\documentclass[12pt]{report}

\DeclareMathAlphabet{\mathpzc}{OT1}{pzc}{m}{it}

\usepackage{psfrag}

\usepackage{relsize}
\usepackage{bbm}

\input epsf

\usepackage{mathrsfs}
\usepackage{amsmath,amssymb}
\usepackage{epsfig}
\input epsf

\def\x'{\mathaccent 19 x}
\def\y'{\mathaccent 19 y}
\def\n'{\mathaccent 19 n}
\def\u'{\mathaccent 19 u}

\newcommand\bp{\hbox{\larger\larger $\pi$}}

\def\et'{\mathaccent 19 \eta}
\def\th'{\mathaccent 19 \theta}
\def\lam'{\mathaccent 19 \lambda}
\def\varet'{\mathaccent 19 \vartheta}
\def\rh'{\mathaccent 19 \rho}
\def\ph'{\mathaccent 19 \phi}
\def\xb'{\mathaccent 19 {\bar{x}}}

\newcommand{\Y}{\mathcal Y}

\def\l{{\lambda}}

\def \N {{\cal N}}

\def \A {{\cal{A}}}

\def\det{\hbox{det}}
\def\be{\begin{equation}}
\def\ee{\end{equation}}

\newcommand{\bea}{\begin{eqnarray}}
\newcommand{\eea}{\end{eqnarray}}

\def\r {\rho}
\def\a {\alpha}
\def\b {\beta}
\def\s {\sigma}
\def\pa {\partial}

\newcommand{\alg}[1]{\mathfrak{#1}}
\newcommand{\su}{\alg{su}}
\newcommand{\sls}{\alg{sl}}
\newcommand{\psu}{\alg{psu}}
\newcommand{\un}{\alg{u}}

\newcommand{\AdS}{{\rm  AdS}_5\times {\rm S}^5}

\renewcommand{\L}{{\mathscr L}}

\newcommand{\atopfrac}[2]{\genfrac{}{}{0pt}{}{#1}{#2}}
\newcommand{\sfrac}[2]{{\textstyle\frac{#1}{#2}}}
\newcommand{\half}{\sfrac{1}{2}}

\newcommand{\so}{\alg{so}}

\renewcommand{\v}{{\varphi}}
\renewcommand{\o}{{\omega}}

\def \i {\text i}

\def \S {\text S}

\def\bS{{\bf S}}
\def\hJ{{\bf J}}
\def\ba{{\bf a}}
\def\bb{{\bf b}}
\def\hP{\hat{P}}
\def\vp{\varphi}
\def\ov{\over}

\def \nn {\nonumber}
\def\la{\label}

\newcommand{\sn}{\mathop{\mathrm{sn}}\nolimits}
\newcommand{\cn}{\mathop{\mathrm{cn}}\nolimits}
\newcommand{\dn}{\mathop{\mathrm{dn}}\nolimits}

\def\et'{\mathaccent 19 \eta}
\def\th'{\mathaccent 19 \theta}
\def\lam'{\mathaccent 19 \lambda}
\def\varet'{\mathaccent 19 \vartheta}
\def\rh'{\mathaccent 19 \rho}
\def\ph'{\mathaccent 19 \phi}
\def\xb'{\mathaccent 19 {\bar{x}}}

\def\cex{{_{\cal C}}}

\def\bQ{\overline{Q}}
\def\bC{C^\dagger}

\def\hP{\hat{P}}

\def\cA{{\cal A}}

\def \ss {\alg{s}}

\def\l{{\lambda}}

\def\N{${\cal N}=4$ }

\def\sl(2){\alg{sl}(2)}

\def\det{\hbox{det}}
\def\be{\begin{equation}}
\def\ee{\end{equation}}

\newcommand{\bei}{\begin{itemize}}
\newcommand{\eei}{\end{itemize}}

\def\GS{Green-Schwarz }

\def\r {\rho}
\def\a {\alpha}
\def\b {\beta}
\def\s {\sigma}
\def\pa {\partial}

\def\g {\gamma}
\def\om {\omega}
\def\p{\phi}
\def\la{\label}
\def\e{\epsilon}
\def\ov{\over}
\def\tr{{\rm tr}}

\def\str{{str}}

\def\S{\Sigma}
\def\H{{\cal H}}
\def\dM{\dot{M}}
\def\dN{\dot{N}}
\def\dK{\dot{K}}
\def\dL{\dot{L}}
\def\dx{\dot{x}}

\def\hP{{\bf P}}
\def\hJ{{\bf J}}

\def\bJ{{\bf J}}
\def\bJ{{\bf J}}
\def\bS{{\bf S}}
\def\bR{{\bf R}}
\def\bL{{\bf L}}
\def\bQ{{\bf Q}}
\def\tbQ{{\widetilde{ \bf Q}}}
\def\bC{{\bf C}}
\def\tbC{{\widetilde{ \bf C}}}

\def\bH{{\bf H}}
\def\tbH{{\widetilde {\bf H}}}

\def\ba{{\bf a}}
\def\bb{{\bf b}}

\def\vp{\varphi}

\def\hP{{\mathbb P}}
\def\hJ{{\mathbb J}}

\def\bI{{\mathbb I}}
\def\bJ{{\mathbb J}}
\def\bJ{{\mathbb J}}
\def\bS{{\mathbb S}}
\def\bR{{\mathbb R}}
\def\bL{{\mathbb L}}
\def\bQ{{\mathbb Q}}
\def\tbQ{{\widetilde{ \mathbb Q}}}
\def\bC{{\mathbb C}}
\def\tbC{{\widetilde{ \mathbb C}}}

\def\bH{{\mathbb H}}
\def\tbH{{\widetilde {\mathbb H}}}

\newcommand{\ag}{\alg{g}}
\def \A {A}
\newcommand{\Z}{\mathcal Z}

\def\de{\delta}
\def\De{\Delta}
\newcommand{\Gtt}{G_{tt}}
\newcommand{\Gpp}{G_{\phi\phi}}
\def\str{{\rm str}}
\def\Kk{{\cal K}}
\def\ws{{\rm ws}}
\def\pws{p_{{\rm ws}}}
\def\dz{\dot{z}}

\def\eps{\epsilon}
\def\da{{\dot a}}
\def\dal{{\dot \a}}
\def\db{{\dot b}}
\def\dbe{{\dot \b}}
\def\dc{{\dot c}}
\def\dga{{\dot \g}}

\def\mV{\mathscr{V}}

\def\V{{\mathscr V}}

\usepackage{hyperref}

\renewcommand{\L}{{\mathscr L}}
\renewcommand{\l}{{\lambda}}

\def\bN{{\mathbb{N}}}
\def\bA{{\mathbb{A}}}
\newcommand{\gr}{\epsilon}

\def\Om{\Omega}

\def\mI{\mathbbm{1}}

\begin{document}


\null\vskip-40pt
 \vskip-5pt \hfill
{\tt \footnotesize ITP-UU-09-05} \vskip-5pt \hfill
{\tt\footnotesize SPIN-09-05} \vskip-5pt \hfill {\tt\footnotesize
TCD-MATH-09-06}
 \vskip-5pt \hfill {\tt\footnotesize
HMI-09-03}

\vskip 1cm \vskip0.2truecm
\begin{center}
\begin{center}
\vskip 0.8truecm {\Large\bf Foundations of the $\AdS$ Superstring
\vskip 0.2cm Part I}
\end{center}

\renewcommand{\thefootnote}{\fnsymbol{footnote}}

\vskip 0.9truecm
Gleb Arutyunov$^a$\footnote[1]{Email: G.E.Arutyunov@uu.nl,
frolovs@maths.tcd.ie}{}\footnote[2]{ Correspondent fellow at
Steklov Mathematical Institute, Moscow.} \  and Sergey
Frolov$^{b\, * \dagger}$
 \\
\vskip 0.5cm

 $^{a}$ {\it Institute for Theoretical Physics and Spinoza
Institute,\\ ~~Utrecht University, 3508 TD
Utrecht, The Netherlands} \\
\vskip 0.1cm

$~$~~~$^b$ {\it Hamilton Mathematics Institute and School of Mathematics, \\
Trinity College, Dublin 2,
Ireland}


\end{center}
\vskip 1cm \noindent\centerline{\bf Abstract} \vskip 0.2cm We
review the recent advances towards  finding the spectrum of the
$\AdS$ superstring. We thoroughly explain the theoretical
techniques which should be useful for the ultimate solution of the
spectral problem. In certain cases our exposition is original and
cannot be found in the existing literature. The present Part I
deals with foundations of classical string theory in $\AdS$,
light-cone perturbative quantization and derivation of the exact
light-cone world-sheet scattering matrix.

\newpage

\tableofcontents

\newpage

\chapter*{Introduction}
\vspace{-0.5cm} \addcontentsline{toc}{chapter}{Introduction}
Already in the mid seventies it became clear that the existing
theoretical tools are hardly capable of providing an ultimate
solution to the theory of strong interactions --  Quantum
Chromodynamics (QCD). At small distances quarks interact weakly
and the physical properties of the theory can be well described by
perturbative expansion based on Feynman diagrammatics. However, at
large separation, forces between quarks become strong and this
precludes the usage of perturbation theory. Understanding the
strong coupling dynamics of quantum Yang-Mills theories remains
one of the daunting challenges of theoretical particle physics.

A spectacular new insight into dynamics of non-abelian gauge
fields has recently been offered by the AdS/CFT
(Anti-de-Sitter/Conformal Field Theory) duality conjecture also
known under the name of the ``gauge-string correspondence"
\cite{M}. This conjecture states that certain four-dimensional
quantum gauge theories could be alternatively described in terms
of closed strings moving in a ten-dimensional curved space-time.

The prime example of the gauge-string correspondence involves the
four-dimen\-sional maximally-supersymmetric ${\cal N}=4$
Yang-Mills theory with gauge group ${\rm SU}(N)$ and type IIB
superstring theory defined in an ${\rm AdS}_5\times {\rm S}^5$
space-time, which is the product of a five-dimensional
Anti-de-Sitter space (the maximally symmetric space of constant
negative curvature) and a five-sphere. Since no candidate for a
string dual of QCD is presently known, the ${\cal N}=4$ theory
together with its conjectured string partner offers a unique
playground for testing the correspondence between strings and
quantum field theories, as well as for understanding
strongly-coupled gauge theories in general. The success of the
whole gauge-string duality program relies on our ability to
quantitatively verify this prime example of the correspondence
and, more importantly, to clarify the physical principles at work.

The ${\cal N}=4$ super Yang-Mills theory has a vanishing
beta-function and, for this reason, is an exact conformal field
theory in four dimensions. The algebra of conformal
transformations  coincides with $\alg{so}(4,2)$ which, in addition
to the Poincar\'e algebra, includes the generators of scale
transformations (dilatation) and conformal boosts. The
supersymmetry generators extend the conformal algebra to the
superconformal algebra $\psu(2,2|4)$, the latter being the full
algebra of global symmetries of the ${\cal N}=4$ theory.
Simultaneously, $\psu(2,2|4)$ plays the role of the symmetry
algebra of type IIB superstring in the ${\rm AdS}_5\times {\rm
S}^5$ background. Thus, the gauge and string theory at hand share
the same kinematical symmetry. This, however, does not a priory
imply their undoubted equivalence.

To solve a conformal field theory, one has to identify the
spectrum of primary operators (forming irreducible representations
of the conformal group) and to compute their three-point
correlation functions. Scaling (conformal) dimensions of primary
operators and the three-point correlators encode all the
information   about the theory since all higher-point correlation
functions can in principle be found  by using the Operator Product
Expansion. The ${\cal N}=4$ theory has two parameters: the
coupling constant $g_{_{\rm YM}}$ and the rank $N$ of the gauge
group, and it admits a well-defined \mbox{'t Hooft} expansion in
powers of $1/N$ with the \mbox{'t Hooft} coupling
$\lambda=g_{_{\rm YM}}^2N$ kept fixed. The AdS/CFT duality
conjecture relates these parameters to the string coupling
constant $g_s$ and the string tension $g$ as follows:
$g_s=\lambda/4\pi N$ and $g=\sqrt{\lambda}/2\pi$. Scaling
dimensions $\Delta$ of composite gauge invariant primary operators
are eigenvalues of the dilatation operator and they depend on the
couplings: $\Delta\equiv \Delta(\lambda,1/N)$. Scaling dimension
is the only label of a (super)-conformal representation which is
allowed to continuously depend on the parameters of the model. In
spite of the finiteness of the ${\cal N}=4$ theory, composite
operators undergo non-trivial renormalization which explains the
appearance of coupling-dependent anomalous dimensions.
Alternatively, in string theory on $\AdS$ energies $E$ of string
states are functions of the couplings: $E\equiv E(g,g_s)$. In the
most general setting, the gauge-string duality conjecture implies
that physical states of gauge and string theories are organized in
precisely the same set of $\psu(2,2|4)$-multiplets. In particular,
energies of string states measured in the global AdS coordinates
must coincide with scaling dimensions of gauge theory primary
operators, both regarded as non-trivial functions of their
couplings. Exhibiting this fact would be the first important step
towards proving the conjecture.

The initial research on the ${\cal N}=4$ gauge-string duality was
concentrated on deriving scaling dimensions/correlation functions
of primary operators in the supergravity approximation
\cite{GKP1,W}. This corresponds to the strongly-coupled planar
regime in the gauge theory where $\lambda$ is infinite and $N$ is
large. Only rather special states -- those which are protected
from renormalization by a large amount of supersymmetry -- could
be a subject of comparison here.

The next important step has been undertaken in \cite{BMN}, where a
special scaling limit was introduced. This work initiated
intensive studies of unprotected operators with large $R$-charge
which eventually led to the discovery of integrable structures in
the gauge theory \cite{MZ}-\cite{Beisert:2003yb}. This discovery
marked a new phase in the research on the fundamental model of
AdS/CFT.

\begin{figure}
\begin{minipage}{\textwidth}
\begin{center}
\includegraphics[width=0.70\textwidth]{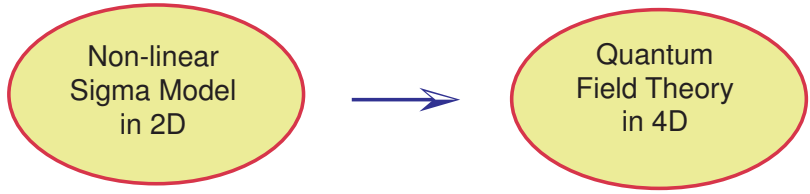}
\end{center}
\begin{center}
\parbox{5in}{\footnotesize{\caption{\label{ADSCFT} The AdS/CFT correspondence: The spectrum of a
2d non-linear sigma-model describing string theory on a curved
background is expected to be equivalent to the spectrum of a 4d
quantum non-abelian gauge theory in the large $N$ limit.}}}
\end{center}
\end{minipage}
\end{figure}

In the limit where the rank of the gauge group becomes infinite,
one can neglect string interactions and consider free string
theory. Free strings propagating in a non-trivial gravitational
background such as ${\rm AdS}_5\times {\rm S}^5$ are described by
a two-dimensional quantum non-linear sigma model. Finding the
spectrum of the sigma model will determine the spectrum of scaling
dimensions of composite operators in the dual gauge theory, Figure
\ref{ADSCFT}.

In general, to solve a non-linear quantum sigma model would be a
hopeless enterprise. Remarkably, it appears, however, that
classical strings in ${\rm AdS}_5\times {\rm S}^5$ are described
by an integrable model \cite{Bena:2003wd}. Integrable models
constitute a special class of dynamical systems with an infinite
number of conservation laws which in many cases hold the key to
their exact solution.  If string integrability continues to exist
for the corresponding quantum theory then we are facing a
breathtaking possibility to solve the string model exactly and,
via the gauge-string duality, to find an exact solution of an
interacting quantum field theory in four dimensions.

In recent years there has been a lot of exciting progress towards
understanding integrable properties of both the string sigma model
and the dual gauge theory. Not all this progress is yet logically
deducible from the first principles and in certain cases it is
based on new assumptions or clever guesses. Nevertheless, we feel
that a clear and self-contained picture starts to emerge of how to
obtain a solution (spectrum) of quantum strings in $\AdS$. It is
the scope of this review to explain this picture and to provide
all the necessary technical tools in its support.

The review should be accessible to PhD students. It is certainly
desirable to have a prerequisite knowledge of string theory
\cite{Green:1987sp}. The review might also be useful for
specialists: as a handbook and as a source of formulae. In order
not to distract the reader's attention with references, we comment
on the literature in a special section concluding each chapter.
Further, we emphasize that this review is most exclusively about
string theory. To get more familiar with gauge theory
constructions, the reader is invited to consult the original
literature and reviews \cite{Belitsky:2004cz,Beisert:2004ry}.

As is seen for the moment, solving the string sigma-model is a
complicated multi-step procedure. In view of this, before we start
our actual journey, we would like to briefly describe the
corresponding steps and to summarize the most relevant current
progress in the field. This will also help the reader to get
familiar with the content of the review.

\vspace{-0.4cm}
\subsubsection*{Light-cone gauge}
The starting point is the Green-Schwarz action for strings in
$\AdS$ which defines a two-dimensional non-linear sigma model of
Wess-Zumino type \cite{Metsaev:1998it}. The isometries of the
$\AdS$ space-time constitute the global symmetry algebra of the
sigma model and string states are naturally characterized by the
charges (representation labels) they carry under this symmetry
algebra. Among all representation labels two charges, $J$ and $E$,
are of particular importance for the light-cone gauge fixing. The
charge $J$ is the angular momentum carried by the string due to
its rotation around the equator of ${\rm S}^5$ and $E$ is the
string energy, the latter corresponds to the symmetry of the
Green-Schwarz action under constant shifts of the global time
coordinate of the AdS space. It is the energy spectrum of string
states that we would like to determine and subsequently compare to
the spectrum of scaling dimensions of primary operators in the
gauge theory.

To describe the physical states, it is advantageous to fix the
so-called generalized light-cone gauge. In this gauge the
world-sheet Hamiltonian is equal to $E-J$, while the light-cone
momentum $P_+$ is another global charge which, generically, is a
linear combination of $J$ and $E$. Physical states should satisfy
the level-matching condition: the total world-sheet momentum
carried by a state must vanish. Solving the model is then
equivalent to computing the physical spectrum of the (quantized)
light-cone Hamiltonian for a fixed value of $P_+$.

 Fixing the
light-cone gauge for the Green-Schwarz string  in a curved
background is subtle because of a local fermionic  symmetry. This
question has been studied in \cite{AF0,FPZ} where the exact
gauge-fixed classical Hamiltonian was found.  This Hamiltonian is
non-polynomial in the world-sheet fields and, as such, can hardly
be quantized in a straightforward manner.

\vspace{-0.4cm}
\subsubsection*{From cylinder to plane: Decompactification and symmetries}
 In the light-cone gauge the world-sheet action
depends explicitly on the light-cone momentum $P_+$. By
appropriately rescaling the world-sheet coordinates, the theory
becomes defined on a cylinder of circumference $P_+$. At this
stage, one can consider the decompactification limit, {\it i.e.}
the limit where $P_+$ and therefore the radius of the cylinder go
to infinity, while keeping the string tension fixed. In this limit
one is left with a theory on a plane which leads to significant
simplifications. Most importantly, the world-sheet theory has a
massive spectrum and the notion of asymptotic states (particles)
is well defined, calling for an application of scattering theory.
Quantum integrability should then imply the absence of particle
production and factorization of multi-particle scattering into a
sequence of two-body events.

Thus, assuming quantum integrability, the next step is to find the
dispersion relation for elementary excitations and the S-matrix
describing their pairwise scattering. To deal with particles with
arbitrary world-sheet momenta, one has to give up the
level-matching condition. This leads to an important modification
of the global symmetry algebra of the model. Namely, the manifest
$\psu(2|2) \oplus \psu(2|2) \subset \psu(2,2|4)$ symmetry algebra
of the light-cone string theory gets enhanced by two central
charges \cite{AFPZ}. The central charges vanish on physical states
satisfying the level-matching condition but they play a crucial
role in fixing the structure of the world-sheet S-matrix. The same
centrally-extended algebra also appears in the dual gauge theory
\cite{B}.

\vspace{-0.4cm}
\subsubsection*{Dispersion relation and scattering matrix}
Insights coming from both gauge and string theory \cite{BMN} led
to a conjecture for the dispersion relation \cite{BDS}. It has the
following unusual form \bea
\nonumber \epsilon(p) = \sqrt{1 + 4g^2\sin^2 \frac{p}{2}} \, ,\eea
where $g$ is the string tension, $\epsilon$ and $p$ are the energy
and the momentum of an elementary excitation.

An important observation made in \cite{B} is that  the dispersion
relation  is uniquely  determined by the symmetry algebra of the
model provided its central charges are known as exact functions of
the string tension and the world-sheet momentum. The dispersion
relation is non-relativistic although it reveals the usual square
root dependence of relativistic field theory. On the other hand,
the sine function under the square root is a common feature of
lattice theories, and its appearance here is rather surprising,
given that the string world-sheet is continuous.

The various pieces of the two-body scattering matrix were
conjectured in \cite{AFS,S,BS} based on the analysis of the
integral equations \cite{KMMZ} describing classical spinning
strings \cite{Gubser:2002tv,FT,Frolov:2003qc} and insights from
gauge theory \cite{BDS}. Later, it was found that the matrix
structure of this S-matrix is uniquely fixed by the
centrally-extended $\psu(2|2) \oplus \psu(2|2)$ symmetry algebra,
the Yang-Baxter equation and the generalized physical unitarity
condition \cite{B, AFZzf, AFtba}.

\vspace{-0.4cm}
\subsubsection{Dressing factor}
The S-matrix is thus determined up to an overall  scalar function
$\sigma(p_1,p_2)$ -- the so-called dressing factor \cite{AFS}.
Ideally, one would hope that further physical requirements would
allow for complete determination of this factor. In relativistic
integrable quantum field theories implementation of Lorentz
invariance together with crossing symmetry exchanging particles
with anti-particles imposes  an additional crossing relation on
the S-matrix \cite{ZZ}.

The light-cone gauge-fixed sigma model is not Lorentz invariant.
However, as was argued in \cite{Janik}, some version of the
crossing relation might hold for the corresponding S-matrix; the
crossing relation then implies a non-trivial functional equation
for the dressing factor. This crossing equation is rather
complicated; it is unclear how to solve it in full generality and
how to single out the physically relevant solution.

Luckily, the logarithm of the dressing factor turns out to be a
two-form on the vector space of local conserved charges of the
model which severely constraints  its functional form \cite{AFS}.
The dressing factor explicitly depends on the string tension $g$
and admits a ``strong coupling'' expansion in powers of $1/g$ that
corresponds to an asymptotic perturbative expansion of the string
sigma model.

Combining the functional form of the dressing factor together with
the first two known orders in the strong coupling expansion
\cite{AFS,HL}, a set of solutions to the crossing equation in
terms of an all-order strong coupling asymptotic series has been
proposed \cite{BHL}. A particular solution was conjectured to
correspond to the actual string sigma model perturbative
expansion. This solution was shown to agree with the explicit
two-loop sigma model result \cite{RTT,KlMMZ}. It should be
stressed, however, that all these solutions are only asymptotic
and, therefore, they do not define the dressing factor as a
function of $g$.

In contrast to the strong coupling expansion, gauge theory
perturbative expansion of the dressing factor is in powers of $g$
and it has a finite radius of convergence. As a result, the
dressing factor can be defined as a function of $g$. An
interesting proposal for the exact dressing factor has been put
forward in \cite{BES}. On the one hand, it agrees with the
explicit four-loop gauge theory computation \cite{Bern,BMR}. On
the other hand, it was argued to have the same strong coupling
asymptotic expansion as the particular solution by \cite{BHL}
corresponding to the string sigma model. Taking all this into
account, one can adopt the working assumption that the exact
dressing factor and, therefore, the S-matrix are established.
However, a word of caution to bear in mind -- there is no unique
solution to the crossing equation; additional yet to be found
physical constraints should be used to single out the right
solution unambiguously.

\vspace{-0.4cm}
\subsubsection{Bound states}
Having found the exact dispersion relation and the S-matrix, the
next step is to determine the complete asymptotic spectrum of the
model. This amounts to finding all bound states of the elementary
excitations and  bound states of the bound states, etc. This
problem can be solved by analyzing the pole structure of the
S-matrix. The analysis reveals that all bound states are those of
elementary particles \cite{D1}. More explicitly, $Q$-particle
bound states comprise into the tensor product of two $4Q$-dim
atypical totally symmetric multiplets of the centrally-extended
symmetry algebra $\su(2|2)$ \cite{B2}. Since the light-cone string
sigma model is not Lorentz-invariant, the identification of what
is called  the ``physical region" of the S-matrix is very subtle
and it affects the counting of bound states \cite{AFtba}.

The problem of computing a bound state S-matrix is rather
non-trivial and  reduces to finding its dressing factor and fixing
its matrix structure. The dressing factor can be computed by using
the fusion procedure for the $\su(2)$ sector S-matrix
\cite{D2,Roiban}, and appears to be of the same universal form as
the one for the elementary particles S-matrix \cite{AFS}. As to
the  matrix structure, it can be found by using the superfield
approach by \cite{AFbs}.

\subsubsection{Back from plane to cylinder: Finite $P_+$ spectrum}
Having understood the spectrum of the light-cone string sigma
model on a plane,   one has to  ``upgrade'' the findings to a
cylinder. All physical string configurations (and dual gauge
theory operators) are characterized by a finite value of $P_+$,
and as such they are excitations of a theory on a cylinder.

The first step in determining the finite-size spectrum of a
two-dimensional integrable model is to consider the model on a
cylinder of a very large but finite circumference $P_+$. In this
case integrability implies that a multi-particle state can be
approximately described  by the wave function of the Bethe-type
\cite{ZZ}. Factorizability of the multi-particle scattering matrix
together with the periodicity condition for the Bethe wave
function leads to a system of equations on the particle momenta
known as the Bethe-Yang equations. In the AdS/CFT context these
equations are usually referred to as the asymptotic Bethe
ansatz\footnote{In the theory of integrable models the asymptotic
Bethe ansatz is known for a long time \cite{Sutherland}. }
\cite{BS}. The $\AdS$ string S-matrix has a complicated matrix
structure which results at the end in a set of nested Bethe
equations \cite{BS,B,Martins:2007hb}.

The Bethe-Yang equations determine any power-like $1/P_+$
corrections to the energy of multi-particle states. It is known,
however, that for large $P_+$  there are also exponentially small
corrections. To compute the leading exponential corrections, one
can adapt L\"uscher's formulae \cite{Lu,Lu1} for the
non-Lorentz-invariant case at hand \cite{Janik:2007wt}. This
computation has been done for some string states at strong
coupling.

Remarkably,  L\"uscher's approach could be also applied to find
perturbative scaling dimensions of gauge theory operators up to
the first order where the Bethe-Yang description breaks down
\cite{BJ}. The corresponding computation has been done \cite{BJ}
for the simplest case of the so-called Konishi operator and
stunning agreement with a very complicated four-loop result based
on the standard Feynman diagrammatics \cite{Zanon} has been found.
String theory starts to reveal its extreme power, elegance and
simplicity  in comparison to the conventional perturbative
approach!

\vspace{-0.4cm}
\subsubsection{Thermodynamic Bethe Ansatz}
The success in computing gauge theory perturbative anomalous
dimensions is very encouraging. However, one is really interested
in non-perturbative gauge theory, ${\it i.e.}$ in the exact
spectrum for finite values of the gauge coupling (or equivalently
for finite string tension and finite $P_+$). One tempting
possibility is to generalize the thermodynamic Bethe ansatz (TBA),
originally developed for relativistic integrable models \cite{za},
to the light-cone string theory at hand.

The TBA approach would be based on the following construction.
Consider a closed string of length $L\equiv P_+$  which wraps  a
loop of ``time" length $R$. The topology of the corresponding
surface spanned by the string is a torus, {\it i.e.} the Cartesian
product of two orthogonal circles with circumferences $L$  and $R$
, respectively. According to the imaginary time formalism of
Statistical Mechanics, the circumference of any of these two
circles can be treated as the inverse temperature for a
statistical field theory with the Hilbert space of
quantum-mechanical states defined on the complementary circle.
Thus, there are two models related to one and the same torus: the
original theory of strings with length $L$  at temperature $1/R$
and the ``mirror" model defined on a circle of length $R$ at
temperature $1/L$. The smaller and the colder the original theory,
the hotter and the bigger its mirror. In particular, the ground
state energy of the original string model in a finite
one-dimensional volume $L$  is equal to the Gibbs free energy (or
Witten's index in the case of periodic fermions) of the mirror
model in infinite volume, {\it i.e.} for infinite $R$. It should
be also possible to relate the whole string spectrum to the proper
thermodynamic quantities of the mirror model defined for infinite
$R$, a problem which is not well understood at present.

Since the light-cone string sigma model is not Lorentz-invariant,
the mirror model is governed by a different Hamiltonian and
therefore has very different dynamics.  Thus, to implement the TBA
approach one has to study the mirror theory in detail. The first
step  in this direction has been already done in \cite{AFtba},
where the Bethe-Yang equations for the mirror model were derived.
Another result obtained in \cite{AFtba} was a classification of
the mirror bound states according to which they comprise the
tensor product of two $4Q$-dim atypical totally anti-symmetric
multiplets\footnote{Notice the difference with the bound states in
the original model which transform in symmetric representations!}
of the centrally-extended algebra $\su(2|2)$. This observation was
of crucial importance for the derivation \cite{BJ} of the scaling
dimension of the Konishi operator.  We consider this derivation as
prime evidence for the validity of the mirror theory approach.
Recently two interesting conjectures has been made: one concerns
the classification of states contributing in the thermodynamic
limit of the mirror theory \cite{Arutyunov:2009zu}, another
formulates the so-called Y-system
\cite{Zamolodchikov:1991vg,Kuniba:1993cn} which is supposed to
encode the finite-size string spectrum \cite{Gromov:2009tv}.

Because of a large amount of necessary material, we decided to
split the review into two parts. The present Part I deals with
foundations of classical string theory in $\AdS$, the light-cone
perturbative quantization and derivation of the light-cone
world-sheet scattering matrix. Part II will include the derivation
of the Bethe-Yang equations, the discussion of bound states and
the progress in understanding the finite-size spectrum of the
string sigma model, both in L\"uscher's and in the TBA setting. We
will also present yet ``phenomenological" arguments which led to
the determination of the dressing phase. In the last chapter of
Part II we plan to list the important topics which were uncovered
in the present review.

This concludes our brief description of a possible approach to
find the spectrum of quantum strings in $\AdS$. At present we do
not know if the route we follow is the unique or the simplest one.
Time will tell. In any case, the success we encounter underway
makes us believe that the first ever exact solution of a
four-dimensional interacting quantum field theory is within our
reach.

\chapter{String sigma model}
In addition to the flat ten-dimensional Minkowski space, type IIB
supergravity admits another maximally supersymmetric solution
which is  product of the five-dimensional Anti-de-Sitter space
${\rm AdS}_5$ and the five-sphere ${\rm S}^5$. This solution is
supported by the self-dual Ramond-Ramond five-form flux. The
presence of this background flux precludes the usage of the
standard NSR approach to build up the action for strings
propagating in this geometry. Indeed, the Ramond-Ramond vertex
operator is known to be non-local in terms of the world-sheet
fields and, for this reason, it is unclear how to couple it to the
string world-sheet.

\smallskip

There exists another approach to define string theory for a
background geometry supported by Ramond-Ramond fields -- the
so-called Green-Schwarz formalism. This formalism has a further
advantage, namely, it allows one to realize the space-time
supersymmetry in a manifest way. The Green-Schwarz approach can be
used for any background obeying the supergravity equations of
motion to guarantee the invariance of the corresponding string
action with respect to the local fermionic symmetry
($\kappa$-symmetry), the latter being responsible for the
space-time supersymmetry of the physical spectrum. In practice,
construction of the Green-Schwarz action for an arbitrary
supergravity solution faces a serious difficulty. Namely, starting
from a given bosonic solution, one has to determine the full
structure of the type IIB superfield, a problem that has not been
solved so far for a generic background.

\smallskip

Fortunately, there is an alternative approach to define the
Green-Schwarz superstring which makes use of the special symmetry
properties of the background solution. This approach has already
been shown to work nicely in the case of a flat background, where
it amounts to defining the Green-Schwarz string as a WZNW-type
non-linear sigma model on the coset superspace being a quotient of
the ten-dimensional super-Poincar\'e group over its Lorentz
subgroup ${\rm SO}(9,1)$. The super-Poincar\'e group acts
naturally on this coset space and it is a manifest symmetry of the
corresponding sigma model action. The Wess-Zumino term guarantees
invariance of the full action under $\kappa$-symmetry
transformations.

\smallskip

Remarkably, a similar sigma model approach can be developed in the
\mbox{$\AdS$} case. Namely, we define type IIB Green-Schwarz
superstring in the $\AdS$ background as a non-linear sigma-model
with target space being the following coset \bea \frac{{\rm
PSU}(2,2|4)}{{\rm SO(4,1)}\times {\rm SO}(5)}\, . \label{sAdS}
\eea The supergroup ${\rm PSU}(2,2|4)$ contains the bosonic
subgroup ${\rm SU(2,2)}\times {\rm SU}(4)$ which is locally
isomorphic to ${\rm SO(4,2)}\times {\rm SO}(6)$; the quotient of
the latter over ${\rm SO(4,1)}\times {\rm SO}(5)$ provides a model
of the $\AdS$ manifold with ${\rm SO(4,1)}\times {\rm SO}(5)$
being the group of local Lorentz transformations. Correspondingly,
the coset (\ref{sAdS}) can be regarded as a model of the $\AdS$
superspace. The group ${\rm PSU}(2,2|4)$ which acts on the coset
by left multiplications plays the role of the isometry group of
the $\AdS$ superspace. Thus, considering a non-linear sigma-model
with target superspace (\ref{sAdS}) provides a natural way to
couple the string world-sheet to the background Ramond-Ramond
fields.

\smallskip

In this chapter we will describe the corresponding sigma-model in
detail. We will discuss its global and local symmetries and show
that it can be embedded into the standard framework of classical
integrable systems.

\section{Superconformal algebra}
The construction of the coset sigma-model essentially relies on
the properties of the superconformal algebra $\psu(2,2|4)$. Here
we will summarize the necessary facts about this algebra and
introduce our notation.

\subsection{Matrix realization of \texorpdfstring{$\su(2,2|4)$}{\su(2,2|4)} }
\label{subMatrixrealization} We start our discussion with the
definition of the superalgebra $\sls(4|4)$ considered over the
field ${\mathbb C}$. As a matrix superalgebra, $\sls(4|4)$ is
spanned by $8\times 8$ matrices $M$, which we write in terms of
$4\times 4$ blocks as \bea M=\left(
\begin{array}{cc}
  m & \theta \\
  \eta & n
\end{array} \right)\, .
\eea These matrices are required to have vanishing supertrace
${\rm str}M\equiv {\rm tr}\, m-{\rm tr}\, n=0$. The superalgebra
$\sls(4|4)$ carries the structure of a ${\mathbb Z}_2$-graded
algebra: the matrices $m$ and $n$ are regarded as even, and
$\theta,\eta$ as odd, respectively. The entries of $\theta$ and
$\eta$ can be thought of as grassmann (fermionic) anti-commuting
variables.

\smallskip

The superalgebra $\su(2,2|4)$ is a non-compact real form of
$\sls(4|4)$. It is identified with a set of fixed points
$M^{\star}=M$ of $\sls(4|4)$ under the Cartan involution\footnote{
 It is worthwile to note that our definition of the Cartan
involution is different but equivalent to the standard one:
\mbox{$M^{\star}=-i^{\eps_M}HM^{\dagger}H^{-1}$}, where $\eps_M=0$
for even and $\eps_M=1$ for odd elements respectively.}
\mbox{$M^{\star}=-HM^{\dagger}H^{-1}$.} In other words, a matrix
$M$ from $\su(2,2|4)$ is subject to the following reality
condition
 \bea \label{real} M^{\dagger}H+ H M=0\, . \eea Here the
adjoint of the supermatrix $M$ is defined as $M^{\dagger}=(M^t)^*$
and the hermitian matrix $H$ is taken to be
 \bea  H=\left(
\begin{array}{cc}
  \Sigma & 0 \\
  0 & \mI_{4}
\end{array} \right)\, ,
\eea where $\Sigma$ is the following $4\times 4$ matrix \bea
\Sigma=\, {\footnotesize{\left(
\begin{array}{cc}
 \mI_2  & 0 \\
  0 & -\mI_2 \\
\end{array} \right)}} \eea
and $\mI_n$ denotes the $n\times n$ identity matrix. We further
note that for any odd element $\theta$ the conjugation acts as a
${\mathbb C}$-anti-linear anti-involution:
$$(c\,\theta)^*=\bar{c}\, \theta^*\, ,~~~~~
\theta^{\ast\ast}=\theta\, , ~~~~~~~
(\theta_1\theta_2)^*=\theta_2^*\theta_1^*\, ,$$ which guarantees,
in particular,  that
$(M_1M_2)^{\dagger}=M_2^{\dagger}M_1^{\dagger}$, {\it i.e.} that
anti-hermitian supermatrices form a Lie superalgebra.

\medskip

Condition (\ref{real}) implies that \bea\label{rc}
m^{\dagger}=-\Sigma\,  m\, \Sigma\, , ~~~~~~n^{\dagger}=-n\, ,
~~~~~~~~\eta^{\dagger}=-\Sigma\,  \theta\, . \eea Thus, $m$ and
$n$ span the unitary subalgebras $\un(2,2)$ and $\un(4)$
respectively. The algebra $\su(2,2|4)$ also contains the
$\un(1)$-generator $i\mI$, as the latter obeys eq.(\ref{real}) and
has vanishing supertrace. Thus, the bosonic subalgebra of
$\su(2,2|4)$ is \bea \su(2,2)\oplus \su(4)\oplus \un(1)\, . \eea
The superalgebra $\psu(2,2|4)$ is defined as a {\it quotient
algebra} of $\su(2,2|4)$ over this $\un(1)$-factor. It is
important to note that $\psu(2,2|4)$, as the quotient algebra, has
no realization in terms of $8\times 8$ supermatrices.

\smallskip

It is convenient to fix a basis for the bosonic subalgebra
$\su(2,2)\oplus \su(4)$. Throughout this work we will use the
following representation of Dirac matrices
\begin{eqnarray*}
\gamma^1&=&{\scriptsize\left(
\begin{array}{cccc}
  0 & 0 & 0 & -1 \\
  0 & 0 & 1 & 0 \\
  0 & 1 & 0 & 0 \\
 -1 & 0 & 0 & 0
\end{array} \right)},\hspace{0.3in}
\gamma^2={\scriptsize\left(
\begin{array}{cccc}
  0 & 0 & 0 & i \\
  0 & 0 & i & 0 \\
  0 & -i & 0 & 0 \\
 -i & 0 & 0 & 0
\end{array} \right)},\hspace{0.3in}
\gamma^3={\scriptsize\left(
\begin{array}{cccc}
  0 & 0 & 1 & 0 \\
  0 & 0 & 0 & 1 \\
  1 & 0 & 0 & 0 \\
  0 & 1 & 0 & 0
\end{array} \right)}, \\
\nonumber \gamma^4&=&{\scriptsize\left(
\begin{array}{cccc}
  0 & 0 & -i & 0 \\
  0 & 0 & 0 & i \\
  i & 0  & 0 & 0\\
  0 & -i & 0 & 0
\end{array} \right)},\hspace{0.25in}~\gamma^5={\scriptsize \left(
\begin{array}{cccc}
  1 & 0 & 0 & 0 \\
  0 & 1 & 0 & 0 \\
  0 & 0 & -1 & 0 \\
   0 & 0 & 0 & -1
\end{array} \right)}=\Sigma\, ,
\end{eqnarray*}
satisfying the SO(5) Clifford algebra relations
$$
\gamma^i\gamma^j+\gamma^j\gamma^i=2\delta^{ij}\, ,
~~~~~~~i,j=1,\ldots, 5\, .
$$
Note that $\gamma^5=-\gamma^1\gamma^2\gamma^3\gamma^4$. All these
matrices are hermitian: $(\gamma^{i})^*=(\gamma^i)^t$, so that
$i\gamma^i$ belongs to $\alg{su}(4)$. The spinor representation of
$\so(5)$ is spanned by the generators
$n^{ij}=\frac{1}{4}[\gamma^i,\gamma^j]$  satisfying the relations
\bea \label{so5}
[n^{ij},n^{kl}]=\delta^{jk}n^{il}-\delta^{ik}n^{j\,l}-\delta^{j\,l}n^{ik}+\delta^{il}n^{jk}\,
, ~~~~~n^{ij}=-n^{j\,i}\, . \eea Adding
$n^{i6}=\frac{i}{2}\gamma^i$, one can verify that $n^{ij}=-n^{j\,
i}$ generate an irreducible (Weyl) spinor representation of
$\alg{so}(6)\sim \alg{su}(4)$ with defining relations (\ref{so5})
where now $i,j=1,\ldots , 6$. The other Weyl representation would
correspond to choosing $n^{i6}=-\frac{i}{2}\gamma^i$.

\smallskip

Analogously, a set $\{i\gamma^5,\gamma^i\}$ with $i=1,\ldots ,4$
generates the Clifford algebra for SO(4,1). Indeed, if we
introduce $\gamma^0\equiv i\gamma^5$, then $m^{ij}=
\frac{1}{4}[\gamma^i,\gamma^j]$ with $i,j=0,\ldots,4$ satisfy the
$\alg{so}(4,1)$ algebra relations \bea \label{so41}
[m^{ij},m^{kl}]=\eta^{jk}m^{il}-\eta
^{ik}m^{j\,l}-\eta^{j\,l}m^{ik}+\eta^{il}m^{jk}\, ,
~~~~~m^{ij}=-m^{j\,i}\, ,
 \eea
where $\eta={\rm diag}(-1,1,1,1,1)$. Enlarging this set of
generators by $m^{i5}=\frac{1}{2}\gamma^{i}$, $i=0,\ldots, 4$, we
obtain a realization of $\alg{so}(4,2)\sim \alg{su}(2,2)$ with the
same defining relations (\ref{so41}) where this time $\eta={\rm
diag}(-1,1,1,1,1,-1)$ and $i,j=0,\ldots, 5$.

\smallskip

Thus, we regard $\su(2,2)$ and $\su(4)$ as real vector spaces
spanned by the following set of generators
 \bea
\label{basisbosons}
 \begin{aligned} \su(2,2)&\sim \, {\rm span}_{\mathbb
R}\big\{\sfrac{1}{2}\gamma^i ,\, \sfrac{i}{2}\gamma^5, \,
\sfrac{1}{4}[\gamma^i,\gamma^j], \,
\sfrac{i}{4}[\gamma^5,\gamma^j]\big\}\, ,~~~~~~~ & i,j=1,\ldots, 4,\\
\su(4)&\sim\,  {\rm span}_{\mathbb R}\big\{\sfrac{i}{2}\gamma^i ,
\, \sfrac{1}{4}[\gamma^i,\gamma^j]\big\}\, ,~~~~~~~ &
i,j=1,\ldots, 5.
\end{aligned} \eea
Together with the central element $i\mI$, this set of generators
provides an explicit basis for the bosonic subalgebra of
$\su(2,2|4)$.

\smallskip

Our next goal is to elaborate more on the structure of the
conformal algebra $\su(2,2)$. Introduce the notation
$\gamma^{ij}=\frac{1}{4}[\gamma^i,\gamma^j]$. First, we note that
the matrices $i\gamma^{15},i\gamma^{25},i\gamma^{35},i\gamma^{45}$
together with $\gamma^{1,2,3,4}$ are block off-diagonal, {\it
i.e.} in terms of $2\times 2$ blocks they span the (real)
8-dimensional space
$$
\left(\begin{array}{rr} 0 & \bullet \\
\bullet & 0  \end{array}\right)\subset \su(2,2)\, .
$$
On the other hand, the matrices $\gamma^{ij}$ with $i,j=1,\ldots
4$ span the $\alg{so}(4)$ subalgebra embedded into the conformal
algebra diagonally as two copies of $\su(2)$:
$$
\left(\begin{array}{cc} \su(2) & 0 \\
0 & \su(2)  \end{array}\right)\subset \su(2,2)\, .
$$
Finally,  $\frac{i}{2}\gamma^5$ is diagonal and its centralizer in
$\su(2,2)$ coincides with the maximal compact subalgebra
$\su(2)\oplus \su(2)\oplus \alg{u}(1)\subset \su(2,2)$. Sometimes
the generator $\frac{1}{2}\gamma^5$ is referred to as the
``conformal Hamiltonian".

\smallskip

Second, consider the one-dimensional subalgebra generated by
$\frac{1}{2}\gamma^3\equiv -i{\rm D}$. It is usually called the
``dilatation subalgebra". Evidently, in addition to $\gamma^3$,
the centralizer of $\gamma^3$ in $\su(2,2)$ is generated by
$\gamma^{12},\gamma^{14},\gamma^{24}$ and
$i\gamma^{15},i\gamma^{25},i\gamma^{45}$. The first three matrices
generate $\alg{so}(3)$, while, all together, the six matrices
generate the Lorentz subalgebra  $\alg{so}(3,1)$. The orthogonal
complement to $\alg{so}(3,1)\oplus i{\rm D}$ is the 8-dimensional
real space. The basis in this space can be chosen from
eigenvectors of $i{\rm D}$. The eigenvectors ${\rm K}_i$,
$i=1,\ldots, 4$ with negative eigenvalues form the subalgebra of
{\it special conformal transformation}, while the eigenvectors
${\rm P}_i$ with positive eigenvalues form the subalgebra of {\it
translations}.

\smallskip

Finally, we note that the matrices $\gamma^3$ and $\gamma^5$ are
related by an orthogonal transformation \bea
e^{-\frac{\pi}{4}\gamma^3\gamma^5}\,\gamma^3\,
e^{+\frac{\pi}{4}\gamma^3\gamma^5}=\gamma^5\,  \eea implying
thereby the well-known relation between the dilatation generator
${\rm D}$ and the conformal Hamiltonian. In unitary
representations the operator ${\rm D}$ must be hermitian: ${\rm
D}^{\dagger}={\rm D}$. Here ${\rm D}=\frac{i}{2}\gamma^3$ is
anti-hermitian which is compatible with the fact that we are
dealing with the finite-dimensional and, therefore, non-unitary
representation of the non-compact algebra $\su(2,2)$.

\medskip

The following matrix $K$
\begin{equation}
\label{K} K=-\gamma^2\gamma^4={\scriptsize\left(
\begin{array}{cccc}
  0 & -1 & 0 & 0 \\
  1 & 0 & 0 & 0 \\
   0 & 0 & 0 & -1 \\
   0 & 0 & 1 & 0
\end{array} \right)\, }\, ,
\end{equation}
will play a distinguished role in our subsequent discussion. One
can check that for all Dirac matrices the following relation is
satisfied \bea\hskip 1.5cm (\gamma^{i})^t=K\gamma^iK^{-1}\, ,
~~~~~~~i=1,\ldots, 5. \label{pK}\eea Also we define the charge
conjugation matrix $C=\gamma^1\gamma^3$ which commutes with $K$
and has  the following properties
$$
C\gamma^i C^{-1}=-(\gamma^i)^t\, , ~~~~C\gamma^5
C^{-1}=(\gamma^5)^t\, ,~~~~C^2=-\mI\, , ~~~~i=1,\ldots ,4.
$$

\subsection{\texorpdfstring{$\mathbb{Z}_4$}{{\mathbb{Z}4}}-grading}
\label{Z4grad} The outer automorphism group of a Lie algebra plays
an important role in the corresponding representation theory.  It
appears that for $\sls(4|4)$ the outer automorphism group
$\rm{Out}(\sls(4|4))$ contains continuous and finite subgroups.

\medskip

Consider the continuous group $\{\delta_{\rho},\rho\in {\mathbb
C}^*\}$ which acts on $M$ in the following way \bea
\delta_{\rho}(M)=\left(
\begin{array}{cc}
  m & \rho\, \theta \\
  \frac{1}{\rho}\, \eta & n
\end{array} \right)\, ,
\eea {\it i.e.} it leaves the bosonic elements untouched and acts
on the fermionic elements as a dilatation. In fact, this
transformation is generated by the so-called hypercharge \bea
\label{hyper}
\Upsilon= \left(
\begin{array}{cc}
  \mI_4 & 0 \\
  0 & -\mI_4
\end{array} \right)\,
\eea and can be formally written in the form $
\delta_{\rho}(M)=e^{\frac{1}{2}\Upsilon\log\rho }M
e^{-\frac{1}{2}\Upsilon\log\rho}\, $. Of course, the hypercharge
is not an element of $\sls(4|4)$ as it has non-vanishing
supertrace. On the other hand, \bea e^{\frac{1}{2}
\Upsilon\log\rho}=\left(
\begin{array}{cc}
 \rho^{\frac{1}{2}} \, \mI_4 & 0 \\
  0 & \rho^{-\frac{1}{2}}\, \mI_4
\end{array} \right)\, .
\eea The superdeterminant of this matrix is equal to $\rho^4$.
Thus, for $\rho$ satisfying the relation $\rho^4=1$, the
corresponding automorphisms $\delta_{\rho}$ are, in fact, inner.
Hence, the continuous family of outer automorphisms of $\sls(4|4)$
coincides with the factor-group $\delta_{\rho}/\{\delta_{\rho}:
\rho^4=1\}$. We further note that the automorphism group
$\delta_{\rho}$ admits a restriction to $\su(2,2|4)$ provided the
parameter $\rho$ lies on a circle $|\rho|=1$.

\medskip

The finite subgroup of $\rm{Out}(\sls(4|4))$ coincides with the
Klein four-group ${\mathbb Z}_2\times {\mathbb Z}_2$. The first
factor is generated by the transformation \bea
 M=\left(
\begin{array}{cc}
  m & \theta \\
  \eta & n
\end{array} \right)\,\to\, \left(
\begin{array}{cc}
  n & \eta \\
  \theta & m
\end{array} \right)\, ,
\eea while the second one is generated by \bea M\to -M^{st} \,
,\eea where the supertranspose $M^{st}$ is defined as \bea M^{st}=
\left(
\begin{array}{cc}
  m^t & -\eta^t \\
  \theta^t & n^t
\end{array} \right)\, .\eea
The ``minus supertransposition" is an automorphism of order four.
We see, however, that \bea\label{inneraut} (M^{st})^{st}=\left(
\begin{array}{cc}
  m & -\theta \\
  -\eta & n
\end{array} \right)=\delta_{-1}(M)\, ,
\eea which, according to the discussion above,  is an inner
automorphism. Thus, in the group of outer automorphisms the order
of ``minus supertransposition" is indeed two, while in the group
of all automorphisms its order is equal to four.

\smallskip

The fourth order automorphism $M\to -M^{st}$ allows one to endow
$\sls(4|4)$ with the structure of a ${\mathbb Z}_4$-graded Lie
superalgebra. For our further purposes it is important, however,
to choose an equivalent automorphism\footnote{ Although the
actions of these two automorphisms are related by the similarity
transformation, they introduce inequivalent $\mathbb{Z}_4$-graded
structures on $\sls(4|4)$.} \bea M\to \Omega(M)= -{\cal K}M^{
st}{\cal K}^{-1}\, , \eea where ${\cal K}$ is the $8\times
8$-matrix, ${\cal K}={\rm diag}(K,K)$, and the $4\times 4$ matrix
$K$ is given in eq.(\ref{K}).  On the product of two supermatrices
one has $ \Omega(M_1M_2)=-\Omega(M_2)\Omega(M_1)\, $.

\smallskip

Introducing the notation ${\mathscr G}=\sls(4|4)$, let us define
\bea {\mathscr G}^{(k)}=\Big\{M\in {\mathscr G},\, ~~
\Omega(M)=i^k M\Big\}\, . \eea
 Then, as a vector space, ${\mathscr
G}$ can be decomposed into a direct sum of graded subspaces \bea
{\mathscr G}={\mathscr G}^{(0)}\oplus{\mathscr G}^{(1)}\oplus
{\mathscr G}^{(2)}\oplus {\mathscr G}^{(3)}\, \label{Z4}\eea where
$[{\mathscr G}^{(k)},{\mathscr G}^{(m)}]\subset {\mathscr
G}^{(k+m)}$ modulo $\mathbb{Z}_4$. For any matrix $M\in {\mathscr
G}$ its projection $M^{(k)}\in {\mathscr G}^{(k)}$ is given by
\bea
M^{(k)}=\frac{1}{4}\Big(M+i^{3k}\Omega(M)+i^{2k}\Omega^2(M)+i^k\Omega^3(M)\Big)\,
. \label{Z4proj} \eea It is easy to see that the projections
$M^{(0)}$ and $M^{(2)}$ are even, while $M^{(1)}$ and $M^{(3)}$
are odd.

\medskip

While $[K,\Sigma]=[\gamma^5,\gamma^2\gamma^4]=0$, in general
$(M^{st})^{\dagger}\neq (M^{\dagger})^{st}$. As a result, one
finds that the action of $\Omega$ (anti-) commutes with the Cartan
involution: \bea
\begin{aligned}
\Omega(M)^{\dagger}&=\Omega(M^{\dagger})\,
~~~~~~~~~~&{\rm for}~M~~{\rm even} \, , \\
\Omega(M)^{\dagger}&=-\Omega(M^{\dagger})\, ~~~~&{\rm for}~M~~{\rm
odd} \, .
\end{aligned}
\eea In fact, these two formulae can be concisely written as a
single expression \bea\label{daggerM}
\Omega(M)^{\dagger}=\Upsilon\,
\Omega(M^{\dagger})\Upsilon^{-1}=-(\Upsilon H)\, \Omega(M)(
\Upsilon H)^{-1}\, , \eea where $\Upsilon$ is hypercharge
(\ref{hyper}) and we assumed that $M\in \su(2,2|4)$. Thus,
$\Omega$ admits a restriction to the bosonic subalgebra of the
real form $\su(2,2|4)$. On the whole $\su(2,2|4)$ the map $\Omega$
is not diagonalizable, since two eigenvalues of $\Omega$ are
imaginary: for the projections $M^{(k)}$ with $k=1,3$ we have
$\Omega(M^{(k)})=\pm i M^{(k)}$, while $\su(2,2|4)$ is a Lie
superalgebra over real numbers. Nevertheless, any matrix $M\in
\su(2,2|4)$ can be uniquely decomposed into the sum (\ref{Z4}),
where each component $M^{(k)}$ takes values in $\su(2,2|4)$. To
make this point clear, we compute the hermitian-conjugate of
$M^{(k)}$ given by eq.(\ref{Z4proj})
$$
M^{(k)\dagger}=-\frac{1}{4}H\Big[M+i^{k}\Upsilon\,
\Omega(M)\Upsilon^{-1} +i^{2k}\Omega^2(M)+ i^{3k} \Upsilon\,
\Omega^3(M)\Upsilon^{-1}\Big]H^{-1}
$$
where we made use of eqs.(\ref{daggerM}) and (\ref{real}). It
remains to note that according to eq.(\ref{inneraut}) one has
$\Upsilon\,\Omega(M)\Upsilon^{-1}=\Omega^3(M)$ so that
$M^{(k)\dagger}=-HMH^{-1}$, {\it i.e.} $M^{(k)}$ belongs to
$\su(2,2|4)$ for any $k$. Thus, denoting now
$\mathscr{G}=\su(2,2|4)$, in what follows we will refer to
eq.(\ref{Z4}) as the $\mathbb{Z}_4$-graded decomposition of
$\su(2,2|4)$, where the individual subspaces are defined by means
of eq.(\ref{Z4proj}).

\medskip

According to our discussion, with respect to the action of
$\Omega$ the bosonic subalgebra $\su(2,2)\oplus \su(4)\oplus
\alg{u}(1)\subset \alg{su}(2,2|4)$ is decomposed into the direct
sum of two graded components. Working out explicitly the
projection $M^{(0)}$, one finds \bea
M^{(0)}&=&\frac{1}{2}\left(\begin{array}{cc} m-Km^tK^{-1} &
0 \\
0 & n-Kn^tK^{-1} \end{array}\right) \, . \eea Analogously, for
$M^{(2)}$ one obtains \bea
M^{(2)}&=&\frac{1}{2}\left(\begin{array}{cc} m+Km^tK^{-1} &
0 \\
0 & n+Kn^tK^{-1} \end{array}\right) \, . \eea At this point it is
advantageous to make use of the explicit bases (\ref{basisbosons})
for $\su(2,2)\oplus \su(4)$ introduced in the previous section.
According to the discussion there,
$\sfrac{1}{4}[\gamma^i,\gamma^j]$ with $i,j=1,\ldots, 5$ generate
the subalgebra $\alg{so}(5)\subset \alg{su}(4)$, while the
commutators $\sfrac{1}{4}[\gamma^i,\gamma^j]$ and
$\sfrac{i}{4}[\gamma^i,\gamma^5]$ with $i,j=1,\ldots, 4$ generate
$\alg{so}(4,1)\subset \su(2,2)$. Further, the matrix $K$ was
chosen such that the following relations are satisfied \bea
\gamma^i=K(\gamma^i)^tK^{-1}\, ,
~~~~~~~[\gamma^i,\gamma^j]=-K[\gamma^i,\gamma^j]^tK^{-1}\,,\quad
i,j=1,\ldots,5\, . \label{KG} \eea These formulae reveal that the
space ${\mathscr G}^{(0)}$ in the $\mathbb{Z}_4$-graded
decomposition of $\alg{psu}(2,2|4)$ coincides with the subalgebra
$\alg{so}(4,1)\oplus \alg{so}(5)\subset \alg{su}(2,2)\oplus
\alg{su}(4)$.

\smallskip

Similarly, comparing the structure of $M^{(2)}$ with
eqs.(\ref{KG}), one finds that the space $\mathscr{G}^{(2)}$ is
spanned by the matrices $\{\gamma^{1,2,3,4},i\gamma^5\}\in
\alg{su}(2,2)$ and $\{i\gamma^i\}\in \alg{su}(4)$, where
$i=1,\ldots,5$. As we will see in section \ref{cpgs}, these are
the Lie algebra generators along the directions corresponding to
the coset space ${\rm SU}(2,2)\times {\rm SU}(4)/{\rm
SO}(4,1)\times {\rm SO}(5)=\AdS$. The central element $i\mI\in
\alg{su}(2,2|4)$ also occurs in the projection $M^{(2)}$.

\smallskip
To complete the discussion of the $\mathbb{Z}_4$-graded
decomposition, we also give the explicit formulae for the odd
projections
 \bea
\begin{aligned}
 M^{(1)}&=\frac{1}{2}\left(\begin{array}{cc} 0 &
\theta-i K\eta^t K^{-1} \\
\eta+i K \theta^tK^{-1} & 0 \end{array}\right) \, ,~~~~~\\
M^{(3)}&=\frac{1}{2}\left(\begin{array}{cc} 0 &
\theta+i K\eta^t K^{-1} \\
\eta-i K \theta^tK^{-1} & 0 \end{array}\right) \, .
\end{aligned}\eea


\section{Green-Schwarz string as coset model}
For our further discussion, it is convenient to introduce an
effective dimensionless string tension $g$, which for strings in
$\AdS$ is expressed through the radius $R$ of ${\rm S}^5$ and
string slope $\a'$ as $g = R^2/2\pi\a'$. In the AdS/CFT
correspondence this tension is related to the `t Hooft coupling
constant $\l$ as \bea g = \frac{\sqrt{\lambda}}{2\pi}\, . \eea We
will consider a single closed string propagating in the $\AdS$
space. Let coordinates $\s$ and $\tau$ parametrize the string
world-sheet which is a cylinder of circumference $2r$. For later
convenience we assume the range of the world-sheet spatial
coordinate $\s$ to be $- r\le \s\le  r$, where $r$ is an arbitrary
constant. The standard choice for a closed string is $r=\pi$. The
string action is then \bea S=\int {\rm d}\tau{\rm d}\sigma\,
\mathscr{L} \, ,\eea where $\mathscr{L}$ is the Lagrangian density
and the integration range for $\sigma$ is assumed from $-r$ to
$r$. In this section we outline the construction of the string
Lagrangian and also analyze its global and local symmetries.

\subsection{Lagrangian}
\label{Sect:Lagrangian} Let $\ag$ be an element of the supergroup
${\rm SU}(2,2|4)$. Introduce the following one-form with values in
$\alg{su}(2,2|4)$ \bea \label{la} A=-\ag^{-1}{\rm
d}\ag=A^{(0)}+A^{(2)}+A^{(1)}+A^{(3)}\, . \eea Here on the right
hand side of the last formula we exhibited the
$\mathbb{Z}_4$-decomposition of $A$, {\it c.f.} eq.(\ref{Z4}). By
construction,  $A$ has vanishing curvature $F={\rm d }A-A\wedge
A=0$ or, in components, \bea \label{zeroc} \pa_{\a}
A_{\b}-\pa_{\b} A_{\a}-[A_{\a},A_{\b}]=0\, . \eea

\smallskip

Now we postulate the following Lagrangian density describing a
superstring in the $\AdS$ background \bea \label{sLag} \L
=-\frac{g}{2}\Big[\gamma^{\a\b}{\rm
str}\big(A^{(2)}_{\a}A^{(2)}_{\b}\big)+\kappa\,
\epsilon^{\a\beta}{\rm str}\big(A^{(1)}_{\a}A^{(3)}_{\beta}\big)
\Big]\, , \eea which is the sum of the kinetic and the Wess-Zumino
term. Here we use the convention $\epsilon^{\tau\sigma}=1$ and
$\gamma^{\a\b}= h^{\a\b} \sqrt {-h}$ is the Weyl-invariant
combination\footnote{Note the following formula for the inverse
metric $$ \gamma^{\a\b}=\left(\begin{array}{rr}-\gamma^{22} ~&~
\gamma^{12} \\
\gamma^{21} ~&~ -\gamma^{11}
\end{array}\right)\, . $$} of the world-sheet
metric $h_{\a\beta}$ with $\det\gamma=-1$.  In the conformal gauge
$\g^{\a\b} = {\mbox{diag}}(-1,1)$. The parameter $\kappa$ in front
of the Wess-Zumino term has to be a real number to guarantee that
the Lagrangian is a real (even) Grassmann element.\footnote{ As we
will see shortly, the requirement of $\kappa$-symmetry leaves two
possibilities $\kappa=\pm 1$.} Indeed, assuming $\kappa=\kappa^*$
and taking into account the conjugation rule for the fermionic
entries: $(\theta_1\theta_2)^*=\theta^*_{2}\theta^*_{1}$, as well
as the cyclic property of the supertrace, we see that
$$
\L^*=-\frac{g}{2}\Big[\gamma^{\a\b}{\rm
str}\big(A^{(2)\dagger}_{\a} A^{(2)\dagger}_{\b}\big)+\kappa \,
\epsilon^{\a\beta}{\rm
str}\big(A^{(3)\dagger}_{\b}A^{(1)\dagger}_{\a}\big) \Big]=\L\, ,
$$
because all the projections $A^{(i)}$ are pseudo-hermitian
matrices obeying (\ref{real}). Thus, the Lagrangian (\ref{sLag})
is real.

\smallskip

Before we motivate formula (\ref{sLag}), we would like to comment
on the Wess-Zumino term. Originally, this term can be thought of
as entering the action in the usual non-local fashion, {\it i.e.}
as the following ${\rm SO(4,1)}\times {\rm SO(5)}$-invariant
closed three-form \bea \Theta_3={\rm str}\Big(\A^{(2)}\wedge
\A^{(3)}\wedge \A^{(3)}-\A^{(2)}\wedge \A^{(1)}\wedge
\A^{(1)}\Big) \eea integrated over a three-cycle with the boundary
being a two-dimensional string world-sheet. The fact that
$\Theta_3$ is closed can be easily derived from the flatness
condition for $\A$. However, since the third cohomology group of
the superconformal group is trivial the form $\Theta_3$ appears to
be exact \bea 2\Theta_3={\rm d}~{\rm str}\Big(\A^{(1)}\wedge
\A^{(3)}\Big)
\eea and, as a consequence, the Wess-Zumino term can
be reduced to the two-dimensional integral, {\it c.f.}
eq.(\ref{sLag}).

\smallskip

Consider a transformation
\begin{equation} \la{rmult}
 \ag\to \ag\alg{h}\,,
\end{equation}
where $\alg{h}$ belongs to ${\rm SO}(4,1)\times {\rm SO}(5)$.
Under this transformation the one-form transforms as \bea A\to
\alg{h}^{-1}A\alg{h} -\alg{h}^{-1}{\rm d}\alg{h}\, . \eea It is
easy to see that for the $\mathbb{Z}_4$-components of $A$ this
transformation implies \bea A^{(1,2,3)}\to
\alg{h}^{-1}A^{(1,2,3)}\alg{h}\, , ~~~~~A^{(0)}\to
\alg{h}^{-1}A^{(0)}\alg{h} -\alg{h}^{-1}{\rm d}\alg{h}\, . \eea
Thus, the component $A^{(0)}$ undergoes a gauge transformation,
while all the other homogeneous components transform by the
adjoint action.

\smallskip

By construction, the Lagrangian (\ref{sLag}) depends on the group
element $\ag$. However, as was shown above, under the right
multiplication of $\alg{g}$ with a local, {\it i.e.} $\sigma$- and
$\tau$-dependent element $\alg{h}\in {\rm SO}(4,1)\times {\rm
SO}(5)$, the homogeneous components $A^{(1)}$, $A^{(2)}$ and
$A^{(3)}$ undergo a similarity transformation leaving  the
Lagrangian (\ref{sLag}) invariant. Thus, the Lagrangian actually
depends on a coset element from ${\rm SU}(2,2|4)/{\rm
SO}(4,1)\times {\rm SO}(5)$, rather than on $\alg{g}\in {\rm
SU}(2,2|4)$.

\smallskip

Recall that in the $\mathbb{Z}_4$-decomposition of $A\in
\alg{su}(2,2|4)$ the central element $i\mI$ occurs in the
projection $A^{(2)}$. As  a result, under the right multiplication
of $\ag$ with a group element from ${\rm U}(1)$ corresponding to
$i\mI$, the component $A^{(2)}$ undergoes a shift
$$
A^{(2)}\to A^{(2)}+c\cdot i \mI\, .
$$
Since the supertrace of both the identity matrix and $A^{(2)}$
vanishes, this transformation leaves the Lagrangian (\ref{sLag})
invariant. Thus, in addition to $\alg{so}(4,1)\times \alg{so}(5)$,
we have an extra local $\alg{u}(1)$-symmetry induced by the
central element $i\, \mI$. Clearly, this symmetry can be used to
gauge away the trace part of $A^{(2)}$. Thus, in what follows we
will assume that $A^{(2)}$ is chosen to be traceless, which can be
viewed as the gauge fixing condition for these
$\alg{u}(1)$-transformations.

\smallskip

The group of global symmetry transformations of the Lagrangian
(\ref{sLag}) coincides with ${\rm PSU}(2,2|4)$. Indeed, ${\rm
PSU}(2,2|4)$ acts on the coset space (\ref{sAdS}) by
multiplication from the left. If $\ag \in {\rm PSU}(2,2|4)$ is a
coset space representative and $G$ is an arbitrary group element
from ${\rm PSU}(2,2|4)$, then the action of $G$ on $\ag$ is as
follows \bea G:~\ag\to \ag'\, , \label{actionG}\eea where $\ag'$
is determined from the following equation \bea G\cdot \ag=\ag'\,
\alg{h}\, . \label{actionG1}\eea Here $\ag'$ is a new coset
representative and $\alg{h}$ is a "compensating" local element
from ${\rm SO}(4,1)\times {\rm SO}(5)$. Because of the local
invariance under ${\rm SO}(4,1)\times {\rm SO}(5)$ the Lagrangian
(\ref{sLag}) is also invariant under global ${\rm
PSU}(2,2|4)$-transformations. The detailed discussion of these
global symmetry transformations will be postponed till section
\ref{cpgs}.

\smallskip

Further justification of the Lagrangian (\ref{sLag}) comes from
the fact that when restricted to bosonic variables only, it
reproduces the usual Polyakov action for bosonic strings
propagating in the $\AdS$ geometry. We will present the
corresponding derivation in section \ref{Subsect:altLag}.

\smallskip

Our next goal is to derive the equations of motion following from
eq.(\ref{sLag}). We first note that if $M_1$ and $M_2$ are two
supermatrices then \bea {\rm str}(\Omega^k(M_1)M_2)= {\rm
str}(M_1\Omega^{4-k}(M_2))\, \eea for $k=1,2,3$. By using this
property, the variation of the Lagrangian density can be cast in
the form \bea \delta \L=- {\rm str}(\delta \A_{\a}\, \Lambda^{\a}
)\, , \eea where \bea
\Lambda^{\a}=g\Big[\gamma^{\a\b}\A^{(2)}_{\b}
-\sfrac{1}{2}\kappa\,
\epsilon^{\a\beta}(\A^{(1)}_{\beta}-\A^{(3)}_{\beta}) \Big]\, .
\label{cS} \eea Taking into account that the variation of $A_{\a}$
is
$$
\delta A_{\a}=-\delta(\ag^{-1}\pa_{\a}\ag)=-\ag^{-1}\delta\ag\,
A_{\a}-\ag^{-1}\pa_{\a}(\delta \ag)\, ,
$$
we obtain
$$
\delta\L={\rm str}\Big[\ag^{-1}\delta \ag\,
A_{\a}\Lambda^{\a}+\ag^{-1}\pa_{\a}(\delta \ag) \Lambda^{\a}
\Big]\, .
$$
Finally, integrating the last term by parts and omitting the total
derivative contribution, we arrive at the following expression for
the variation of the Lagrangian density  \bea \delta \L=-{\rm
str}\Big[\ag^{-1}\delta \ag\,
(\pa_{\a}\Lambda^{\a}-[\A_{\a},\Lambda^{\a}] )\Big]\, . \eea Thus,
if we regard $\pa_{\a}\Lambda^{\a}-[\A_{\a},\Lambda^{\a}]$ as an
element of $\alg{su}(2,2|4)$,\ then the equations of motion read
as \bea \label{seom0}
\pa_{\a}\Lambda^{\a}-[\A_{\a},\Lambda^{\a}]=\varrho\cdot \mI\, ,
\eea where the coefficient $\varrho$ is found by taking the trace
of both sides of the last equation. Since $\psu(2,2|4)$ is
understood as the quotient of $\su(2,2|4)$ over its
one-dimensional center, in $\psu(2,2|4)$ the equations of motion
take the form \bea \label{seom}
\pa_{\a}\Lambda^{\a}-[\A_{\a},\Lambda^{\a}]=0\, .\eea

The single equation (\ref{seom}) can be projected on various
$\mathbb{Z}_4$-components. First, one finds that the projection on
$\mathscr{G}^{(0)}$ vanishes. Second, for the projection on
$\mathscr{G}^{(2)}$ we get \bea \label{Eqb}
\pa_{\a}(\gamma^{\a\b}\A_{\b}^{(2)})
-\gamma^{\a\b}[\A_{\a}^{(0)},\A_{\b}^{(2)}] +\sfrac{1}{2}\kappa
\epsilon^{\a\beta}\big([\A_{\a}^{(1)},\A_{\b}^{(1)}]-[\A_{\a}^{(3)},\A_{\b}^{(3)}]\big)=0\,
, \eea while the for projections on $\mathscr{G}^{(1,3)}$ one
finds
\begin{equation}
\label{Eqf1} \begin{aligned}
 & \gamma^{\a\b}[\A_{\a}^{(3)},\A_{\b}^{(2)}] +\kappa
\epsilon^{\a\beta}[\A_{\a}^{(2)},\A_{\b}^{(3)}]=0\, , \\
& \gamma^{\a\b}[\A_{\a}^{(1)},\A_{\b}^{(2)}] -\kappa
\epsilon^{\a\beta}[\A_{\a}^{(2)},\A_{\b}^{(1)}]=0\,
.\end{aligned}\end{equation} In deriving these equations we  also
used the condition of zero curvature for the connection $\A_{\a}$.
Introducing the tensors \bea {\rm P}_{\pm}^{\a\b}=\sfrac{1}{2}(
\gamma^{\a\b}\pm \kappa \epsilon^{\a\b} )\, ,\eea equations
(\ref{Eqf1})  can be written as
\begin{equation}
\label{Eqf} \begin{aligned}
&{\rm P}_{-}^{\a\b}[\A_{\a}^{(2)},\A_{\b}^{(3)}] = 0 \, ,\\
&{\rm P}_{+}^{\a\b}[\A_{\a}^{(2)},\A_{\b}^{(1)}] = 0 \, .
\end{aligned}\end{equation} We further note that for $\kappa=\pm
1$ the tensors ${\rm P}_{\pm}$ are orthogonal projectors: \bea
{\rm P}_{+}^{\a\b}+{\rm P}_{-}^{\a\b}=\gamma^{\a\b}\, ,~~~~~{\rm
P}_{\pm}^{\a\delta}{\rm P}_{\pm \delta}^{~~\b}={\rm
P}_{\pm}^{\a\b} \, ,~~~~~~\, {\rm P}_{\pm}^{\a\delta}{\rm P}_{\mp
\delta}^{~~\b}=0\, .\eea Further, we emphasize the relation
between the equations of motion and the global symmetries of the
model (the Noether theorem). Consider the following current \bea
J^{\a}=\ag\, \Lambda^{\a} \ag^{-1}\, .\label{gcurrent}\eea Due to
eq.(\ref{seom}), this current is conserved: \bea
\pa_{\a}J^{\a}=0\, . \eea In fact, $J^{\a}$ is nothing else but
the Noether current corresponding to global PSU(2,2$|$4)-symmetry
transformations. The corresponding conserved charge ${\rm Q }$ is
given by the following integral of the $J^{\tau}$ component \bea
\label{NC} {\rm Q}=\int_{-r}^{r}{\rm d}\sigma\,
J^{\tau}=g\int_{-r}^{r}{\rm d}\sigma\, \ag\Big[
\gamma^{\tau\tau}A_{\tau}^{(2)}+\gamma^{\tau\sigma}A_{\sigma}^{(2)}-\frac{\kappa}{2}(A^{(1)}_{\sigma}
-A^{(3)}_{\sigma})\Big]\ag^{-1}\, .\eea It is worth pointing out
that in the matrix representation the current $J^{\a}$ is an
element of $\su(2,2|4)$ and, for this reason, only its traceless
part is conserved.

\smallskip

Finally, we also have equations of motion for the world-sheet
metric which are equivalent to vanishing the world-sheet
stress-tensor \bea \label{Vir} {\rm
str}(\A^{(2)}_{\a}\A^{(2)}_{\b})
-\sfrac{1}{2}\gamma_{\a\b}\gamma^{\rho\delta}{\rm
str}(\A^{(2)}_{\rho}\A^{(2)}_{\delta})=0\, .
 \eea
These equations are known as the Virasoro constraints and they
reflect the two-dimensional reparametrization invariance of the
string action.

\smallskip

In summary, we presented a construction of the superstring
Lagrangian based on the flat connection $A$. The Lagrangian
comprises degrees of freedom corresponding to the coset space
(\ref{sAdS}) and it is invariant with respect to the global ${\rm
PSU}(2,2|4)$-symmetry transformations. The flat connection $A$
allows one to introduce a new current $J^{\a}$ which is conserved
due to the superstring equations of motion; the corresponding
conserved charge is a generator of these global symmetry
transformations.

\subsection{Parity transform and time reversal}\la{sect:parity}
In section (\ref{Z4grad}) we introduced a continuous group
$\{\delta_{\rho}\}$ of automorphisms of $\sls(4|4)$. For $\rho$
restricted to the unit circle this group also becomes an
automorphism group of $\psu(2,2|4)$. In particular, the
automorphisms $\delta_{-1}$ and $\delta_{\pm i}$ are inner. Here
we will argue that the action of the elements $\delta_{\pm i}$ on
the string Lagrangian (\ref{sLag}) can be essentially viewed as
the parity transformation or, equivalently, as the time reversal
operation.

\smallskip

More generally, we start our analysis by considering the following
transformation \bea \label{Utr} \ag'=U\, \ag\,  U^{-1}\, , \eea
where $\ag\in {\rm PSU}(2,2|4)$ and $U$ is some global (constant)
bosonic matrix. The matrix $U$ should not however belong to ${\rm
SO}(4,1)\times {\rm SO}(5)$, as in the opposite case we have
already established the invariance of the string Lagrangian: it is
separately invariant under multiplication of $\ag$ by a global
element $U$ from the left and by a local element $V$ from the
right. Under transformation (\ref{Utr}) the connection
$A=-\ag^{-1}{\rm d}\ag$ undergoes a change
$$
A\to A'=U\, A \,  U^{-1}\, .
$$
Imposing an extra requirement that $U$ commutes with
$\mathcal{K}$, we obtain \bea\label{trcon} \Omega(A')=
-\mathcal{K}(U\, A \, U^{-1})^{st} \mathcal{K}^{-1}=(U^{t})^{-1}\,
\Omega(A) \, U^t\, . \eea This formula allows us to construct the
$\mathbb{Z}_4$-graded decomposition of the transformed connection
$A'$. First, we look at the projection $A'^{(2)}$ \bea
A'^{(2)}=\frac{1}{4}\Big[A'-\Omega(A')+\Omega^2(A')-\Omega^3(A')\Big]\,
, \eea which, upon the usage of eq.(\ref{trcon}), takes the form
\bea A'^{(2)}=\frac{1}{4}\Big[ U\big(
A+\Omega^2(A)\big)U^{-1}-(U^t)^{-1}\big(\Omega(A)+\Omega^3(A)\big)U^t\Big]\,
. \eea Substituting here the $\mathbb{Z}_4$-graded decomposition
(\ref{la}) of $A$, we see that \bea \label{A2tr1} A'^{(2)}=
\frac{1}{2}\Big[U\big(
A^{(0)}+A^{(2)}\big)U^{-1}-(U^t)^{-1}\big(A^{(0)}-A^{(2)}\big)U^t\Big]\,
. \eea Analogous considerations allow one to establish the
formulae for the odd components of the transformed connection \bea
\label{A13}
\begin{aligned}
A'^{(1)}&=\frac{1}{2}\Big[ U\big(
A^{(1)}+A^{(3)}\big)U^{-1}+(U^t)^{-1}\big(A^{(1)}-A^{(3)})\big)U^t\Big]\,
,
\\
 A'^{(3)}&=\frac{1}{2}\Big[ U\big(
A^{(1)}+A^{(3)}\big)U^{-1}-(U^t)^{-1}\big(A^{(1)}-A^{(3)})\big)U^t\Big]\,
. \end{aligned}\eea These expressions suggest to consider the
following two cases. The first one corresponds  to taking $U$ such
that \bea\label{trivial} \hskip 1cm U^tU=\mI\,\ ,
~~~~~~~~~~[U,\mathcal{K}]=0\, . \eea With this choice  the
transformation formulae (\ref{A2tr1}) and (\ref{A13}) simplify to
\bea A^{'(2)}=UA^{(2)}U^{-1}\, , ~~~~~ A^{'(1)}=UA^{(1)}U^{-1}\,
,~~~~~ A^{'(3)}=UA^{(3)}U^{-1}\, . \eea We thus see that the
Lagrangian (\ref{sLag}) remains invariant\footnote{In fact, the
Lagrangian remains invariant under a milder assumption on $U$,
namely, $U^tU=e^{i\a}\mI$, where $e^{i\a}$ is an arbitrary phase.
However, this phase plays no role -- being absorbed into $U$, it
drops out of the similarity transformation (\ref{Utr}). The matrix
$U$ corresponding to $\delta_{-1}$ is $U=i\Upsilon$, so that it
commutes with $\mathcal{K}$ and obeys $U^tU=-\mI$. Thus, the
action of $\delta_{-1}$ leaves the Lagrangian invariant. },
however, there is nothing new here because the group singled out
by the requirements (\ref{trivial}) is just a subgroup of ${\rm
SO}(4,1)\times {\rm SO}(5)$.

\smallskip

The second case  corresponds to imposing the following
requirements \bea\label{chcon} U^tU=\Upsilon \,
,~~~~~[U,\mathcal{K}]=0\, ,\eea where we omitted an unessential
overall phase in front of $\Upsilon$, see the footnote 5. Since
for any odd matrix $M$ one has $\Upsilon M \Upsilon^{-1}=-M$,
expressions (\ref{A2tr1}) and (\ref{A13}) reduce to
$$
A^{'(2)}=UA^{(2)}U^{-1}\, , ~~~~~ A^{'(1)}=UA^{(3)}U^{-1}\, ,~~~~~
A^{'(3)}=UA^{(1)}U^{-1}\, .
$$
Thus, in essence, the transformation above exchanges the
projections $A^{(1)}$ and $A^{(3)}$. For this reason, it does not
leave the Lagrangian (\ref{sLag}) invariant, rather it changes the
sign in front of the Wess-Zumino term.

\smallskip

As the simplest solutions to eqs.(\ref{chcon}), we can take
\bea\label{cp} U=\left(\begin{array}{cc} i^{\frac{1}{2}}\,\mI_4 & 0 \\
0 & i^{-\frac{1}{2}}\,\mI_4
\end{array}\right)=e^{i\frac{\pi}{4}\Upsilon}\, ,
\eea which corresponds to the action of $\delta_i$.  We identify
$U$ as a matrix corresponding to the parity transformation
$\mathscr{P}\equiv U$. Indeed, under the map $\sigma\to -\sigma$
the Wess-Zumino term changes its sign.\footnote{The pseudo-tensor
$\epsilon^{\a\b}$ does not change its sign under $\sigma\to\sigma$
or $\tau\to -\tau$.} This sign change can be then compensated by
transformation (\ref{Utr}) with $U$ given by (\ref{cp}). Thus,
under the combined transformation \bea \sigma\to -\sigma\,
,~~~~~~\ag \to \mathscr{P}\ag \mathscr{P}^{-1} \eea the action
remains invariant. Under $\mathscr{P}$ a supermatrix $M$
transforms as follows \bea
 M=\left(
\begin{array}{cc}
  m & \theta \\
  \eta & n
\end{array} \right)\,\to\, \mathscr{P}M\mathscr{P}^{-1}= \left(
\begin{array}{cc}
  m & i\theta \\
  -i\eta & n
\end{array} \right)\, ,
\eea {\it i.e.} fermions are multiplied by $\pm i$ which can be
identified as their intrinsic parity.

\smallskip

Before the gauge fixing, $\sigma$ and $\tau$ variables enter the
sigma model action on equal footing. Therefore, one can equally
regard the action of $U$ together with the change $\tau\to-\tau$
as the time reversal operation. In the gauge-fixed theory the time
reversal operation acts differently. We will return to this issue
in chapter 3.

\subsection{Kappa-symmetry}
\label{Sect:kappa} Kappa-symmetry is a local fermionic symmetry of
the Green-Schwarz superstring. It generalizes the local fermionic
symmetries first discovered for massive and massless
superparticles and its presence is crucial to ensure the
space-time supersymmetry of the physical spectrum.  Here we
establish $\kappa$-symmetry transformations associated with the
Lagrangian (\ref{sLag}).

\subsubsection*{Deriving $\kappa$-symmetry}

\noindent Recall that the action of the global symmetry group
${\rm PSU}(2,2|4)$ is realized on a coset element by
multiplication from the left. In this respect, $\kappa$-symmetry
transformations can be viewed as the {\it right local} action of
$G=\exp \epsilon$ on the coset representative $\ag$: \bea
\label{ks} \ag\, \cdot G=\ag'\, \alg{h}\, , \eea where
$\epsilon\equiv \epsilon(\tau,\sigma)$ is a local fermionic
parameter taking values in $\psu(2,2|4)$. Here $\alg{h}$ is a
compensating element from ${\rm SO(4,1)}\times {\rm SO(5)}$. The
main difference with the case of global symmetry is that for
arbitrary $\epsilon$ the action is not invariant under
transformation (\ref{ks}). Below we find the conditions on
$\epsilon$ which guarantee the invariance of the action.

\smallskip First, we note that under local multiplication from
the right the one-form $\A$ transforms as follows \bea
\delta_{\epsilon}\A=-{\rm d}\epsilon +[\A,\epsilon]\, . \eea The
$\mathbb{Z}_4$-decomposition of this equation gives
\bea\label{kappa}
\begin{aligned} \delta_{\epsilon}\A^{(1)}&=-{\rm d}\epsilon^{(1)}
+[\A^{(0)},\epsilon^{(1)}]+[\A^{(2)},\epsilon^{(3)}]\,
,\\\delta_{\epsilon}\A^{(3)}&=-{\rm d}\epsilon^{(3)}
+[\A^{(2)},\epsilon^{(1)}]+[\A^{(0)},\epsilon^{(3)}]\, ,\\
 \delta_{\epsilon}\A^{(2)}&=
[\A^{(1)},\epsilon^{(1)}]+[\A^{(3)},\epsilon^{(3)}]\,  ,\\
 \delta_{\epsilon}\A^{(0)}&=
[\A^{(3)},\epsilon^{(1)}]+[\A^{(1)},\epsilon^{(3)}]\,  ,
\end{aligned}\eea where we have assumed that
$\epsilon=\epsilon^{(1)}+\epsilon^{(3)}$. By using these formulae,
we find for the variation of the Lagrangian density  \bea
\nonumber -\frac{2}{g}\,
\delta_{\epsilon}\L&=&\delta\gamma^{\a\b}{\rm
str}\big(\A^{(2)}_{\a}\A^{(2)}_{\b}\big)-2\gamma^{\a\b}{\rm
str}\big([\A^{(1)}_{\a},\A^{(2)}_{\b}]\epsilon^{(1)}
+[\A^{(3)}_{\a},\A^{(2)}_{\b}]\epsilon^{(3)}\big)\\
\nonumber &+&\kappa \epsilon^{\a\beta}{\rm
str}\Big(\pa_{\a}\A^{(3)}_{\b}\epsilon^{(1)}
-\pa_{\a}\A^{(1)}_{\beta}\epsilon^{(3)}+[\A^{(0)}_{\a},\epsilon^{(1)}]\A^{(3)}_{\b}
+[\A^{(2)}_{\a},\epsilon^{(3)}]\A^{(3)}_{\b}
\\
&&~~~~~~~~~~~ +\A^{(1)}_{\a}[\A^{(0)}_{\b},\epsilon^{(3)}]
+\A^{(1)}_{\a}[\A^{(2)}_{\b},\epsilon^{(1)}] \Big)\, . \eea Note
that the derivatives of $\epsilon$ have been eliminated by
integrating by parts and subsequently neglecting the corresponding
total derivatives. The variation of the world-sheet metric is left
unspecified. The flatness condition (\ref{zeroc}) implies \bea
\begin{aligned}
\epsilon^{\a\b} \pa_{\a}
\A_{\b}^{(1)}&=\epsilon^{\a\b}[\A_{\a}^{(0)},\A_{\b}^{(1)}]
+\epsilon^{\a\b}[\A_{\a}^{(2)},\A_{\b}^{(3)}] \, ,\\
\epsilon^{\a\b} \pa_{\a}
\A_{\b}^{(3)}&=\epsilon^{\a\b}[\A_{\a}^{(0)},\A_{\b}^{(3)}]
+\epsilon^{\a\b}[\A_{\a}^{(1)},\A_{\b}^{(2)}]  \, .
\end{aligned}\eea Taking into account these formulae, we obtain
$$
-\frac{2}{g}\, \delta_{\epsilon}\L=\delta\gamma^{\a\b}{\rm
str}\big(\A^{(2)}_{\a}\A^{(2)}_{\b}\big)-4\, {\rm str}\big( {\rm
P}_{+}^{\a\b}[\A^{(1)}_{\b},\A^{(2)}_{\a}]\epsilon^{(1)} +{\rm
P}_{-}^{\a\b}[\A^{(3)}_{\b},\A^{(2)}_{\a}]\epsilon^{(3)} \big) \,
.
$$
For any vector $V^{\a}$ we introduce the projections $V^\a_\pm$:
$$
V^\a_\pm = {\rm P}_{\pm}^{\a\b}V_\b\,
$$
so that the variation of the Lagrangian acquires the form
\bea\hskip -0.3cm
-\frac{2}{g}\,\delta_{\epsilon}\L=\delta\gamma^{\a\b}{\rm
str}\big(\A^{(2)}_{\a}\A^{(2)}_{\b}\big)-4\, {\rm str}\big(
[\A^{(1),\a}_{+},\A^{(2)}_{\a,-}]\epsilon^{(1)}
+[\A^{(3),\a}_{-},\A^{(2)}_{\a,+}]\epsilon^{(3)} \big) \, .\eea We
further note that from the condition ${\rm
P}_{\pm}^{\a\b}\A_{\b,\mp}=0$ the components $\A_{\tau,\pm}$ and
$\A_{\sigma,\pm}$ are proportional: \bea
 \label{prop}
\A_{\tau,\pm}=-\frac{\gamma^{\tau\sigma}\mp\kappa
}{\gamma^{\tau\tau}}\A_{\sigma,\pm}\, . \eea

The crucial point of our construction is the following ansatz for
the $\kappa$-symmetry parameters $\epsilon^{(1)}$ and
$\epsilon^{(3)}$ \bea\label{kappaansatz} \begin{aligned}
\epsilon^{(1)} &=& \A^{(2)}_{\a,-}\kappa^{(1),\a}_{+} +
\kappa^{(1),\a}_{+}\A^{(2)}_{\a,-}\, , \\  \epsilon^{(3)} &=&
\A^{(2)}_{\a,+}\kappa^{(3),\a}_{-} +
\kappa^{(3),\a}_{-}\A^{(2)}_{\a,+}\, . \end{aligned}\eea Here
$\kappa^{(i),\a}_{\pm}$ are new independent parameters of
$\kappa$-symmetry transformation which are homogeneous elements of
degree $i=1,3$ with respect to $\Omega$. The correct degree  of
$\epsilon$ is inherited from the properties of $\Omega$ (see
footnote 2). For instance, one has \bea\nonumber
&&\Omega(\A^{(2)}_{\a,-}\kappa^{(1),\a}_{+} +
\kappa^{(1),\a}_{+}\A^{(2)}_{\a,-})=\\
\nonumber &&~~~~~~~~~~~~~
-\Omega(\kappa^{(1),\a}_{+})\Omega(\A^{(2)}_{\a,-})-
\Omega(\A^{(2)}_{\a,-})\Omega(\kappa^{(1),\a}_{+})=i(\A^{(2)}_{\a,-}\kappa^{(1),\a}_{+}
+ \kappa^{(1),\a}_{+}\A^{(2)}_{\a,-})\, . \eea Finally,
$\epsilon^{(1,3)}\in \su(2,2|4)$ provided the matrices
$\kappa^{(1)}$ and $\kappa^{(3)}$ satisfy the following reality
conditions \bea H\kappa^{(1)}-(\kappa^{(1)})^{\dagger}H=0\, ,
~~~~~H\kappa^{(3)}-(\kappa^{(3)})^{\dagger}H=0\, .\eea

\smallskip

As was explained in section \ref{Z4grad}, the (traceless)
component $\A^{(2)}$ taking values in $\su(2,2|4)$ can be expanded
as \bea \A^{(2)}=\left(
\begin{array}{cc} m^i\gamma^{i} & 0 \\
0 & n^i \gamma^i\end{array}\right)\, ,\eea where $\gamma^i$ are
the ${\rm SO}(5)$ Dirac matrices. The coefficients $n^i$ are all
imaginary, while $m^i$ are real for $i=1,\ldots,4$ and imaginary
for $i=5$.
 Thus, assuming $\A^{(2)}$ to be traceless we can
further write
$$
\A^{(2)}_{\a,\pm} \A^{(2)}_{\b,\pm} =\left(
\begin{array}{cc} m^i_{\a,\pm}m^j_{\b,\pm} \gamma^i\gamma^j & 0 \\
0 & n^i_{\a,\pm}n^j_{\b,\pm} \gamma^i\gamma^j
\end{array}\right) .
$$
  Since the chiral
components $\A_{\tau,\pm}$ and $\A_{\sigma,\pm}$ are proportional
to each other, see eq.(\ref{prop}), we can rewrite the last
formula as \bea \nonumber \A^{(2)}_{\a,\pm} \A^{(2)}_{\b,\pm}
&=&\left(
\begin{array}{cc} m^i_{\a,\pm}m^j_{\b,\pm} \sfrac{1}{2}\{\gamma^i,\gamma^j\} & 0 \\
0 & n^i_{\a,\pm}n^j_{\b,\pm} \sfrac{1}{2}\{\gamma^i,\gamma^j\}
\end{array}\right)\, .
\nonumber\eea This expression can be concisely written as \bea
\label{mf} \A^{(2)}_{\a,\pm} \A^{(2)}_{\b,\pm} =\left(
\begin{array}{cc} m^i_{\a,\pm}m^i_{\b,\pm}  & 0 \\
0 & n^i_{\a,\pm}n^i_{\b,\pm}
\end{array}\right)= \sfrac{1}{8}\,
\Upsilon \, {\rm str}\big( \A^{(2)}_{\a,\pm}
\A^{(2)}_{\b,\pm}\big) + c_{\a\b} \mI_8 \eea where $c_{\a\b}
=\sfrac{1}{2}(m^i_{\a,\pm}m^i_{\b,\pm}+n^i_{\a,\pm}n^i_{\b,\pm})$
and $\Upsilon$ is the hypercharge (\ref{hyper}). In other words,
the product of two $A^{(2)}$'s entering the variation upon
substitution of the ansatz (\ref{kappaansatz}) appears to be just
a linear combination of two matrices, one of them being the
identity matrix and the other being $\Upsilon$.

\smallskip

 Therefore,  for
the variation of the Lagrangian density we find \bea \nonumber &&
-\frac{2}{g}\,\delta_{\epsilon}\L=\delta\gamma^{\a\b}{\rm
str}\big(\A^{(2)}_{\a}\A^{(2)}_{\b}\big)\\\nonumber
&&-\sfrac{1}{2}~{\rm str}\big( \A^{(2)}_{\a,-}
\A^{(2)}_{\b,-}\big) {\rm str}\big(\Upsilon
[\kappa^{(1),\b}_{+},\A^{(1),\a}_{+}] \big) - \sfrac{1}{2}~{\rm
str} \big(\A^{(2)}_{\a,+} \A^{(2)}_{\b,+}\big) {\rm
str}\big(\Upsilon [\kappa^{(3),\b}_{-},\A^{(3),\a}_{-}] \big) \,
,\eea where the contribution of the term in eq.(\ref{mf})
proportional to the identity matrix drops out.
 It is now clear
that this variation vanishes provided we assume the following
transformation rule for the world-sheet metric under
$\kappa$-symmetry transformations \bea \nonumber \delta
\gamma^{\a\b} &=& \frac{1}{4}{\rm str}\Big(\Upsilon
([\kappa^{(1),\a}_{+},\A^{(1),\b}_{+}]
 +
[\kappa^{(1),\b}_{+},\A^{(1),\a}_{+}] +
[\kappa^{(3),\a}_{-},\A^{(3),\b}_{-}]
 +
[\kappa^{(3),\b}_{-},\A^{(3),\a}_{-}]) \Big)\, . \eea This
variation is an even symmetric tensor satisfying the identity
$\gamma_{\a\b}\delta\gamma^{\a\b}=0$. Moreover, the reality
conditions for $A$ and $\kappa$ guarantee that the variation
$\delta\gamma^{\a\b}$ is a tensor with real components.

\smallskip

It is useful to note that the projectors ${\rm P}_{\pm}^{\a\b}$
satisfy the following important identity \bea
 \label{projprop}
 {\rm
P}_{\pm}^{\a\gamma} {\rm P}_{\pm}^{\b\delta}={\rm
P}_{\pm}^{\b\gamma} {\rm P}_{\pm}^{\a\delta}\, . \eea This
identity allows one to rewrite the variation of the metric in a
simpler form \bea  \delta \gamma^{\a\b} &=& \frac{1}{2}{\rm
tr}\Big([\kappa^{(1),\a}_{+},\A^{(1),\b}_{+}]
 +
[\kappa^{(3),\a}_{-},\A^{(3),\b}_{-}] \Big)\, ,\eea where we used
the fact that the supertrace of any matrix with an insertion of
$\Upsilon$ is the same as the usual trace of this matrix. It is
worthwhile to point out that in our derivation of
$\kappa$-symmetry transformations we exploited  the fact that
${\rm P}_{\pm}^{\a\b}$ are orthogonal projectors and, therefore,
the realization of $\kappa$-symmetry requires the parameter
$\kappa$ in front of the Wess-Zumino term to take one of the
values $\kappa=\pm 1$.

\subsubsection{On-shell rank of $\kappa$-symmetry transformations}

\noindent The next important question is to understand how many
fermionic degrees of freedom can be gauged away on-shell by means
of $\kappa$-symmetry. To this end, one can make use of the
light-cone gauge\footnote{String theory in the light-cone gauge
will be treated in great detail in the next chapter.}.
Generically, the light-cone coordinates $x_{\pm}$ are introduced
by making linear combinations of one field corresponding to the
time direction from ${\rm AdS}_5$ and one field from ${\rm S}^5$.
Without loss of generality we can assume that the transversal
fluctuations are all suppressed and the corresponding element
$A^{(2)}$ has the form \bea\label{Alc} A^{(2)}=
\left(\begin{array}{cc} i x\gamma^5 & 0
\\
0 & i y \gamma^5
\end{array} \right)\, .
\eea Indeed, the matrix $i\gamma^5$ corresponds to the time
direction in ${\rm AdS}_5$ and any element from the tangent space
to ${\rm S}^5$ can be brought to $\gamma^5$ by means of an
$\alg{su}(4)$ transformation.

\smallskip

Consider first the $\kappa$-symmetry parameter $\epsilon^{(1)}$.
In the present context, going on-shell means the fulfillment of
the Virasoro constraint ${\rm str}(A^{(2)}_{\a,-}A^{(2)}_{\b,
-})=0$, the latter boils down to $x^2=y^2$, {\it i.e.} to $y=\pm
x$. According to eq.(\ref{kappa}), we have \bea
\epsilon^{(1)}=A_{\tau,-}^{(2)}\varkappa+\varkappa
A_{\tau,-}^{(2)}\, , ~~~~~\varkappa\equiv
\kappa_+^{(1),\tau}-\frac{\gamma^{\tau\tau}}{\gamma^{\tau\sigma}\mp
\kappa }\kappa_{+}^{(1),\sigma}\, . \label{pk}\eea Picking, {\it
e.g.}, the solution $y=x$, we compute the element
$\epsilon^{(1)}$. Plugging eq.(\ref{Alc}) into eq.(\ref{pk}) and
assuming for the moment that $\varkappa$ is generic, {\it i.e.}
that it depends on 32 independent (real) fermionic variables, we
obtain \bea\label{kappaeps1}
\epsilon^{(1)}=2ix\left(\begin{array}{cc} 0 ~&~ \varepsilon
\\
-\varepsilon^{\dagger}\Sigma ~&~ 0
\end{array} \right)\, , \eea
where $\varepsilon$ is the following matrix \bea \label{kappaeps2}
\varepsilon={\small \left(\begin{array}{cccc} \varkappa_{11} &
\varkappa_{12} &
0 & 0 \\
\varkappa_{21} & \varkappa_{22} & 0 & 0 \\
0 & 0 & -\varkappa_{33} & -\varkappa_{34} \\
0 & 0 & -\varkappa_{43} & -\varkappa_{44} \\
\end{array}\right)} \, .
\eea and $\varkappa_{ij}$ are the entries of the matrix
$\varkappa$. As we see, the matrix $\varepsilon$ depends on 8
independent complex fermionic parameters. Now we can account for
the fact that the odd matrix $\varkappa$ belongs to the
homogeneous component $\mathscr{G}^{(1)}$ which reduces the number
of independent fermions by half. As a result, $\epsilon^{(1)}$
depends on 8 real fermionic parameters. A similar analysis shows
that $\epsilon^{(3)}$ will also depend on other 8 real fermions.
Thus, in total $\epsilon^{(1)}$ and $\epsilon^{(3)}$ involve 16
real fermions and these are those degrees of freedom which can be
gauged away by $\kappa$-symmetry. The gauge-fixed coset model will
therefore  involve 16 physical fermions only.

\smallskip

The above analysis, especially eqs.(\ref{kappaeps1}) and
(\ref{kappaeps2}), shows that $\kappa$-symmetry suffice to bring a
generic odd element of $\su(2,2|4)$ to the following form
\bea\label{kappagauge} {\small \left(\begin{array}{rrrr|rrrr} 0 &
0 & 0 & 0 & 0 & 0 & \bullet & \bullet
\\
0 & 0 & 0 & 0 & 0 & 0 & \bullet & \bullet
\\
0 & 0 & 0 & 0 & \bullet & \bullet & 0 & 0
\\
0 & 0 & 0 & 0 & \bullet & \bullet & 0 & 0
\\
\hline 0 & 0 & \bullet & \bullet & 0 & 0 & 0 & 0
\\
0 & 0 & \bullet & \bullet & 0 & 0 & 0 & 0
\\
\bullet & \bullet & 0 & 0 & 0 & 0 & 0 & 0
\\
\bullet & \bullet & 0 & 0 & 0 & 0 & 0 & 0
\\
\end{array}
\right)}\, , \eea where bullets stand for odd elements which
cannot be gauged away by $\kappa$-symmetry transformations. We
thus consider (\ref{kappagauge}) as a convenient $\kappa$-symmetry
gauge choice and we will implement it in our construction of the
light-cone string action in the next chapter.

\section{Integrability of classical superstrings}
\label{Sect:Integrability} In this section we show that the
non-linear sigma model describing strings on $\AdS$ is a classical
two-dimensional integrable system. This opens up the possibility
to analyze it by means of techniques developed in the theory of
integrable models. We start with recalling the general concept of
integrability and then we demonstrate  integrability of the string
sigma model by constructing the zero curvature representation of
the corresponding equations of motion. Finally, we discuss the
relationship between integrability and the local, and global
symmetries of the string model.

\subsection{General concept of integrability}
The classical inverse scattering method, {\it i.e.} the method of
finding a certain class of solutions of non-linear integrable
partial differential equations, is based on a remarkable
observation that a two-dimensional partial differential equation
arises as the consistency condition of the following
overdetermined system of equations
\begin{equation}\label{zcg}
\begin{aligned}
\frac{\pa \Psi}{\pa \sigma}={L}_{\sigma}(\sigma,\tau,z)\Psi \, ,
 \\
\frac{\pa \Psi}{\pa \tau} ={L}_{\tau}(\sigma,\tau,z)\Psi \, ,
\end{aligned}
\end{equation}
which is sometimes referred to as the fundamental linear problem.
Here $\Psi\equiv \Psi(\sigma,\tau,z)$ is a vector of rank
$\alg{r}$ and ${L}_{\sigma}\equiv {L}_{\sigma}(\sigma,\tau,z)$ and
${L}_{\tau}\equiv {L}_{\tau}(\sigma,\tau,z)$ are properly chosen
$\alg{r}\times \alg{r}$ matrices. Both $\Psi$ and
${L}_{\sigma},{L}_{\tau}$ depend on an additional {\it spectral
parameter} $z$ taking values in the complex plane\footnote{In more
complicated situations the spectral parameter can live on a
higher-genus Riemann surface.}. Differentiating the first equation
in (\ref{zcg}) with respect to $\tau$ and the second one with
respect to $\sigma$, we obtain
 \bea \frac{\pa^2 \Psi}{\pa \tau\pa
\sigma}&=&\pa_{\tau}{L}_{\sigma}\Psi+{L}_{\sigma}\pa_{\tau}\Psi
=(\pa_{\tau}{L}_{\sigma}+{L}_{\sigma}L_{\tau})\Psi \, ,
\nonumber \\
\frac{\pa^2 \Psi}{\pa \sigma\pa \tau}&=&\pa_{\sigma}
{L}_{\tau}\Psi+L_{\tau}\pa_{\sigma}\Psi
=(\pa_{\sigma}L_{\tau}+L_{\tau}L_{\sigma})\Psi \, , \nonumber \eea
which implies the fulfilment of the following consistency
condition
$$
\pa_{\tau}L_{\sigma}-\pa_{\sigma}
L_{\tau}+[L_{\sigma},L_{\tau}]=0\,
$$
for all values of the spectral parameter. If we introduce a
two-dimensional non-abelian connection ${L}_\a$ with components
$L_{\tau}$ and $L_{\sigma}$, then the consistency condition
derived above can be reinstated as vanishing of the curvature of
${L}_\a$:
 \bea \pa_{\a}L_{\beta}-\pa_{\b}L_{\a}- [L_{\a},L_{\b}]=0
\, . \label{zcr}\eea The matrices $L_{\tau}$ and $L_{\sigma}$ must
be chosen in such a way that the zero curvature condition above
should imply the fulfilment of the original differential equation
for all values of the spectral parameter. A connection ${L}_\a$
with these properties is known as the Lax connection (or the Lax
pair), while equation (\ref{zcr}) as the zero-curvature (Lax)
representation of an integrable partial differential equation.

\smallskip

For a given integrable partial differential equation the Lax
connection is by no means unique. Even the rank $\alg{r}$ of the
matrices $L_{\a}$ can be different for different Lax
representations. Also, the condition of zero curvature (\ref{zcr})
is invariant with respect to the gauge transformations \bea
L_{\a}\to L'_{\a}=h L_{\a} h^{-1}+\pa_{\a}hh^{-1}\, ,\eea where
$h$ is an arbitrary matrix, in general depending on dynamical
variables of the model and the spectral parameter.

\subsubsection*{Conservation laws}

\noindent The usefulness of the Lax connection lies in the fact
that  for a given integrable model it provides a canonical way to
exhibit the conservation laws (integrals of motion) which is
usually the first step in constructing explicit solutions of the
corresponding equations of motion. Indeed, the one-parameter
family of flat connections allows one to define the monodromy
matrix ${\rm T}(z)$ which is the path-ordered exponential of the
Lax component ${L}_{\sigma}(z)$ \bea {\rm T}(z)\,= \,
\stackrel{\longleftarrow}{\exp}\int_0^{2\pi}{\rm d}\sigma\,
L_{\sigma}(z)\, . \eea For definiteness, we assume that a model is
defined on a circle $0\leq \sigma<2\pi$ and all dynamical
variables are periodic functions of $\sigma$.

\smallskip

 Let us derive the  evolution equation for
this matrix with respect to the parameter $\tau$. We have \bea
\nonumber \pa_{\tau}{\rm T}(z)&=&\int_0^{2\pi}{\rm d}\sigma\,
\Big[\stackrel{\longleftarrow}{\exp}{\int_{\sigma}^{2\pi}
{L}_{\sigma} }\Big]\pa_{\tau}{L}_{\sigma}\Big[
\stackrel{\longleftarrow}{\exp}{\int^{\sigma}_{0} {L}_{\sigma}}\Big]\\
\nonumber &=&\int_0^{2\pi}{\rm d}\sigma~\Big[
\stackrel{\longleftarrow}{\exp}{\int_{\sigma}^{2\pi}{L}_{\sigma}
}\Big]\left(\pa_{\sigma}{L}_{\tau}+[{L}_{\tau},{L}_{\sigma}]\right)\Big[
\stackrel{\longleftarrow}{\exp}{\int^{\sigma}_{0}{L}_{\sigma}
}\Big] \, ,\eea where in the last formula we used the flatness of
$L_{\a}$. Now we notice that the integrand of the expression above
is the total derivative \bea \pa_{\tau}{\rm
T}(z)&=&\int_0^{2\pi}{\rm d}\sigma~\pa_{\sigma}\left[ \Big(
\stackrel{\longleftarrow}{\exp}{\int_{\sigma}^{2\pi}{L}_{\sigma}
}\Big){L}_{\tau}\Big(
\stackrel{\longleftarrow}{\exp}{\int^{\sigma}_{0}{L}_{\sigma}
}\Big) \,\right]\, .\eea Given that the Lax connection is a
periodic function of $\sigma$, for the monodromy we find the
following evolution equation \bea \label{evmon}\pa_{\tau}{\rm
T}(z)&=&[{L}_{\tau}(0,\tau,z),{\rm T}(z)] \, .\eea This important
formula implies that the eigenvalues of ${\rm T}(z)$ defined by
the characteristic equation \bea \label{ac} \Gamma(z,\mu)\equiv
\det({\rm T}(z)-\mu \mI )=0\,  \eea do not depend on $\tau$, in
other words they are integrals of motion. Thus, the spectral
properties of the model are encoded into the monodromy matrix.
Equation (\ref{ac}) defines an algebraic curve in $\mathbb{C}^2$
called the spectral curve.

\medskip

An alternative way to obtain the evolution equation (\ref{evmon})
is to notice that ${\rm T}(\tau)$ introduced above represents the
monodromy of a solution of the fundamental linear problem:
$$
\Psi(2\pi,\tau)={\rm T}(\tau)\Psi(0,\tau)\, .
$$
Indeed, if we differentiate this equation with respect to $\tau$,
we get
$$
\pa_\tau\Psi(2\pi,\tau)=\pa_{\tau}{\rm T}(\tau) \Psi(0,\tau)+{\rm
T}(\tau)\pa_\tau\Psi(0,\tau)\, ,
$$
which, according to the fundamental linear system, gives
$$
{L}_\tau(2\pi,\tau){\rm T}(\tau)\Psi(0,\tau)=\pa_\tau{\rm
T}(\tau)\Psi(0,\tau)+{\rm T}(\tau){L}_\tau(0,\tau)\Psi(0,\tau)\, .
$$
This relation implies the evolution equation (\ref{evmon}).

\subsubsection{ An example: Principal chiral model}

\noindent To familiarize the reader with the concept of
integrability, we consider, as an example, the so-called principal
chiral model. This integrable system is rather similar but much
simpler than the string sigma model introduced in the previous
chapter and, therefore, our discussion here will provide a
necessary warm-up before an actual handling of string
integrability.

\smallskip

The principal chiral model is a non-linear sigma model based on a
field $\ag\equiv \ag(\sigma,\tau)$ with values in a Lie group. The
action reads
$$
S=-\frac{1}{2}\int {\rm d}\tau{\rm d}\sigma\, \gamma^{\a\beta}{\rm
tr}\big(\pa_{\a}\ag\ag^{-1}\pa_{\b}\ag\ag^{-1}\big)\, ,
$$
where $\gamma^{\a\beta}$ is the Weyl-invariant metric introduced
in section \ref{Sect:Lagrangian}. Equations of motion are \bea
\la{eomprinciple} \pa_{\a}(\gamma^{\a\b}\pa_{\b}\ag
\ag^{-1})=\pa_{\a}(\gamma^{\a\beta}\ag^{-1}\pa_{\b}\ag)=0\, , \eea
and they can be conveniently written in terms of the  left and
right currents \bea A_{\it l}^{\a}=-\gamma^{\a\beta}\pa_{\b}\ag
\ag^{-1}\, , ~~~~~~~~A_{\it
r}^{\a}=-\gamma^{\a\beta}\ag^{-1}\pa_{\b}\ag \,  \eea as
$$
\pa_{\a}A_{\it l}^{\a}=0=\pa_{\a}A^{\a}_{\it r}\, .
$$
One can easily see that $A_{\it l}$ and $A_{\it r}$ are the
Noether currents corresponding to multiplications of $\ag$ by a
constant group element from the left $\ag\to \alg{h}\ag$ and from
the right $\ag\to \ag\alg{h}$, respectively. These shifts from the
left and from the right constitute the global symmetries of the
model.

\smallskip

Introduce the following connection
 \bea\label{bc} L_{\a}=\ell_1\, A_{\a}+\ell_2\,
\gamma_{\a\b}\epsilon^{\beta\rho}A_{\rho}\, , \eea where $\ell_1$
and $\ell_2$ are two undetermined parameters and $A$ is either
$A^{\it r}$ or $A^{\it l}$. In two dimensions the zero curvature
condition (\ref{zcr}) can be equivalently written as \bea
\label{zcrm}
2\epsilon^{\a\b}\pa_{\a}L_{\beta}-\epsilon^{\a\b}[L_{\a},L_{\b}]=0\,
. \eea Now we want to determine the coefficients $\ell_1$ and
$\ell_2$ by requiring the fulfillment of eq.(\ref{zcrm}) on-shell.
Taking into account the following identity \bea
\epsilon^{\alpha\beta}\gamma_{\b\rho}\epsilon^{\rho\delta}=\gamma^{\alpha\delta}\,
, \eea a simple computation reveals that eq.(\ref{zcrm}) for the
connection (\ref{bc}) reduces to
$$
2\ell_1\epsilon^{\a\b}\pa_{\a}A_{\beta}-(\ell_1^2-\ell_2^2)\epsilon^{\a\b}[A_{\a},A_{\b}]
+2\ell_2\pa_{\a}A^{\a}=0\, .
$$
The last term vanishes due to the equations of motion. As to the
first two terms, we recall that both $A^{\it r}$ or $A^{\it l}$
are flat, {\it i.e.}
$$
\pa_{\a}A_{\b}-\pa_{\b}A_{\a}\pm [A_{\a},A_{\b}]=0\, ,
$$
where the minus in front of the commutator term is for the right
current and the plus for the left one, respectively. Thus, the
first two terms will vanish due to the flatness of $A$ provided
the parameters $\ell$ are related as \bea
\begin{aligned} &
\ell_1^2-\ell_2^2-\ell_1=0\, ~~~~~~{\rm for}~~A=A^{\it r}\, ,\\
& \ell_1^2-\ell_2^2+\ell_1=0\, ~~~~~~{\rm for}~~A=A^{\it l}\, .
\end{aligned}
\eea Both equations can be resolved in term of a single free
parameter $z$ so that we find two Lax formulations of the
equations of motion of the principal chiral model \bea
\begin{aligned}
L_{\a}^{\it r}&=-\frac{z^2}{1-z^2}\, A_{\a}^{\it
r}+\frac{z}{1-z^2}\,
\gamma_{\a\b}\epsilon^{\beta\rho}A_{\rho}^{\it
r}\, ,\\
L_{\a}^{\it l}&=\frac{z^2}{1-z^2}\, A_{\a}^{\it
l}+\frac{z}{1-z^2}\,
\gamma_{\a\b}\epsilon^{\beta\rho}A_{\rho}^{\it l}\, .
\end{aligned}
\eea The parameter $z$ plays now the role of the spectral
parameter of the model. Both Lax connections exhibit first order
poles at $z=\pm 1$. These poles play a very special role. As we
will see later, expanding the trace of the monodromy matrix around
these poles leads to an explicit construction of ${\it local}$
conserved charges. We also note that the leading term in the
expansion of the $L$'s around $z=\infty$ ($z=0$) coincides with
the Noether current (the Hodge dual of the Noether current)
corresponding to right or left global symmetries of the model.
Finally, we remark that the connections $L^{\it r}$ and $L^{\it
l}$ are related by the gauge transformation
$$
L^{\it l}=hL^{\it r}h^{-1}+{\rm d}hh^{-1}\,
$$
with $h=\ag$, which means that they are essentially equivalent.
\smallskip

Now we turn our attention to the construction of the Lax
representation for our string sigma model.

\subsection{Lax pair}
\label{sec:Lax} To build up the zero curvature representation of
the string equations of motion, we start with the following ansatz
for the Lax connection ${L}_{\a}$
 \bea
 \label{wL}
{L}_{\a}=\ell_0 \A_{\a}^{(0)}+\ell_1 \A_{\a}^{(2)}
+\ell_2\gamma_{\a\b}\epsilon^{\beta\rho}\A_{\rho}^{(2)} +\ell_3
\A_{\a}^{(1)}+\ell_4 \A_{\a}^{(3)}\, , \eea where $\ell_i$ are
undetermined constants and $A^{(k)}$ are the
$\mathbb{Z}_4$-components of the flat connection (\ref{la}). The
connection $L_{\a}$ is required to have zero curvature (\ref{zcr})
as a consequence of the dynamical equations (\ref{seom}) and the
flatness of $\A_{\a}$. This requirement will impose certain
constraints on $\ell_i$, much similar to the case of the principal
chiral model discussed in the previous section.

\smallskip

Computing the curvature of $L_\a$, we expand the resulting
expression into the sum of the homogeneous components
$\mathscr{G}^{(k)}$ under the $\mathbb{Z}_4$-grading. First, the
projection on $\mathscr{G}^{(0)}$ reads
$$
2\ell_0
\epsilon^{\alpha\beta}\pa_{\a}\A_{\beta}^{(0)}-\epsilon^{\alpha\beta}\big(\ell_0^2[\A_{\a}^{(0)},\A_{\beta}^{(0)}]
+(\ell_1^2-\ell_2^2)[\A_{\a}^{(2)},\A_{\beta}^{(2)}]+2\ell_3\ell_4[\A_{\a}^{(1)},\A_{\beta}^{(3)}]\big)=0\,
.
$$
The flatness of $\A^{(0)}$ then implies \bea \label{cc1}
\ell_0=1\, ,~~~~~\ell_1^2-\ell_2^2=1\, , ~~~~~\ell_3\ell_4=1\, .
\eea Second, for the projection on $\mathscr{G}^{(2)}$ we find
\bea\nonumber &&\ell_1
\epsilon^{\alpha\beta}\pa_{\a}\A_{\beta}^{(2)}
+\ell_2\pa_{\a}\big(\gamma^{\a\beta}\A_{\beta}^{(2)}\big)\\
\nonumber
&&~~~~~-(\epsilon^{\alpha\beta}\ell_0\ell_1+\gamma^{\alpha\beta}\ell_0\ell_2)[\A_{\a}^{(0)},\A_{\beta}^{(2)}]
-\frac{1}{2}\epsilon^{\alpha\beta}\ell_3^2[\A_{\a}^{(1)},\A_{\beta}^{(1)}]
-\frac{1}{2}\epsilon^{\alpha\beta}\ell_4^2[\A_{\a}^{(3)},\A_{\beta}^{(3)}]=0\,
. \eea Using the flatness condition for $\A^{(2)}$, we can bring
this equation to the form
$$
\pa_{\a}\big(\gamma^{\a\beta}\A_{\beta}^{(2)}\big)-\gamma^{\alpha\beta}[\A_{\a}^{(0)},\A_{\beta}^{(2)}]
-\epsilon^{\alpha\beta}\frac{(\ell_3^2-\ell_1)}{2\ell_2}[\A_{\a}^{(1)},\A_{\beta}^{(1)}]
-\epsilon^{\alpha\beta}\frac{(\ell_4^2-\ell_1)}{2\ell_2}[\A_{\a}^{(3)},\A_{\beta}^{(3)}]=0\,
.
$$
The last expression coincides with the string equations of motion
(\ref{Eqb}) provided \bea \label{cc2}
\frac{\ell_3^2-\ell_1}{\ell_2}=-\kappa\, ,~~~~~~~
\frac{\ell_4^2-\ell_1}{\ell_2}=\kappa\, .\eea Third, projections
on $\mathscr{G}^{(1)}$ and $\mathscr{G}^{(3)}$ are \bea \nonumber
&& \ell_3\epsilon^{\alpha\beta}\pa_{\a}\A_{\beta}^{(1)}
-\epsilon^{\alpha\beta}\ell_0\ell_3[\A_{\a}^{(0)},\A_{\beta}^{(1)}]
-\epsilon^{\alpha\beta}\ell_1\ell_4[\A_{\a}^{(2)},\A_{\beta}^{(3)}]
+\gamma^{\a\beta}\ell_2\ell_4[\A_{\a}^{(2)},\A_{\beta}^{(3)}]=0\, ,\\
\nonumber && \ell_4\epsilon^{\alpha\beta}\pa_{\a}\A_{\beta}^{(3)}
-\epsilon^{\alpha\beta}\ell_0\ell_4[\A_{\a}^{(0)},\A_{\beta}^{(3)}]
-\epsilon^{\alpha\beta}\ell_1\ell_3[\A_{\a}^{(2)},\A_{\beta}^{(1)}]
+\gamma^{\a\beta}\ell_2\ell_3[\A_{\a}^{(2)},\A_{\beta}^{(1)}]=0\,
. \eea Once again, by invoking the flatness of $\A^{(1,3)}$, we
can rewrite these equations as follows \bea \nonumber
\big(\gamma^{\a\beta}-\frac{\ell_1\ell_4-\ell_3}{\ell_2\ell_4}
\epsilon^{\alpha\beta}\big)[\A_{\a}^{(2)},\A_{\beta}^{(3)}]=0\, ,\\
\nonumber
\big(\gamma^{\a\beta}+\frac{\ell_4-\ell_1\ell_3}{\ell_2\ell_3}
\epsilon^{\alpha\beta}\big)[\A_{\a}^{(2)},\A_{\beta}^{(1)}]=0\, .
\eea These will coincide with the string equations (\ref{Eqf})
provided \bea\label{cc3}
\frac{\ell_1\ell_4-\ell_3}{\ell_2\ell_4}=\kappa\, , ~~~~~~~~
\frac{\ell_4-\ell_1\ell_3}{\ell_2\ell_3}=\kappa\, .\eea Summing up
eqs.(\ref{cc2}), we find \bea \label{con1}
2\ell_1=\ell_3^2+\ell_4^2\, . \eea The same condition follows from
eqs.(\ref{cc3}) upon taking into account that $\ell_3\ell_4=1$.
Furthermore, it appears that the parameter $\kappa$ gets fixed up
to the sign. Indeed, multiplying eqs.(\ref{cc3}) and using
eqs.(\ref{cc1}), (\ref{con1}), we get \bea \label{con2}
\kappa^2=1\, , \eea which is precisely the condition for having
$\kappa$-symmetry! Thus, we have obtained a striking result:
Integrability of the equations of motion for the Lagrangian
(\ref{sLag}), {\it i.e.} existence of the corresponding Lax
connection, implies that the model possesses $\kappa$-symmetry.

\smallskip

Proceeding, we uniformize the parameters $\ell_i$ in terms of a
single variable $z$ taking values on the Riemann sphere:
\bea\label{fp1} \hspace{-0.5cm}\ell_0=1\, ,~~~~
\ell_1=\frac{1}{2}\Big(z^2+\frac{1}{z^2}\Big)\, , ~~~~
\ell_2=-\frac{1}{2\kappa}\Big(z^2-\frac{1}{z^2}\Big)\, ,~~~~
\ell_3=z\, , ~~~~\ell_4=\frac{1}{z}\, . \eea The reader can easily
verify that these $\ell_i$ solve all the constraints imposed by
the zero curvature for $L_{\a}$. For a given $\kappa=\pm 1$, there
is a unique Lax connection which is a meromorphic matrix-valued
function on the Riemann sphere.

\smallskip

Finally, we point out how the grading map $\Omega$ acts on the Lax
connection $L_{\a}$. Since $\Omega$ is an automorphism of
$\sls(4|4)$, the curvature of $\Omega(L_{\a})$ also vanishes. It
can be easily checked that $\Omega(L_{\a})$ is related to $L_{\a}$
by a certain diffeomorphism of the spectral parameter, namely,
$$
\Omega(L_{\a}(z))=L_{\a}(iz)\, ,
$$
{\it i.e.} $z$ undergoes a rotation by the angle $\pi/2$. Using
the explicit form of $\Omega$, we can write the last relation as
\bea\label{symtran} {\cal K}L_{\a}^{st}(z){\cal
K}^{-1}=-L_{\a}(iz)\, . \eea Since  $z$ is complex, the Lax
connection takes values in $\sls(4|4)$ rather than in
$\su(2,2|4)$. Obviously, the action of $\Omega$ on $L_{\a}$ is
compatible with the fact that $\Omega$  is the forth order
automorphism of $\sls(4|4)$.

\smallskip

Finally, we mention the action of the parity transformation
(\ref{cp}) on the Lax connection. Under $\sigma\to-\sigma$ we have
$A_{\tau}\to A_{\tau}$ and $A_{\sigma}\to -A_{\sigma}$. Thus,
$$
\mathscr{P}L_{\tau}(z)|_{\sigma\to
-\sigma}\mathscr{P}^{-1}=L_{\tau}^{\mathscr{P}}(1/z)\, ,
~~~~\mathscr{P}L_{\sigma}(z)|_{\sigma\to
-\sigma}\mathscr{P}^{-1}=-L_{\sigma}^{\mathscr{P}}(1/z)\, ,
$$
where we have taken into account that the parity transformation
exchanges\footnote{Specifying an explicit dependence of the Lax
connection on $\kappa$ as $L_{\a}(z,\kappa)$, we see that without
changing $\sigma\to -\sigma$ the connection enjoys the following
property
$\mathscr{P}L_{\a}(z,\kappa)\mathscr{P}^{-1}=L_{\a}(1/z,-\kappa)$.
} the projections $A^{(1)}$ and $A^{(3)}$. Here
$L_{\a}^{\mathscr{P}}$ is the same connection (\ref{wL}) where
$\ag(\sigma)$ in $A_{\a}=-\ag^{-1}\pa_{\a}\ag$ is replaced by
$\ag^{\mathscr{P}}(\sigma)=\ag(-\sigma)$. Obviously,
$L_{\a}^{\mathscr{P}}$ retains vanishing curvature.

\smallskip

In summary, we have shown that the string equations of motion
admit zero-curvature representation which ensures their
kinematical integrability. We have also seen that inclusion of the
Wess-Zumino term into the string Lagrangian is allowed by
integrability only for $\kappa=\pm 1$, {\it i.e.} only for those
values of $\kappa$ for which the model has $\kappa$-symmetry.

\subsection{Integrability and symmetries}
In the previous section we have shown that string equations of
motion admit the Lax representation provided the parameter
$\kappa$ in the Lagrangian takes values $\pm 1$. It is for these
values of $\kappa$ that the model exhibits the local fermionic
symmetry. In addition to the $\kappa$-symmetry, the string sigma
model has the usual reparametrization invariance. Due to these
local symmetries not all degrees of freedom appearing in the
Lagrangian (\ref{sLag}) are physical. Thus, ultimately we would
like to understand if and how integrability is inherited by the
physical subspace which is obtained by making a gauge choice and
imposing the Virasoro constraints. In this section we will make a
first step in this direction by analyzing in detail the
transformation properties of the Lax connection under the
$\kappa$-symmetry and diffeomorphism transformations. We also
indicate a relation between the Lax connection and the global
$\psu(2,2|4)$ symmetry of the model.

\medskip

We start with the analysis of the relationship between the Lax
connection and $\kappa$-symmetry. Recalling  eqs.(\ref{kappa})
which describe how the $\mathbb{Z}_4$-components of $\A$ transform
under $\kappa$-symmetry, it is straightforward to
find\footnote{For our present purposes it is enough to consider a
 variation with non-trivial $\epsilon^{(1)}$
only.} \bea \nonumber
\delta L_{\a}&=&[L_{\a},\Lambda]-\pa_{\a}\Lambda+\\
\nonumber
&+&(\ell_4-\ell_1\ell_3)[\A_{\a}^{(2)},\epsilon^{(1)}]-\ell_2\ell_3
[\epsilon_{\a\b}\gamma^{\b\delta}\A_{\delta}^{(2)},\epsilon^{(1)}]\\
\nonumber &+&
[(\ell_1-\ell_3^2)\A_{\a}^{(1)}+\ell_2\epsilon_{\a\b}\gamma^{\b\delta}\A_{\delta}^{(1)},\epsilon^{(1)}]
+\ell_2 \epsilon_{\a\b}\delta\gamma^{\b\delta}\A_{\delta}^{(2)}\,
. \eea Here $\Lambda=\ell_3 \epsilon^{(1)}$. Taking into account
the relations between the coefficients $\ell_i$ found in the
previous section from the requirement of integrability, the last
formula can be cast into the form \bea \nonumber \delta
L_{\a}&=&[L_{\a},\Lambda]-\pa_{\a}\Lambda +\ell_2\ell_3\,
\epsilon_{\a\b}[\A^{(2),\beta}_{-},\epsilon^{(1)}] +\ell_2\,
\epsilon_{\a\b}\left(2[\A^{(1),\b}_{+},\epsilon^{(1)}]+
\delta\gamma^{\b\delta}\A_{\delta}^{(2)}\right)\, . \label{fv}
\eea The $\Lambda$-dependent term here is nothing else but an
infinitesimal gauge transformation of the Lax connection. Under
this transformation, the curvature of the transformed connection
retains its zero value. On the other hand, the last two terms
proportional to $\ell_2\ell_3$ and $\ell_2$ would destroy the zero
curvature condition unless they (separately) vanish. It turns out
that vanishing of these terms is equivalent to the requirement of
the Virasoro constraints as well as equations of motion for the
fermions! Consider the first term
$$
I_1\equiv[\A_{\alpha,-}^{(2)},\epsilon^{(1)}]=[\A_{\alpha,-}^{(2)},\A^{(2)}_{\b,-}\kappa^{(1),\b}_{+}
+ \kappa^{(1),\b}_{+}\A^{(2)}_{\b,-}]=\sfrac{1}{8}{\rm str}
(\A^{(2)}_{\a,-} \A^{(2)}_{\b,-})[\Upsilon,\kappa^{(1),\b}_{+}]\,
.
$$
Here we used eq.(\ref{mf}) and also eq.(\ref{prop}) stating that
$\A_{\alpha,-}^{(2)}$ and $\A_{\beta,-}^{(2)}$ with different $\a$
and $\beta$ are proportional to each other. It is not hard to
prove that
$$
{\rm str} (\A^{(2)}_{\a,-} \A^{(2)}_{\b,-})=0
$$
implies fulfilment of the Virasoro constraint (\ref{Vir}) and vice
versa.

\smallskip

The second unwanted term involves an expression
$$
I_2\equiv[\A_{+}^{(1),\a},\epsilon^{(1)}]=[\A_{+}^{(1),\a},\A^{(2)}_{\b,-}\kappa^{(1),\b}_{+}
+
\kappa^{(1),\b}_{+}\A^{(2)}_{\b,-}]=[\A_{+}^{(1),\b},\A^{(2)}_{\b,-}\kappa^{(1),\a}_{+}
+ \kappa^{(1),\a}_{+}\A^{(2)}_{\b,-}]\, ,
$$
where we made use of the identity (\ref{projprop}). Taking into
account the equation of motion for fermions,
$[\A_{+}^{(1),\b},\A^{(2)}_{\b,-}]=0$, the last formula boils down
to
$$
I_2=\A^{(2)}_{\b,-}[\A_{+}^{(1),\b},\kappa^{(1),\a}_{+}]+[\A_{+}^{(1),\b},\kappa^{(1),\a}_{+}]
\A^{(2)}_{\b,-}\, .
$$
Since the commutator $[\A_{+}^{(1),\b},\kappa^{(1),\a}_{+}]$
belongs to the space $\mathscr{G}^{(2)}$, we can write
$$
[\A_{+}^{(1),\b},\kappa^{(1),\a}_{+}]=\left(
\begin{array}{cc} m^a\gamma^a & 0 \\
0 & n^a \gamma^a\end{array}\right)+\frac{1}{8}{\rm str
}(\Upsilon[\A_{+}^{(1),\b},\kappa^{(1),\a}_{+}])~\mI\, ,
$$
for some coefficients $m^a$ and $n^a$. Therefore,
$$
I_2=\{\A^{(2)}_{\b,-},\left(
\begin{array}{cc} m^a\gamma^a & 0 \\
0 & n^a \gamma^a \end{array}\right)\}+\frac{1}{4}{\rm str
}(\Upsilon [\A_{+}^{(1),\b},\kappa^{(1),\a}_{+}])~\A^{(2)}_{\b,-}
$$
Expanding  the element $\A^{(2)}_{\b,-}$ over the Dirac matrices
and substituting it in the anti-commutator above, we see that, due
to the Clifford algebra, $2I_2$ must have the following structure
\bea 2I_2=\rho_1\, \mI+\rho_2\, \Upsilon-\frac{1}{2}{\rm str
}(\Upsilon[\kappa^{(1),\a}_{+},\A_{+}^{(1),\b}])~\A^{(2)}_{\b,-}\,
. \label{I2s}\eea Actually, the coefficient $\rho_2$ must vanish
as the supertrace of $I_2$ equals zero (this follows from the
original definition of $I_2$ as a commutator of two terms). The
last term proportional to $\A^{(2)}_{\b,-}$ will then cancel in
eq.(\ref{fv}) the term containing the variation of the metric
$\delta\gamma^{\a\b}\A_{\b}^{(2)}=\delta\gamma^{\a\b}\A_{\b,-}^{(2)}$.
Finally, the term in eq.(\ref{I2s}) proportional to the identity
matrix is unessential because the Lax representation is understood
modulo an element $i\mI$.

\medskip

Thus, we have obtained the following important result. Although
the Virasoro constraints (\ref{Vir}) do not apparently follow from
the zero curvature condition, we see that upon $\kappa$-symmetry
transformations the Lax connection retains its zero curvature  if
and only if the Virasoro constraints (and equations of motion for
fermions) are satisfied. This shows that the local symmetries of
the model and the existence of the Lax connection are tightly
related to each other.

\medskip
Let us now show that diffeomorphisms also induce gauge
transformations of the Lax connection. Indeed, under a
diffeomorphism $\sigma^{\a}\to \sigma^{\a}+f^{\a}(\sigma)$ any
one-form and, in particular, the Lax connection transforms as
follows \bea \delta
L_{\a}=f^{\beta}\pa_{\beta}L_{\a}+L_{\beta}\pa_{\a}f^{\beta}\, .
\eea Using the zero-curvature condition for $L_{\a}$, we can
rewrite the last formula as \bea \delta
L_{\a}=f^{\beta}\big(\pa_{\a}L_{\b}+[L_{\b},L_{\a}]\big)+L_{\beta}\pa_{\a}f^{\beta}\,=
\pa_{\a}(f^{\b}L_{\b})+[f^{\b}L_{\b},L_{\a}] , \eea which is a
gauge transformation with a parameter $f^{\b}L_{\b}$.

\smallskip

Now we explain the interrelation between the Lax connection and
the generators of the global $\psu(2,2|4)$ symmetry.  So far our
discussion of the Lax connection was based on the one-form
$A=-{\rm d}\ag \ag^{-1}$ which, as the reader undoubtedly noticed,
is analogous to the right connection of the principal chiral
model. At $z=1$ the Lax connection (\ref{fp1}) turns into
$\A_{\a}$. As we have already mentioned, the condition of zero
curvature (\ref{zcr}) is invariant with respect to the gauge
transformations
$$L\to L'=h L h^{-1}+{\rm d}hh^{-1}\, .$$
The inhomogeneous term on the right hand side does not depend on
$z$ and, therefore, this gauge freedom can be used to gauge away
the constant piece $A$ arising at $z=1$. For this one has to take
$h=\ag$. Indeed, define $a^{({i})}=\ag\A^{(i)}\ag^{-1}$ and
represent the "dual" one-form $\tilde{\A}=-{\rm d}\ag\ag^{-1}$ in
the following way
 \bea
 \nonumber
\tilde{\A}=\ag\A
\ag^{-1}=\ag(\A^{(0)}+\A^{(1)}+\A^{(2)}+\A^{(3)})\ag^{-1}=a^{(0)}+
a^{(1)}+a^{(2)}+a^{(3)}\, . \eea Then  the result of the gauge
transformation of $L$ takes the form \bea \label{newLax}
L_{\a}=\ell_0 a_{\a}^{(0)}+\ell_1 a_{\a}^{(2)}
+\ell_2\gamma_{\a\b}\epsilon^{\beta\rho}a_{\rho}^{(2)} +\ell_3
a^{(1)}_{\a}+\ell_4 a^{(3)}_{\a}\, , \eea where $\ell_0=0$ and the
other coefficients $\ell_i$ are given by \bea \nonumber
\ell_1=\frac{(1-z^2)^2}{2z^2}\, , ~~~~
\ell_2&=&-\frac{1}{2\kappa}\Big(z^2-\frac{1}{z^2}\Big)\, ,
~~~~\ell_3=z-1\, ,~~~~\ell_4=\frac{1}{z}-1 \, . \eea Expanding
this connection around $w=1-z$  \bea \label{zero}
L_{\a}=\frac{2w}{\kappa} \, {\cal L}_{\a} +\ldots \, ,\eea we
discover that the leading term ${\cal L}_{\a}$ is \bea\nonumber
 {\cal
L}_{\a}=\gamma_{\a\b}\epsilon^{\beta\rho}a_{\rho}^{(2)}-\frac{\kappa}{2}(a^{(1)}_{\a}-a^{(3)}_{\a})
\, . \eea The zero-curvature condition is satisfied at each order
in $w$;  at first order in $w$ it gives
 \bea
 \nonumber
\pa_{\a}{\cal L}_{\beta}-\pa_{\beta}{\cal L}_{\a}=0 \,
~~~~\Longrightarrow~~~~ \pa_{\a}\big(\epsilon^{\a\beta}{\cal
L}_{\beta}\big)=0\, ,  \eea which is obviously the conservation
law for a non-abelian current
 \bea \label{cc}
J^{\a}=g\, \epsilon^{\a\b}{\cal
L}_{\beta}=g\Big[\gamma^{\a\b}a_{\b}^{(2)}-\frac{\kappa}{2}\epsilon^{\a\b}(a^{(1)}_{\b}-a^{(3)}_{\b})\,
\Big]. \eea Comparing the last expression with eqs.(\ref{cS}),
(\ref{gcurrent}), we conclude that $J^{\a}$
 is just the Noether current corresponding to
the global $\psu(2,2|4)$ symmetry of the model. The dual one-form
$\tilde{A}$ is an analogue of the left connection of the principal
chiral model.

\smallskip
One can analyze the expansion of the Lax connection around $z=-1$
in a similar fashion. Expanded around $z=-1$, the connection
exhibits a constant piece which can be gauged away by a proper
gauge transformation. After this is done, at order $(z+1)$ one
finds a non-abelian conserved current, which is again the Noether
current generating the global $\psu(2,2|4)$-symmetry.

\section{Coset parametrizations}
\label{cpgs} This section is devoted to the discussion of various
embeddings of the coset space (\ref{sAdS}) into the supergroup
${\rm SU}(2,2|4)$. We put an emphasis on a particular embedding
which is most suitable for the light-cone gauge fixing. We also
identify a bosonic subalgebra of the full symmetry algebra which
acts linearly on the coordinates of the coset space.

\subsection{Coset parametrization}
 To give an explicit expression for the Lagrangian
density (\ref{sLag}) in terms of the coset degrees of freedom
varying over the two-dimensional world-sheet, it is necessary to
choose an embedding of the coset element (\ref{sAdS}) into the
supergroup ${\rm SU}(2,2|4)$. This can be done in many different
ways, all of them are related by non-linear field redefinitions.
Before we motivate our preferred coset parametrization, we need to
describe the space $\mathscr{G}^{(2)}$ constituting the orthogonal
complement of $\alg{so}(4,1)\oplus \alg{so}(5)$ in the bosonic
subalgebra of $\psu(2,2|4)$.

\smallskip

The space $\mathscr{G}^{(2)}\subset \mathscr{G}=\psu(2,2|4)$ is
spanned by solutions to the following equation
$$
{\cal K}M^{st}{\cal K}^{-1}=M\, ,
$$
which for $M$ even is equivalent to
$$
M^t={\cal K}M{\cal K}^{-1}\, .
$$
According to the discussion of section \ref{Z4grad}, the matrices
$M=M^{(2)}$ solving the equation above can be parametrized as \bea
\label{LieAdS} M=\frac{1}{2}\left(\begin{array}{cc}
it\gamma^5+z^i\gamma^i & 0 \\
0  & i \phi \gamma^5 + i y^i \gamma^i
\end{array} \right)\, ,
\eea where the summation index $i$ runs from $1$ to $4$. As will
be explained in appendix \ref{App:emcoord}, the coordinates
$t,z^i$ cover the ${\rm AdS}_5$ space, while $\phi,y^i$ cover the
five-sphere. In particular, $\phi$ parametrizes a big circle in
${\rm S}^5$ and it has range $0\leq \phi < 2\pi$. More generally,
since we deal with closed strings, the global coordinates on the
sphere must be periodic functions of $\sigma$. On the other hand,
$\phi$ is an angle and, therefore, one can have configurations
with a non-trivial winding \bea \la{winding}
\phi(2\pi)-\phi(0)=2\pi m \, , \eea where $m$ is an integer. All
the coordinates are assumed to be periodic functions of $\sigma$
(we do not allow winding in the time direction).

\smallskip

One obvious way to define an embedding of the ${\AdS}$ space into
the bosonic subgroup of ${\rm SU}(2,2|4)$ is just to exponentiate
an element (\ref{LieAdS}):
$$
\ag_{\alg{b}}=\exp\frac{1}{2}\left(\begin{array}{cc}
it\gamma^5+z^i\gamma^i & 0 \\
0  & i \phi \gamma^5 + i y^i \gamma^i
\end{array} \right)\, .
$$
The fermionic degrees of freedom can be incorporated in the
following group element \bea \ag_{\alg{f}}=\exp\chi\, , ~~~~~
\chi=\left(\begin{array}{cc} 0 ~&~
\Theta \\
-\Theta^{\dagger}\Sigma & 0
\end{array} \right)\, .
\label{chi} \eea A group element describing an embedding of the
coset space (\ref{sAdS}) into ${\rm SU}(2,2|4)$ can be then
constructed as \bea\label{fb} \ag=\ag_{\alg f}\ag_{\alg{b}}\, .
\eea Clearly, this is just one of infinitely many ways to choose a
coset representative; for instance, one could also define
$\ag=\ag_{\alg{b}}\ag_{\alg f}$.

\smallskip

It appears, however, that the choice (\ref{fb}) is particularly
convenient to manifest the global bosonic symmetries of the model,
because the latter act linearly on fermionic variables. Indeed,
the symmetry group acts on a coset element by multiplication from
the left, see eqs.(\ref{actionG}) and (\ref{actionG1}). If a coset
element is realized as in eq.(\ref{fb}), then the action of $G\in
{\rm SU(2,2)}\times {\rm SU}(4)$ preserves the structure of the
fermionic coset representative:
$$
G\cdot \ag=G\ag_{\alg{f}}G^{-1}\cdot
G\ag_{\alg{b}}=G\ag_{\alg{f}}G^{-1}\cdot \ag_{\alg{b}}'\alg{h}\, ,
$$
where $\alg{h}$ is a compensating element from ${\rm
SO(4,1)}\times {\rm SO}(5)$. From here we deduce that the matrix
$\ag_{\alg{f}}$ transforms as
$$
\ag_{\alg{f}}\to G\, \ag_{\alg{f}}\, G^{-1}=\exp G\chi G^{-1}\, .
$$
Thus, fermions undergo the adjoint (linear) action of $G$, while
bosons generically transform in a non-linear fashion:
$\ag_{\alg{b}}\to \ag_{\alg{b}}'$. In particular, fermions are
charged under {\it all} Cartan generators of $\psu(2,2|4)$, the
latter represents a set of commuting $\alg{u}(1)$-isometries of
the coset space (\ref{sAdS}).

\smallskip
Another reason to choose a coset representative (\ref{fb}) is that
in this case supersymmetry transformations act  on the fermionic
and bosonic variables in a simple way. Indeed, under an
infinitezimal supersymmetry transformation with a fermionic
parameter $\epsilon$ the coset variables undergo the following
transformation \bea \delta_{\epsilon}\chi=\epsilon\, , ~~~~~~~
\delta_{\epsilon}\ag_{\alg{b}}=\frac{1}{2}[\epsilon,\chi]\ag_{\alg{b}}-\ag_{\alg{b}}\,\alg{h}\,
,
 \eea
where $\alg{h}$ is a compensating element from ${\rm
SO(4,1)}\times {\rm SO}(5)$. The last formula makes it obvious
that invariance of the model under supersymmetry transformations
requires fermionic variables in the representation (\ref{fb}) to
be periodic functions of $\sigma$.

\smallskip

As will be discussed in the next chapter, fixing the light-cone
gauge is greatly facilitated by working with fermions which are
{\it neutral} under the isometries corresponding to shifts of the
AdS time $t$ and the sphere angle $\phi$. By the above, fermions
of the coset element (\ref{fb}) do not meet this requirement. The
idea is, therefore, to redefine the original fermionic variables
in such a fashion that they become neutral under the isometries
related to $t$ and $\phi$. This can be understood in the following
way. Introduce a diagonal matrix \bea \label{Lambda}
\Lambda(t,\phi)=\exp\left(\begin{array}{cc}
\sfrac{i}{2}t\gamma^5 & 0 \\
0  & \sfrac{i}{2} \phi \gamma^5
\end{array} \right)\,
\eea with the property
$\Lambda(t_1+t_2,\phi_1+\phi_2)=\Lambda(t_1,\phi_1)\Lambda(t_2,\phi_2)$,
and  the following exponential \bea \label{X} \ag(\mathbb{X})=\exp
\mathbb{X}\, , ~~~~~\mathbb{X}=\left(\begin{array}{cc}
\sfrac{1}{2}z^i\gamma^i & 0 \\
0  &  \sfrac{i}{2} y^i \gamma^i
\end{array} \right)\, .
\eea  Now, instead of (\ref{fb}), consider a new parametrization
of the coset representative \bea \label{basiccoset}
\ag=\Lambda(t,\phi)\, \ag(\chi)\, \ag(\mathbb{X})\, , \eea where
$\ag(\chi)\equiv\ag_{\alg{f}}$. Obviously, an element $G$
corresponding to global shifts $t\to t+a$, $\phi\to \phi+b$ can be
identified with $\Lambda(a,b)$. Thus, under the left
multiplication \bea G\cdot \ag=\Lambda(a,b )\Lambda(t,\phi)\,
\ag(\chi)\, \ag(\mathbb{X})=\Lambda(t+a ,\phi+b)\, \ag(\chi)\,
\ag(\mathbb{X})\, , \label{basiccoset1}\eea {\it i.e.} both $\chi$
and $\mathbb{X}$ remain untouched by this transformation. In other
words, with our new choice (\ref{basiccoset}),  not only the
fermions $\chi$ but also all the remaining  eight bosons $z^i$ and
$y^i$, appear to be neutral under the isometries related to $t$
and $\phi$\,! This property motivates our  choice
(\ref{basiccoset}). In fact, coset representatives (\ref{fb}) and
(\ref{basiccoset}) are related to each other by a non-linear field
redefinition, which for fermionic variables is of the form
$\chi\to \Lambda(t,\phi)\chi\Lambda(t,\phi)^{-1}$.

\smallskip

It should be noted, however, that non-linear field redefinitions
can change the boundary conditions for the world-sheet fields. In
parametrization (\ref{fb}) fermions $\chi$ transform linearly
under all bosonic symmetries and they are periodic functions of
$\sigma$. To pass to parametrization (\ref{basiccoset}), we
redefine \bea \la{ferred} \chi\to \chi'=\Lambda^{-1}\chi\Lambda
~~~~\Longrightarrow~~~~\Theta\to
\Theta'=e^{\frac{i}{2}(\phi-t)\gamma_5}\Theta\, ,
 \eea
where we have invoked the parametrization (\ref{chi}). As a
result, the new fermions satisfy the following boundary conditions
\bea \Theta'(\sigma+2\pi)=e^{i\pi m\gamma_5}\Theta'(\sigma)\, ,
\eea {\it i.e.} they remain periodic for $m$ even (the even
winding number sector) and they become anti-periodic for $m$ odd
(the odd winding number sector).

\smallskip

We conclude our discussion of coset representatives by emphasizing
that given the structure (\ref{basiccoset}), one is entirely free
to choose parametrizations for $\alg{g}(\chi)$ and
$\alg{g}(\mathbb{X})$ different from those in eqs.(\ref{chi}) and
(\ref{X}). In particular, in appendix \ref{Subsect:altLag} we
describe another useful choice for the element $\ag(\mathbb{X})$.

\subsection{Linearly realized bosonic symmetries}
\label{Sect:lrbs} Well adjusted to the light-cone gauge,
parametrization (\ref{basiccoset}) does not allow for a linear
realization of all the bosonic symmetries. Our next task is,
therefore, to determine a maximal subgroup of the bosonic symmetry
group which acts linearly on the dynamical fields $\mathbb{X}$ and
$\chi$. This subgroup will then coincide with the manifest bosonic
symmetry of the light-cone gauge-fixed string Lagrangian.

\smallskip

It is easy to see that the centralizer of the
$\alg{u}(1)$-isometries  corresponding to shifts of $t$ and $\phi$
in the algebra $\su(2,2)\oplus \su(4)$ coincides  with
\bea\label{linearsym} \alg{C}=\alg{so}(4)\oplus
\alg{so}(4)=\su(2)\oplus \su(2)\oplus \su(2)\oplus \su(2)\, , \eea
where the first factor is $
\alg{so}(4)\subset\alg{so}(4,1)\subset\alg{so}(4,2)$ and the
second one $\alg{so}(4)\subset \alg{so}(5)\subset\alg{so}(6)$.
Indeed, both copies of $\alg{so}(4)$ are generated by
$\sfrac{1}{4}[\gamma^i,\gamma^j]$, $i,j=1,\ldots,4$ because the
latter matrices commute with $i\gamma^5$ generating shifts in the
$t$- or $\phi$-directions. Let now $G$ be a group element
corresponding to a Lie algebra element from (\ref{linearsym}).
Then $G\, \Lambda(t,\phi)\, G^{-1}=\Lambda(t,\phi)$. Due this
condition, one gets
$$
G\cdot \ag=\Lambda(t,\phi)\cdot G\ag(\chi)G^{-1}\cdot
G\ag(\mathbb{X})G^{-1}\cdot G \, .
$$
Now one can recognize that the last  $G$ in the right hand side of
this formula is nothing else but the compensating element
$\alg{h}$ from ${\rm SO(4,1)}\times {\rm SO}(5)$: $\alg{h}=G$.
Indeed, the adjoint transformation with $G$ preserves the
structure of the coset element $\ag(\mathbb{X})$, because the
generator $\sfrac{1}{4}[\gamma^i,\gamma^j]$ commutes with
$\gamma^k$ for $j\neq k\neq i$ and is equal to $2\gamma^j$ for
$k=j$. Thus, under the action of $G$ both bosons and fermions
undergo a linear transformation
$$
\chi\to \chi'=G\, \chi\,  G^{-1}\, ,~~~~~ \mathbb{X}\to
\mathbb{X}'=G\,\mathbb{X}\, G^{-1}\, .
$$
To conclude, the centralizer $\alg{C}$ of the isometries related
to $t$ and $\phi$ induces linear transformations of the dynamical
variables.

\smallskip

In terms of 2 by 2 blocks a matrix $G$ from the centralizer can be
represented as follows \bea \la{gec} G={\small \left(
\begin{array}{cccc} \ag_1 & 0
& 0 & 0 \\
0 & \ag_2 & 0 & 0  \\
0 & 0 & \ag_3 & 0 \\
 0 & 0 & 0 & \ag_4
\end{array}\right)}\, .
\eea Here $\ag_1,\ldots,\ag_4$ denote four independent copies of
${\rm SU}(2)$. Analogously, the elements $\mathbb{X}$ and $\chi$
can be represented as \bea \label{block}
\mathbb{X}={\small\left( \begin{array}{cccc} 0 & Z & 0 & 0 \\
Z^{\dagger} & 0 & 0 & 0  \\
 0 & 0 & 0 & i Y \\
 0 & 0 & i Y^{\dagger} & 0
\end{array}\right)}\, ,~~~~~~
\chi={\footnotesize \left( \begin{array}{cccc} 0 & 0 & \Theta_1 & \Theta_2 \\
0 & 0 & \Theta_3^\dagger & \Theta_4  \\
-\Theta_1^{\dagger} & \Theta_3 & 0 & 0 \\
-\Theta_2^{\dagger} & \Theta_4^{\dagger} & 0 & 0
\end{array}\right)}\,
. \eea Here $Z$ and $Y$ are two $2\times 2$ matrices which
incorporate eight bosonic degrees of freedom \bea
Z=\frac{1}{2}\left(\begin{array}{rr} z_3-i z_4 & -z_1+iz_2
\\ z_1+iz_2 & z_3+i z_4\end{array}\right)\, , ~~~~
Y=\frac{1}{2}\left(\begin{array}{rr} y_3-iy_4 & -y_1+iy_2 \\
y_1+iy_2 & y_3+iy_4
\end{array}\right)\, , \eea
while four $2\times 2$ blocks $\Theta_1,\ldots,\Theta_4$ comprise
16 complex fermions. Matrices $Z$ and $Y$ satisfy the following
reality condition \bea\la{conjrule} Z^\dagger=\eps\, Z^t
\eps^{-1}\, , ~~~~~Y^\dagger=\eps\, Y^t \, \eps^{-1}\, ,
~~~~\eps\equiv i\sigma_2\, , \eea where $\sigma_2$ is the Pauli
matrix.

\smallskip

Thus, we deduce that under the action of $G$: \bea \la{lawboson}
\mathbb{X}\to G\,\mathbb{X}\,G^{-1}= {\small \left( \begin{array}{cccc} 0 & \ag_1 Z \ag_2^{-1} & 0 & 0 \\
\ag_2Z^{\dagger}\ag_1^{-1} & 0 & 0 & 0  \\
 0 & 0 & 0 & i\,\ag_3Y\ag_4^{-1} \\
 0 & 0 & i\, \ag_4Y^{\dagger}\ag_3^{-1} & 0
\end{array}\right)}\,
\eea and
 \bea
 \label{lawfermion}
 \chi\to G\,\chi\, G^{-1}=
{\small \left( \begin{array}{cccc} 0 & 0 & \ag_1\Theta_1\ag_3^{-1} & \ag_1\Theta_2\ag_4^{-1} \\
0 & 0 & \ag_2\Theta_3^{\dagger}\ag_3^{-1} & \ag_2 \Theta_4\ag_4^{-1}  \\
-\ag_3\Theta_1^{\dagger}\ag_1^{-1} & \ag_3\Theta_3\ag_2^{-1} & 0 & 0 \\
-\ag_4\Theta_2^{\dagger}\ag_1^{-1} &
\ag_4\Theta_4^{\dagger}\ag_2^{-1} & 0 & 0
\end{array}\right)}\,
. \eea Before we proceed with discussing these symmetry
transformations, let us note that, according to
eq.(\ref{kappagauge}), one can implement the $\kappa$-symmetry
gauge by requiring the absence of fermionic blocks
$\Theta_1,\Theta_1^{\dagger}$ and $\Theta_4,\Theta_4^{\dagger}$.
As we now see, under the action of $\alg{C}$ the block structure
(\ref{block}) is preserved and, therefore, it is indeed consistent
to put $\Theta_1,\Theta_1^{\dagger}$ and
$\Theta_4,\Theta_4^{\dagger}$ to zero. In what follows we will
assume this gauge choice for the odd part of the coset
representative.

\smallskip

Now we will introduce a convenient two-index notation which allows
us to naturally treat the dynamical variables of the model as
transforming in bi-fundamental representations of $\su(2)$. To
this end, we recall that any SU(2)-matrix $\ag$ obeys the
following property \bea\ag^{\dagger}=\ag^{-1}=\eps\, \ag^t
\eps^{-1}\, ~~\Longrightarrow~~ \ag^*=\eps\, \ag\, \eps^{-1},
\la{su2conj}\eea which provides an equivalence between an irrep of
${\rm SU}(2)$ and its complex conjugate.

\smallskip

Consider, {\it e.g.}, matrix $Z$ and multiply it by $\eps$ from
the right. According to eq.(\ref{lawboson}), $Z\eps$ transforms
under $\alg{C}$ as follows \bea \la{su(2)law} Z\eps\to \ag_1
Z\ag_2^{-1}\eps= \ag_1 Z\eps\, \ag_2^t\, . \eea If we now
associate the index $\a=3,4$ to the fundamental irrep of $\ag_1$
and the index $\dot{\a}=\dot{3},\dot{4}$ to the fundamental irrep
of $\ag_2$ , then $Z\eps$ can be regarded as the matrix with
entries $Z^{\a\dot{\a}}$: \bea
Z\eps=\left(\begin{array}{rr} Z^{3\dot{3}} & Z^{3\dot{4}}\\
Z^{4\dot{3}} & Z^{4\dot{4}}
\end{array}\right)\, .\eea
Then formula (\ref{su(2)law}) written in components takes the form
$$
Z'^{\a\dot{\a}}=\ag^{\a}_{~\beta}\ag^{\dot{\a}}_{~\dot{\b}}Z^{\b\dot{\b}}\,
,
$$
which shows that $Z\eps$ transforms in the bi-fundamental representation
of $\su(2)$. The matrix $Z$ itself is
expressed via the entries of $Z\eps$ as \bea
Z=\left(\begin{array}{rr} Z^{3\dot{4}} & -Z^{3\dot{3}}\\
Z^{4\dot{4}} & -Z^{4\dot{3}}
\end{array}\right)\, . \eea
Analogously, we associate the indices $a=1,2$ and
$\dot{a}=\dot{1},\dot{2}$
 with the third and the fourth copies of
$\su(2)$ in eq.(\ref{linearsym}), respectively. Then,
parametrization (\ref{block}) of the bosonic Lie algebra element
$\mathbb{X}$ in terms of the entries $Z^{\a\dot{\a}}$ and
$Y^{a\dot{a}}$ reads as follows
 \bea\la{Xlc}
\mathbb{X}=\left(\begin{array}{cccc|cccc} 0 & 0 & Z^{3\dot{4}}
& -Z^{3\dot{3}} & 0 & 0 & 0 & 0 \\
0 & 0 & Z^{4\dot{4}} & -Z^{4\dot{3}} & 0 & 0 & 0 & 0 \\
-Z^{4\dot{3}} & Z^{3\dot{3}} & 0 & 0 & 0 & 0 & 0 & 0 \\
-Z^{4\dot{4}} & Z^{3\dot{4}} & 0 & 0 & 0 & 0 & 0 & 0 \\
\hline
0 & 0 & 0 & 0 & 0 & 0 & iY^{1\dot{2}} & -iY^{1\dot{1}}\\
0 & 0 & 0 & 0 & 0 & 0 & iY^{2\dot{2}} & -iY^{2\dot{1}} \\
0 & 0 & 0 & 0 & -iY^{2\dot{1}} & iY^{1\dot{1}} & 0 & 0 \\
0 & 0 & 0 & 0 & -iY^{2\dot{2}} & iY^{1\dot{2}} & 0 & 0
\end{array}\right)\, .
\eea To obtain this formula, we have replaced the matrices
$Z^{\dagger}$ and $Y^{\dagger}$ in eq.(\ref{block}) via $Z$ and
$Y$ by using the reality condition (\ref{conjrule}). In a similar
fashion we deduce the following parametrization of the fermionic
Lie algebra element $\chi$:
 \bea\la{chilc}
\chi={\small \left(\begin{array}{cccc|cccc} 0 & 0 & 0 & 0 & 0 & 0
&
\eta^{3\dot{2}} & -\eta^{3\dot{1}}\\
 0 & 0 & 0 & 0 & 0 & 0 & \eta^{4\dot{2}}
& -\eta^{4\dot{1}} \\
 0 & 0 & 0 & 0 & ~~\theta^\dagger_{1\dot{4}} & ~~\theta^\dagger_{2\dot{4}} & 0
& 0 \\
 0 & 0 & 0 & 0 & -\theta^\dagger_{1\dot{3}} & -\theta^\dagger_{2\dot{3}} & 0
& 0 \\ \hline
 0 & 0 & \theta^{1\dot{4}} & -\theta^{1\dot{3}} & 0 & 0 & 0 & 0
\\
 0 & 0 & \theta^{2\dot{4}}
& -\theta^{2\dot{3}} & 0 & 0 & 0 & 0\\
-\eta^\dagger_{3\dot{2}} & -\eta^\dagger_{4\dot{2}} & 0 & 0 & 0 &
0 & 0 & 0
\\
~~ \eta^\dagger_{3\dot{1}} &~~ \eta^\dagger_{4\dot{1}} & 0 & 0 & 0
& 0 & 0 & 0
\end{array}
\right) }\eea Here, by definition,  $\theta_{\a\da}^\dagger$ and
$\eta_{a\dal}^\dagger$ are understood as complex conjugate of
$\theta^{\a\da}$ and $\eta^{a\dal}$, respectively, \bea
(\theta^{a\a})^*\equiv\theta_{a\a}^{\dagger}\, ,~~~~~~ (\eta^{\a
a})^*\equiv \eta_{\a a}^\dagger\,  . \eea

\smallskip

In summary, we have shown that the bosonic symmetry algebra
$\alg{G}$ which commutes with an element $\Lambda(t,\phi)$
coincides with four copies of $\su(2)$. The corresponding group
acts linearly on the remaining dynamical variables. We choose to
parametrize these dynamical variables by fields
$$
Z^{\a\dot{\a}}\, , ~~~~Y^{a\dot{a}}\, ,~~~~\theta^{a\dot{\a}}\, ,
~~~~~\eta^{a\dot{\a}}\, ,
$$
which transform in the bi-fundamental representation of $\su(2)$.

\section{Appendix}
\subsection{Embedding coordinates} \la{App:emcoord}

The bosonic coset element $\ag_{\alg{b}}$ provides a
parametrization of the $\AdS$ space in terms of $5+5$
unconstrained coordinates $z^i$ and $y^i$. Sometimes it is however
more convenient to work with constrained $6+6$ coordinates which
describe the embedding of the ${\rm AdS}_5$ and the five-sphere
into ${\mathbb R}^{4,2}$ and ${\mathbb R}^6$, respectively.

\smallskip The
 embedding coordinates are defined in the following way.
 For the five-sphere we introduce six real coordinates
 $Y_A$, $A=1,\ldots, 6$ obeying the
condition $Y_A^2=1$. These coordinates are related to five
unconstrained variables $\phi,y^a$ as follows
 \bea
\begin{aligned}
\Y_1\equiv Y_1+iY_2=\frac{y_1+iy_2}{1+\frac{y^2}{4}}\, &,~~~~~
\Y_2\equiv Y_3+iY_4=\frac{y_3+iy_4}{1+\frac{y^2}{4}}\, ,~~~~~\label{paramsphere}\\
&\hspace{-2.5cm} \Y_3\equiv
Y_5+iY_6=\frac{1-\frac{y^2}{4}}{1+\frac{y^2}{4}} \exp(i\phi) \, .
\end{aligned}
\eea Here we used the shorthand notation $y^2=y^iy^i$. The metric
induced on S$^5$ from the flat metric of the embedding space is
\bea \label{metricsphere}
dY_AdY_A=\left(\frac{1-\frac{y^2}{4}}{1+\frac{y^2}{4}}
\right)^2d\phi^2\, +\frac{dy_idy_i}{(1+\frac{y^2}{4})^2}\, . \eea

\smallskip

Analogously, to describe the five-dimensional AdS space we
introduce the embedding coordinates $Z_A$. These coordinates  are
constrained to obey $\eta_{AB}Z^AZ^B=-1$ with the metric
$\eta_{AB}=(-1,1,1,1,1,-1)$ and are related to $t,z^a$ as \bea
\begin{aligned}
 \Z_1\equiv
Z_1+iZ_2=\frac{z_1+iz_2}{1-\frac{z^2}{4}}\, ,&~~~~~
\Z_2\equiv Z_3+iZ_4=\frac{z_3+iz_4}{1-\frac{z^2}{4}}\, ,~~~~~\label{paramAdS}\\
&\hspace{-2.5cm} \Z_3\equiv Z_0+iZ_5
=\frac{1+\frac{z^2}{4}}{1-\frac{z^2}{4}} \exp(it)\, ,
\end{aligned}\eea
where $z^2=z^iz^i$. For the induced metric one obtains
\bea\label{metricAdS} \eta_{AB}dZ^AdZ^B
=-\left(\frac{1+\frac{z^2}{4}}{1-\frac{z^2}{4}} \right)^2dt^2
+\frac{dz_idz_i}{(1-\frac{z^2}{4})^2} \, . \eea In the last
formula we assume $-\infty <t <\infty$, which corresponds to
considering the universal cover of the AdS space without closed
time-like curves. In what follows we do not distinguish between
the lower and upper indices for the $z$ and $y$ coordinates, that
is $z_i\equiv z^i$, $y_i\equiv y^i$. For future convenience we
combine the coordinates $z^i,y^i$ into a single vector $x^{\mu}$
with $\mu=1,\ldots, 8$ for which $x^{i}=z^{i}, x^{i+4}=y^i$.

\smallskip

Thus, the metric of the $\AdS$ space takes the following diagonal
form \bea\begin{aligned} ds^2 &= -G_{tt}\, dt^2\, +\, G_{\p\p}\,
d\p^2\, +\, G_{zz}\,dz_idz_i \, +\, G_{yy}\,dy_idy_i
\la{metricadss}
\\ &= -G_{tt}\, dt^2\, +\, G_{\p\p}\, d\p^2\, +\,
G_{\mu\mu}\,dx^\mu dx^\mu  \, , \end{aligned} \eea where
\bea\nonumber \Gtt=\left(\frac{1+\frac{z^2}{4}}{1-\frac{z^2}{4}}
\right)^2\, , \quad
\Gpp=\left(\frac{1-\frac{y^2}{4}}{1+\frac{y^2}{4}} \right)^2\, ,
\quad G_{zz}=\frac{1}{(1-\frac{z^2}{4})^2}\, , \quad
G_{yy}=\frac{1}{(1+\frac{y^2}{4})^2}   \, ,\eea  and
$G_{ii}=G_{zz}$, $G_{4+i,4+i}=G_{yy}$ for $i=1,\ldots ,4$.

\smallskip

Having introduced the embedding coordinates, we would like to ask
whether there exists a bosonic coset representative
$\ag_{\alg{b}}$ such that the bilinear form ${\rm
str}(\ag_{\alg{b}}^{-1}\, {\rm d}\ag_{\alg{b}})^2$ would coincide with the metric
(\ref{metricadss}). Introduce the following matrices \bea
\label{someg(X)}
\ag_{\alg{b}} = \Lambda(t,\p)\,\ag(\mathbb{X})\,,\quad \ag(\mathbb{X})=\left(\frac{\mI+\mathbb{X}}{\mI-\mathbb{X}}\right)^{\frac{1}{2}}\,
, \eea where $\mathbb{X}$ is the Lie algebra element (\ref{X}).
Substituting here the matrix representation for $\mathbb{X}$, we
find the the following result \bea \label{anotherg}
 \ag(\mathbb{X})=\small{\left(
\begin{array}{cc}
\frac{1}{\sqrt{1-\sfrac{z^2}{4}}}\Big[\mI+\frac{1}{2}z^i\gamma^i\Big]
& 0
\\
0 &
\frac{1}{\sqrt{1+\sfrac{y^2}{4}}}\Big[\mI+\frac{i}{2}y^i\gamma^i\Big]
\end{array}
\right)}\, . \eea One can easily verify that\footnote{The map
$\mathbb{X}\to \frac{\mI+\mathbb{X}}{\mI-\mathbb{X}}$ is the
Cayley transform which maps a (pseudo-) anti-hermitian matrix into
a (pseudo-) unitary one. In eq.(\ref{someg(X)}) one can replace
the square root by any real function $f(x)$ which admits a power
series expansion around $x=0$. Also, note that
$\ag^{-1}(\mathbb{X})=\ag(-\mathbb{X})$. }
$\ag(\mathbb{X})^{\dagger}H\ag(\mathbb{X})=H$, {\it i.e.} $\ag_{\alg{b}}$
belongs to the bosonic subgroup of $\su(2,2|4)$. It depends
on $t$, $\p$ and $\mathbb{X}$, {\it i.e.} it comprises the coset degrees of
freedom corresponding to the $\AdS$ space. Thus, we can consider
$\ag_{\alg{b}}$ as another embedding of the coset element into
${\rm SU}(2,2)\times {\rm SU}(4)$, alternative to the exponential
map (\ref{X}). Finally, computing the metric ${\rm str}(\ag_{\alg{b}}^{-1}\,
{\rm d}\ag_{\alg{b}})^2$, we see that it reproduces (\ref{metricadss}).

\subsection{Alternative form of the string Lagrangian}
\label{Subsect:altLag} We start with an alternative description of
the bosonic coset element and further use it to construct another
convenient representation of the string Lagrangian (\ref{sLag}).

\smallskip

 Let $\ag$ be an arbitrary matrix from ${\rm
SU(2,2)}\times {\rm SU}(4)$.  Construct the following matrix \bea
{\rm G}=\ag\, {\cal K}\, \ag^t\, . \label{altG}\eea Obviously,
${\rm G}$ is skew-symmetric: ${\rm G}^t=-{\rm G}$. It is also
pseudo-unitary: ${\rm G}^{\dagger}H{\rm G}=H$. Let $\alg{h}\in
{\rm SO(4,1)}\times {\rm SO}(5)$. Then $\alg{h}$ leaves the matrix
${\cal K}$ invariant: $\alg{h}\, {\cal K}\, \alg{h}^t={\cal K}$.
Therefore, under the right multiplication $\ag\to \ag \alg{h}$ the
matrix ${\rm G}$ remains unchanged:
$$
\ag\, {\cal K}\, \ag^t ~~\to~~ \ag\,\alg{h}\, {\cal K}\,
\alg{h}^t\, \ag^t={\rm G}\, .
$$
Thus, ${\rm G}$ depends solely on the coset degrees of freedom
comprising the $\AdS$ space. This space itself can be thought of as
an intersection of (even) pseudo-unitary and skew-symmetric
matrices.

\smallskip

Computing now ${\rm G}$ corresponding to the bosonic element
$\ag=\Lambda(t,\phi)\, \ag(\mathbb{X})$ with $\ag(\mathbb{X})$
given by eq.(\ref{anotherg}), we find \bea {\rm G}=\Lambda\,
\frac{\mI+\mathbb{X}}{\mI-\mathbb{X}}{\mathcal K}\, \Lambda
=\Big[\Lambda^2\frac{\mI+\mathbb{X}^2}{\mI-\mathbb{X}^2}-2\frac{\mathbb{X}}{\mI-\mathbb{X}^2}
\Big]\mathcal{K} \, ,\eea where we have used the property
$\mathbb{X}\Lambda=\Lambda^{-1}\mathbb{X}$. We see that, opposite
to $\ag$, the element ${\rm G}$ depends on $\Lambda^2$ rather than
on $\Lambda$. Thus, ${\rm G}$ is a periodic function of $\sigma$
irrespective of a winding sector. Another way to see this is to
write ${\rm G}$ in terms of global embedding coordinates.
Representing \bea {\rm G}=\left(\begin{array}{cc} {\rm G}_{\alg{ads}} & 0 \\
0 & {\rm G}_{\alg{sphere}}
\end{array}\right) \, ,
\eea we obtain
 \bea {\rm G}_{\alg{ads}}=\footnotesize{\left(\begin{array}{rrrr} 0 & -\Z_3 &
\Z_1^* &
\Z_2^* \\
\Z_3 & 0 & -\Z_2 & \Z_1 \\
-\Z_1^*  & \Z_2 & 0 &-\Z_3^* \\
-\Z_2^* & -\Z_1 & \Z_3^* & 0
\end{array}\right)} \, ,~~~~~ {\rm G}_{\alg{sphere}}=\footnotesize{\left(\begin{array}{rrrr}
0 & -\Y_3 & -i\Y_1^* & -i\Y_2^* \\
\Y_3 & 0 & i\Y_2 & -i\Y_1 \\
i\Y_1^* & -i\Y_2 & 0 & -\Y_3^* \\
i\Y_2^* & i\Y_1 & \Y_3^* & 0 \end{array}\right)}\, , \nonumber\eea
where the entries above are written in terms of complex embedding
coordinates given by eqs.(\ref{paramsphere}) and (\ref{paramAdS}).

\smallskip

We point out that the actual convenience of the embedding
coordinates is explained by the fact that, in opposite to
$x^{\mu}$, under the action of the whole bosonic symmetry group
they transform linearly. Indeed, if $G\in {\rm SU(2,2)}\times {\rm
SU}(4)$, then
$$
G\cdot \ag=\ag'\,\alg{h}_{c} ~~~\Longrightarrow~~~ {\rm G}\to {\rm
G}'= G\cdot {\rm G} \cdot G^t\, ,
$$
because the compensating element $\alg{h}$ from ${\rm
SO}(5,1)\times{\rm SO}(5)$ decouples from ${\rm G}'$ as a
consequence of definition (\ref{altG}).

\subsubsection*{\bf String Lagrangian}

\noindent Starting with the coset parametrization (\ref{fb}), we
write down the corresponding one-form $A$ \bea A=-\ag^{-1}{\rm
d}\ag= -\ag^{-1}_{\alg{b}}\big(\ag_{\alg{f}}^{-1}{\rm
d}\ag_{\alg{f}}\big)\ag_{\alg{b}}-\ag^{-1}_{\alg{b}}{\rm
d}\ag^{-1}_{\alg{b}}\, .\eea The element $\ag_{\alg{f}}^{-1}{\rm
d}\ag_{\alg{f}}$ takes values in $\su(2,2|4)$ and it is the sum of
even and odd components denoted by  ${\rm B}$ and ${\rm F}$,
respectively, \bea\nonumber \ag_{\alg{f}}^{-1}\, {\rm
d}\ag_{\alg{f}}={\rm B}+{\rm F} \, .\eea Hence, $A=A_{\rm
e}+A_{\rm o}$, where the even, $A_{\rm e}$, and odd, $A_{\rm o}$,
components are
 \bea\la{Aeo}
A_{\rm e}=-\ag^{-1}_{\alg{b}}\,{\rm
B}\,\ag_{\alg{b}}-\ag^{-1}_{\alg{b}}{\rm d}\ag^{-1}_{\alg{b}} \, ,
~~~~A_{\rm o}= -\ag^{-1}_{\alg{b}}\,{\rm F}\,\ag_{\alg{b}}\, .
\eea It is interesting to note that with this choice of the coset
parametrization the even component of the flat current is a gauge
transformation of the even element ${\rm B}$, while the odd one is
the adjoint transform of ${\rm F}$ with the bosonic matrix
$\ag_{\alg{b}}$.

\smallskip

As the next step, we compute the ${\mathbb Z }_4$-projections
$A^{(k)}$ of the connection (\ref{Aeo}). Straightforward
application of the formulae (\ref{Z4proj}) together with the
definition (\ref{altG}) gives \bea
\begin{aligned}
2A^{(0)}&=A_{\rm e}-{\cal K}A^t_{\rm e}{\cal
K}^{-1}=-2\ag_{\alg{b}}^{-1} {\rm
d}\ag-\ag_{\alg{b}}^{-1}\big({\rm B}-{\rm G}{\rm B}^t{\rm G}-{\rm
d}{\rm G}{\rm G}^{-1}\big)\ag_{\alg{b}}\, ,
\\
 2A^{(2)}&=A_{\rm e}+{\cal K}A^t_{\rm e}{\cal K}^{-1}=
-\ag_{\alg{b}}^{-1}\big({\rm B}+{\rm G}{\rm B}^t{\rm G}+{\rm
d}{\rm G}{\rm G}^{-1}\big)\ag_{\alg{b}}\, , \end{aligned}\eea for
the even components of $A$, and \bea
\begin{aligned}
2A^{(1)}&=A_{\rm o}+i\, {\cal K}A^{st}_{\rm o}{\cal
K}^{-1}=-\ag_{\alg{b}}^{-1}\big({\rm F}+i\, {\rm G}{\rm F
}^{st}{\rm G }^{-1}\big)\ag_{\alg{b}}\, ,
\\
 2A^{(3)}&=A_{\rm o}-i\, {\cal K}A^{st}_{\rm o}{\cal K}^{-1}=
-\ag_{\alg{b}}^{-1}\big({\rm F}-i\, {\rm G}{\rm F }^{st}{\rm G
}^{-1}\big)\ag_{\alg{b}}\, , \end{aligned}\eea for the odd ones.
Substituting these projections into the Lagrangian density
(\ref{sLag}), we obtain \bea\label{altLag}\begin{aligned}
\mathscr{L}=&-\frac{g}{8}{\rm str}\Big[ \gamma^{\a\b}({\rm
B}_{\a}+{\rm G}{\rm B}_{\a}{\rm G}^{-1}+\pa_{\a}{\rm G}{\rm
G}^{-1})({\rm B}_{\b}+{\rm G}{\rm
B}_{\b}{\rm G}^{-1}+\pa_{\b}{\rm G}{\rm G}^{-1}) -\\
&-2i\, \kappa \, \eps^{\a\beta}\, {\rm F}_{\a}{\rm G}{\rm
F}^{st}_{\beta}\, {\rm G}^{-1}\Big]\, . \end{aligned}\eea The nice
feature of this Lagrangian is that it only involves the fields
which carry the linear representation of the bosonic symmetry
algebra.

\smallskip

Finally, the form (\ref{altLag}) provides a shortcut to reproduce
the Polyakov Lagrangian for strings on $\AdS$,  when fermions are
switched off. Indeed, putting fermions to zero reduces expression
(\ref{altLag}) to \bea \mathscr{L}= -\frac{g}{8}\,
\gamma^{\a\b}{\rm str}\big(\pa_{\a}{\rm G}{\rm G}^{-1}\pa_{\b}{\rm
G}{\rm G}^{-1}\big)\, ,\eea which is the Lagrangian density for a
non-linear sigma-model with bosonic fields taking value in the
$\AdS$ space described by a group element ${\rm G}$.

\section{Bibliographic remarks}

{\small A manifestly supersymmetric covariant flat space
superstring action has been found in \cite{Green:1983wt} based on
the covariant action for superparticles \cite{Brink:1981nb}. This
action exhibits $\kappa$-symmetry \cite{Green:1983wt} which
generalizes the local fermionic symmetries first discovered for
massive and massless superparticles
\cite{deAzcarraga:1982dw,Siegel:1983hh}. For an introduction to
the Green-Schwarz formalism and further references on the
covariant quantization issue we refer the reader to the book
\cite{Green:1987sp}. Interpretation of Green-Schwarz string  as a
coset sigma-model of the Wess-Zumino type has been proposed by
Henneaux and Mezincescu \cite{Henneaux:1984mh}. It was shown in
\cite{Grisaru:1985fv} that type IIB superstring can be
consistently coupled  to a generic supergravity background with
preservation of $\kappa$-symmetry gauge invariance, see also an
earlier work \cite{Witten:1985nt} on the same subject for the
ten-dimensional superstring with $N=1$ target space supersymmetry.

The action for type IIB superstring on $\AdS$ was constructed by
Metsaev and Tseytlin \cite{Metsaev:1998it} along the lines of the
Henneaux-Mezincescu approach \cite{Henneaux:1984mh}. Various
aspects of this action, alternative formulations and related
models have been discussed in
\cite{Kallosh:1998nx}-\cite{Roiban:2000yy}. In
\cite{Berkovits:1999zq} it was found that the Wess-Zumino term
entering the sigma model action is ${\rm d}$-exact and can be
written in the local fashion provided the subgroup $H$ defining
the coset space $G/H$ is the invariant locus of a
$\mathbb{Z}_4$-automorphism of $G$. Our exposition of the string
sigma model based on the coset space (\ref{sAdS}) follows closely
\cite{Alday:2005gi}.

There is a vast literature on Lie superalgebras.  The reader is
invited to consult \cite{Kac:1977em,Bars:1982ps,Frappat:1996pb}.
Automorphisms of simple Lie superalgebras have been classified in
\cite{serganova} and we mention the corresponding classification
for $\sls(4|4)$ in section \ref{Z4grad}.

Our treatment of $\kappa$-symmetry in section \ref{Sect:kappa} is
based on the observation that this symmetry can be understood as
the right local action on the coset space supplied with the proper
transformation of the two-dimensional world-sheet metric
\cite{McArthur:1999dy}. The reader might also find some
similarities with the corresponding discussion in
\cite{Roiban:2000yy}.

Concerning the general concept of integrability and conservation
laws, we refer the reader to the books \cite{Faddeev:1987ph,BBT}.
Dynamics of bosonic strings propagating in the $\AdS$ geometry is
described by the corresponding non-linear sigma model. This model
inherits its classical and quantum integrability from the
principal chiral model based on the group ${\rm SO}(4,2)\times
{\rm SO}(6)$. Classically the model is conformally invariant but
at the quantum level it develops a mass gap.

Integrability of classical superstring theory on $\AdS$ has been
established for the first time in \cite{Bena:2003wd} by exhibiting
the zero curvature representation of the string equations of
motion. The corresponding (full and bosonic) Lax pair and the
associated conservation laws have been further studied in many
papers, see e.g. \cite{Alday:2005gi},
\cite{Kazakov:2004nh}-\cite{Magro:2008dv}. The relation between
$\kappa$-symmetry and integrability was emphasized in
\cite{Bena:2003wd,Polyakov:2004br}.

Coset parametrization of the type $\ag=\ag_{\alg{f}}\ag_{\alg{b}}$
has been introduced in \cite{Alday:2005jm}. Also, the action of
the global symmetry algebra on a coset representative was analyzed
there. Representation (\ref{basiccoset}),  suitable for the
light-cone gauge fixing, appeared in \cite{FPZ}. The
$\kappa$-symmetry gauge choice (\ref{kappagauge}) was pointed out
in \cite{Alday:2005jm,FPZ}. Two-index notation to encode the
transformation properties of the world-sheet fields with respect
to the linearly realized bosonic subgroup ${\rm SU}(2)^4$ was
introduced in \cite{KMRZ}. For the alternative parametrization of
the coset space discussed in appendix 2 we refer to the work
\cite{Alday:2005jm} and \cite{Alday:2005ww}. The latter paper also
contains the alternative form of the string Lagrangian --
eq.(\ref{altLag}).

}

\chapter{Strings in light-cone gauge}

\def\z{\zeta}

\def\Texp{{\cal T}{\rm exp}}

\def\cS{{\cal S}}
\def\dcS{\dot{{\cal S}}}
\def\cT{{\cal T}}
\def\dcT{\dot{{\cal T}}}

\def\dK{\dot{K}}
\def\dL{\dot{L}}
\def\dP{\dot{P}}
\def\dQ{\dot{Q}}

\def\bP{{\mathbb P}}
\def\bV{{\mathbb V}}
\def\bT{{\mathbb T}}
\def\Bp{{\rm  p}}
\def\Bp{{\hat{\rm P}}}

\def\de{\delta}
\def\De{\Delta}

\def\str{{\rm str}}

\def\Kk{{\cal K}}
\def\ws{{\rm ws}}
\def\pws{p_{{\rm ws}}}
\def\dz{\dot{z}}

\def\eps{\epsilon}
\def\da{{\dot a}}
\def\dal{{\dot \a}}
\def\db{{\dot b}}
\def\dbe{{\dot \b}}
\def\dc{{\dot c}}
\def\dd{{\dot d}}
\def\dga{{\dot \g}}
\def\dr{{\dot r}}
\def\drh{{\dot \rho}}

\def\bX{\mathbb{X}}
\def\bU{\mathbb{U}}
\newcommand{\sQ}{\rm Q}
\newcommand{\sH}{\rm H}
\newcommand{\eqn}[1]{(\ref{#1})}

To fix the reparametrization freedom of the string sigma model, in
this chapter we introduce a special one-parameter class of gauges.
They are usually called the uniform light-cone gauges. In the
light-cone gauge
 the string sigma model is a two-dimensional
field theory defined on a cylinder of circumference $P_+$ with the
light-cone Hamiltonian depending on the string tension $g$ and
$P_+$. It describes a sector of string states, all carrying the
same space-time light-cone momentum $P_+$. Not all of these states
are considered to be physical -- a physical state should satisfy
the level-matching condition that is its {\it total world-sheet
momentum} must vanish.

Quantization of the light-cone string sigma model simplifies
greatly in the so-called decompactification limit where the
light-cone momentum tends to infinity, while the string tension is
kept fixed. In the decompactification limit  the gauge-fixed model
is defined on the plane and has massive excitations. Giving up the
level-matching condition defines the theory {\it off-shell}. In
the off-shell theory world-sheet excitations (particles) carry
non-trivial world-sheet momenta and can scatter among themselves.
Their pairwise scattering is encoded into the two-body {\it
world-sheet S-matrix}.

In this chapter the  light-cone model is quantized perturbatively
in the large string tension expansion. At the leading order the
model is nothing else but a massive relativistic  two-dimensional
theory with eight bosons and eight fermions. Developing the
expansion in powers of $1/g$, one can compute the corresponding
perturbative world-sheet S-matrix. We present here the
corresponding calculation in the tree-level (Born) approximation.
We also study the symmetry algebra of the light-cone model and
show that in the off-shell theory it undergoes a central
extension; the latter turns out to be crucial for fixing the
matrix structure of the exact world-sheet S-matrix.

\section{Light-cone gauge}

In this section we introduce the first-order formalism for the \GS
superstring in $\AdS$. Then we impose the uniform light-cone gauge
and fix $\kappa$-symmetry. The uniform light-cone gauge
generalizes the standard phase-space light-cone gauge to a curved
background, and it is distinguished from other possible light-cone
gauges by the choice of the light-cone coordinates and
$\kappa$-symmetry fixing. To make the discussion clearer, we start
by considering bosonic strings, and then include fermions and fix
$\kappa$-symmetry.

\subsection{Bosonic strings in light-cone gauge}

We consider strings propagating in a target manifold possessing
(at least) two abelian isometries realized by shifts of the time
coordinate of the manifold denoted by $t$, and a space coordinate
denoted by $\phi$. If the variable $\phi$ is an angle then the
range of $\phi$ is chosen to be from $0$ to $2\pi$.

\medskip

To impose a uniform gauge, we also assume that the string
sigma-model action is invariant under shifts of  $t$ and $\phi$,
all the other bosonic and fermionic fields being invariant under
the shifts. This means that the string action does not have an
explicit dependence on $t$ and $\p$ and depends only on the
derivatives of the fields. An example of such a string action is
provided by the Green-Schwarz superstring in $\AdS$ where the
metric can be written in the form, see (\ref{metricadss}) \bea
\la{metrgen} ds^2 = -G_{tt}\, dt^2\, +\, G_{\p\p}\, d\p^2\, +\,
G_{\mu\nu}\,dx^\mu dx^\nu \, . \eea Here $t$ is the global time
coordinate of $\rm{AdS}_5$, $\p$ is an angle parametrizing the
equator of ${\rm S}^5$, and $x^\mu$, $\mu=1,\ldots ,8$,  are the
remaining ``transversal'' coordinates of  $\AdS$.

\medskip

In this subsection we consider only the bosonic part of a string
sigma model action, and assume that the B-field vanishes.

The corresponding part of the string action is of the
following form \bea \la{S1} S = -{g\over 2}\int_{-r}^{ r}\, {\rm
d}\s{\rm d}\tau\, \g^{\a\b}\partial_\a X^M\partial_\b X^N\,
G_{MN}\,, \eea where $g$ is the effective dimensionless string
tension, $X^M = \{t,\p,x^\mu\}$ are string coordinates and
$G_{MN}$ is the target-space metric independent of $t$ and $\p$.

\medskip

The simplest way to impose a uniform light-cone gauge is to
introduce momenta canonically-conjugate to the coordinates $X^M$
 \bea\nonumber p_M ={\de S\ov \de \dot{X}^M} = -g\, \g^{0\b}\partial_\b X^N\,
G_{MN}\,,\quad \dot{X}^M\equiv \pa_0 X^M \,, \eea and rewrite the
string action (\ref{S1}) in the first-order form \bea \la{S2} S=
\int_{- r}^{ r}\, {\rm d}\s{\rm d}\tau\, \left( p_M \dot{X}^M +
{\g^{01}\ov\g^{00}} C_1+ {1\ov 2g\, \g^{00}}C_2\right)\,.  \eea
The reparametrisation invariance of the string action leads to the
two Virasoro constraints \bea \nonumber C_1=p_MX'^M\,,\quad
C_2=G^{MN} p_M p_N + g^2\, X'^M X'^N G_{MN}\,,\qquad X'^M\equiv
\pa_1 X^M\,, \eea which are to be solved after imposing a gauge
condition.

The invariance of the string action under the shifts of the time
and space coordinates, $t$ and $\p$,  of the manifold leads to the
existence of two conserved charges \bea\la{charges} E = - \int_{-
r}^{ r}\, {\rm d}\s\, p_t\ \, , \qquad J= \int_{- r}^{ r}\, {\rm
d}\s\, p_\p\ . \eea It is clear that the charge $E$ is the target
space-time energy, and $J$ is the U(1) charge of the string equal
to the total (angular) momentum of the string in the
$\p$-direction.

To impose a uniform gauge we introduce the ``light-cone''
coordinates and momenta: \bea \la{lcc} &&x_- =\p \,-\,t\ , \,\,
x_+ =(1-a)\,t\, +\,a\,\p\ ,\,\, p_- = p_\p\,+\,p_t\ ,\,\, p_+ =
(1-a) p_\p \,-\,a\,p_t\ ,\nonumber \\ &&t = x_+ \,-\,a\,x_-\ ,\,
\, \p = x_+ \,+\, (1-a) x_-\ , \,\, p_t = (1-a)\,p_- - p_+ \ ,\,\,
p_\p = p_+ \,+\, a\,p_- \ . \nonumber \eea Here $a$ is an
arbitrary number which parametrizes the most general uniform gauge
(up to some trivial rescaling of the light-cone coordinates) such
that the light-cone momentum $p_-$ is equal to $p_\p\,+\,p_t$.
This choice of gauge is natural in the AdS/CFT context because, as
we will see in a moment, in such a uniform gauge the world-sheet
Hamiltonian is equal to $E-J$.

Taking into account (\ref{charges}), we get the following
expressions for the light-cone momenta \bea\la{charges2}\nonumber
P_- \,=\,\int_{- r}^{ r}\, {\rm d}\s\, p_-\,= \, J\,-\,E\ \, ,
\qquad P_+ \,=\, \int_{- r}^{ r}\, {\rm d}\s\, p_+\,=\, (1-a)\, J
\,+\, a\,E\, . \eea In terms of the light-cone coordinates the
action (\ref{S2}) takes the form \bea \la{S3} S = \int_{- r}^{
r}\, {\rm d}\s{\rm d}\tau\, \left( p_-\dot{x}_++ p_+ \dot{x}_- +
p_\mu \dot{x}^\mu + {\g^{01}\ov\g^{00}} C_1+ {1\ov 2g\,
\g^{00}}C_2\right)\, , \eea where \bea \la{C1b} C_1\,=\,p_+x_-'
\,+\, p_-x_+' \,+\, p_\mu x'^\mu\,  . \eea The second Virasoro
constraint is a quadratic polynomial in $p_-$ which can be cast in
the following form \bea\nonumber C_2 &=& \left( a^2\,
G_{\p\p}^{-1} - (a-1)^2\, G_{tt}^{-1} \right) p_-^2\,  +\, 2\left(
a\, G_{\p\p}^{-1} - (a-1)\, G_{tt}^{-1} \right) p_-p_+ \, +\,
\left(G_{\p\p}^{-1} - G_{tt}^{-1} \right) p_+^2\\\nonumber &+&g^2
\left(  (a-1)^2\, G_{\p\p} -a^2\, G_{tt} \right) x_-'^2\,  -\,
2g^2 \left(  (a-1)\, G_{\p\p} -a\, G_{tt} \right) x_-'x_+'
\\\la{c2} &+& g^2 \left(  G_{\p\p} -G_{tt} \right) x_+'^2 \, + \,
\H_x\,, \eea where $\H_x$ is the part of the constraint which
depends only on the transversal fields $x^\mu$ and $p_\mu$
\bea\nonumber \H_x = G^{\mu\nu}p_\mu p_\nu + g^2\, x'^\mu x'^\nu\,
G_{\mu\nu}\,, \eea and we assume that the target space-time metric
is of the form (\ref{metrgen}).

We then fix the uniform light-cone gauge by imposing the
conditions \bea \la{ulc} x_+ \,=\, \tau \,+\, a\, m\,\s\
,\quad\quad p_+ \,=\, 1\ . \eea The condition $p_+\,=\,1$ means
that the light-cone momentum is distributed uniformly along the
string, and this explains the word ``uniform'' in the name of the
gauge. The integer number $m$ is the winding number which
represents the number of times the string winds around the circle
parametrized by $\phi$. The winding number appears because we
consider closed strings and  the coordinate $\p$ is an angle
variable with the range $0\le\p\le 2\pi$ and, therefore, it  has
to satisfy the constraint:
 \bea \label{period}
\phi(r)-\phi(-r)=2\pi  m\, , ~~~~~m\in {\mathbb Z}\, . \eea The
consistency of this gauge choice also fixes  the constant $r$ to
be equal to \bea\nonumber r = {1\ov 2}P_+\,, \eea which means that the
light-cone string sigma model is defined on a cylinder of
circumference equal to the total light-cone momentum $P_+$.

To find the gauge-fixed action, we first solve the Virasoro
constraint $C_1$ for $x_-'$ \bea\nonumber C_1\,=\,x_-' \,+\,  a m p_- \,+\,
p_\mu x'^\mu\,=\,0\, \quad \Longrightarrow \quad x_-'\,=\, - a m
p_- \,-\, p_\mu x'^\mu\,, \eea then we substitute the solution
into $C_2$ and solve the resulting quadratic equation for $p_-$.
Substituting all these solutions into the string action
(\ref{S3}) and omitting the total derivative $\dx_-^{(0)}$ of the zero mode of $x_-$, we end up with the gauge-fixed action
\bea \la{S4} S
=\int_{- r}^{ r}\, {\rm d}\s{\rm d}\tau\, \left( p_\mu \dot{x}^\mu
\,-\, \H \right)\, ,
 \eea
  where \bea \la{denH} \H \,=\, -p_-(p_\mu, x^\mu ,x'^\mu ) \eea is the
density of the world-sheet Hamiltonian which depends only on the
physical
 fields $p_\mu, x^\mu$. It is worth noting that $\H$ has no
dependence on $P_+$, and the dependence of the gauge-fixed action
and the world-sheet Hamiltonian $H \,=\, \int_{- r}^{ r}\, {\rm
d}\s\, \H$ on $P_+$ comes only through the integration bounds $\pm
r$.

\medskip

Since we consider closed strings, the transversal fields $x^\mu$
are periodic: $x^\mu(r) = x^\mu(-r)$. Therefore, the gauge-fixed
action defines a two-dimensional model on a cylinder of
circumference $2 r =P_+$.  In addition, the physical states should
also satisfy the level-matching condition \bea\la{LM} \De x_-=
\int_{- r}^{ r}\, {\rm d}\s\, x_-' = am H - \int_{- r}^{ r}\, {\rm
d}\s\, p_\mu x'^\mu = 2\pi m\,, \eea that follows by integrating
the Virasoro constraint $C_1$ (\ref{C1b}) over $\s$ and taking
into account that $\p$ is an angle variable.

The gauge-fixed action is obviously invariant under the shifts of
the world-sheet coordinate $\s$.  This leads to the existence of
the conserved charge \bea\la{pws} p_{{\rm ws}} = -\int_{- r}^{
r}\, {\rm d}\s\, p_\mu x'^\mu\,, \eea which is just the total
world-sheet momentum of the string. In what follows we will be
mostly interested in the zero-winding number case, $m=0$ because
only in this case the large tension perturbative expansion is
well-defined.  Then the level-matching condition simply states
that the total world-sheet momentum vanishes for physical
configurations \bea\la{LM2} \De x_-= p_{{\rm ws}} =0\,,\quad
m=0\,.  \eea

It is worth  stressing  that  to quantize the light-cone string
sigma model and to also identify its symmetry algebra, one has to
consider all states with periodic $x^\mu$ and to impose the
level-matching condition singling out the physical subspace only
at the very end. In a uniform light-cone gauge one has a
well-defined model on a cylinder. However, if a string
configuration does not satisfy the level-matching condition then
its target space-time image is an open string with end points
moving in unison so that $\De x_-$ remains constant. Another
peculiarity is related  to the fact that the gauge-fixed string
sigma models are equivalent for different choices of a uniform
gauge, {\it i.e.} for different values of $a$, provided the
level-matching condition is satisfied. String configurations which
violate the level-matching condition may depend on $a$. This
gauge-dependence makes the problem of quantizing string theory in
a uniform gauge very subtle. On the other hand, the requirement
that physical states are gauge-independent should impose severe
constraints on the structure of the theory. It may also happen
that for finite $J$ there is a preferred choice of the parameter
$a$ simplifying the exact quantization of the model. In fact, we
will see that for finite $J$ the choice $a=0$  seems to be the
most natural one, at least in the AdS/CFT context. For example,
 only in the $a=0$ uniform gauge
 one can study string configurations with an arbitrary
winding number in one go.

Let us now consider in more detail bosonic strings in $\AdS$ where the metric  takes the
form (\ref{metricadss}).
We consider string states with zero winding number $m=0$ and
impose the uniform light-cone gauge (\ref{ulc}) $x_+=\tau\,,\
p_+=1$. Solving the first Virasoro constraint $C_1$ (\ref{C1b}) for $x_-'$, we
get
\bea\nonumber x_-'= -p_\mu x'^\mu\,,
\eea while the second constraint
(\ref{c2}) takes the following form \bea\nonumber C_2 &=& \left(
a^2\, G_{\p\p}^{-1} - (a-1)^2\, G_{tt}^{-1} \right) p_-^2\,  +\,
2\left( a\, G_{\p\p}^{-1} - (a-1)\, G_{tt}^{-1} \right) p_- \, +\,
G_{\p\p}^{-1} - G_{tt}^{-1} \\\la{c2b} &+&g^2 \left(  (a-1)^2\,
G_{\p\p} -a^2\, G_{tt} \right) x_-'^2 \, + \, \H_x\,. \eea There
are two solutions of the constraint equation $C_2=0$, and one
should keep those that leads to a positive definite Hamiltonian
density through the relation $\H=-p_-$. A simple computation shows
that the solution is given by the following expression
\bea\nonumber
\H &=& {\sqrt{G_{\p\p}G_{tt}\left( 1+ \left((a-1)^2G_{\p\p} - a^2G_{tt}\right) \H_x  + g^2\left((a-1)^2G_{\p\p} - a^2G_{tt}\right)^2 x_-'^2  \right)}\ov(a-1)^2G_{\p\p} - a^2G_{tt}}~~~\\
\la{Hb} &+&  {(a-1)G_{\p\p} - a G_{tt} \ov(a-1)^2G_{\p\p} -
a^2G_{tt}}\,. \eea The world-sheet light-cone Hamiltonian has a
very complicated non-linear dependence on the physical coordinates
and momenta, and it  could hardly be used to perform a direct
canonical quantization of the model.

The gauge-fixed action corresponding to the world-sheet
Hamiltonian\footnote{ The  action is written in the first-order
formalism. It is not difficult to see, however, that one can
eliminate the momenta from the action by using their equations of
motion, and get an action which depends only on $x^\mu$ and their
first derivatives.} can be used to analyze string theory in
various limits. One well-known limit is the BMN limit in which one
takes $g\to\infty$ and $P_+ \rightarrow \infty$, while keeping
$g/P_+$ fixed. In this case it is useful to rescale $\s$ so that
the range of $\s$ will be from $-\pi$ to $\pi$. The gauge-fixed
action then admits a well-defined expansion in powers of $1/g$ (or
equivalently $1/P_+$), with the leading part being just a
quadratic action for eight massive bosons (and eight fermions for
the full model). The action can be easily quantized perturbatively
and subsequently used to compute $1/P_+$ corrections to the energy
of string states.

Another interesting limit is the decompactification  limit where
$P_+\to\infty$ with $g$ kept fixed.  In this limit the
circumference $2r$ goes to infinity and we get a two-dimensional
massive model defined on a plane. Since the gauge-fixed theory is
defined on a plane the asymptotic states and S-matrix are
well-defined.
 An important observation is that in the limit
 the light-cone string sigma model admits one- and multi-soliton solutions.
   The corresponding one-soliton solutions were named giant
   magnons because  they are dual to field theory spin chain magnons and also because
generically their size is of order of the radius of ${\rm S}^5$.
Since for a giant magnon $\De x_-$ is not an integer multiple of
$2\pi$, such a soliton configuration does not describe a closed
string. We will discuss giant magnons in the next section in
detail.

Let us also mention that the world-sheet Hamiltonian in the
light-cone gauge is related to the target space-time energy $E$
and the U(1) charge $J$ as follows \bea \la{He} H \,=\, \int_{-
r}^{ r}\, {\rm d}\s\, \H = E-J\,. \eea According to the AdS/CFT
correspondence, the space-time energy $E$ of a string state is
identified with the conformal dimension $\De$ of the dual CFT
operator: $E\equiv \De$. Since the Hamiltonian $H$ is a function
of $P_+=(1-a)J + aE$, for generic values of $a$ the relation
(\ref{He}) gives us a nontrivial equation on the energy $E$.
Computing the spectrum of $H$ and solving the equation (\ref{He})
would allow one to find conformal dimensions of dual CFT
operators.

There are three natural choices of the parameter $a$. If $a=0$ we
have the temporal gauge $t = \tau\,,\ P_+ = J$.  In this gauge the
world-sheet Hamiltonian depends on $J$ only and therefore its
spectrum immediately determines the space-time energy $E$.   If
$a={1\ov 2}$, we obtain the usual light-cone gauge $x_+ ={1\ov
2}(t +\p) =\tau\,,\ P_+ = {1\ov 2}(E+J)$. The light-cone gauge
appears to drastically simplify  perturbative computations  in the
large tension limit, as we will demonstrate later in this chapter.
Finally, one can also set $a=1$. In this case, the uniform gauge
reduces to $x_+=\p =\tau\,,\ P_+=E$, where the angle variable $\p$
is identified with the world-sheet time $\tau$, and the energy $E$
is distributed uniformly along the string. String theory in $\AdS$
has not been analyzed in this gauge yet.

\subsection{First-order formalism}

To generalize the discussion of the previous subsection to the
Green-Schwarz superstring in $\AdS$, one should use the
parametrization (\ref{basiccoset}) of the coset element that
ensures that all fermions are neutral under the U$(1)$ isometries
generated by shifts of $t$ and $\p$. Then, to impose the
light-cone gauge in the Hamiltonian setting, one should first
determine the momenta canonically-conjugate to the coordinates $t$
and $\phi$ (or, equivalently, to the light-cone coordinates
$x_\pm$). Because of non-trivial interactions between bosonic and
fermionic fields, to find the momenta is not straightforward. A
better way to proceed is to introduce a Lie-algebra valued
auxiliary field $\bp$, and rewrite the superstring Lagrangian
(\ref{sLag}) in the form
\begin{eqnarray}\nonumber
\L =  -\str \bigg( \bp \, A_0^{(2)}\, + \, \kappa{g\over
2}\epsilon^{\a\b}A_\a^{(1)} A_\b^{(3)}\, +\,  {\gamma^{01}\over
\gamma^{00}} \, \bp \, A_1^{(2)}\,
 - \, {1\over 2 g \gamma^{00}} \left( \bp^2 + g^2 (A_1^{(2)})^2 \right)
\bigg)\,. \\\label{Lang}
\end{eqnarray}
It is easy to see that if one solves the equations of motion for
$\bp$ and substitutes the solution back into (\ref{Lang}) one
obtains (\ref{sLag}). The last two terms in (\ref{Lang}) yield the
Virasoro constraints \bea \la{C1}
C_1 &=& \str\,\bp \, A_1^{(2)}=0\,,\\
\la{C2} C_2 &=& \str\left( \bp^2 + g^2 (A_1^{(2)})^2 \right) =0\,,
\eea which are to be solved after imposing the light-cone gauge
and fixing the $\kappa$-symmetry.

It is clear that without loss of generality we can assume that
$\bp$ belongs to the subspace $M^{(2)}$ of $\su(2,2|4)$, as the
other components in the $\mathbb{Z}_4$ grading decouple. It
therefore admits the following decomposition {\small \bea
\label{bpexp} \bp = \bp^{(2)} = {i\ov 2}\bp_+ \Sigma_+ +
 {i\over 4}\bp_- \Sigma_- + {1\over 2}\bp_\mu \Sigma_\mu
 + \bp_\mI i\mI_8 \, .
\eea where $\Sigma$'s are $8\times 8$ matrices defined as follows
 \bea\la{Sigpm} \Sigma_+= \left(\begin{matrix}
\S & 0 \cr 0 & \S\cr \end{matrix}\right )   \,, \ \
\Sigma_-=  \left(\begin{matrix}
-\S & 0 \cr 0 & \S\cr \end{matrix}\right ) \,,\ \ \S_k= \left(
\begin{array}{cc}
 \gamma_k& 0  \\
  0 & 0
\end{array} \right)  \,,\ \ \S_{4+k}=  \left(
\begin{array}{cc}
 0& 0  \\
  0 & i  \gamma_k
\end{array} \right) .
\eea } \normalsize \noindent \hskip -0.1cm Since $A_\a^{(2)}$
belongs to the superalgebra $\su(2,2|4)$, $\str A_\a^{(2)} =0$,
and the quantity $\bp_\mI$ does not contribute to the Lagrangian.

It is worth stressing that the fields $\bp_\pm$ do not coincide
with the momenta $p_\pm$ canonically conjugate to $x_\mp$ but they
can be expressed in terms of $p_\pm$. Before doing this, we impose
the $\kappa$-symmetry gauge, which will dramatically simplify our
further treatment.

\subsection{Kappa-symmetry gauge fixing}

As was discussed in the previous chapter, the key property of the
\GS action is its invariance under the fermionic $\kappa$-symmetry
that halves the number of fermionic degrees of freedom. A
$\kappa$-symmetry gauge should be compatible with the bosonic
gauge imposed, and the analysis of the $\kappa$-symmetry
transformations (\ref{kappa})  for the \GS  superstring  action
(\ref{Lang}) performed in subsection \ref{Sect:kappa} revealed
that for the uniform light cone gauge the $\kappa$-symmetry could
be fixed by choosing the fermion field $\chi$ (\ref{chi}) to be of
the form (\ref{chilc}). It is not difficult to check that the
gauge-fixed fermion field $\chi$ satisfies the following important
relations \bea\la{Schi} \S_+\chi = -\chi\S_+\,,\quad \S_-\chi =
\chi\S_-\,. \eea In fact these relations may be considered as the
defining ones for the $\kappa$-symmetry gauge we have chosen and
can be used instead of specifying the explicit form of $\chi$.
Taking into account that $\ag^{-1}(\chi)= \ag(-\chi)$ and these
identities, one can then easily show that
\begin{align}\nonumber
\ag^{-1}(\chi)\S_+&=\S_+\ag(\chi)\quad \Rightarrow \quad
\ag^{-1}(\chi)\S_+\ag(\chi)=
\S_+\ag(\chi)^2\,,\\\nonumber
\ag^{-1}(\chi)\S_-&=\S_-\ag^{-1}(\chi)  \Rightarrow \quad
\ag^{-1}(\chi)\S_-\ag(\chi)=\S_-\,.
\end{align}
The perturbative expansion of the light-cone Lagrangian in powers of the fields simplifies if one uses $\ag(x)\equiv \ag(\bX)$  of the form (\ref{someg(X)}), and the matrix $g(\chi)$ of the form
\bea\la{chi2}
g(\chi) = \chi + \sqrt{1+\chi^2}\,.
\eea
The  standard exponential parametrization (\ref{chi}) can be obtained from (\ref{chi2}) by means of the following fermion redefinition
 $\chi \to \sinh \chi$.

Now it is straightforward to use the coset parametrization
(\ref{basiccoset}) to compute the current (\ref{la}) $$ A =
A_{even}+A_{odd}\,,$$ where \bea\nonumber
 A _{even}&=&-\ag^{-1}(x)\Big[
{i\ov 2} \left(dx_+ + \Big({1\ov 2} -a\Big) dx_-\right)\S_+(1+
2\chi^2) + {i\ov 4} dx_- \S_- \Big]\ag(x)\\\la{Aeven}
&~&-\ag^{-1}(x)\Big[ \sqrt{1+\chi^2}d\sqrt{1+\chi^2} - \chi d\chi
+d\ag(x)\ag^{-1}(x)\Big]\ag(x)\,, \\\nonumber
A_{odd}&=&-\ag^{-1}(x)\Big[ i  \left(dx_+ + \Big({1\ov 2} -a\Big)
dx_-\right) \S_+ \chi\sqrt{1+\chi^2}\\\la{Aodd}
&&~~~~~~~~~~~~~~~~~~~~~~~~~~~~+ \sqrt{1+\chi^2}d\chi-\chi
d\sqrt{1+\chi^2} \Big]\ag(x)\,. \eea These formulae clearly
demonstrate that the currents acquire the simplest form if the
parameter $a$ of the uniform light-cone gauge is equal to $1/2$.
For $a=1/2$ the odd part of the current $A$ does not depend on the
light-cone coordinate $x_-$! This explains the drastic
simplifications that occur for the $a=1/2$ gauge in comparison to
the general uniform gauge. For $a=1/2$ and in the gauge $x_+ =
\tau$ the odd part of $A$ depends on the derivatives of the
fermion $\chi$ only. In what follows we restrict our discussion of
the fermionic part of the light-cone \GS action to the simplest
case $a=1/2$.

\subsection{Light-cone gauge fixing}

Now we  can use the formulae established above to express
$\bp_\pm$ in terms of $p_\pm$. To this end, omitting the Virasoro
constraints, we can rewrite the Lagrangian (\ref{Lang}) as follows
\bea\label{Lang2} \L = p_+\dot{x}_- +{\bf p}_-\dot{x}_+ -\str
\bigg( \bp A_{even}^\perp + \kappa{g\over
2}\epsilon^{\a\b}A_\a^{(1)} A_\b^{(3)}\bigg)\,,
 \eea where
 \bea\nonumber A_{even}^\perp =
-\ag^{-1}(x)\Big[ \sqrt{1+\chi^2}\pa_\tau\sqrt{1+\chi^2} - \chi
\pa_\tau\chi +\pa_\tau \ag(x)\ag^{-1}(x)\Big]\ag(x)\,, \eea and
the momentum $p_+$, canonically conjugate to $x_-$, is shown to be
equal to \bea\label{p+explicit} p_+ = {i\ov 4}\, \str \left( \bp
\Sigma_- \ag(x)^2 \right) = G_+ \bp_+ - {1\ov 2}G_- \bp_-\,,\qquad
G_\pm &=&{1\over 2}(G_{tt}^{1\over 2} \pm G_{\phi\phi}^{1\over 2})
\,.~~~~~~~~
\end{eqnarray}
The variable ${\bf p}_-$ is not equal to the momentum $p_-$
canonically conjugate to $x_+$. It differs from $p_-$ by a
contribution coming from the Wess-Zumino term  in (\ref{Lang2}) ,
and is defined as follows \bea \la{bpm} {\bf p}_- ={i\ov 2}\, \str
\left( \bp \Sigma_+ \ag(x)(1+2\chi^2)\ag(x) \right) \,. \eea Now
having identified the light-cone momentum $p_+$, we can impose the
uniform light-cone gauge (\ref{ulc}) with $a=1/2$ \bea \la{ulch}
x_+ \,=\, \tau \,+\, {1\ov 2}\, m\,\s\ ,\quad\quad p_+ \,=\, 1\ .
\eea
 Let us stress again that the density $\H$ of the world-sheet light-cone Hamiltonian is  equal to $-p_-$ but not to  $-{\bf p}_-$.

It is also important to recall that to impose the light-cone gauge
we had to make all the fermions of the string sigma model neutral
with respect to the two U$(1)$ isometry groups generated by the
shifts of $t$ and $\p$. As a result, in  the light-cone gauge  the
fermions are periodic in the even winding sector and they are
anti-periodic in the odd winding sector.

In what follows we will be interested in the decompactification
and large string tension limits, and, therefore, we  set $m=0$.

\subsection{Gauge-fixed Lagrangian}\la{sec:gfL}

Now we are ready to find the light-cone gauge-fixed Lagrangian.
This is a multi-step procedure. First, we solve equation
(\ref{p+explicit})  determining $p_+$ for $\bp_+=\bp_+(p_+,\bp_-)$
and set $p_+=1$ in the solution. Second, we solve the Virasoro
constraint $C_1$ of equation (\ref{C1}) to find $x_-'$. Finally,
we determine $\bp_-$ from the second Virasoro constraint $C_2$
eq.(\ref{C2}). Substituting all the solutions into the Lagrangian
of equation (\ref{Lang2}), we end up with the total gauge-fixed
Lagrangian. The explicit derivation is rather involved and we
refer the reader to the original literature for details, see
section \ref{sec:biblight-cone}.

The upshot is a Lagrangian which can be written in the standard
form as the difference of a kinetic term and the Hamiltonian
density \bea \label{Lgf} &&\L_{gf} = \L_{kin} - \H \,.~~~~~~~ \eea
The kinetic term $\L_{kin}$ depends on the time derivatives of the
physical fields, and determines the Poisson structure of the
theory. It can be cast in the form \bea \label{Lkin} \L_{kin} &=&
p_\mu\dot{x}_\mu -
\frac{i}{2}\str\left(\Sigma_+\chi\pa_\tau\chi\right)
+\frac{1}{2}\ag_\nu\bp_\mu\, \str\left(\left[\S_\nu,\Sigma_\mu\right] B_\tau\right)\\
\nonumber &-& i\kappa{g\over 2}(G_+^2-G_-^2)\, \str\left( F_\tau
\Kk F_\s^{st} \Kk\right) +i\kappa{g\over 2} G_\mu G_\nu\,
\str\left(\S_\nu F_\tau\S_\mu \Kk F_\s^{st} \Kk\right) \,,~~~~~~~
\eea where  we  use the following definitions
 \bea\la{g}\nonumber \ag(x)
= \ag_+ I_8 + \ag_-\Upsilon + \ag_\mu \Sigma_\mu\, ,\quad
 \ag(x)^2 = G_+I_8 +G_-\Upsilon +G_\mu\S_\mu \, ,
\eea and the functions $B_\alpha$ and $F_\alpha$ refer to the even
and odd components of $\ag^{-1}(\chi)\,\partial_\alpha  \ag(\chi)$
 \bea\la{BF}
&&\ag^{-1}(\chi)\pa_\a \ag(\chi) = B_\a +
F_\a\,,~~~~~~~\\\nonumber &&B_\a =  -{1\ov 2}\chi\pa_\a\chi +
{1\ov 2}\pa_\a\chi\chi + {1\ov
2}\sqrt{1+\chi^2}\pa_\a\sqrt{1+\chi^2} - {1\ov
2}\pa_\a\sqrt{1+\chi^2}\sqrt{1+\chi^2}\, ,~~~~~\\\nonumber &&F_\a
=
  \sqrt{1+\chi^2}\pa_\a\chi - \chi\pa_\a\sqrt{1+\chi^2}\, .
\eea As one can see, the kinetic term is highly nontrivial and
leads to a complicated Poisson structure. To quantize the theory
perturbatively, {\it e.g.} in the large string tension limit, one
would need to redefine the fields  so that the kinetic term
acquires the conventional form \bea\la{kincan} \L_{kin}\rightarrow
p_\mu\dot{x}_\mu -
\frac{i}{2}\str\left(\Sigma_+\chi\pa_\tau\chi\right)\,, \eea and,
therefore, the redefined fields would satisfy the canonical
commutation relations. This will be done in the next section up to
the quartic order in the fields.

The density $\H$ of the Hamiltonian is given by the sum of $- {\bf
p_-}$ and the Wess-Zumino term \bea \label{H} \H &=& - {\bf p_-} +
\H_{WZ}\, ,\la{Hwz} \eea where \bea \nonumber
 \H_{WZ}&=&-\kappa{g\over 2}(G_+^2-G_-^2)\,
\str\left(\S_+ \chi\sqrt{1+\chi^2}\Kk F_\s^{st} \Kk\right)
\\
\nonumber &~&~~~~~~- \kappa{g\over 2}G_\mu G_\nu\,
\str\left(\S_+\S_\nu\chi\sqrt{1+\chi^2} \S_\mu \Kk F_\s^{st}
\Kk\right) \,.~~~~~~~ \eea Let us stress that, in this way, we
find the gauge-fixed Lagrangian as an {\it exact} function of  the
string tension $g$. The corresponding light-cone gauge-fixed
action is written in the standard form \bea\la{sgf} S_{gf} =
\int_{-r}^{ r}\, {\rm d}\s{\rm d}\tau\, \L_{gf}\,,\quad r =
P_+/2\,, \eea and its dependence on the total light-cone momentum
$P_+$ comes only through the integration bounds, as it was in the
bosonic case discussed in the previous subsection. Then it is
straightforward  to take the decompactification limit and get a
two-dimensional model on the plane. This will be discussed in
detail in the next section.

The gauge-fixed Lagrangian and Hamiltonian are obviously invariant
under the transformations generated by the SU(2)$^4$ bosonic
subgroup of the PSU(2,2$|$4) supergroup discussed in subsection
\ref{Sect:lrbs}  because  the subgroup commutes with the
$\alg{u}(1)$-isometries  corresponding to shifts of $t$ and
$\phi$, and, therefore, preserves the light-cone and
$\kappa$-symmetry gauge-fixing conditions.

Finally, the physical states should satisfy the level matching
condition which is obtained by integrating the Virasoro constraint
$C_1$ (\ref{C1}) over $\s$ \bea \nonumber &&\hspace{-1.0cm}\De
x_-= \int_{-r}^{ r}\, {\rm d}\s\, x_-'= - \int_{-r}^{ r}\, {\rm
d}\s\, \left( p_\mu x_\mu' -
\frac{i}{2}\,\str\left(\Sigma_+\chi\chi'\right)
+\frac{1}{2}g_\nu\bp_\mu\,\str\left(\left[\S_\nu,\Sigma_\mu\right]
B_\s\right) \right)\\\la{lmcf}
\eea The right hand side of the equation is equal to the
world-sheet momentum carried by the string, and, since we consider
the zero-winding number sector, it must vanish for  closed strings
\bea\nonumber \De x_-= p_\ws = 0\,. \eea

\section{Decompactification limit}

In this section we discuss  properties of the
light-cone string theory in the decompactification limit where the
total light-cone momentum $P_+$ goes to infinity, and one gets a
massive two-dimensional model defined on the plane. The resulting
model possesses multi-soliton solutions, and we construct the
simplest one-soliton solution and find its dispersion relation.
Then, we study the structure of the model in the large tension
perturbative expansion, perform its perturbative quantization, identify closed
sectors, and construct a perturbative world-sheet S-matrix which satisfies the
classical Yang-Baxter equation.

\subsection{From cylinder to plane}
The light-cone string sigma model Hamiltonian constructed in the
previous section describes a highly nonlinear two-dimensional
model defined on a cylinder, and it is obviously too complicated
to be quantized and solved exactly by using canonical methods. A
better way to address the spectral problem is to first consider
the states carrying very large light-cone momentum $P_+$, and then
to take into account the finite $P_+$ effects.

As was shown in the previous section, the light-cone string sigma
model action is  of the following form \bea\nonumber S = \int_{-r}^{ r}\,
{\rm d}\s\, {\rm d}\tau\, \L\,, \eea where $r=P_+/2$, and the
Lagrangian density $\L$ depends on the string tension $g$, but it
has no dependence on the light-cone momentum $P_+$. The light-cone
model is defined on a cylinder, and this is reflected in  the
periodic boundary conditions imposed  on the bosonic and fermionic
fields entering the Lagrangian. A physical configuration
corresponding to a closed string must satisfy the level-matching
condition which is equivalent to the vanishing of its world-sheet
momentum.

The specific dependence of the action on the light-cone momentum
$P_+$ allows one to define the decompactification or infinite
light-cone momentum limit. In this limit the circumference of the
cylinder tends to infinity, while the string tension is kept
fixed, and one is left with a non-trivial interacting model
defined on the plane. The periodic boundary conditions turn into
the vanishing ones because one is interested in string states with
finite world-sheet energy. Since $H=E-J$ is finite, and
$P_+=(1-a)J + a E \to\infty$, the charge $J$ also goes to infinity
in the decompactification limit.

\vskip 0.7cm \noindent
\begin{figure}
\begin{minipage}{\textwidth}
\begin{center}
\includegraphics[width=0.70\textwidth]{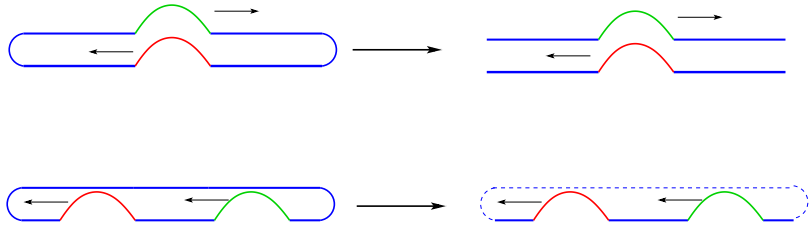}
\end{center}
\begin{center}
\parbox{5in}{\footnotesize{\caption{\label{SolDec} Solitonic excitations of a closed string in the decompactification limit.}}}
\end{center}
\end{minipage}
\end{figure}

\vskip -1.4cm

The resulting model appears to be non-Lorentz invariant but it has
massive spectrum, and, therefore, the notion of particles and
their scattering matrix is well-defined. Moreover, this model is
expected to be integrable at the quantum level, and hence
multi-body interactions should factorize into a sequence of
two-body events. Thus, in the decompactification limit the problem
of solving the theory reduces to three steps: first identify the
asymptotic spectrum, {\it i.e.} elementary excitations and their
bound states, second find the dispersion relations for all the
particles and, finally, determine the two-body S-matrices. It is
worth stressing, however, that in order to be able to consider
particles with arbitrary momenta, one should go off-shell, {\it
i.e.} to give up the level-matching condition and allow for
unphysical configurations that do not correspond to closed
strings. As a result, some quantities, {\it e.g.} the world-sheet
S-matrix, acquire a mild gauge dependence.

At leading order in the large tension expansion the light-cone
model describes eight free bosons and eight free fermions of equal
mass. The quadratic action, in fact, coincides with the light-cone
action for superstrings in the plane-wave background and, for this
reason, the large tension expansion is sometimes called the near
plane wave one. This expansion is rather peculiar because one can
easily perform the perturbative quantization of the light-cone
model and perturbatively compute the world-sheet S-matrix that
describes scattering of massive bosons and fermions.  In order to
determine the exact S-matrix, one has to use more sophisticated
methods to be developed in chapter 3.

An interesting feature of the light-cone string sigma model in the
decompactification limit is that it admits (multi-)soliton
solutions, see Figure \ref{SolDec}. Below we discuss the simplest
one-soliton solution called the giant magnon.

\subsection{Giant magnon}\la{magn}

To construct classical solutions of the light-cone string sigma
model  it is sufficient to consider only  its bosonic part. In
general a solution may involve several fields from both the AdS
and S$^5$ parts of the background geometry. Simplest solutions
would obviously depend only on one field, and one can show that a
solution with a finite energy which can therefore be called a
soliton exists only if one takes a field from the five-sphere.

The corresponding part of the gauge-fixed string action is
obtained from the Hamiltonian (\ref{Hb}) by setting to zero all
the fields but one, say $y_1$, from the S$^5$ part of the action.
One can easily check that it is a consistent reduction of the
light-cone model.  Then, it is convenient to make the following
change of variables \bea\nonumber z = {y_1\ov 1 + {y_1^2\ov 4}}\,.
\eea In the conformal gauge the corresponding reduction of the
string sigma model is to strings moving in the $\mathbb{R}\times
{\rm S}^2$ part of the $\AdS$ background. In terms of the angle
coordinate $\p$ and $z$ the metric of S$^2$ takes the form
\bea\nonumber ds^2_{{\rm S}^2} = {dz^2\ov 1-z^2} + (1-z^2)
d\p^2\,. \eea The coordinate $z$ is related to the angle $\theta$
as $z = \cos \theta$. The values $z=\pm 1$ correspond to the north
and south poles of the sphere,  and at $z=0$ the angle $\phi$
parametrizes the equator.

The light-cone Hamiltonian depends on the string tension, and it
is convenient to rescale the world-sheet coordinate $\s$ as $\s\to
g\s$. Then, the light-cone action takes the following form \bea
\la{Su1} S = g\int_{- \infty}^{ \infty}\, {\rm d}\s{\rm d}\tau\,
\left( p_z \dot{z} \,-\, \H \right)\, , \eea where the density of
the gauge-fixed Hamiltonian is a function of the coordinate $z$
and its canonically conjugate momentum $p_z$, but it has no
dependence on the string tension $g$. Explicit expressions for the
Hamiltonian and other quantities computed in this subsection can
be found in appendix \ref{app:magnon} where we also present their
forms for the three simplest cases $a=0,1/2,1$.

\medskip

To find soliton solutions of the gauge-fixed string theory, it is
convenient to go to the Lagrangian description by eliminating the
momentum $p_z$. Solving the equation of motion for $p_z$ that
follows from the action (\ref{Su1}), we determine the momentum as
a function of $\dz$ and $z$. Then substituting the solution into
(\ref{Su1}), we obtain the action in the Lagrangian form:
$S=S(z,z',\dz)$. The explicit form of the action is given in
Appendix \ref{app:magnon}, and it is of the Nambu-Goto form. We
will see in a moment that this leads to the existence of
finite-energy singular solitons.

To find a one-soliton solution, we make the most general ansatz
describing a wave propagating along the string \bea \la{ansatz} z
= z(\s - v\,\tau)\,, \eea where $v$ is the velocity of the
soliton. Substituting the ansatz into the action (\ref{Su2}) from
Appendix
 \ref{app:magnon} , we derive the Lagrangian, $L_{red}=L_{red}(z,z')$, of a reduced
model which defines a one-particle system if we regard $\s$ as a
time variable.  The $\s$-evolution of this system can be easily
determined by introducing the ``momentum'' conjugate to $z$ with
respect to ``time'' $\s$ \bea\nonumber \pi_z = {\pa L_{red}\ov \pa
z'} \,, \eea and computing the reduced Hamiltonian \bea\nonumber
H_{red} =\pi_z z' - L_{red} \,. \eea The reduced Hamiltonian is a
conserved quantity with respect to $\s$. Since the coordinate $z$
of the soliton satisfies vanishing boundary conditions,
$z(\pm\infty)=z'(\pm\infty)=0$,  we conclude that \bea\nonumber
H_{red} = 0\, .  \eea Solving this equation with respect to $z'$,
we get the following basic equation \bea\la{zp} z'^2 =
\left(\frac{1-z^2}{1- (1-a)z^2}\right)^2\, \frac{ z^2}{1-v^2-z^2}
\ , \eea which can be easily integrated in terms of elementary
functions.

It is not difficult to see that  a solution with finite energy
exists if the following inequalities hold \bea\la{rang} 0\le a\le
1\,,\quad 0\le |v| \le 1\,. \eea Then, assuming for definiteness
that $z\ge 0$, the corresponding solution of the  equation
(\ref{zp}) lies between $0$ and $z_{max} = \sqrt{1-v^2}$. One can
easily see from equation (\ref{zp}) that in the range of
parameters (\ref{rang}) the shape of the soliton is similar for
any values of $a$ and $v$. The allowed values of $z$ are $0\le
z\le z_{max}$, and $z'$ vanishes at $z=0$, and goes to infinity at
$z=z_{max}$.

The corresponding solution is, as we see, not smooth at $z=
z_{max}$. The energy of this soliton is however  finite. To
compute the energy, we need to evaluate $\H/|z'|$ on the solution:
\bea\nonumber {\H\ov |z'|} = { z\ov\sqrt{z_{max}^2 - z^2}}\,. \eea
Then the soliton energy  is given by the following integral
\bea\la{energya} E-J = g\int_{- \infty}^{ \infty}\, {\rm d}\s\, \H
= 2g\int_{0}^{z_{max}}\, {\rm d}z {\H\ov |z'|} =
2g\,\sqrt{1-v^2}\,. \eea Finally, to find the dispersion relation
we also need to compute the world-sheet momentum (\ref{pws})
\bea\la{pwsu} \pws = -\int_{-\infty}^{\infty}  {\rm d}\s p_z z' =
2\int_{0}^{z_{max}}\, {\rm d}z |p_z|\,, \eea where we have assumed
that $v>0$, and took into account that for the soliton we consider
the expression $-p_z z'$ is positive. The following explicit
formula for the momentum $p_z$ canonically conjugate to $z$ can be
easily found by using eqs. (\ref{pza}),(\ref{ansatz}) and
(\ref{zp}) \bea\la{pzu} p_z = \frac{v
z}{(1-z^2)\sqrt{z_{max}^2-z^2}}\,. \eea Computing the world-sheet
momentum \bea \la{pQ0}\nonumber \pws =2\arccos v\,, \eea and
expressing $v$ in terms of $\pws$, we derive the giant magnon
dispersion relation \bea\la{gmdr} E-J = 2g\left|\sin {\pws\ov 2}
\right|\,. \eea The dispersion relation explicitly shows  that the
light-cone model is not Lorentz invariant.  It appears to be
independent of the gauge parameter $a$. Note also the appearance
of trigonometric functions which are usually associated with a
lattice structure, while here the dispersion relation was obtained
for a continuum model. The dispersion relation was derived in
classical theory, {\it i.e.} in the limit of large string tension
$g$ and finite world-sheet momentum $\pws$. In the quantum theory
it gets modified, and we will discuss the exact dispersion
relation in the next chapter.

\medskip

Let us finally mention that in the case of a one-soliton solution
the world-sheet momentum (\ref{pwsu}) is just equal to the
canonical momentum carried by the center of mass of the soliton.
To see that we just need to plug the ansatz (\ref{ansatz}) into
the string action (\ref{Su1}), and integrate over $\s$. Then we
obtain the following action for a point particle
$$
S = g \int {\rm d}\tau\, \left( \pws\, v \,-\, \H \right)\,,
$$
that explicitly shows  that $\pws$ is the soliton momentum.

\subsection{Large string tension expansion}\la{sec:lste}

In this subsection we discuss the large string tension expansion
and perturbative quantization of the light-cone gauge-fixed action
(\ref{sgf}) in the decompactification limit. To develop the
expansion, we first note that the string tension $g$ in the
gauge-fixed Lagrangian (\ref{Lgf}) is always accompanied by a
$\s$-derivative of a field. Thus, rescaling\footnote{Performing
the  rescaling with finite $P_+$ changes the integration bounds in
(\ref{sgf})  as $r\to  P_+/2g$.} the coordinate $\s$ as $\s\to
g\s$ removes the $g$-dependence from the Lagrangian, and the
light-cone action takes the form \bea\la{sgf2} S_{gf} = g \int
{\rm d}\s  {\rm d}\tau\, \L_{gf}\,, \eea where $\L_{gf}$ is given
by (\ref{Lgf}) with $g=1$. Finally, one rescales all the fields
appearing in (\ref{sgf2}) as  follows \bea\la{resc} x_\mu\to
x_\mu/\sqrt{g}\,,\quad p_\mu\to p_\mu/\sqrt{g}\,,\quad \chi\to
\chi/\sqrt{g}\,, \eea and expands the action (\ref{sgf2}) in
powers of $1/g$ \bea\la{sgf3} S_{gf} = \int  {\rm d}\s  {\rm
d}\tau\,\left( \L_2 + {1\ov g} \L_4  + {1\ov g^2} \L_6+\cdots
\right)\,, \eea where $\L_2$ is quadratic in the fields,  $\L_4$
is quartic, and so on.

It is worth mentioning that the rescaling (\ref{resc}) implies the
following rescaling of the world-sheet momentum of a state
\bea\nonumber
\pws = -  \int  {\rm d}\s  \left( p_\mu x_\mu' + \cdots\right)= {1\ov g}\,p\,,
\eea
where $p$ is the rescaled world-sheet momentum given by the same formula $p=-  \int  {\rm d}\s  \left( p_\mu x_\mu' + \cdots\right)$ in terms of the rescaled coordinates and momenta. It is kept fixed in the large tension expansion and, therefore,  one considers states with very small
world-sheet momenta $\pws$ of order $1/g$.

In principle it is straightforward to expand the light-cone
Lagrangian (\ref{Lgf}) and find the quadratic and quartic
Lagrangians.  The quadratic  Lagrangian appears to be of the
following simple form \bea\la{L2} \L_2 = p_\mu\dot{x}_\mu -
\frac{i}{2}\str\left(\Sigma_+\chi\dot\chi\right) - \H_2\,, \eea
where the first two terms with time derivatives define the
standard Poisson structure for the bosons and fermions, and $\H_2$
is the density of the quadratic Hamiltonian \bea\la{Hquadr2} \H_2
= {1\ov 2} p_\mu^2 + {1\ov 2}x_\mu^2 +{1\ov 2}x_\mu'^2  -
{\kappa\ov 2} \str \left(\S_+ \chi\Kk\chi'^{st} \Kk\right) + {1\ov
2}\str\,\chi^2\, . \eea The quadratic  Lagrangian obviously
describes a Lorentz-invariant theory of eight massive bosons and
eight massive fermions with masses equal to unity. It can be
easily canonically quantized as we describe in the next
subsection.

\medskip

The quartic Lagrangian obtained just by expanding (\ref{Lgf}),
however, has two unpleasant properties. First of all, it contains
terms depending on the time derivatives of the fields which come
from the interacting part of the kinetic Lagrangian (\ref{Lkin}).
These terms modify the Poisson structure and make quantizing the
model more complicated. One should remove these terms by
redefining the fields.

To find the necessary field redefinition,  we notice that the
kinetic Lagrangian (\ref{Lkin}) can be written in the following
form \bea \label{Lkin2} \L_{kin} &=& p_\mu\dot{x}_\mu -
\frac{i}{2}\str\left(\Sigma_+\chi\dot\chi\right)
+\frac{i}{g}\str\left(\Sigma_+\Phi(p,x,\chi)\dot\chi\right)
\,.~~~~~~~ \eea where $\Phi$ is a function of at least cubic order
in physical fields. It is then clear that the last term can be
removed by the following redefinition of $\chi$ \bea\la{chired}
\chi\rightarrow \chi + \frac{1}{g}\Phi(p,x,\chi)\,. \eea This
redefinition casts the kinetic term (\ref{Lkin2}) into the form
(up to a total derivative) \bea \label{Lkin3} &&\L_{kin} = p_\mu
\dot{x}_\mu  -
\frac{i}{2}\str\left(\Sigma_+\chi\dot\chi\right)\\\nonumber
&&+\frac{i}{g}\str\left(\Sigma_+\left(\Phi(p,x,\chi+\frac{1}{g}\Phi)
- \Phi(p,x,\chi)\right)\dot\chi\right)
+\frac{i}{2g^2}\str\left(\Sigma_+\Phi(p,x,\chi)\dot\Phi(p,x,\chi)\right)
. \eea Since $\Phi$ is at least of cubic order in the fields, the
terms on the second line of (\ref{Lkin3}) are at least of  sixth
order. These terms can be also removed by a similar field
redefinition. However, this time one would need to redefine not
only the fermions but also the bosonic coordinates $x_\mu$ and
$p_\mu$. For our purposes here it is sufficient to perform only
the simplest redefinition (\ref{chired}), and just drop the terms
on the second line of (\ref{Lkin3}). This reduces the kinetic term
to the canonical quadratic form which enters the quadratic
Lagrangian (\ref{L2}). Since the redefinition removes all the time
derivative terms from the quartic Lagrangian, the latter becomes
equal up to the minus sign to the quartic Hamiltonian: $\L_4 = -
\H_4$.

\medskip

It is also necessary to mention an important and nice property of
the redefinition (\ref{chired}). One can check that up to sixth
order in fields, the formula (\ref{lmcf}) for $x_-'$ takes the
form \bea \la{xmpred} x_-' = - {1\ov g}\left( p_\mu x_\mu' -
\frac{i}{2}\,\str\left(\Sigma_+\chi\chi'\right) +\pa_\s
f(p,x,\chi)\right) \, , \eea where $f(p,x,\chi)$ is a function of
the momenta and coordinates of at least fourth order in the
fields. Thus, we see that integrating (\ref{xmpred}) over $\s$, we
get the usual ``flat space'' form of the level-matching condition
and world-sheet momentum \bea \De x_-= \int_{-\infty}^{ \infty}\,
{\rm d}\s\, x_-'=\pws =  {p\ov g}
 =  - {1\ov g}
\int_{-\infty}^{\infty}\, {\rm d}\s\, \left( p_\mu x_\mu' -
\frac{i}{2}\,\str\left(\Sigma_+\chi\chi'\right)
 \right)\,. \la{lmcf2}
\eea Let us stress again that even though for physical states the
total world-sheet momentum must vanish, to define asymptotic
states and the scattering matrix we should consider states with
arbitrary world-sheet momenta.

\medskip

The second unpleasant property of the quartic Lagrangian (and
Hamiltonian) is that it contains bosonic terms of the form $p^2
x^2$ which do not depend on the space derivatives. These terms,
however, can be removed by a proper canonical transformation. The
final form of the quartic Hamiltonian is
 \bea\nonumber
\H_4&=&{1\ov 4}\Big[\, 2 \left(y'^2z^2-z'^2y^2+ z'^2 z^2 - y'^2
y^2\right)\\\nonumber
 &-&\str\left({1\ov 2}\chi\chi'\chi\chi' +{1\ov 2} \chi^2\chi'^2+
{1\ov 4}\left[\chi,\chi'\right]\Kk \left[\chi,\chi'\right]^t\Kk
 +
\chi\Kk\chi'^{st} \Kk \chi\Kk\chi'^{st} \Kk \right) \\\nonumber
&+& \str\left( (z^2-y^2)\chi'\chi' +{1\ov 2} x_\mu' x_\nu
\left[\S_\mu,\S_\nu\right]\left[\chi,\chi'\right] -2 x_\mu
x_\nu\S_\mu\chi'\S_\nu\chi'\right)
\\\la{H41}
&+&{i\kappa\ov 4}\,x_\nu p_\mu\str\left(
\left[\S_\nu,\Sigma_\mu\right]\left[\Kk\chi^{st}
\Kk,\chi\right]'\right)  \Big]  \, . \eea The computation of the
quartic Hamiltonian is rather involved, and we refer the reader to
the original literature for details.   The quadratic and the
quartic Hamiltonians can also be written in terms of the bosonic
and fermionic matrices $\bX$ and $\chi$, see   (\ref{Xlc}) and
(\ref{chilc}), as follows \bea \la{H2matr} \H_2 &=& {1\ov 2}
\str\left(\bp^\perp\bp^\perp + \bX\bX +\bX'\bX'  -
\kappa \S_+ \chi\Kk\chi'^{st} \Kk+ \chi\chi\right)\, ,\\
\la{H4matr}
\H_4&=&{1\ov 2}\,\str\, \Upsilon\bX\bX\,  \str\,  \bX' \bX' + {1\ov 4}\,\str\, \Upsilon\bX\bX\,  \str\, \chi'\chi' \\\nonumber
&-&
\str\left({1\ov 2}
\left[\bX,\bX'\right]\left[\chi,\chi'\right] +2\bX\,\chi'\,\bX\,\chi' -{i\kappa\ov 4}
\left[\bX,\bp^\perp\right]\left[\Kk\chi^{st} \Kk,\chi\right]' \right)  \\\nonumber
 &-&\str\left({1\ov 8}\chi\chi'\chi\chi' +{1\ov 8} \chi^2\chi'^2+
{1\ov 16}\left[\chi,\chi'\right]\Kk \left[\chi^t,\chi'^t\right]\Kk
 +
{1\ov 4}\chi\Kk\chi'^{st} \Kk \chi\Kk\chi'^{st} \Kk \right) \,
,\eea where $\Upsilon={\rm diag}(\mI_4, -\mI_4)$, and the momentum
$\bp^\perp= {1\ov 2}p_\mu\Sigma_\mu$  has the following form in
terms of  two-index fields
 \bea\la{Plc}
\bp^\perp= {1\ov 2}\left(\begin{array}{cccc|cccc} 0 & 0 &- P_{4\dot{3}}
& -P_{4\dot{4}} & 0 & 0 & 0 & 0 \\
0 & 0 & P_{3\dot{3}}& P_{3\dot{4}}  & 0 & 0 & 0 & 0 \\
P_{3\dot{4}} & P_{4\dot{4}} & 0 & 0 & 0 & 0 & 0 & 0 \\
-P_{3\dot{3}} & -P_{4\dot{3}} & 0 & 0 & 0 & 0 & 0 & 0 \\
\hline
0 & 0 & 0 & 0 & 0 & 0 &- iP_{2\dot{1}} & -iP_{2\dot{2}}\\
0 & 0 & 0 & 0 & 0 & 0 & iP_{1\dot{1}} & iP_{1\dot{2}} \\
0 & 0 & 0 & 0 & iP_{1\dot{2}} & iP_{2\dot{2}} & 0 & 0 \\
0 & 0 & 0 & 0 & -iP_{1\dot{1}} & -iP_{2\dot{1}} & 0 & 0
\end{array}\right)\, .
\eea The momenta $P_{a\da}$ and $P_{\a\dal}$ are canonically
conjugate to $Y^{a\da}$ and $Z^{\a\dal}$, and $\bp^\perp$
satisfies the relation $\str\, \bp^\perp \dot{\bX}=p_\mu
\dot{x}^\mu  = P_{a\da}\dot{Y}^{a\da} +
P_{\a\dal}\dot{Z}^{\a\dal}$. This form also makes  the invariance
of the Hamiltonians  under the transformations generated by the
SU(2)$^4$ subgroup of PSU(2,2$|$4) manifest.

\medskip

Summarizing the discussion in this subsection, we conclude that by
means of proper field redefinitions at each order of the large $g$
expansion the light-cone gauge-fixed Lagrangian (\ref{Lgf}) can be
brought to the following canonical form \bea\la{Lexp} \L_{gf} =
 \str\left(\bp^\perp \dot{\bX} -
\frac{i}{2}\Sigma_+\chi\dot\chi\right) - \H_2 - {1\ov g}\H_4-
{1\ov g^2}\H_6 - \cdots\,, \eea where the interaction part does
not contain terms with the time derivatives, and also terms which
do not depend on the space derivatives. Perturbative quantization
of the model can be performed in the canonical way by using the
quadratic part of the Lagrangian which describes eight massive
bosons and eight massive fermions. The quartic Hamiltonian can be
then used to compute the tree-level two-particle world-sheet
scattering matrix.

\subsection{Quantization}
\label{subsect:quantization}

We now turn to the perturbative quantization of the light-cone
$\AdS$ superstring in the large tension expansion. We start with
rewriting the quadratic Lagrangian density in terms of the
two-index fields, see eqs.(\ref{Xlc}),  (\ref{chilc}) and
(\ref{Plc}). \bea\la{L2b} \L_2 = P_{a\da}\dot{Y}^{a\da}  +
P_{\a\dal}\dot{Z}^{\a\dal} +
i\,\eta_{\a\da}^\dagger\dot{\eta}^{\a\da}+
i\,\theta_{a\dal}^\dagger\dot{\theta}^{a\dal} - \H_2\,, \eea where
the density of the quadratic Hamiltonian is given by \bea\la{dH2}
\H_2 &=&{1\ov 4}P_{a\da} P^{a\da} +Y_{a\da} Y^{a\da} +
Y'_{a\da}Y'^{a\da} + {1\ov 4}P_{\a\dal} P^{\a\dal} +Z_{\a\dal}
Z^{\a\dal} +Z'_{\a\dal}Z'^{\a\dal} \\\nonumber &+&
\eta_{\a\da}^\dagger\eta^{\a\da} + {\kappa\ov
2}\eta^{\a\da}\eta'_{\a\da} -{\kappa\ov 2}\eta^{\dagger\a\da}
\eta_{\a\da}'^\dagger + \theta_{a\dal}^\dagger\theta^{a\dal}  +
{\kappa\ov 2}\theta^{a\dal}\theta'_{a\dal} -{\kappa\ov
2}\theta^{\dagger a\dal} \theta_{a\dal}'^\dagger\,. ~~~~~~~\eea
Here $\theta_{a\dal}^\dagger$ and $\eta_{\a\da}^\dagger$ are
complex conjugate of $\theta^{a\dal}$ and $\eta^{\a\da}$,
respectively, and we lower and raise the indices by using the
$\eps$-tensor \bea Y_{a\da}
=\eps_{ab}\eps_{\da\db}Y^{b\db}\,,\quad P^{a\da}
=\eps^{ab}\eps^{\da\db}P_{b\db}\,,\quad \eta_{\a\da}
=\eps_{\a\b}\eps_{\da\db}\eta^{\b\db}\,,\quad \eta^{\dagger\a\da}
=\eps^{\a\b}\eps^{\da\db}\eta^\dagger_{\b\db}\,, \eea and similar
formulae for $Z_{\a\dal}\,,\  P^{\a\dal}\,,\  \theta_{a\dal}\,,\
\theta^{\dagger a\dal} $. The reality condition for these bosonic
and fermionic fields then takes the following form
$$
\left(Y^{a\da}\right)^\dagger = Y_{a\da}\,,\quad \left(P_{a\da}\right)^\dagger = P^{a\da}\,,\quad \left(\eta_{\a\da}\right)^\dagger = \eta^{\dagger \a\da}\,.
$$
The canonical equal-time (anti)commutation relations for the fields can be
now easily read off from
 (\ref{L2b})
 \bea\nonumber
&&\hspace{-0.7cm}[\, Y^{a\da}(\s,\tau)\,  ,\, P_{b\db}(\s',\tau) \, ] = i\, \delta^a_{b}
\delta^\da_{\db}\delta(\s-\s') \,,\quad
[\, Z^{\a\dal}(\s,\tau)\,  ,\, P_{\b\dbe}(\s',\tau) \, ] = i\, \delta^\a_{\b} \delta^\dal_{\dbe}\delta(\s-\s') \,,\\\nonumber
&&\hspace{-0.7cm} \{\, \theta^{a\dal}(\s,\tau)\, ,\, \theta_{b\dbe}^\dagger(\s',\tau)\,
\}=\delta^a_{b} \delta^\dal_{\dbe}\delta(\s-\s')\,,\quad  \{\,
\eta^{\a\da}(\s,\tau)\, ,\, \eta_{\b\db}^\dagger(\s',\tau)\,
\}=\delta^\a_{\b} \delta^\da_{\db}\delta(\s-\s') \, ,~~~ \eea and we
just  need to establish a mode decomposition of the bosonic and
fermionic fields which renders the quadratic Lagrangian
(\ref{L2b}) in a diagonal form.

We set  $\kappa=1$  for definiteness, and
 choose the following mode decompositions for the bosonic fields
\bea\la{bosrep}\begin{aligned} &~~~~Y^{a\da}(\s,\tau) = {1\ov
\sqrt{2\pi}}\int\,{\rm d}p\,{1\ov 2\sqrt{\om_p}}\left( e^{ip\s}
a^{a\da}(p,\tau) + e^{-ip\s}\eps^{ab}\eps^{\da\db}
a_{b\db}^\dagger(p,\tau)\right)\,,~~~~~~~~
\\ &~~~~P_{a\da}(\s,\tau) = {1\ov \sqrt{2\pi}}\int\,{\rm d}p\,\,i\,
\sqrt{\om_p}\left( e^{-ip\s} a_{a\da}^\dagger(p,\tau)-
e^{ip\s}\eps_{ab}\eps_{\da\db} a^{b\db}(p,\tau)\right)\,,\\
&~~~~Z^{\a\dal}(\s,\tau) = {1\ov \sqrt{2\pi}}\int\,{\rm d}p\,{1\ov
2\sqrt{\om_p}}\left( e^{ip\s} a^{\a\dal}(p,\tau) +
e^{-ip\s}\eps^{\a\b}\eps^{\dal\dbe}
a_{\b\dbe}^\dagger(p,\tau)\right)\,, \\ &~~~~P_{\a\dal}(\s,\tau) =
{1\ov \sqrt{2\pi}}\int\,{\rm d}p\,i\, \sqrt{\om_p}\left( e^{-ip\s}
a_{\a\dal}^\dagger(p,\tau)- e^{ip\s}\eps_{\a\b}\eps_{\dal\dbe}
a^{\b\dbe}(p,\tau)\right)\,,
\end{aligned}\eea and similarly for fermionic ones
\bea\la{ferrep}\begin{aligned} &~~~~~\theta^{a\dal}(\s,\tau)=
{e^{-i\pi/4}\ov \sqrt{2\pi}}\int\,{{\rm
d}p\ov\sqrt{\om_p}}\,\left( e^{ip\s}\,f_p\, a^{a\dal}(p,\tau)
+e^{-ip\s}\,h_p\,\eps^{ab}\eps^{\dal\dbe}
a_{b\dbe}^\dagger(p,\tau)\right)\,,\\
 &~~~~~\eta^{\a\da}(\s,\tau)=
{e^{-i\pi/4}\ov \sqrt{2\pi}}\int\,{{\rm
d}p\ov\sqrt{\om_p}}\,\left( e^{ip\s}\,f_p\, a^{\a\da}(p,\tau)
+e^{-ip\s}\,h_p\,\eps^{\a\b}\eps^{\da\db}
a_{\b\db}^\dagger(p,\tau)\right) \,. ~~~~~~~~\end{aligned}\eea
Here the creation $a_{M\dot{M}}^\dagger$ and annihilation
$a^{M\dot{M}}$ operators are conjugate to each other:
$\big(a^{M\dot{M}}\big)^\dagger = a_{M\dot{M}}^\dagger$, where
$M=1,\ldots,4$ and ${\dot M} ={\dot 1},\ldots,{\dot 4}$;   the
frequency is $\om_p = \sqrt{1+p^2}$, and the quantities
 \bea\nonumber f_p = \sqrt{{\om_p+1\ov 2}}\,,\quad h_p =
{ p\ov 2f_p}\,,\qquad f_p^2-h_p^2=1\,,\quad f_p^2+h_p^2=\om_p \,,
\eea
play the role of the fermion wave functions.
In what follows we always use capital Latin letters $M,N,\ldots$ and ${\dot M},{\dot N},\ldots$  to denote superindices  $M=( a|\a )$,
and  $\dM=( \da|\dal )$, where the lower-case Latin indices are even and the greek indices are odd. Thus, the grading of $M$, $\dM$ is defined to be
$\eps_a =\eps_\da = 0$ and $\eps_\a=\eps_\dal = 1$.

For the sake of simplicity, we will not explicitly show  the time
dependence in all the operators everywhere where it cannot lead to
any confusion. Then, omitting total derivative terms, the
quadratic Lagrangian indeed takes the diagonal form \bea\nonumber
L_2 =\int\,{\rm d}\s\, \L_2 =\int\,{\rm d}p\,
\sum_{M,\dot{M}}\left(
i\,a_{M\dot{M}}^\dagger(p)\dot{a}^{M\dot{M}}(p)  -\om_p\,
a_{M\dot{M}}^\dagger(p)a^{M\dot{M}}(p)\right)\,, \eea which shows
explicitly that the creation and annihilation operators satisfy
the canonical equal-time (anti-)commutation relations
\bea\la{comrel} [\, a^{M\dot{M}}(p,\tau)\,  ,\,
a_{N\dot{N}}^\dagger(p',\tau) \, \} = \delta^M_{N}\,
\delta^{\dot{M}}_{\dot{N}}\, \delta(p-p') \, , \eea where we take
the commutator for bosons, and the anti-commutator for fermions.

\medskip

The quadratic Hamiltonian is, therefore, of the standard harmonic
oscillator form \bea\la{hquad} \bH_2 = \int\,{\rm d}p\,
\sum_{M,\dot{M}}\, \om_p\,
a_{M\dot{M}}^\dagger(p)a^{M\dot{M}}(p)\,, \eea and its generic
$Q$-particle state  can now be created by acting with  creation
operators on the vacuum \bea\la{gstate} |\Psi \rangle
=a_{M_1\dot{M}_1}^\dagger(p_1)\, a_{M_2\dot{M}_2}^\dagger(p_2)\,
\cdots\,  a_{M_Q\dot{M}_Q}^\dagger(p_Q)\,
 |0 \rangle\,,
\eea where we may assume that the momenta are ordered as follows
$$
p_1> p_2 >\cdots> p_{Q-1}> p_Q \,.
$$
The energy of this state is obviously equal to \bea\nonumber
\bH_2|\Psi \rangle = E|\Psi \rangle\,,\quad E= \sum_{i}
\om_{p_i}\,. \eea This state is also an eigenvector of the
world-sheet momentum operator which takes the following form
\bea\nonumber \bP\, \equiv\pws &=&  -{1\ov g} \int\, {\rm d}\s\,
\left( P_{a\da}Y'^{a\da} + P_{\a\dal}Z'^{\a\dal} + i
\theta_{\a\da}^\dagger\theta'^{\a\da} +
i\eta_{a\dal}^\dagger\eta'^{a\dal}
 \right)\\\la{pwsm}
  &=& {1\ov g} \int\,{\rm d}p\, \sum_{M,\dot{M}}\, p\, a_{M\dot{M}}^\dagger(p)a^{M\dot{M}}(p) \,.
\eea A physical string state must also satisfy the level-matching
condition implying that its world-sheet momentum vanishes
\bea\nonumber \bP\, |\Psi \rangle = 0\ \Rightarrow \sum_{i}p_i =
0\, . \eea Nevertheless, to understand the general properties of
the scattering matrix we would need to consider states with
arbitrary momenta.

\medskip

The tree-level two-particle scattering matrix is determined by the
quartic Hamiltonian $\bH_{4}$ that we take to be normal-ordered
with respect to these bosonic and fermionic oscillator modes. Its
expression in terms of the two-index fields is given in appendix
\ref{H4double}.

\subsection{Closed sectors}

It is clear that there are 16 one-particle states of different
flavors, and, therefore, the two-particle scattering matrix is a
$(16\times 16)\times(16\times 16)$ matrix. The S-matrix is not
diagonal, and in the scattering process particles can exchange
their momenta and flavors.  The model is believed to be
integrable, and the multi-particle scattering matrix can be
expressed through a product of the two-particle ones. There are,
however, groups of particles of definite flavors which can scatter
only among themselves.  They are said to form closed sectors.

The simplest way to identify closed sectors is to use the fact
that all the 16 particles are charged under the bosonic $\su(2)\oplus \su(2)\oplus \su(2)\oplus \su(2)$ subalgebra
of the symmetry algebra of the
light-cone model, and the total charges carried by the scattering
states are preserved in the scattering process. Let us recall that
 two $\su(2)$'s belong to
$\su(4)\subset\psu(2,2|4)$,  and act on the undotted and dotted
lower-case latin indices $a,b,\da,\db,\ldots$ which take the
values $1, 2$ and $\dot{1},\dot{2}$, and  that the other two
$\su(2)$'s belong to $\su(2,2)\subset \psu(2,2|4)$,  and act on
the undotted and dotted greek indices $\a,\b,\dal,\dbe,\ldots$
which take the values $3, 4$ and $\dot{3},\dot{4}$. Thus, the
bosonic fields with all latin indices come from the five-sphere,
and those with all greek  indices come from the AdS part of the
string sigma model. Below we describe some closed sectors.

\subsubsection*{\bf $\su(2)$ sector}
\noindent The  $\su(2)$ sector is a rank-one sector which consists
of bosonic particles of  type $a_{1\dot{1}}^\dagger$ originating
from the five-sphere of $\AdS$, and a generic  $Q$-particle state
from the sector is of the form \bea\la{gstatesu2} |\Psi_{\su(2)}
\rangle =a_{1\dot{1}}^\dagger(p_1)\, a_{1\dot{1}}^\dagger(p_2)\,
\cdots\, a_{1\dot{1}}^\dagger(p_Q)\,
 |0 \rangle\,.
\eea These states can obviously scatter only among themselves
because they carry the maximum charges $Q/2,Q/2$ with respect to
$\su(2)\oplus\su(2)\subset \su(4)$. The $\su(2)\oplus\su(2)$
algebra is isomorphic to the $\so(4)$ that rotates the four
coordinates $y_i$ from S$^5$, see section \ref{App:emcoord} for
detail, and a $a_{1\dot{1}}^\dagger$ particle carries  charge $1$
with respect to the $\alg{o}(2)\sim \alg{u}(1)$ which rotates the
$y_1y_2$-plane, and  charge 0 with respect to the $\alg{o}(2)\sim
\alg{u}(1)$ which rotates the $y_3y_4$-plane.

Field theory operators dual to the states (\ref{gstatesu2}) with
vanishing total world-sheet momentum can be easily identified. To
this end we assume that the light-cone momentum $P_+ = {1\ov
2}(E+J)$ is very large but not infinite. Recall that $J$ is the
charge associated to the U(1) generating  shifts of the angle
$\phi$ of ${\rm S}^5$. Then,
 the charge $J$ is also large, and it is assigned to
the light-cone vacuum and {\it no} creation and annihilation
operator carries charges under this U(1). Thus, the states
(\ref{gstatesu2}) are the lightest states which  only carry the
two charges $J$ and $Q$, and they should be dual to the \N SYM
operators  of the form \bea\la{su2oper} O_{\su(2)} = \tr \left(
Z^J X^Q + \mbox{permutations} \right)\,, \eea where $Z$ and $X$
are the two complex gauge theory scalars which carry one unit of
the charges $J$ and $Q$, respectively. Note that there is an
$\su(2)$ algebra which rotates the two complex scalars $Z,X$, and
this explains why the sector is called the $\su(2)$ sector. It is
clear that the particles of  type $a_{2\dot{2}}^\dagger$ form
another closed $\su(2)$ sector.

\subsubsection*{\bf $\sl(2)$ sector}
\noindent The  $\sl(2)$ sector is a rank-one sector consisting of
bosonic particles of  type $a_{3\dot{3}}^\dagger$  from the AdS
part of $\AdS$, and a generic  $\sl(2)$ sector $Q$-particle state
is \bea\la{gstatesl2} |\Psi_{\sl(2)} \rangle
=a_{3\dot{3}}^\dagger(p_1)\, a_{3\dot{3}}^\dagger(p_2)\, \cdots\,
a_{3\dot{3}}^\dagger(p_Q)\,
 |0 \rangle\,.
\eea These states scatter only among themselves because they carry
the maximum charges $Q/2,Q/2$ with respect to
$\su(2)\oplus\su(2)\subset \su(2,2)$. The $\su(2)\oplus\su(2)\sim
\so(4)$ rotates the four coordinates $z_i$ from AdS$_5$, and a
$a_{3\dot{3}}^\dagger$ particle carries  charges 1 and 0 with
respect to the two $\alg{o}(2)\sim \alg{u}(1)$'s which rotate the
$z_1z_2$- and $z_3z_4$-planes, respectively.

Thus, the states (\ref{gstatesl2}) are the lightest states which
only carry the two charges $J$ and $Q$, and they are dual to the
\N SYM operators  of the form \bea\la{sl2oper} O_{\sl(2)} = \tr
\left( D_-^Q Z^J + \mbox{permutations} \right)\,, \eea where $D_-$
is the covariant derivative in a light-cone direction carrying
unit charge under the $\alg{u}(1)$ subalgebra of $\su(2,2)$, that
in the string picture corresponds to the $\alg{o}(2)$ which
rotates the $z_1z_2$-plane.  The particles of  type
$a_{4\dot{4}}^\dagger$ obviously form another closed $\sl(2)$
sector.

\subsubsection*{\bf $\su(1|1)$ sector}
\noindent The  $\su(1|1)$ sector is a rank-one sector consisting
of fermionic particles of  type $a_{3\dot{1}}^\dagger$, and a
generic $\su(1|1)$ sector $Q$-particle state  is
\bea\la{gstatesu11} |\Psi_{\su(1|1)} \rangle
=a_{3\dot{1}}^\dagger(p_1)\, a_{3\dot{1}}^\dagger(p_2)\, \cdots\,
a_{3\dot{1}}^\dagger(p_Q)\,
 |0 \rangle\,.
\eea These states scatter only among themselves, and are dual to
the \N SYM operators  of the form \bea \label{gssu11} O_{\su(1|1)}
= {\rm tr}\big(Z^{J-\frac{Q}{2}} \Psi^Q  + \mbox{permutations}
\big)\,. \eea The fermion $\Psi$ is the highest weight component
of the gaugino from the vector multiplet of the gauge theory. The
gaugino $\Psi_{\a}$ belongs to the vector multiplet, it is neutral
under $\su(3)$ which rotates the three gauge theory complex
scalars among themselves, and it carries the same charge $1/2$
under any of the three ${\rm U}(1)$  subgroups of SU(4). Note also
that there is the second equivalent $\su(1|1)$ sector consisting
of fermionic particles of  type $a_{1\dot{3}}^\dagger$.

\subsubsection*{\bf $\su(1|2)$ sector}
\noindent The $\su(1|2)$ sector can be considered as the union of
the $\su(2)$ and $\su(1|1)$ sectors, because it consists of
particles of  types $a_{1\dot{1}}^\dagger$ and
$a_{3\dot{1}}^\dagger$, and a generic  $\su(1|2)$ sector
$Q$-particle state  is \bea\la{gstatesu12} |\Psi_{\su(1|2)}
\rangle =a_{3\dot{1}}^\dagger(p_1)\, a_{3\dot{1}}^\dagger(p_2)\,
\cdots\, a_{3\dot{1}}^\dagger(p_M)a_{1\dot{1}}^\dagger(k_1)\,
a_{1\dot{1}}^\dagger(k_2)\, \cdots\,  a_{1\dot{1}}^\dagger(k_K)\,
 |0 \rangle\,.
\eea Counting the charges carried by these states shows that they
scatter only among themselves, and the number of bosons and
fermions is unchanged in the scattering process.

These states obviously are dual to the \N SYM operators  of the
form \bea\la{su12oper} O_{\su(1|2)} = \tr \left( Z^{J-\frac{M}{2}}
\Psi^M X^K + \mbox{permutations}
 \right)\,,
\eea because, as was discussed above, the gauge theory fields $X$
and $\Psi$ correspond to the creation operators
$a_{1\dot{1}}^\dagger$ and $a_{3\dot{1}}^\dagger$, respectively.

\subsubsection*{ \bf $\su(2|3)$ sector}
\noindent The $\su(2|3)$ sector is the largest closed sector, and
it is an extension of the  $\su(1|2)$ sector. It involves two
bosonic particles of  types $a_{1\dot{1}}^\dagger$ and
$a_{2\dot{1}}^\dagger$, and two fermionic particles of  types
$a_{3\dot{1}}^\dagger$ and $a_{4\dot{1}}^\dagger$. A generic
$Q$-particle state in the $\su(2|3)$ sector  is
\bea
\nonumber ~~~~~a_{3\dot{1}}^\dagger(p_1)\cdots
a_{3\dot{1}}^\dagger(p_{M_+}) a_{4\dot{1}}^\dagger({\bar
p}_1)\cdots  a_{4\dot{1}}^\dagger({\bar p}_{M_-})
a_{1\dot{1}}^\dagger(k_1) \cdots  a_{1\dot{1}}^\dagger(k_{J_1})
a_{2\dot{1}}^\dagger({\bar k}_1) \cdots a_{2\dot{1}}^\dagger({\bar
k}_{J_2})
 |0 \rangle\,.
\eea We see that the left $\su(2|2)$ subalgebra of the symmetry
algebra $\su(2|2)\oplus \su(2|2)$ acts on the states of the
sector.

The $\su(2|3)$ sector exhibits the following new feature. One can
easily check  that the operators $a_{1\dot{1}}^\dagger
a_{2\dot{1}}^\dagger$ and $a_{3\dot{1}}^\dagger
a_{4\dot{1}}^\dagger$ have the same charges, and, therefore,  the
scattering of two bosons can result into two fermions. Thus, the
number of bosons and fermions is not preserved in the scattering
process involving particles from this sector.

These states can be shown to be dual to the \N SYM operators  of
the form \bea\la{su23oper} O_{\su(2|3)} = \tr \left(
Z^{J-\frac{M_+}{2} -\frac{M_-}{2}}X^{J_1} Y^{J_2} \Psi_+^{M_+}
\Psi_-^{M_-}  + \mbox{permutations} \right)\,, \eea where $\Psi_+$
is the highest weight component of the gaugino $\Psi_{\a}$ from
the vector multiplet that was denoted as $\Psi$ previously, and
$\Psi_-$ is the lowest weight component.

\section{Perturbative world-sheet S-matrix}\la{PSM}

\subsection{Generalities}

In scattering theory the S-matrix is a unitary operator, which we
denote by $\bS$, mapping free particle $out$-states to free
particle $in$-states in the Heisenberg picture. Both $in$- and
$out$-states belong to the same Hilbert space of the model, and
are eigenvectors of the full Hamiltonian $\bH$  with the same
eigenvalue $E$ \bea\la{hfinout} \bH\,  |p_1,p_2, \ldots , p_n
\rangle^{(in/out)}_{i_1,...,i_n}  = E\, |p_1,p_2, \ldots , p_n
\rangle^{(in/out)}_{i_1,...,i_n}\, .\eea Here $i_1,...,i_n$ are
flavor indices used to account for different kinds of particles in
the model, and $p_k$ is the momentum carried by the particle with
the flavor $i_k$ either at $t=-\infty$ for $in$-states or at
$t=\infty$ for $out$-states. The eigenvalue $E$ is given by \bea
\label{eigenenergy} E= \sum_{k=1}^n \om_{p_k}^{(i_k)}\, , \eea
where $\om_p^{(i)}$ is the energy (the dispersion relation) of a
particle of type $i$ with the momentum $p$. Recall that in
relativistic theory the dispersion relation is of the form $\om_p
= \sqrt{m^2+p^2}$, where $m$ is the mass of the particle which may
depend on coupling constants of the model and may receive quantum
corrections; momentum $p$ can take any real value. In a lattice
discretization of a relativistic model the dispersion relation
appears in the form $\om_p = \sqrt{m^2+{4\ov {\ell }^2}\sin^2{
p\ov 2}}$, where ${\ell}$ is a lattice step and $p$ changes from
$-\pi$ to $\pi$. As we will see in the next chapter, the exact
dispersion relation for particles of the light-cone string theory
in the decompactification limit is $\om_p = \sqrt{1+4g^2\sin^2{
p\ov 2}}$ where $g$ is the string tension, and therefore the
quantum light-cone string sigma model can be regarded as a lattice
model with the lattice step ${\ell} =1/g$. In general, in
non-relativistic theory $\om_p$ can be an arbitrary function of
$p$. It is worthwhile stressing that the dispersion relations
(\ref{eigenenergy}) entering in the eigenvalue problem
(\ref{hfinout}) are {\it exact}, {\it i.e.} they include all
quantum corrections. In this subsection, to avoid discussing
subtleties related to ultra-violet divergencies, we assume that we
are dealing with a lattice model.

\smallskip

To describe the $in$- and $out$-states, we introduce creation and
annihilation $in$- and $out$-operators acting in the same Hilbert
space and satisfying the canonical commutation relations
(\ref{comrel}). The Hilbert space has a state $|\Om\rangle$, called
vacuum, which is annihilated by all annihilation operators
$a_{{\rm in}}(p,t)|\Om\rangle =a_{{\rm out}}(p,t)|\Om\rangle =0$. The
$in$- and $out$-states corresponding to free fields are obtained
by applying creation $in$-operators $a^{{\rm
in}\,\dagger}_{k}(p)\equiv a^{{\rm in}\,\dagger}_{k}(p,0)$ and
$out$-operators $a^{{\rm out}\,\dagger}_{k}(p)\equiv a^{{\rm
out}\,\dagger}_{k}(p,0)$ to the vacuum state, respectively,
 \bea\label{inoutstates}\begin{aligned}
 |p_1,p_2, \ldots , p_n \rangle^{(in)}_{i_1,...,i_n} &=a^{{\rm in}\,\dagger}_{i_1}(p_1)\cdots
 a^{{\rm in}\,\dagger}_{i_n}(p_n)|\Om \rangle    \,,\\
  |p_1,p_2, \ldots , p_n \rangle^{(out)}_{i_1,...,i_n} &=
 a^{{\rm out}\,\dagger}_{i_1}(p_1)\cdots  a^{{\rm out}\,\dagger}_{i_n}(p_n)|\Om \rangle
 \, .\end{aligned}
\eea In the Heisenberg picture the time evolution of $in$- and
$out$-operators is governed by the free Hamiltonians $\bH^{{\rm
in}}$ and $\bH^{{\rm out}}$: \bea\la{hinout}\begin{aligned}
\bH^{{\rm in}} &= \int{\rm d}p \sum_{i} \om_p^{(i)} a_{i}^{{\rm
in}\,\dagger}(p)a^{i}_{\rm in}(p)\,,
\\
\bH^{{\rm out}} &= \int{\rm d}p \sum_{i} \om_p^{(i)} a_{i}^{{\rm
out}\,\dagger}(p)a^{i}_{\rm out}(p)\,.
\end{aligned}
 \eea
By construction, $in/out$-states (\ref{inoutstates}) are the
eigenstates of $\bH_0^{{\rm in/out}}$
with the same eigenvalue (\ref{eigenenergy}).

\smallskip

The $in$- and $out$-operators satisfy the canonical commutation
relations, and therefore, by virtue of the Stone -– von Neumann
theorem,
 they are related by a unitary operator
$\bS$ \bea\la{inout} a_{{\rm in}}^\dagger(p,t) = \bS\cdot a_{{\rm
out}}^\dagger(p,t)\cdot \bS^\dagger\,,\quad a_{{\rm in}}(p,t) =
\bS\cdot a_{{\rm out}}(p,t)\cdot \bS^\dagger\,,\quad \bS\,
|\Om\rangle =|\Om\rangle\,, \eea which is the S-matrix operator. The
S-matrix is time-independent because the $in$- and $out$-operators
have the same free field time dependence which factors out from
eq.(\ref{inout}). Therefore, $in$ and $out$ states are related as
follows \bea\la{insout} |p_1, \ldots
,p_n\rangle^{(in)}_{i_1,...,i_n}= \bS\cdot |p_1, \ldots
,p_n\rangle^{(out)}_{i_1,...,i_n}\,, \eea and we can expand
initial states on a basis of final states and vise versa.
 In particular, for the two-particle $in$
and $out$ states we get either \bea\la{sijkl}
|p_1,p_2\rangle^{(in)}_{i,j} = \bS\cdot
|p_1,p_2\rangle^{(out)}_{i,j}=\bS^{kl}_{ij}(p_1,p_2)
|p_1,p_2\rangle^{(out)}_{k,l} \,, \eea or equivalently, by
multiplying (\ref{sijkl}) by $\bS$ and using (\ref{insout})
\bea\la{sijkl2}
 \bS\cdot
|p_1,p_2\rangle^{(in)}_{i,j}=\bS^{kl}_{ij}(p_1,p_2)
|p_1,p_2\rangle^{(in)}_{k,l} \,. \eea Here we take into account
that in one-dimensional space  the set of momenta of the two
scattering particles does not change in the scattering process,
and
 we also order the particle momenta in decreasing order
$p_1>p_2>\cdots >p_n$ to take into account the particle's
statistics.
 It is clear
 that the S-matrix commutes with the full Hamiltonian
 $$\bS\cdot\bH = \bH\cdot\bS\,,$$ and
 that in the absence of interaction $\bS =\mI$, and $\bS^{kl}_{ij}(p_1,p_2) = \de_i^k\de_j^l$.
 According to eqs. (\ref{sijkl}) and (\ref{sijkl2}),
it does not matter whether one computes the matrix elements
$\bS^{kl}_{ij}(p_1,p_2)$ by using the basis of $in$ or $out$
states.

If there is no external field, and the particles interact only
among themselves, then a one-particle $in$-state coincides with
its $out$-state, and therefore the S-matrix must also satisfy the
following condition
 \bea\la{sinout1}
\bS\, |p\rangle^{(in)}_{k} =|p\rangle^{(in)}_{k} \ \ \Longleftrightarrow\ \ \bS\, |p\rangle^{(out)}_{k} =|p\rangle^{(out)}_{k} \,. \eea
This condition can be used to determine dispersion relations.

\medskip

To compute the S-matrix in perturbation theory one splits the full
Hamiltonian into  free and interaction parts \bea\nonumber \bH = \bH_0 +
\bV \,, \eea
 and introduces creation and annihilation operators $a, a^\dagger$ satisfying the canonical commutation relations
 (\ref{comrel}). In terms of these operators the free Hamiltonian $\bH_0$ takes the form
\bea\la{hfree} \bH_0 = \int\,{\rm d}p\, \sum_{k}\, \om_p^{(k)}\,
a_{k}^{\dagger}(p,t)a^{k}(p,t)\,. \eea
 The operators $a, a^\dagger$ (and  $\bH_0$) are interacting Heisenberg
 fields obeying the following equations of motion
 \bea\la{eoma}
 \dot{a}^{k}(p,t) = i\, [\bH\,,\,a^{k}(p,t)\,] = -i\,\om_p^{(k)}\,a^k(p,t) +  i\, [\bV\,,\,a^{k}(p,t)\,]  \,,
 \eea
 where $\bV = \bV(a^\dagger\,,\, a)$ is a function of $a^\dagger_k$ and $a^k$.
Note that if the dispersion relation receives quantum
corrections then the interaction Hamiltonian $\bV$ contains terms
quadratic in $a, a^\dagger$.

\smallskip

Since the  creation and annihilation operators $a, a^\dagger$
satisfy the canonical commutation relations,  they are related to
the $in$- and $out$-operators by  unitary transformations
\bea
\la{ain} &&\hspace{-1.7cm}a^\dagger(p,t) = \bU_{{\rm
in}}^\dagger(t)\cdot a_{{\rm in}}^\dagger(p,t)\cdot \bU_{{\rm
in}}(t)\,,\quad~~~~ a(p,t) = \bU_{{\rm in}}^\dagger(t)\cdot
a_{{\rm in}}(p,t)\cdot \bU_{{\rm
in}}(t)\,, \\
 \la{aout} &&\hspace{-1.7cm}a^\dagger(p,t) = \bU_{{\rm out}}(t)\cdot
a_{{\rm out}}^\dagger(p,t)\cdot \bU_{{\rm out}}^\dagger(t)\,,~~~
a(p,t) = \bU_{{\rm out}}(t)\cdot a_{{\rm out}}(p,t)\cdot \bU_{{\rm
out}}^\dagger(t)\,.
\eea The unitary operators $\bU_{{\rm in}}\,,\ \bU_{{\rm out}}$
are determined up to constant unitary transformations, which we
fix by imposing the following boundary conditions\footnote{It
would be sufficient for our purposes to impose a weaker condition
$\bU_{{\rm in}}(-\infty)\cdot \bU_{{\rm out}}(\infty) =\mI$.}
\bea\la{bcu} \bU_{{\rm in}}(-\infty) =\mI \,,\quad \bU_{{\rm
out}}(\infty)=\mI\,, \eea and up to multiplication by a phase
$\bU(t)\to e^{i\vp(t)}\bU(t)$, where $\vp(t)$ is an arbitrary real
function independent of the creation and annihilation operators.
In fact, the conditions (\ref{bcu}) imply that the interacting
Heisenberg field $a(p,t)$ tends to the free operators $a_{\rm
in}(p,t)$ and $a_{\rm out}(p,t)$ in the asymptotic past $t\to -
\infty$ and the asymptotic future $t\to +\infty$, respectively.

Comparing formulae (\ref{ain}) and (\ref{aout}) with
eqs.(\ref{inout}) defining the S-matrix, we find the following
expression for $\bS$ in terms of $\bU_{{\rm in}}\,,\ \bU_{{\rm
out}}$ \bea\la{suu} \bS = \bU_{{\rm in}}(t)\cdot \bU_{{\rm
out}}(t)\,. \eea The S-matrix is time-independent and, therefore,
in the above formula we can put $t$ to any desired value. Choosing
$t=\infty$ or $t=-\infty$ and taking into account the boundary
conditions (\ref{bcu}), we get the following two convenient
representation for the S-matrix
\bea\la{suu2}
\bS = \bU_{{\rm in}}(\infty) = \bU_{{\rm out}}(-\infty)\,.
\eea

To find $\bU_{{\rm in}}$, we differentiate (\ref{ain}) with
respect to $t$, and use the equations of motion for the operators
involved. After simple algebra, we get the following equalities
\bea \la{eoma2} \left[\, \dot{\bU}_{{\rm in}}\bU_{{\rm
in}}^\dagger + i\bV(a_{{\rm in}}^\dagger, a_{{\rm in}})\,, a_{{\rm
in}}^\dagger(p,t)\,\right] =0\,,\quad \left[\, \dot{\bU}_{{\rm
in}}\bU_{{\rm in}}^\dagger + i\bV(a_{{\rm in}}^\dagger, a_{{\rm
in}})\,, a_{{\rm in}}(p,t)\,\right] =0\,,~~~~~ \eea
 where the interaction Hamiltonian is now a function of the $in$-operators
 \bea\nonumber
 \bV(a_{{\rm in}}^\dagger, a_{{\rm in}})=  \bH(a_{{\rm in}}^\dagger, a_{{\rm in}}) - \bH_0^{{\rm in}}  = \bU_{{\rm in}}\, \bH(a^\dagger_k\,,\, a^k)\, \bU_{{\rm in}}^\dagger - \bH_0^{{\rm in}} \,.
 \eea
 Eqs.(\ref{eoma2}) imply that
$
  \dot{\bU}_{{\rm in}}\bU_{{\rm in}}^\dagger + i\bV(a_{{\rm in}}^\dagger, a_{{\rm in}}) = c(t)\, \mI\,,
$
 where $c(t)$ does not depend on $a_{{\rm in}}^\dagger\,,\  a_{{\rm in}}$.
 By properly choosing  the phase $\vp(t)$, we can always ensure vanishing of $c(t)$,
 so that $\bU_{{\rm in}}$ will be then determined unambiguously by the following equation
\bea\nonumber
  \dot{\bU}_{{\rm in}}\bU_{{\rm in}}^\dagger + i\bV(a_{{\rm in}}^\dagger, a_{{\rm in}}) = 0\,.
 \eea
The equation can be solved in terms of  the time-ordered exponential function
$\Texp$
\bea\la{uin}
\bU_{{\rm in}}(t) =  \Texp\left(- i \int_{-\infty}^t \,{\rm d}\tau\, \bV\big(a_{{\rm in}}^\dagger(\tau), a_{{\rm in}}(\tau)\big)\right)\,,
\eea
where we have taken into account the boundary condition (\ref{bcu}) for $\bU_{{\rm in}}$.

\smallskip

The operator $\bU_{{\rm out}}$ can be found in the same way. It
satisfies the equation
 $$\bU_{{\rm out}}^\dagger\dot{\bU}_{{\rm out}} - i\bV(a_{{\rm out}}^\dagger, a_{{\rm out}}) = 0\,,$$  whose solution is given by the following formula
\bea\la{uout}
\bU_{{\rm out}}(t) =  \Texp\left(- i \int^{\infty}_t \,{\rm d}\tau\, \bV\big(a_{{\rm out}}^\dagger(\tau), a_{{\rm out}}(\tau)\big)\right)\,.
\eea

Thus, we have derived the following two explicit expressions for the S-matrix
\bea\nonumber
\bS &=&  \Texp\left(- i \int_{-\infty}^\infty \,{\rm d}\tau\, \bV\big(a_{{\rm in}}^\dagger(\tau), a_{{\rm in}}(\tau)\big)\right)\\&=&  \Texp\left(- i \int^{\infty}_{-\infty} \,{\rm d}\tau\, \bV\big(a_{{\rm out}}^\dagger(\tau), a_{{\rm out}}(\tau)\big)\right)\,.\la{suu3}
\eea
 Expanding the formula in powers of $\bV$, one develops the standard perturbation theory.
 We will need only the leading term in the expansion
 \bea\la{tm}
 \bS = \mI + i\,{1\ov g}\, \bT\,,\quad \bT = - g \int_{-\infty}^\infty \,{\rm d}\tau\, \bV(\tau) +\cdots \,,
 \eea
 where $1/g$ is an expansion parameter of the perturbation theory.

 This formula allows one to compute the world-sheet two-particle S-matrix for the
 light-cone string sigma model to the leading order in the $1/g$ expansion.
    To this end, one has to use
    the quadratic Hamiltonian (\ref{hquad}) and the quartic Hamiltonian  (\ref{H41})
    as the free and interaction ones, respectively.

\medskip

To complete our discussion of the general scattering theory, we
note that $in$ and $out$ states can be also constructed in terms
of the oscillators $a^\dagger(p)=a^\dagger(p,0)$ and
$a(p)=a(p,0)$. Indeed, these oscillators are related to $in$ and
$out$ operators through eqs.(\ref{ain}),(\ref{aout}) \bea\la{ain0}
\hspace{-1cm}&&a^\dagger(p) = \bU_{{\rm in}}^\dagger(0)\cdot
a_{{\rm in}}^\dagger(p)\cdot \bU_{{\rm in}}(0)\,,\quad~~~~
a(p) = \bU_{{\rm in}}^\dagger(0)\cdot a_{{\rm in}}(p)\cdot \bU_{{\rm in}}(0)\,,\\
\la{aout0} \hspace{-1cm}&&a^\dagger(p) = \bU_{{\rm out}}(0)\cdot
a_{{\rm out}}^\dagger(p)\cdot \bU_{{\rm out}}^\dagger(0)\,,\quad
a(p) = \bU_{{\rm out}}(0)\cdot a_{{\rm out}}(p)\cdot \bU_{{\rm
out}}^\dagger(0)\,.~~~~~~~\eea As a result, we can write $in$ and
$out$ states as follows
 \bea\nonumber
 &&|p_1,p_2, \ldots , p_n \rangle^{(in)}_{i_1,...,i_n} = \bU_{{\rm in}}(0)\cdot a^{\dagger}_{i_1}(p_1)\cdots  a^{\dagger}_{i_n}(p_n)|0 \rangle =  \bU_{{\rm in}}(0)|\Phi_{\a} \rangle  \,,\\\nonumber &&|p_1,p_2, \ldots , p_n \rangle^{(out)}_{i_1,...,i_n} = \bU_{{\rm out}}^{\dagger}(0)\cdot a^{\dagger}_{i_1}(p_1)\cdots  a^{\dagger}_{i_n}(p_n)|0 \rangle  = \bU_{{\rm out}}^{\dagger}(0)|\Phi_{\a} \rangle \,,
\eea where $|0  \rangle=   \bU_{{\rm in}}^\dagger(0)|\Om \rangle =
\bU_{{\rm out}}(0)|\Om \rangle$ is the state annihilated by all
operators $a(p)$: $a_k(p) |0  \rangle = 0$, and $\a$ is a
multi-index including all momenta and flavours of the scattering
particles.

It is not difficult to find explicit formulae for the operators
$\bU_{{\rm in}}(0)$ and $ \bU_{{\rm out}}(0)$. To this end we
introduce free time-dependent operators which have the same
time-dependence as the $in$ and $out$ operators \bea\nonumber
a^\dagger_{{\rm fr},\, k}(p,t) = e^{i\,\om_p^{(k)}\, t}\,
a_k^\dagger(p)\,,\quad a_{{\rm fr}}^k(p,t) = e^{-i\,\om_p^{(k)}\,
t}\, a^k(p)\,. \eea The new oscillators are obviously related to
$in$ and $out$ operators through the same
eqs.(\ref{ain0}),(\ref{aout0}) \bea
\nonumber \hspace{-0.2cm}&&a_{{\rm fr}}^\dagger(p,t) = \bU_{{\rm
in}}^\dagger(0)\cdot a_{{\rm in}}^\dagger(p,t)\cdot \bU_{{\rm
in}}(0)\,,\quad~~~~
a_{{\rm fr}}(p,t) = \bU_{{\rm in}}^\dagger(0)\cdot a_{{\rm in}}(p,t)\cdot \bU_{{\rm in}}(0)\,,\\
\nonumber \hspace{-0.2cm}&&a_{{\rm fr}}^\dagger(p,t) = \bU_{{\rm
out}}(0)\cdot a_{{\rm out}}^\dagger(p,t)\cdot \bU_{{\rm
out}}^\dagger(0)\,,\quad a_{{\rm fr}}(p,t) = \bU_{{\rm
out}}(0)\cdot a_{{\rm out}}(p,t)\cdot \bU_{{\rm
out}}^\dagger(0)\,.~~~~~~~\eea Thus, taking into account
eqs.(\ref{uin}) and (\ref{uout}), we get the following formulae
\bea\la{uint} \bU_{{\rm in}}(t) &=& \bU_{{\rm in}}(0)\cdot
\Texp\left(- i \int_{-\infty}^t \,{\rm d}\tau\, \bV\big(a_{{\rm
fr}}^\dagger(\tau), a_{{\rm fr}}(\tau)\big)\right)\cdot  \bU_{{\rm
in}}^\dagger(0)\,,
\\\la{uoutt}
\bU_{{\rm out}}(t) &=&  \bU_{{\rm out}}^\dagger(0)\cdot
\Texp\left(- i \int^{\infty}_t \,{\rm d}\tau\, \bV\big(a_{{\rm
fr}}^\dagger(\tau), a_{{\rm fr}}(\tau)\big)\right)\cdot \bU_{{\rm
out}}(0)\,. \eea From these expressions we can read off $\bU_{{\rm
in}}(0)$ and $ \bU_{{\rm out}}(0)$ in terms of the free
oscillators $a_{{\rm fr}}^\dagger(\tau), a_{{\rm fr}}(\tau)$
\bea\la{uin0} \bU_{{\rm in}}(0) &=& \Texp\left(- i
\int_{-\infty}^0 \,{\rm d}\tau\, \bV\big(a_{{\rm
fr}}^\dagger(\tau), a_{{\rm fr}}(\tau)\big)\right)\,,
\\\la{uout0}
\bU_{{\rm out}}(0) &=&   \Texp\left(- i \int^{\infty}_0 \,{\rm d}\tau\, \bV\big(a_{{\rm fr}}^\dagger(\tau), a_{{\rm fr}}(\tau)\big)\right)\,.
\eea
Then we can easily find the overlap between $in$ and $out$ states, that is the S-matrix elements
\bea\nonumber
 {}_\b\langle out|in \rangle_\a = \langle \Phi_\b| \bU_{{\rm out}}(0)\bU_{{\rm in}}(0) |\Phi_{\a} \rangle = \langle \Phi_\b| \check{\bS} |\Phi_{\a} \rangle\,,
\eea where $\check{\bS}$ is the following operator \bea\nonumber
 \check{\bS} =\bU_{{\rm out}}(0)\bU_{{\rm in}}(0) = \Texp\left(- i \int_{-\infty}^\infty \,{\rm d}\tau\, \bV\big(a_{{\rm fr}}^\dagger(\tau), a_{{\rm fr}}(\tau)\big)\right)\,,
 \eea
Note that the operator $ \check{\bS}$ differs from the S-matrix
operator $\bS$ in eq.(\ref{suu}) by the opposite order of
$\bU_{{\rm in}}(0)$ and $ \bU_{{\rm out}}(0)$.

It is not difficult to show that the operators  $\bU_{{\rm in}}(0)$ and $ \bU_{{\rm out}}(0)$ have the following commutation relations with $\bH$ and $\bH_0(0)$
\bea\nonumber
\bH\cdot \bU_{{\rm in}}(0) = \bU_{{\rm in}}(0)\cdot\bH_0(0)\,,\quad \bH_0(0)\cdot \bU_{{\rm out}}(0) = \bU_{{\rm out}}(0)\cdot\bH\,,
\eea
and, therefore, the operator $\bH_0(0)$ commutes with  $\check{\bS}$
\bea\nonumber
\bH_0(0)\cdot \check{\bS}=  \check{\bS}\cdot\bH_0(0)\,.
\eea
\subsection{A sample computation of  perturbative S-matrix}

To illustrate how the formulae above can be used, let us compute
the perturbative S-matrix for the $Y^{a\da}$ bosons from the
five-sphere. The relevant part of the T-matrix operator is given
by \bea\la{TY} \bT_Y = - g \int_{-\infty}^\infty \,{\rm d}\tau\,
\bV(\tau) = 2\int_{-\infty}^\infty \,{\rm d}\tau\,{\rm d}\s\,
Y^{a\da}Y_{a\da} Y'^{b\db}Y'_{b\db}\,, \eea where we used
eq.(\ref{H4matr}) for the quartic Hamiltonian, and lowered the
indices by means of the $\eps$-tensor  $Y_{a\da} =
\eps_{ab}\eps_{\da\db}Y^{b\db}$. We  use the mode decomposition
(\ref{bosrep}) with the creation and annihilation operators having
the free-field time dependence \bea\nonumber a^{a\da}(p,t) =
e^{-i\,\om_p\, t}\, a^{a\da}(p)\,,\quad a^\dagger_{a\da}(p,t) =
e^{i\,\om_p\, t}\, a^\dagger_{a\da}(p)\,. \eea The creation and
annihilation operators are either $in$ or $out$-operators
depending on the basis we use for the S-matrix computation.

Substituting the mode
decomposition into (\ref{TY}), and integrating over $\tau$ and
$\s$, one gets a sum of terms of the form \bea\nonumber &&\de(\om_1
+\om_2+\om_3+\om_4)\, \de(k_1+k_2+k_3+k_4)\,
a^\dagger(k_1)a^\dagger(k_2)a^\dagger(k_3)a^\dagger(k_4)
\\\nonumber &&+\de(\om_1 +\om_2+\om_3-\om_4)\,
\de(k_1+k_2+k_3-k_4)\,
a^\dagger(k_1)a^\dagger(k_2)a^\dagger(k_3)a(k_4)
\\\nonumber
&&+\de(\om_1 +\om_2-\om_3-\om_4)\, \de(k_1+k_2-k_3-k_4)\,
a^\dagger(k_1)a^\dagger(k_2)a(k_3)a(k_4) + h.c. \eea One can
easily check that due to the energy/momentum conservation
delta-functions only the terms with equal number of creation and
annihilation operators do not vanish. Then, a simple computation
gives \bea\nonumber \bT_Y &=& \int{ {\rm d}k_1{\rm d}k_2{\rm
d}k_3{\rm d}k_4\ov 4\sqrt{\om_1 \om_2\om_3\om_4}}\, \de(\om_1
+\om_2-\om_3-\om_4)\, \de(k_1+k_2-k_3-k_4)~~~~~~\\\nonumber
&&~~~~~~~\times\Big[ (2k_2 k_4-k_1k_2-k_3k_4)
a^\dagger_{b\db}(k_4)a^\dagger_{a\da}(k_3)a^{b\db}(k_2)a^{a\da}(k_1)\\\nonumber
&&~~~~~~~~~~+(k_1k_2+k_3k_4)
a^\dagger_{a\db}(k_4)a^\dagger_{b\da}(k_3)a^{b\db}(k_2)a^{a\da}(k_1)\Big]\,.
\eea The $\de$-functions can be used to integrate over $k_3$ and
$k_4$ because they imply that either $k_3 =k_1, k_4=k_2$ or $k_3
=k_2, k_4=k_1$, and taking into account that the Jacobian of
$\de(\om_1 +\om_2-\om_3-\om_4)$ equals to $\om_1\om_2/|k_1
\om_2-k_2\om_1|$, one gets the T-matrix \bea\nonumber \bT_Y &=&
\int{ {\rm d}k_1{\rm d}k_2\ov 2|k_1 \om_2-k_2\om_1|}\, \Big[{1\ov
2} (k_1-k_2)^2
a^\dagger_{b\db}(k_2)a^\dagger_{a\da}(k_1)a^{b\db}(k_2)a^{a\da}(k_1)\\\nonumber
&&~~~~~~~~~~+2k_1k_2\,
a^\dagger_{a\db}(k_2)a^\dagger_{b\da}(k_1)a^{b\db}(k_2)a^{a\da}(k_1)\Big]\,.
\eea Finally, acting by the T-matrix operator on a two-particle
state, one derives \bea\la{TY2} \bT_Y\cdot
|a^\dagger_{a\da}(p_1)a^\dagger_{b\db}(p_2)\rangle&=&{ (p_1-p_2)^2
\ov 2(p_1 \om_2-p_2\om_1)}\,
|a^\dagger_{a\da}(p_1)a^\dagger_{b\db}(p_2)\rangle\\\nonumber &&+{
p_1p_2 \ov p_1 \om_2-p_2\om_1}\,\big(
|a^\dagger_{b\da}(p_1)a^\dagger_{a\db}(p_2)\rangle
+|a^\dagger_{a\db}(p_1)a^\dagger_{b\da}(p_2)\rangle \big)\, , \eea
where we have assumed that $p_1>p_2$.

The action of  the T-matrix operator on an arbitrary two-particle state is given in appendix \ref{app:T}.
\subsection{S-matrix factorization}\la{sec:Sfact}
The formula (\ref{TY2}) for the T-matrix shows that it has the
following factorized form \bea\nonumber \bT_Y = \cT_Y\otimes \mI +
\mI \otimes \dcT_Y\,, \eea where the operators $\cT_Y$ and
$\dcT_Y$ act only  on the undotted and dotted  indices,
respectively. Moreover, analyzing the formulae from  appendix
\ref{app:T}, one can show that the same factorization also holds
for the full T-matrix: $\bT = \cT\otimes \mI + \mI \otimes \dcT$.
This factorization in fact follows from the corresponding
factorization of the S-matrix operator \bea\nonumber \bS = \cS
\otimes \dcS\,, \eea which is a consequence of the integrability
of the model, as will be discussed in the next chapter in detail.

The simplest way to describe the factorization is  to think about
the two-index creation operators $a^\dagger_{M\dM}$ as a product
of two one-index operators $a^\dagger_{M}$ and $a^\dagger_{\dM}$,
that is  $a^\dagger_{M\dM}(p) \sim
a^\dagger_{M}(p)\,a^\dagger_{\dM}(p)$.  Since the lower-case latin
indices are even, and the greek indices are odd, the operators
$a^\dagger_{a}$, $a^\dagger_{\da}$ are bosonic, and
$a^\dagger_{\a}$, $a^\dagger_{\dal}$ are fermionic, and they
commute or anti-commute depending on their statistics.

We see, therefore, that  one-particle states can be identified with the following tensor product
$$
|a^\dagger_{M\dM}(p)\rangle \sim |a^\dagger_{M}(p)\rangle\otimes |a^\dagger_{\dM}(p)\rangle\,,
$$
and two-particle states with
\bea\la{2ps}
|a^\dagger_{M\dM}(p_1)a^\dagger_{N\dN}(p_2)\rangle\sim (-1)^{\eps_{\dM}\eps_N}|a^\dagger_{M}(p_1)a^\dagger_{N}(p_2)\rangle\otimes |a^\dagger_{\dM}(p_1)a^\dagger_{\dN}(p_2)\rangle\,,
\eea
where the extra sign may appear because one permutes the operators
$a^\dagger_{\dM}$ and $a^\dagger_{N}$.

Then, $\cS$ and $\dcS$ act in the space of
the $|a^\dagger_{M}(p_1)a^\dagger_{N}(p_2)\rangle$ and $|a^\dagger_{\dM}(p_1)a^\dagger_{\dN}(p_2)\rangle$ states, respectively, and their S-matrix elements are defined in the usual way
\bea\la{cS2}
\cS\cdot |a^\dagger_{M}(p_1)a^\dagger_{N}(p_2)\rangle = \cS_{MN}^{PQ}(p_1,p_2)|a^\dagger_{P}(p_1)a^\dagger_{Q}(p_2)\rangle\,,
\eea
and a similar formula for $\dcS$. In particular, we find from (\ref{TY2}) the action of $\cT_Y$ on the states
\bea
\nonumber
 \cT_Y\cdot
|a^\dagger_{a}(p_1)a^\dagger_{b}(p_2)\rangle ={(p_1-p_2)^2 \ov
4(p_1 \om_2-p_2\om_1)}\,
|a^\dagger_{a}(p_1)a^\dagger_{b}(p_2)\rangle + { p_1p_2 \ov p_1
\om_2-p_2\om_1}\, |a^\dagger_{b}(p_1)a^\dagger_{a}(p_2)\rangle \,.
\eea By using  (\ref{2ps})  and (\ref{cS2}), one can easily derive
the following relation between the  elements of the scattering
matrix $\bS$, and those of the auxiliary S-matrices $\cS$ and
$\dcS$ \bea\la{Sfact} \bS_{M\dM, N\dN}^{P\dP,Q\dQ}  (p_1,p_2) =
(-1)^{\eps_{\dM}\eps_N +
\eps_{\dP}\eps_{Q}}\,\cS_{MN}^{PQ}(p_1,p_2)\dcS_{\dM\dN}^{\dP\dQ}(p_1,p_2)\,.
\eea Taking into account that \bea\nonumber \bS=\mI + i{1\ov
g}\bT\,,\quad \cS=\mI + i{1\ov g}\cT\,,\quad \dcS=\mI + i{1\ov
g}\dcT\,, \eea one finds the following relation \bea\la{Tfact}
\bT_{M\dM, N\dN}^{P\dP,Q\dQ}  = (-1)^{\eps_{\dM}(\eps_N +
\eps_{Q})}\cT_{MN}^{PQ}\,\de_{\dM}^{\dP}\,\de_{\dN}^{\dQ} +
(-1)^{(\eps_{\dM} +
\eps_{\dP})\eps_{Q}}\,\de_{M}^{P}\,\de_{N}^{Q}\,
\dcT_{\dM\dN}^{\dP\dQ} \eea for the T-matrix elements. The matrix
elements for $\cT$ and $\dcT$ can be chosen to be equal to each
other, and can be extracted from the formulae in appendix
\ref{app:T}. The result can be written in the following form
\bea\nonumber &&\cT_{ab}^{cd}=
A\,\de_a^c\de_b^d+B\,\de_a^d\de_b^c\,,\qquad ~ \cT_{ab}^{\g\de}=
C\,\eps_{ab}\eps^{\g\de}\,,\\\nonumber &&\cT_{\a\b}^{\g\de}=
D\,\de_\a^\g\de_\b^\de+E\,\de_\a^\de\de_\b^\g\,,\qquad
\cT_{\a\b}^{cd}= F\,\eps_{\a\b}\eps^{cd}\,,\\\nonumber
&&\cT_{a\b}^{c\de}= G\,\de_a^c\de_\b^\de\,,\qquad ~~~~~~~~~~~~~
\cT_{\a b}^{\g d}= L\,\de_\a^\g\de_b^d\,,\\\nonumber
&&\cT_{a\b}^{\g d}= H\,\de_a^d\de_\b^\g\,,\qquad ~~~~~~~~~~~~~
\cT_{\a b}^{\g d}= K\,\de_\a^\g\de_b^d\, \eea where the
coefficients are given by \bea\la{Tmatrcoef}
&&A(p_1,p_2)={(p_1-p_2)^2 \ov 4(p_1 \om_2-p_2\om_1)} + {1\ov
4}(1-2a)( p_1 \om_2-p_2\om_1)\,,\\\nonumber
&&B(p_1,p_2)=-E(p_1,p_2)={p_1p_2\ov p_1
\om_2-p_2\om_1}\,,\\\nonumber &&C(p_1,p_2)=F(p_1,p_2)={1\ov
2}{\sqrt{(\om_1+1)(\om_2+1)}(p_1 \om_2-p_2\om_1+p_2-p_1\ov p_1
\om_2-p_2\om_1}\,,~~~~~\\\nonumber &&D(p_1,p_2)=-{(p_1-p_2)^2 \ov
4(p_1 \om_2-p_2\om_1)}+ {1\ov 4}(1-2a)( p_1
\om_2-p_2\om_1)\,,\\\nonumber
&&G(p_1,p_2)=-L(p_2,p_1)=-{p_1^2-p_2^2 \ov 4(p_1 \om_2-p_2\om_1)}+ {1\ov 4}(1-2a)( p_1 \om_2-p_2\om_1)\,,\\
\nonumber
&&H(p_1,p_2)=K(p_1,p_2)={1\ov 2}{p_1p_2\ov p_1 \om_2-p_2\om_1}{(\om_1+1)(\om_2+1)-p_1p_2\ov \sqrt{(\om_1+1)(\om_2+1)}}\,,
\eea
where we have also added the additional contribution which vanishes in the $a=1/2$ gauge.
The T-matrix $\cT$ is covariant under the SU(2)$\times$SU(2) transformations that reflect the manifest SU(2)$^4$ symmetry of the light-cone string sigma model. The factorization of the T-matrix is a nontrivial test of the integrability of the model.
\section{Symmetry algebra}

In this section we show that the symmetry algebra of the
light-cone  string sigma model in the decompactification limit
gets enlarged by two additional central charges which vanish on
the physical subspace of the model.

\subsection{General structure of symmetry generators}

The invariance of the \GS action under the group PSU(2,2$|$4)
leads to the existence of conserved currents and charges. As was
shown in the previous chapter, see eq.(\ref{gcurrent}), the
conserved currents can be written in terms of $A_\a$ as follows
\bea\la{Ccurr} J^\a = g\, \ag(x,\chi)\left( \g^{\a\b} A_\b^{(2)} -
{\kappa\ov 2} \epsilon^{\a\b}(A_\b^{(1)} -
A_\b^{(3)})\right)\ag(x,\chi)^{-1}\,. \eea The $8\times 8$
supermatrix ${\rm Q}$ of conserved charges is then given by the
integral over $\s$ of $J^\tau$, eq.(\ref{NC}).
For our purposes it is convenient to express the charges in terms
of the momentum $\bp$. To this end, we notice that, as follows
from (\ref{Lang}), $\bp$ satisfies the following equation of
motion \bea\la{eqm} \bp = g\, \g^{\tau\b}A_\b^{(2)} =g\,
\g^{\tau\tau}\left( A_\tau^{(2)} + {\g^{\tau\s}\ov
\g^{\tau\tau}}A_\s^{(2)}\right)\, . \eea Therefore, we can express
$A_\tau^{(2)}$ in terms of $\bp$, and substitute it into the
expression for $ {\rm Q}$. After some simple algebra we get
\bea\nonumber  {\rm Q} =  \int_{-r}^{r} {\rm d}\s\,
\ag(x,\chi)\left(\bp - g{\kappa\ov 2} \big(A_\s^{(1)} -
A_\s^{(3)}\big)\right)\ag(x,\chi)^{-1}\,. \eea The formula can be
written in a more explicit form if we take into account that
\bea\nonumber A_\s^{(1)} - A_\s^{(3)} = i \ag(x)\Kk F_\s^{st} \Kk
\ag(x)^{-1}\,, \eea where $ F_\s $ is an odd component of the
current $g^{-1}(\chi)\pa_\s g(\chi)$ defined in (\ref{BF}). Then,
the $\psu(2,2|4)$ charges are \bea\la{Charges}  {\rm Q}
=\int_{-r}^{r} {\rm d}\s\,  \Lambda\, \ag(\chi) \ag(x)\left(\bp
-ig{\kappa\ov 2}\ag(x)\Kk F_\s^{st} \Kk \ag(x)^{-1} \right)
\ag(x)^{-1}\ag(\chi)^{-1}\Lambda^{-1}\,.~~~~~ \eea The expression
is very simple and  it has an important property of being
explicitly  independent of the world-sheet metric.

We also see that the matrix $\rm Q$ can be
schematically written as follows
\bea\la{Qcharges} {\rm Q} =\int_{-r}^{r} {\rm d}\sigma~ \Lambda\,
U\, \Lambda^{-1}\, , ~~~~~ \eea
where  $U$ depends on
physical fields $(x,p,\chi)$ but not on $x_{\pm}$ and, therefore, is a local function of $\s$. The only
dependence of ${\rm Q}$ on $x_{\pm}$ occurs through the matrix
$\Lambda$ (\ref{Lambda})  which has the following form in the $a=1/2$ light-cone gauge
\be \label{lambdadef}
\Lambda=e^{\frac{i}{2}x_+\Sigma_+ +\frac{i}{4}x_-\Sigma_-}\, ,\ee
where $\Sigma_{\pm}$ are defined in (\ref{Sigpm}), and $x_+=\tau$ due to the light-cone gauge condition.

We recall that the field $x_-$ is unphysical and can be solved in
terms of physical excitations through the equation
\bea\label{eqxmin} x'_-=-{1\ov g}\Big( p_Mx'_M-\frac{i}{2}
\str(\Sigma_+\chi\chi') \Big) +\cdots\, , \eea where $\cdots$
denote terms which are of higher-order in the fields. This
equation determines $x_-$ up to a function of $\tau$ which is the
zero mode of $x_-$ canonically conjugated to $P_+$. The
$\tau$-dependence of the zero mode can be determined from the
evolution equation for $x_-$. In what follows we need to know the
symmetry algebra generators in the decompactification limit only.
In this limit the Hamiltonian and the symmetry generators do not
depend on $P_+$ and, for this reason, the zero mode becomes a
central element.

Linear combinations of components of the matrix ${\rm Q}$ produce
charges which generate rotations, dilatation, supersymmetry and so
on. To single them out one should multiply ${\rm Q}$ by a
corresponding $8\times 8$ matrix ${\cal M}$, and take the
supertrace \bea \label{QM} {\rm Q}_{\cal M} = \str\, ({\rm Q}{\cal
M}) \,. \eea It is clear that the diagonal and off-diagonal
$4\times 4$ blocks of ${\cal M}$ single out bosonic and fermionic
charges of $\psu(2,2|4)$, respectively. In particular, one can
check that the light-cone Hamiltonian can be obtained from $\sQ$
as follows \bea H =- {i\ov 2}\str\, (\sQ\,\S_+)\,, \label{Ham}
\eea and the light-cone momentum  $P_+$  is given by
\begin{equation}
\label{HH} P_+=\frac{i}{4}\str(\sQ\,\S_-)\, . \
\end{equation}

Depending on the choice of ${\cal M}$ the charges ${\rm Q}_{\cal
M}\equiv {\rm Q}_{\cal M}(x_+,x_-)$ can be naturally classified
according to their dependence on $x_{\pm}$. Firstly, with respect
to $x_-$ they are divided into {\it kinematical} (independent of
$x_-$) and {\it dynamical} (dependent on $x_-$). Kinematical
generators do not receive quantum corrections, while the dynamical
generators do. Secondly, the charges, both kinematical and
dynamical, may or may not explicitly depend on $x_+=\tau$.

\medskip
In the Hamiltonian setting the conservation laws  have the
following form
$$
\frac{d \sQ_{\cal M}}{d \tau}=\frac{\pa \sQ_{\cal M}}{\pa \tau}+\{
H,\sQ_{\cal M}\}=0\, .
$$
Therefore, the generators which do not have explicit dependence on
$x_+=\tau$ Poisson-commute with the Hamiltonian. As follows from
the Jacobi identity, they must form an algebra which contains $H$
as central element.

\smallskip
\vskip 0pt \noindent
\begin{figure}
\begin{minipage}{\textwidth}
\begin{center}
\includegraphics[width=0.6\textwidth]{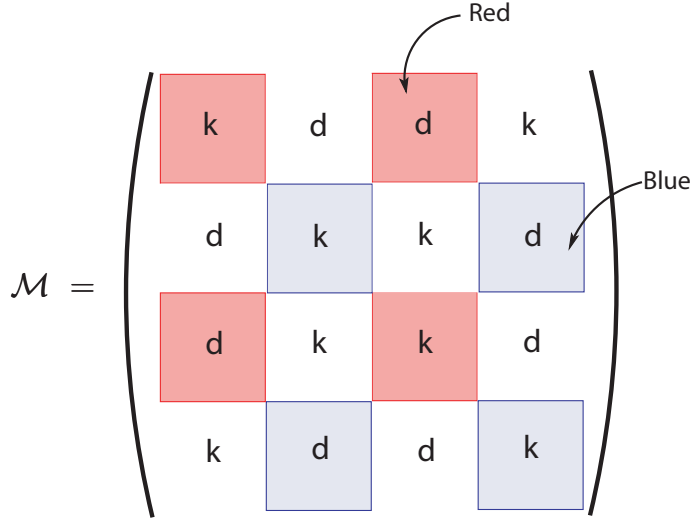}
\vskip 5pt
\parbox{5in}{\footnotesize{\caption{\label{flag} The distribution of the kinematical and
dynamical charges in the ${\cal M}$ supermatrix. The red (dark)
and blue (light) blocks correspond to the subalgebra ${\cal J}$ of
$\psu(2,2|4)$ which leaves the Hamiltonian invariant. } }}
\end{center}
\end{minipage}
\end{figure}

\vskip -0.7cm Analyzing the structure of $\sQ$ one can establish
how a generic matrix ${\cal M}$ is split into $2\times 2$ blocks
each of them giving rise either to kinematical or dynamical
generators. This splitting of ${\cal M}$ is shown in Figure
\ref{flag}, where the kinematical blocks are denoted by ${\bf k}$
and the dynamical ones by $\bf d$ respectively. Furthermore, one
can see that the blocks which are colored in red and blue give
rise to charges which are independent of $x_+=\tau$; by this
reason these charges commute with the Hamiltonian and form the
manifest symmetry algebra of the gauge-fixed string sigma model.
Complementary, we note that the uncolored both kinematical
(fermionic)  and dynamical (bosonic) generators do depend on
$x_+$.

\smallskip

These conclusions about the structure of ${\cal M}$ can be easily
drawn by noting that $\Lambda$ in eq.(\ref{lambdadef}) is built
out of two commuting matrices $\Sigma_{+}$ and $\Sigma_-$. For
instance, leaving in ${\cal M}$ the kinematical blocks only, {\it
i.e.} ${\cal M}\equiv {\cal M}_{\rm kin}$, we find that
$[\Sigma_-, {\cal M}_{\rm kin}]$=0 and, therefore, due to the
structure of $\sQ_{\cal M}$, see eq.(\ref{QM}), the variable $x_-$
cancels out in $\sQ_{\cal M}$. On the other hand, any matrix from
the red-blue submatrix ${\cal J}$ of ${\cal M}$ in Figure
\ref{flag}  commutes with  the element $\Sigma_+$ in $\psu(2,2|4)$
$$
[\Sigma_+,{\cal M}]=0\, , ~~~~~~{\cal M}\in {\cal J}\, ,
$$
leading to a charge $\sQ_{\cal M}$  independent of $x_+=\tau$.
Thus, for $P_+$  finite we  obtain the following vector space decomposition of ${\cal J}$
$$
{\cal J}=\psu(2|2)\oplus\psu(2|2)\oplus \Sigma_+\oplus \Sigma_-\,.
$$
The rank of the latter subalgebra is six and it coincides with
that of $\psu(2,2|4)$. In the case of infinite $P_+$ the last
generator decouples.

Conjugating with $\Lambda$ of \eqn{lambdadef}, one finds
\begin{align}
\Lambda^{-1}\, {\cal M}_{\rm dyn}^{\rm odd}\, \Lambda & = e^{
-\frac{i}{2} x_- \Sigma_-}\,
 {\cal M}_{\rm dyn}^{\rm odd} \, ,\qquad
&\Lambda^{-1}\, {\cal M}_{\rm dyn}^{\rm even}\, \Lambda & = \Lambda^2\,  {\cal M}_{\rm dyn}^{\rm even} \, , \nn \\
\Lambda^{-1}\, {\cal M}_{\rm kin}^{\rm odd}\, \Lambda & = e^{i\,
x_+\, \Sigma_+}\,  {\cal M}_{\rm kin}^{\rm odd} \, ,\qquad
&\Lambda^{-1}\, {\cal M}_{\rm kin}^{\rm even}\, \Lambda & = {\cal
M}_{\rm kin}^{\rm even}\, ,  \la{dynakin}
\end{align}
which shows that the $x_+=\tau$ independent matrices are indeed
given by ${\cal M}_{\rm dyn}^{\rm odd}$ and ${\cal M}_{\rm
kin}^{\rm even}$, {\it i.e.} by the red and blue entries in Figure
\ref{flag}.  We see from Figure \ref{flag} and  (\ref{dynakin})
that in the symmetry algebra all bosonic  charges are kinematic,
and all supercharges are dynamical.

The structure of $\sQ$ discussed above is found for finite $r$ and it also remains
valid in the decompactification limit $r\to \infty$.

\subsection{Centrally extended  \texorpdfstring{$\su(2|2)$}{su(2|2)} algebra}
\label{ceasu}

It is clear that the $\psu(2,2|4)$ charges (\ref{Charges})
transform linearly under the bosonic subalgebra $\alg{C}$ defined
in (\ref{linearsym}) because  $\Lambda$ commutes with any element
of this subalgebra. Therefore, to encode the transformation
properties of the charges under $\alg{C}$, it is convenient to use
the two-index notation introduced in section 1.4. The
time-dependent charges in the white blocks of Figure \ref{flag}
have the same indices as the bosonic and fermionic fields
$Z^{\a\dot{\a}}\, , \ Y^{a\dot{a}}\, ,\ \theta^{a\dot{\a}}\, , \
\eta^{a\dot{\a}}$. The time-independent charges which commute with
the Hamiltonian and form the symmetry algebra can be represented
in terms of $2\times 2$ blocks as follows \bea \la{qsym} {\rm
Q}_{\rm sym} ={\small \left(
\begin{array}{cccc} \bR & 0
&- \bQ^\dagger & 0 \\
0 &\mathring{\bR} & 0 &  \mathring{\bQ}   \\
\bQ & 0 & \bL & 0 \\
 0 & \mathring{\bQ}^\dagger & 0 & \mathring{\bL}
\end{array}\right)}\, .
\eea Here $\bR,\mathring{\bR} \in \su(2,2)$, and
$\bL,\mathring{\bL}\in \su(4)$  are the bosonic charges which
generate the transformations under $\alg{C}$, and  $\bQ,
\bQ^\dagger, \mathring{\bQ}, \mathring{\bQ}^\dagger$ are the eight
complex supercharges. The bosonic charges
 satisfy the usual reality
and tracelessness conditions \bea\la{conjrule2}
\begin{aligned}&\bR^\dagger= - \bR\, ,\ \
\mathring{\bR}^\dagger=-\mathring{\bR}\,,\ \ \bL^\dagger= - \bL\,
,\ \ \mathring{\bL}^\dagger=-\mathring{\bL}\, ,
\\ &\tr\, \bR = \tr\,  \mathring{\bR} = \tr\, \bL = \tr\,
\mathring{\bL} =0 \, .~~~ \end{aligned}\eea These charges should
be complemented by the matrices representing the Hamiltonian and
the light-cone momentum which are of the form \bea \la{HPmatr}
{\rm Q}_{\bH} =-{i\ov 4}\bH {\small \left(
\begin{array}{cccc}- \mI & 0
& 0 & 0 \\
0 &\mI & 0 &0  \\
0& 0 & \mI & 0 \\
 0 & 0& 0 &-\mI
\end{array}\right)}\,,\quad {\rm Q}_{\bP_+} ={i\ov 2}\bP_+ {\small \left(
\begin{array}{cccc} \mI & 0
& 0 & 0 \\
0 &-\mI & 0 &0  \\
0& 0 & \mI & 0 \\
 0 & 0& 0 &-\mI
\end{array}\right)}\, .
\eea In our analysis the light-cone momentum will not play any
role because we will only discuss  the  decompactification limit
where $P_+\to\infty$.

Under the action of the group element (\ref{gec}) the matrix
(\ref{qsym}) transforms as follows
 \bea \la{lawboson2}
{\rm Q}_{\rm sym} \to G\,{\rm Q}_{\rm sym} \,G^{-1}= {\small \left( \begin{array}{cccc} \ag_1 \bR\, \ag_1^{-1}  & 0&- \ag_1 \bQ^\dagger\, \ag_3^{-1} & 0 \\
0& \ag_2\mathring{\bR}\, \ag_2^{-1}  & 0 &  \ag_2\mathring{\bQ}\, \ag_4^{-1}   \\
 \ag_3\bQ\, \ag_1^{-1} & 0 & \ag_3 \bL\, \ag_3^{-1}   &0\\
 0 &  \ag_4\mathring{\bQ}^\dagger\, \ag_2^{-1} & 0 &  \ag_4\mathring{\bL}\, \ag_4^{-1}
\end{array}\right)}\,.
\eea
Since the charges $\bR,\bL,\bQ,\bQ^\dagger$ transform under one $\su(2)\in \su(2,2)$ and one  $\su(2)\in \su(4)$, and the charges $ \mathring{\bR}, \mathring{\bL}, \mathring{\bQ}, \mathring{\bQ^\dagger}$ transform under another $\su(2)\in \su(2,2)$ and another  $\su(2)\in \su(4)$, the charges from the first group must (anti-)commute with the ones from the second group.

Repeating the considerations in subsection \ref{Sect:lrbs}, we
find that the $2\times 2$ blocks $\bR,\bL,\bQ,\bQ^\dagger$  are
expressed via covariant two-index entries $\bL^{ab},
\bR^{\a\b},\bQ^{\a b}, \bQ^\dagger_{a\b}$ as \bea\nonumber
\begin{aligned}
&\bL =-i \left(\begin{array}{rr} \bL^{12} & -\bL^{11}\\
\bL^{22} & -\bL^{21}\end{array}\right)\,,\quad &&
\bR =i \left(\begin{array}{rr} \bR^{34} & -\bR^{33}\\
\bR^{44} & -\bR^{43}\end{array}\right)\,,\quad\\ \nonumber
&\bQ = e^{i\pi/4} \left(\begin{array}{rr} \bQ^{41} & -\bQ^{31}\\
\bQ^{42} & -\bQ^{32}\end{array}\right)\,,\quad &&
\bQ^\dagger = e^{-i\pi/4} \left(\begin{array}{rr} \bQ_{14}^\dagger & \bQ_{24}^\dagger\\
-\bQ_{13}^\dagger & -\bQ_{23}^\dagger\end{array}\right)
\, ,
\end{aligned}
\eea
 and $ \mathring{\bR}, \mathring{\bL}, \mathring{\bQ}, \mathring{\bQ^\dagger}$ are expressed through $\bL^{\da\db}, \bR^{\dal\dbe},\bQ^{ \dal\db}, \bQ^\dagger_{\da\dbe}$  as
\bea\nonumber
\begin{aligned}
~~~~~&\mathring{\bL} =- i\left(\begin{array}{rr} \bL^{\dot{1}\dot{2}} & -\bL^{\dot{1}\dot{1}}\\
\bL^{\dot{2}\dot{2}} & -\bL^{\dot{2}\dot{1}}\end{array}\right)\,,\quad
&&\mathring{\bR} =i \left(\begin{array}{rr} \bR^{\dot{3}\dot{4}} & -\bR^{\dot{3}\dot{3}}\\
\bR^{\dot{4}\dot{4}} & -\bR^{\dot{4}\dot{3}}\end{array}\right)\,,\quad\\ \nonumber
&\mathring{\bQ} =-e^{i\pi/4} \left(\begin{array}{rr} \bQ^{\dot{3}\dot{2}} & -\bQ^{\dot{3}\dot{1}}\\
\bQ^{\dot{4}\dot{2}} & -\bQ^{\dot{4}\dot{1}}\end{array}\right)\,,\quad
&&\mathring{\bQ}^\dagger = -e^{-i\pi/4}\left(\begin{array}{rr} \bQ_{\dot{2}\dot{3}}^\dagger & \bQ_{\dot{2}\dot{4}}^\dagger\\
-\bQ_{\dot{1}\dot{3}}^\dagger & -\bQ_{\dot{1}\dot{4}}^\dagger\end{array}\right)
\, .
\end{aligned}
\eea Here, by definition,  $\bQ_{a \b}^\dagger$ and
$\bQ^\dagger_{\da\dbe}$ are understood as hermitian  conjugate of
$\bQ^{\b a}$ and $\bQ^{\dbe\da}$, respectively, \bea\nonumber
\big(\bQ^{\b a}\big)^\dagger = \bQ_{a \b}^\dagger\,
,\quad\big(\bQ^{\dbe\da}\big)^\dagger =\bQ_{\da\dbe}^\dagger\, ,
\eea and the tracelessness condition for bosonic charges implies
that they are symmetric: $\bL^{ab}=\bL^{ba}$ and so on. Note also
that according to the transformation rule (\ref{lawboson2}) for
$\bQ$,
 it would be more consistent to write the entries
of $\bQ$ as $\bQ^{ b\a}$ rather than $\bQ^{ \a b}$. However, the
order of the indices does not matter because the transformations
by the group elements $\ag_1$ and $\ag_3$ are independent, and
with the choice we made many formulae for the dotted operators are
obtained from the undotted ones by replacing correspondingly the
indices. The phases $e^{\pm i\pi/4}$ in the expressions of the
supercharges are introduced to simplify their representation in
terms of creation and annihilation operators, see appendix
\ref{app:Q}.

We can lower the indices by using the skew-symmetric tensor, and
in what  follows we find it sometimes convenient to lower the
first index and use the following charges \bea\nonumber \bL_a{}^b
= \eps_{ac}\, \bL^{cb}\,,\quad \bR_\a{}^\b = \eps_{\a\g}\,
\bR^{\g\b}\,,\quad \bQ_\a{}^b = \eps_{\a\g}\, \bQ^{\g b}\,,\quad
\bQ_b^{\dagger}{}^{\a} = \eps^{\a\g}\, \bQ^\dagger_{b\g}\,. \eea
One can check that these charges satisfy the following conditions
 \bea\nonumber
\left(\bL_{a}{}^b\right)^\dagger =\bL_{b}{}^a\, ,\ \ \bL_{1}{}^1 + \bL_{2}{}^2 = 0\,,\ \
\left(\bR_{\a}{}^\b\right)^\dagger =\bR_{\b}{}^\a\, ,\ \ \bR_{3}{}^3 + \bR_{4}{}^4 = 0\,,\ \ \left(\bQ_\a{}^b\right)^\dagger =\bQ_{b}^{\dagger}{}^{\a}\,.~~~ \eea

We show in the next subsection that  the bosonic rotation
generators $\bL_a{}^b\,,\ \bR_\a{}^\b$, the supersymmetry
generators $\bQ_\a{}^a,\,\ \bQ_a^{\dagger}{}^\a$, and three
central elements $\bH$, $\bC$ and $\bC^\dagger$ form the centrally
extended $\su(2|2)$ algebra which we will denote $\su(2|2)_\cex$.
The $\su(2|2)_\cex$ algebra relations can be written in the
following form \bea \label{su22} &&
\left[\bL_a{}^b,\bJ_c\right]=\delta_c^b \bJ_a - {1\ov 2}\delta_a^b
\bJ_c\,,\qquad~~~~ \left[\bR_\a{}^\b,\bJ_\g\right]=\delta^\b_\g
\bJ_\a - {1\ov 2}\delta^\b_\a \bJ_\g\,,
 \nonumber \\
&& \left[\bL_a{}^b,\bJ^c\right]=-\delta_a^c \bJ^b + {1\ov 2}\delta_a^b \bJ^c\,,\qquad
~~\left[\bR_\a{}^\b,\bJ^\g\right]=-\delta_\a^\g \bJ^\b + {1\ov 2}\delta_\a^\b \bJ^\g\,, \nonumber \\
&& \{ \bQ_\a{}^a, \bQ_b^\dagger{}^\b\} = \delta_b^a \bR_\a{}^\b + \delta_\a^\b \bL_b{}^a +{1\ov 2}\delta_b^a\delta^\b_\a  \bH\,, \nonumber \\
&& \{ \bQ_\a{}^a, \bQ_\b{}^b\} =
\epsilon_{\a\b}\epsilon^{ab}~\bC\, , ~~~~~~~~ \{
\bQ_a^\dagger{}^\a, \bQ_b^\dagger{}^\b\} =
\epsilon_{ab}\epsilon^{\a\b}~\bC^\dagger \,. \eea Here the first
two lines indicate how the indices $c$ and $\gamma$ of any Lie
algebra generator transform under the action of $\bL_a{}^b$ and
$\bR_\a{}^\b$. Unitarity of the string sigma model requires the
world-sheet light-cone Hamiltonian $\bH$ to be hermitian, and the
supersymmetry generators $\bQ_\a{}^a$ and $\bQ_a^{\dagger}{}^\a$,
and the central elements $\bC$ and $\bC^\dagger$ to be hermitian
conjugate to each other: $\left(
\bQ_\a{}^a\right)^\dagger=\bQ_a^{\dagger}{}^\a$. If one gives up
the hermiticity conditions then all the generators are considered
as independent.

 As we argue in the next subsection,  the
central elements $\bC$ and $\bC^\dagger$  are expressed through
the world-sheet momentum $\pws\equiv \hP$ as follows \bea
\label{Cc} \bC={i\ov 2}g\,(e^{i\hP}-1)e^{2i\xi}\,,\quad\quad
\bC^\dagger=-{i\ov 2}g\,(e^{-i\hP}-1)e^{-2i\xi}\, . \eea In
general, the phase $\xi$ is an arbitrary function of the central
elements. Its presence reflects the obvious fact that the algebra
(\ref{su22}) admits a ${\rm U(1)}$ outer automorphism: $\bQ\to
e^{i\xi}\bQ\,,\ \bC\to e^{2i\xi}\bC$. In perturbative string
theory the phase $\xi$ vanishes, as we will see shortly, and we
find it convenient to set $\xi = 0$ for any value of the string
tension $g$. It is important to realize that the central charges
$\bC$ and $\bC^\dagger$ vanish on the physical subspace
$\bP|\Psi\rangle=0$ where the usual $\su(2|2)$ algebra is
restored.

The remaining generators $\bL_\da{}^\db\,,\ \bR_\dal{}^\dbe$,
$\bQ_\dal{}^\da,\,\ \bQ_\da^{\dagger}{}^\dal$  form another copy
of  $\su(2|2)_\cex$ with the same three central elements $\bH$,
$\bC$ and $\bC^\dagger$.  Thus, the manifest symmetry algebra of
the light-cone $\AdS$ string sigma model coincides  with the sum
of two copies of $\su(2|2)_\cex$ sharing the same set of  central
elements. Because of the location of the generators in the charge
matrix (\ref{qsym}) we will often refer to the algebras generated
by undotted and dotted charges as to the left and right
$\su(2|2)_\cex$ algebras, respectively.

\subsection{Deriving the central charges}


Given the complexity  of the supersymmetry generators
(\ref{Charges}) in the light-cone gauge as well as the
corresponding Poisson structure of the theory, computation of the
exact classical and quantum supersymmetry algebra is difficult.
Hence, simplifying perturbative methods need to be applied. The
perturbative expansion of the supersymmetry generators in powers
of $1/g$ or, equivalently, in the number of fields defines a
particular expansion scheme. Since in the large string tension
expansion one keeps $\Bp=g\,\bP$ fixed, the corresponding
expansion of the central charges starts with $-\Bp/2$, and can be
seen already at the quadratic order.
 This
expansion, however, does not allow one to determine the exact form
of the central charges (\ref{Cc})  because they are non-trivial
functions of $1/g$. To overcome this difficulty,  in this
subsection we describe a ``hybrid'' expansion scheme which can be
used to determine the exact form of the central charges. To be
precise we determine only the part of the central charges which is
independent of fermionic fields. We find that this part depends
solely on the piece of the world-sheet momentum which involves the
bosonic fields. Since the central charges must vanish  if the
world-sheet momentum does, the exact form of the central charges
is, therefore, unambiguously fixed by its bosonic part.

\medskip

More precisely, as can be seen from (\ref{Charges}) and
(\ref{qsym}), a dynamical supersymmetry generator has the
following generic structure \bea \bQ_{A}{}^B=\int {\rm d}\s
~e^{i{\a} x_-} \Omega( x,p,\chi; g)\,, \label{sch} \eea where the
parameter $\a$ in the exponent of (\ref{sch}) is equal to
$\a=1/2(\eps_A - \eps_B)$, and, therefore,  $\a= 1/2$ for
supercharges $\bQ$ and $ \mathring{\bQ}$, and $\a= -1/2$ for
supercharges $\bQ^\dagger$ and $ \mathring{\bQ}^\dagger$. Then,
the function $\Omega(x,p,\chi; g)$ is a {\it local} function of
transversal bosonic fields and fermionic variables. It depends on
$g$ and can be expanded, quite analogously to the Hamiltonian, in
power series
$$ \Omega(x,p,\chi;g)=
\Omega_{2}(x,p,\chi)+{1\ov g} \Omega_{4}(x,p,\chi)+\cdots \,.
$$
Here $\Omega_{2}(x,p,\chi)$ is quadratic in fields,
$\Omega_{4}(x,p,\chi)$ is quartic and so on. Clearly, every term
in this series also admits a finite expansion in the number of
fermions. In the usual perturbative expansion we would also have
to expand the non-local ``vertex'' $e^{i\a x_-}$ in powers of
$1/g$ because $x'_- \sim -px'/g +\cdots$. In the hybrid expansion
we do not expand $e^{i\a x_-}$ but rather treat it as a rigid
object.

\medskip

The complete expression for a supercharge is rather cumbersome.
However, we see that the supercharges and their algebra can be
studied perturbatively: first by expanding up to a given order in
$1/g$ and then by truncating the resulting polynomial up to a
given number of fermionic variables. Then, as was discussed above
the exact form of the central charges is completely fixed by their
parts which depend only on bosons. Thus, to determine these
charges it is sufficient to consider the terms in $\bQ_A{}^B$
which are linear in fermions, and compute their Poisson brackets
(or anticommutators in quantum theory) keeping only terms
independent of fermions. This is, however, a complicated problem
because the Poisson brackets of fermions appearing in
(\ref{Charges}) have a highly non-trivial dependence on bosons as
have been discussed in subsection \ref{sec:gfL}.  We have  shown
in  subsection \ref{sec:lste} that to have the canonical Poisson
brackets one should perform a field redefinition which can be
determined up to any given order in $1/g$. Taking into account the
field redefinition, integrating by parts if necessary, and using
the relation $x'_- \sim -px'/g +\cdots$, one can cast any
supercharge (\ref{sch}) in the following symbolic form \bea
{\bQ}_A{}^B=\int {\rm d}\s ~e^{i{\a} x_-}\,\chi\cdot\big(
\Upsilon_1(x,p) +{1\ov g} \Upsilon_3(x,p) + \cdots \big) + {\cal
O}(\chi^3)\,, \label{sch2} \eea where $\Upsilon_1$ and
$\Upsilon_3$ are linear and cubic in bosonic fields, respectively.
The explicit form of the supercharges expanded up to the leading
order in $1/g$ can be found in appendix \ref{app:Q}.

\medskip

It is clear now that the bosonic part of the Poisson bracket of
two supercharges is of the form
\bea\label{pbsch}
&&\{{\bQ}_1,
{\bQ}_2\}\sim \int_{-\infty}^\infty {\rm d}\s ~ e^{i(\a_1+\a_2)
x_-}\Big(
\Upsilon_1^{(1)}(x,p) \Upsilon_1^{(2)}(x,p)\\
 &&~~~~~~~~~~~~~~~~~~~~~~~~~+ {1\ov g} \big( \Upsilon_1^{(1)}(x,p) \Upsilon_3^{(2)}(x,p)
+ \Upsilon_3^{(1)}(x,p) \Upsilon_1^{(2)}(x,p)\big)+  \cdots
\Big)\,, \nonumber \eea where $\bQ_{1,2}\equiv
\bQ_{{A}_{1,2}}^{~{B}_{1,2}}$. Computing the product
$\Upsilon_1^{(1)}(x,p) \Upsilon_1^{(2)}(x,p)$ in the case
$\a_1=\a_2 =\pm 1/2$, we find that it is given by
\bea\label{uusch} \Upsilon_1^{(1)}(x,p) \Upsilon_1^{(2)}(x,p)\sim
g\, x_-' + {d\ov d\s}f(x,p)\,, \eea where $f(x,p)$ is a local
function of transversal coordinates and momenta. The first term in
(\ref{uusch}) nicely combines with $e^{\pm i x_-}$ to give ${d\ov
d\s}e^{\pm i x_-}$, and integrating this expression over $\s$, we
obtain the sought for central charges \bea\la{Ccg}
\int_{-\infty}^\infty {\rm d}\s ~{d\ov d\s}e^{\pm i x_-} = e^{\pm
i x_-(\infty)} - e^{\pm i x_-(-\infty)} = e^{\pm i
x_-(-\infty)}\left(e^{\pm i \pws} -1\right)\,, \eea where we take
into account that $x_-(\infty)-x_-(-\infty)=\pws$.

\medskip

Making use of the explicit expressions for the supercharges from
appendix \ref{app:Q},  and identifying $x_-(-\infty)\equiv\xi$,
one can easily confirm that the central charges $\bC$ and
$\bC^\dagger$ are given by eqs.(\ref{Cc}). Thus, the phase $\xi$
in the central charges determines the boundary conditions for the
light-cone coordinate $x_-$. As was mentioned above, in what
follows we choose $\xi=0$. It is worth noting however that there
is another natural choice of the boundary conditions for the
light-cone coordinate $x_-$: \bea\nonumber
x_-(+\infty)=-x_-(-\infty) = {\pws\ov 2}\,. \eea This is the
symmetric condition which treats both boundaries on equal footing,
and leads to a real central charge \bea \label{Cc2}
\bC=\bC^\dagger=-g\sin({\pws\ov 2})\, . \eea

\medskip

Since we already obtained the expected central charges, the
contribution of all the other terms in (\ref{pbsch}) should
vanish. Indeed, the second term in (\ref{uusch}) contributes to the
order $1/g$ in the expansion as can be easily seen integrating by parts
and using the relation $x'_- \sim -px'/g +\cdots$. Taking into
account the additional contribution to the terms of order $1/g$ in
(\ref{pbsch}), one can check that the total contribution is given by
a $\s$-derivative of a local function of $x$ and $p$, and, therefore,
only contributes to terms of order $1/g^2$.

It is also not difficult to verify up to the quartic order in fields that the
Poisson bracket of supercharges with $\a_1=-\a_2$ gives the
Hamiltonian and the kinematic generators in complete agreement
with the centrally extended $\su(2|2)$ algebra (\ref{su22}).

The next step is to show that the Hamiltonian commutes with all
dynamical supercharges. As was already mentioned, this can be done
order by order in perturbation theory in powers of the inverse
string tension $1/g$ and in number of fermionic variables. One can
demonstrate that up to the first non-trivial order $1/g$ the
supercharge ${\bQ}$ truncated to the terms linear in fermions
indeed commutes with ${\bH}$. To do this, one needs to keep in
${\bH}$ all quadratic and quartic bosonic terms, and quadratic and
quartic terms which are quadratic in fermions.

The computation we described above was purely classical, and one
may want to know if quantizing the model could lead to some
anomaly in the symmetry algebra. One can compute the symmetry
algebra in the plane-wave limit where one keeps only quadratic
terms in all the symmetry generators, and show that all
potentially divergent terms cancel out and no quantum anomaly
arises. The simplest way to do the computation is to use the form
of the symmetry algebra generators in terms of the creation and
annihilation operators from appendix (\ref{app:Q}).

Another quantum effect might be a modification of the functional
dependence of the central charges on the string tension and the
world-sheet momentum. It is believed, however, that the form
(\ref{Cc}) remains unmodified by quantum corrections, as it is
consistent with both string (large $g$) and field (small $g$)
theory computations of the dispersion relation.

Thus, we have shown that in the decompactification limit and for physical fields
chosen to rapidly decrease at infinity the corresponding string model
enjoys the symmetry which coincides with two copies of the centrally-extended $\su(2|2)$ algebra (\ref{su22})
sharing the same Hamiltonian and central charges.

\section{Appendix}
\subsection{Giant magnon: Explicit formulae} \la{app:magnon}
Here we unwrap some formulae from subsection \ref{magn} and
specify them for the three simplest cases $a=0,1/2,1$.

The density of the gauge-fixed Hamiltonian $\H$ appearing in
(\ref{Su1}) as a function of the coordinate $z$ and the momentum
$p_z$ canonically conjugate to $z$ is \bea \la{denHu1} \H &=&
-\frac{1-(1-a)z^2}{1-2a-(1-a)^2z^2} \\\nonumber &~& +
\frac{\sqrt{1+\left(1 - z^2\right) \left(1-2a-(1 - a)^2
z^2\right)p_z^2} \sqrt{1-z^2 + \left(1-2a-(1-a)^2z^2\right)
z'^2}}{1-2a-(1-a)^2z^2 }\, . \eea The density of the Hamiltonian
(\ref{denHu1}) for the three simplest cases: \bea \la{denHa0}
\nonumber a=0:\quad \H &=&  -1+\sqrt{\frac{1+z'^2}{1 -
z^2}}\sqrt{1+p_z^2 \left(1 - z^2\right)^2} \,,
 \\\la{denHa12} \nonumber
a={1\ov 2}:\quad \H &=& -2 + \frac{4}{z^2} -\frac{1}{z^2}
\sqrt{4(1- z^2)-z^2z'^2}\sqrt{4-p_z^2  z^2\left(1-z^2\right)}\,,
 \\\la{denHa1}\nonumber
a=1:\quad \H &=& 1 -\sqrt{1-z^2 -
\left(z'\right)^2}\sqrt{1-\left(1-z^2\right) p_z^2 }\,. \eea
Solving the equation of motion for $p_z$ that follows from the
action (\ref{Su1}), we determine the momentum as a function of
$\dz$ and $z$ \bea\la{pza}
p_z=\frac{\dz}{\sqrt{\left(1-z^2\right)}
\sqrt{\left(1-z^2\right)^2-\left(1-2 a-(1-a)^2 z^2\right)
   \left(\dz^2-\left(1-z^2\right) \left(z'\right)^2\right)}}\,.
\eea The momentum $p_z$ as a function of $\dz$ and $z$ for the
three simplest cases: \bea\la{pza0}\nonumber a=0:\quad p_z&=& {\dz
\ov (1-z^2)\sqrt{1-z^2-\dz^2 + (1-z^2) z'^2}}  \,,
 \\\la{pza12} \nonumber
a={1\ov 2}:\quad p_z&=&  \frac{2 \dz}{\sqrt{1 - z^2} \sqrt{4
\left(1 - z^2\right)^2+ z^2\left(\dz^2 - \left(1-z^2\right)
z'^2\right) }}\,,
 \\\la{pza1} \nonumber
a=1:\quad p_z&=&  \frac{\dz}{\sqrt{1 - z^2}\sqrt{ \left(1 -
z^2\right)^2 + \dz^2-\left(1 - z^2\right)z'^2 }} \,. \eea
Substituting the solution (\ref{pza}) into the action (\ref{Su1}),
we obtain the action in the Lagrangian form \bea\la{Su2} &&S =g
\int_{- r}^{ r}\, {\rm d}\s{\rm d}\tau\, \left( \frac{1 - (1 - a)
z^2}{1 - 2 a -(1 - a)^2 z^2} \right.\\\nonumber
&&~~~~~~~~~~~~~~~~\left. -\frac{\sqrt{\left(1 - z^2\right)^2 -
\left( 1- 2 a -(1 - a)^2 z^2\right) \left(\dz^2 - \left(1 -
z^2\right) z'^2\right)}}{\sqrt{1 - z^2} \left( 1- 2 a -(1 - a)^2
z^2\right)} \ \right)\,. \eea The action (\ref{Su1}) in the
Lagrangian form for the three simplest cases:
\bea\la{Su2a0}\nonumber a=0:\quad &S& =g \int_{- r}^{ r}\, {\rm
d}\s{\rm d}\tau\, \left( 1-{\sqrt{1-z^2-\dz^2 + (1-z^2) z'^2}\ov
1-z^2} \right)\,, \\\la{Su2a12}\nonumber a={1\ov 2}:\quad &S& =g
\int_{- r}^{ r}\, {\rm d}\s{\rm d}\tau\, \left( 2 -\frac{4}{z^2} +
\frac{2 \sqrt{4 \left(1 - z^2\right)^2+z^2\left(\dz^2 -\left(1 -
z^2\right) z'^2\right)}} {z^2 \sqrt{1 - z^2}}\ \right)\,,
\\\la{Su2a1} \nonumber a=1:\quad &S& =g \int_{- r}^{ r}\, {\rm
d}\s{\rm d}\tau\, \left(-1 + \frac{\sqrt{\left(1 - z^2\right)^2  +
\dz^2- \left(1 -z^2\right)z'^2}}{\sqrt{1 - z^2}} \right)\,. \eea
Substituting the ansatz (\ref{ansatz}) into the action
(\ref{Su2}), we get the following Langrangian of the reduced model
\begin{eqnarray*}
L_{red} = \frac{1 - (1 - a) z^2}{1 - 2 a
-(1 - a)^2 z^2} -\frac{\sqrt{\left(1 - z^2\right)^2 + \left( 1- 2
a -(1 - a)^2 z^2\right) \left(1-v^2 - z^2\right) z'^2}}{\sqrt{1 -
z^2} \left( 1- 2 a -(1 - a)^2 z^2\right)} \,. \, \,
\end{eqnarray*}
The Hamiltonian of the reduced one-dimensional model is
\bea
\begin{aligned}
\nonumber
&H_{red} = \pi_z z' - L_{red}= -\frac{1 - (1 - a) z^2}{ 1- 2 a-(1 - a)^2 z^2} \\
&~~+ \frac{\left(1 -z^2\right)^{3/2}}{\left( 1- 2 a -(1 - a)^2 z^2\right)
\sqrt{\left(1 - z^2\right)^2 + \left( 1- 2 a -(1 - a)^2 z^2\right)
\left(1-v^2 - z^2\right) z'^2} }\,.
\end{aligned}
\eea

\subsection{Quartic Hamiltonian in two-index fields}\la{H4double}

We use eq.(\ref{H4matr}) to find the following expressions for the density of the quartic Hamiltonian in terms of the  two-index fields
\bea\nonumber
\H_4 = \H_4^b + \H_4^f + \H_4^{bf}\,,
\eea
where
\bea\la{H4b} \H_4^b  =
 -2 \big(Y^{a\da}Y_{a\da} - Z^{\a\dal}Z_{\a\dal} \big)  \big(Y'^{b\db}Y'_{b\db} +Z'^{\b\dbe}Z'_{\b\dbe}\big)
\eea
is the bosonic Hamiltonian,
\bea\nonumber\hspace{-0.8cm} \H_4^f  &=&
{1\ov 4}\Big( \eta^{\a \da}\eta^{\b \db}  \eta^{\prime}_{\a \da}\eta'_{\b \db} +\eta^{\a \da}\eta^{\prime\b \db}  \eta_{\a \db}\eta'_{\b \da} +\eta^{\dagger\a \da}\eta^{\dagger\b \db}  \eta^{\dagger\prime}_{\a \da}\eta^{\dagger\prime}_{\b \db} + \eta^{\dagger\a \da}\eta^{\dagger\prime\b \db}  \eta^\dagger_{\a \db}\eta^{\dagger\prime}_{\b \da}~~~\\ \nonumber
&&~+\eta^{\a \da}\eta^{\b \db}  \eta^{\dagger\prime}_{\a \da}\eta^{\dagger\prime}_{\b \db} +\eta^\dagger_{\a \da}\eta^\dagger_{\b \db}  \eta^{\prime\a \da}\eta^{\prime\b \db} - \eta^{\a \da}\eta_{\a \db}  \eta^{\dagger\prime}_{\b \da}\eta^{\dagger\prime\b \db}- \eta^\dagger_{\a \da}\eta^\dagger_{\a \db}  \eta^{\prime\b \da}\eta^{\prime}_{\b \db}\\\nonumber
&&~+
\theta^{a \dal}\theta^{b \dbe}  \theta^{\prime}_{a \dal}\theta'_{b \dbe} +\theta^{a \dal}\theta^{\prime b \dbe}  \theta_{b\dal }\theta'_{a\dbe } + \theta^{\dagger a \dal}\theta^{\dagger b \dbe}  \theta^{\dagger\prime}_{a \dal}\theta^{\dagger\prime}_{b \dbe} + \theta^{\dagger a \dal}\theta^{\dagger\prime b \dbe}  \theta^\dagger_{b\dal }\theta^{\dagger\prime}_{a\dbe }~~~\\
&&~+\theta^{a \dal}\theta^{b \dbe}  \theta^{\dagger\prime}_{a \dal}\theta^{\dagger\prime}_{b \dbe} +\theta^\dagger_{a \dal}\theta^\dagger_{b \dbe}  \theta^{\prime a \dal}\theta^{\prime b \dbe} - \theta^{a \dal}\theta_{b\dal }  \theta^{\dagger\prime}_{a\dbe }\theta^{\dagger\prime b \dbe}-\theta^\dagger_{a \dal}\theta^{\dagger}_{ b\dal }  \theta^{\prime a\dbe }\theta^{\prime}_{b \dbe}\Big)\la{H4f}
\eea
is the fermionic Hamiltonian, and
\bea\nonumber&&\H_4^{bf}  =
  \big(  Z^{\a\dal}Z_{\a\dal} -Y^{a\da}Y_{a\da}\big)  \big(\eta^{\dagger\prime}_{\b \db} \eta^{\prime\b \db} + \theta^{\dagger\prime}_{b \dbe} \theta^{\prime b \dbe} \big) -
  4i\big( \eta'_{\a \da} \theta^{\prime}_{a \dal} +   \eta^{\dagger\prime}_{\a \da}\theta^{\dagger\prime}_{a\dal }\big) Y^{a \da}Z^{\a\dal}~~~~~~~~~~~ \\\nonumber
&&~~~~~~-{1\ov 2}  \big( \eta^{\a \da} \eta^{\dagger\prime}_{\a \da} +   \eta^{\dagger}_{\a \da}\eta^{\prime \a\da } +
 \theta^{a \dal} \theta^{\dagger\prime}_{a \dal} +   \theta^{\dagger}_{a \dal}\theta^{\prime a\dal }\big) \big(Y^{b \db}Y'_{b\db}+Z^{\b \dbe}Z'_{\b\dbe}\big) \\\nonumber
 &&~~~~~~+ \big( \eta^{\a \da} \eta^{\dagger\prime}_{\a \db} +   \eta^{\dagger}_{\a \db}\eta^{\prime \a\da }\big) Y_{a \da}Y'^{a\db} + \big( \theta^{a \dal} \theta^{\dagger\prime}_{b \dal} +   \theta^{\dagger}_{b \dal}\theta^{\prime a\dal }\big) Y_{a \da}Y'^{b\da}\\\nonumber
&&~~~~~~+ \big( \eta^{\b \da} \eta^{\dagger\prime}_{\a \da} +   \eta^{\dagger}_{\a \da}\eta^{\prime \b\da } \big) Z^{\a \dal}Z'_{\b\dal}+  \big( \theta^{a \dbe} \theta^{\dagger\prime}_{a \dal} +   \theta^{\dagger}_{a \dbe}\theta^{\prime a\dal }\big) Z^{\a \dal}Z'_{\a\dbe}\\\nonumber
&&~~~~~~+ {i\kappa\ov 4} \Big( \big( \eta^{\a \da} \eta_{\a \db} + \eta^{\dagger\a \da} \eta^{\dagger}_{\a \db}    \big)\big(P_{a\da}   Y^{a\db}\big)^\prime +  \big(  \theta^{a \dal} \theta_{b \dal} + \theta^{\dagger a \dal} \theta^\dagger_{b \dal}  \big)\big(P_{a\da}   Y^{b\da}\big)^\prime \\
&&~~~~~~
+ \big( \eta_{\b \da} \eta^{\a \da} +  \eta^{\dagger}_{\b \da} \eta^{\dagger\a \da}  \big)\big(P_{\a\dal}Z^{\b\dal}\big)^\prime +  \big(  \theta_{a \dbe} \theta^{a \dal} + \theta^{\dagger}_{a \dbe} \theta^{\dagger a \dal}  \big)(P_{\a\dal}Z^{\a\dbe}\big)^\prime
\Big)\la{H4bf}
\eea
is the mixed Hamiltonian.

\subsection{T-matrix}\la{app:T}
\newcommand{\cpp}{p\,\om'  - p'\,\om }
\newcommand{\ppcpp}{\tfrac{pp'}{\cpp}}
\newcommand{\ket}[1]{\mathopen{|}#1\mathclose{\rangle}}
\newcommand{\lrbrk}[1]{\left(#1\right)}


Here we list the full T-matrix in the uniform $a=1/2$  light-cone
gauge.  To simplify the notations and for visual clarity we use the following notations
\bea\nonumber
&&a^\dagger_{a\da}(p)\to Y_{a\da}\,,\quad a^\dagger_{a\da}(p')\to Y'_{a\da}\,,\quad
a^\dagger_{\a\dal}(p)\to Z_{\a\dal}\,,\quad a^\dagger_{\a\dal}(p')\to Z'_{\a\dal}\,,\\\nonumber
&&a^\dagger_{\a\da}(p)\to \eta_{\a\da}\,,\quad a^\dagger_{\a\da}(p')\to \eta'_{\a\da}\,,\quad
a^\dagger_{a\dal}(p)\to \theta_{\a\dal}\,,\quad a^\dagger_{a\dal}(p')\to \theta'_{a\dal}\,,
\eea
so that we have, in particular
$$|Y_{a\da}\eta'_{\b\db}\rangle\equiv |a^\dagger_{a\da}(p)a^\dagger_{\b\db}(p')\rangle\,,\quad
|\theta_{a\dal}Z'_{\b\dbe}\rangle\equiv |a^\dagger_{a\dal}(p)a^\dagger_{\b\dbe}(p')\rangle\,.
$$
Then we introduce the rapidity $\theta$ related to the momentum $p$ and energy $\om$ as follows
$$
p=\sinh\theta\,,\quad \om=\cosh\theta\,.
$$
Since the model is not Lorentz-invariant, the T-matrix does not
depend only on the difference $\theta -\theta'$, and one may find
the following
 identities useful
\begin{align}
  & p\,\om'  - p'\, \om = \sinh(\theta-\theta') \,,\quad (p-p') \cosh\tfrac{\theta-\theta'}{2} = (\om+\om') \sinh\tfrac{\theta-\theta'}{2} \nn \\
  & \sinh\tfrac{\theta}{2} = \half \sqrt{\om + p} - \half \sqrt{\om - p}\,,\quad
  \cosh\tfrac{\theta}{2} = \half \sqrt{\om + p} + \half \sqrt{\om - p} \nn \\
  & \sinh\tfrac{\theta-\theta'}{2} = \half \sqrt{(\om + p)(\om' - p')} - \half \sqrt{(\om - p)(\om' + p')} \nn \\
  & \cosh\tfrac{\theta-\theta'}{2} = \half \sqrt{(\om + p)(\om' - p')} + \half \sqrt{(\om - p)(\om' + p')} \nn
\end{align}
The two momenta $p$ and $p'$ satisfy $p>p'$.


\subsubsection*{Boson-Boson}
\[
\begin{split}
\bT\cdot \ket{Y_{{a}{\da}} Y'_{{b}{\db}}} = \ &
  + \half \tfrac{(p-p')^2}{\cpp}
  \ket{Y_{{a}{\da}} Y'_{{b}{\db}}}
  + \ppcpp\lrbrk{
  \ket{Y_{{a}{\db}} Y'_{{b}{\da}}} +
  \ket{Y_{{b}{\da}} Y'_{{a}{\db}}} } \\ &
  - \ppcpp \sinh\tfrac{\theta-\theta'}{2}\lrbrk{
  \eps_{{\da}{\db}} \eps^{{\dal}{\dbe}} \ket{\eta_{{a}{\dal}}\eta'_{{b}{\dbe}}} +
  \eps_{{a}{b}} \eps^{{\a}{\b}} \ket{\theta_{{\a}{\da}}\theta'_{{\b}{\db}}} } \\[1mm]
\bT\cdot  \ket{Z_{{\a}{\dal}} Z'_{{\b}{\dbe}}} = \ &
  - \half \tfrac{(p-p')^2}{\cpp}
  \ket{Z_{{\a}{\dal}} Z'_{{\b}{\dbe}}}
  - \ppcpp\lrbrk{
  \ket{Z_{{\a}{\dbe}} Z'_{{\b}{\dal}}} +
  \ket{Z_{{\b}{\dal}} Z'_{{\a}{\dbe}}} } \\ &
  + \ppcpp \sinh\tfrac{\theta-\theta'}{2}\lrbrk{
  \eps_{{\dal}{\dbe}} \eps^{{\da}{\db}} \ket{\theta_{{\a}{\da}}\theta'_{{\b}{\db}}} +
  \eps_{{\a}{\b}} \eps^{{a}{b}} \ket{\eta_{{a}{\dal}}\eta'_{{b}{\dbe}}} } \\[1mm]
\bT\cdot  \ket{Y_{{a}{\da}} Z'_{{\a}{\dal}}} = \ &
  - \half \tfrac{p^2-p'^2}{\cpp} \ket{Y_{{a}{\da}} Z'_{{\a}{\dal}}}
  + \ppcpp \cosh\tfrac{\theta-\theta'}{2}
  \lrbrk{ \ket{\theta_{{\a}{\da}}\eta'_{{a}{\dal}}}
        - \ket{\eta_{{a}{\dal}}\theta'_{{\a}{\da}}} } \\[1mm]
\bT\cdot  \ket{Z_{{\a}{\dal}} Y'_{{a}{\da}}} = \ &
  + \half \tfrac{p^2-p'^2}{\cpp} \ket{Z_{{\a}{\dal}} Y'_{{a}{\da}}}
  - \ppcpp \cosh\tfrac{\theta-\theta'}{2}
  \lrbrk{ \ket{\eta_{{a}{\dal}}\theta'_{{\a}{\da}}}
        - \ket{\theta_{{\a}{\da}}\eta'_{{a}{\dal}}} }
\end{split}
\]

\subsubsection*{Fermion-Fermion}
\[
\begin{split}
\bT\cdot   \ket{\eta_{{a}{\dal}}\eta'_{{b}{\dbe}}} = \ &
  + \ppcpp\lrbrk{
  \ket{\eta_{{b}{\dal}}\eta'_{{a}{\dbe}}} -
  \ket{\eta_{{a}{\dbe}}\eta'_{{b}{\dal}}} } \\ &
  - \ppcpp \sinh\tfrac{\theta-\theta'}{2}\lrbrk{
  \eps_{{\dal}{\dbe}} \eps^{{\da}{\db}} \ket{Y_{{a}{\da}} Y'_{{b}{\db}}} -
  \eps_{{a}{b}} \eps^{{\a}{\b}} \ket{Z_{{\a}{\dal}} Z'_{{\b}{\dbe}}} } \\[1mm]
\bT\cdot   \ket{\theta_{{\a}{\da}}\theta'_{{\b}{\db}}} = \ &
  - \ppcpp\lrbrk{
  \ket{\theta_{{\b}{\da}}\theta'_{{\a}{\db}}} -
  \ket{\theta_{{\a}{\db}}\theta'_{{\b}{\da}}} } \\ &
  + \ppcpp \sinh\tfrac{\theta-\theta'}{2}\lrbrk{
  \eps_{{\da}{\db}} \eps^{{\dal}{\dbe}} \ket{Z_{{\a}{\dal}} Z'_{{\b}{\dbe}}} -
  \eps_{{\a}{\b}} \eps^{{a}{b}} \ket{Y_{{a}{\da}} Y'_{{b}{\db}}} } \\[1mm]
\bT\cdot   \ket{\eta_{{a}{\dal}}\theta'_{{\b}{\db}}} = \ &
  - \ppcpp \cosh\tfrac{\theta-\theta'}{2}\lrbrk{
  \ket{Y_{{a}{\db}} Z'_{{\b}{\dal}}} +
  \ket{Z_{{\b}{\dal}} Y'_{{a}{\db}}} } \\[1mm]
\bT\cdot   \ket{\theta_{{\a}{\da}}\eta'_{{b}{\dbe}}} = \ &
  + \ppcpp \cosh\tfrac{\theta-\theta'}{2}\lrbrk{
  \ket{Z_{{\a}{\dbe}} Y'_{{b}{\da}}} +
  \ket{Y_{{b}{\da}} Z'_{{\a}{\dbe}}} }
\end{split}
\]

\vspace{-0.3cm}
\subsubsection*{Boson-Fermion}
\[
\begin{split}
\bT\cdot   \ket{Y_{{a}{\da}}\eta'_{{b}{\dbe}}} = \ &
  + \half \tfrac{(p'-p)p'}{\cpp} \ket{Y_{{a}{\da}}\eta'_{{b}{\dbe}}}
  + \ppcpp \ket{Y_{{b}{\da}}\eta'_{{a}{\dbe}}} \\ &
  + \ppcpp \cosh\tfrac{\theta-\theta'}{2} \ket{\eta_{{a}{\dbe}} Y'_{{b}{\da}}}
  - \ppcpp \sinh\tfrac{\theta-\theta'}{2} \eps_{{a}{b}} \eps^{{\a}{\b}} \ket{\theta_{{\a}{\da}} Z'_{{\b}{\dbe}}} \\[1mm]
\bT\cdot   \ket{Y_{{a}{\da}}\theta'_{{\b}{\db}}} = \ &
  + \half \tfrac{(p'-p)p'}{\cpp} \ket{Y_{{a}{\da}}\theta'_{{\b}{\db}}}
  + \ppcpp \ket{Y_{{a}{\db}}\theta'_{{\b}{\da}}} \\ &
  + \ppcpp \cosh\tfrac{\theta-\theta'}{2} \ket{\theta_{{\b}{\da}} Y'_{{a}{\db}}}
  + \ppcpp \sinh\tfrac{\theta-\theta'}{2} \eps_{{\da}{\db}} \eps^{{\dal}{\dbe}} \ket{\eta_{{a}{\dal}} Z'_{{\b}{\dbe}}} \\[1mm]
\bT\cdot   \ket{\eta_{{a}{\dal}} Y'_{{b}{\db}}} = \ &
  + \half \tfrac{(p-p')p}{\cpp} \ket{\eta_{{a}{\dal}} Y'_{{b}{\db}}}
  + \ppcpp \ket{\eta_{{b}{\dal}} Y'_{{a}{\db}}} \\ &
  + \ppcpp \cosh\tfrac{\theta-\theta'}{2} \ket{Y_{{a}{\db}}\eta'_{{b}{\dal}}}
  + \ppcpp \sinh\tfrac{\theta-\theta'}{2} \eps_{{a}{b}} \eps^{{\a}{\b}} \ket{Z_{{\a}{\dal}}\theta'_{{\b}{\db}}} \\[1mm]
\bT\cdot   \ket{\theta_{{\a}{\da}} Y'_{{b}{\db}}} = \ &
  + \half \tfrac{(p-p')p}{\cpp} \ket{\theta_{{\a}{\da}} Y'_{{b}{\db}}}
  + \ppcpp \ket{\theta_{{\a}{\db}} Y'_{{b}{\da}}} \\ &
  + \ppcpp \cosh\tfrac{\theta-\theta'}{2} \ket{Y_{{b}{\da}}\theta'_{{\a}{\db}}}
  - \ppcpp \sinh\tfrac{\theta-\theta'}{2} \eps_{{\da}{\db}} \eps^{{\dal}{\dbe}} \ket{Z_{{\a}{\dal}}\eta'_{{b}{\dbe}}}
\end{split}
\]

\[
\begin{split}
\bT\cdot   \ket{Z_{{\a}{\dal}}\eta'_{{b}{\dbe}}} = \ &
  - \half \tfrac{(p'-p)p'}{\cpp} \ket{Z_{{\a}{\dal}}\eta'_{{b}{\dbe}}}
  - \ppcpp \ket{Z_{{\a}{\dbe}}\eta'_{{b}{\dal}}} \\ &
  - \ppcpp \cosh\tfrac{\theta-\theta'}{2} \ket{\eta_{{b}{\dal}} Z'_{{\a}{\dbe}}}
  - \ppcpp \sinh\tfrac{\theta-\theta'}{2} \eps_{{\dal}{\dbe}} \eps^{{\da}{\db}} \ket{\theta_{{\a}{\da}} Y'_{{b}{\db}}} \\[3mm]
\bT\cdot   \ket{Z_{{\a}{\dal}}\theta'_{{\b}{\db}}} = \ &
  - \half \tfrac{(p'-p)p'}{\cpp} \ket{Z_{{\a}{\dal}}\theta'_{{\b}{\db}}}
  - \ppcpp \ket{Z_{{\b}{\dal}}\theta'_{{\a}{\db}}} \\ &
  - \ppcpp \cosh\tfrac{\theta-\theta'}{2} \ket{\theta_{{\a}{\db}} Z'_{{\b}{\dal}}}
  + \ppcpp \sinh\tfrac{\theta-\theta'}{2} \eps_{{\a}{\b}} \eps^{{a}{b}} \ket{\eta_{{a}{\dal}} Y'_{{b}{\db}}} \\[3mm]
\bT\cdot   \ket{\eta_{{a}{\dal}} Z'_{{\b}{\dbe}}} = \ &
  - \half \tfrac{(p-p')p}{\cpp} \ket{\eta_{{a}{\dal}} Z'_{{\b}{\dbe}}}
  - \ppcpp \ket{\eta_{{a}{\dbe}} Z'_{{\b}{\dal}}} \\ &
  - \ppcpp \cosh\tfrac{\theta-\theta'}{2} \ket{Z_{{\b}{\dal}}\eta'_{{a}{\dbe}}}
  + \ppcpp \sinh\tfrac{\theta-\theta'}{2} \eps_{{\dal}{\dbe}} \eps^{{\da}{\db}} \ket{Y_{{a}{\da}}\theta'_{{\b}{\db}}} \\[3mm]
\bT\cdot   \ket{\theta_{{\a}{\da}} Z'_{{\b}{\dbe}}} = \ &
  - \half \tfrac{(p-p')p}{\cpp} \ket{\theta_{{\a}{\da}} Z'_{{\b}{\dbe}}}
  - \ppcpp \ket{\theta_{{\b}{\da}} Z'_{{\a}{\dbe}}} \\ &
  - \ppcpp \cosh\tfrac{\theta-\theta'}{2} \ket{Z_{{\a}{\dbe}}\theta'_{{\b}{\da}}}
  - \ppcpp \sinh\tfrac{\theta-\theta'}{2} \eps_{{\a}{\b}} \eps^{{a}{b}} \ket{Y_{{a}{\da}}\eta'_{{b}{\dbe}}}
\end{split}
\]


\subsection{Symmetry algebra generators}\la{app:Q}

The generators of the centrally-extended $\su(2|2)\oplus\su(2|2)$
symmetry algebra up to quadratic order in the fields are given by
the following expressions \bea\nonumber \bL^{ab}= \int {\rm d}\s\,
\Big[{i\ov 2}\big(\eps^{ac}P_{c\dc}Y^{b\dc} +
\eps^{bc}P_{c\dc}Y^{a\dc} \big)- {1\ov
4}\big(\eps^{ac}\theta^\dagger_{c\dga}\theta^{b\dga} +
\eps^{bc}\theta^\dagger_{c\dga}\theta^{a\dga}\big)\Big] \,, \eea
\bea\nonumber \bR^{\a\b}= \int {\rm d}\s\,\Big[ {i\ov
2}\big(\eps^{\a\g}P_{\g\dga}Z^{\b\dga} +
\eps^{\b\g}P_{\g\dga}Z^{\a\dga}  \big) - {1\ov
4}\big(\eps^{\a\g}\eta^\dagger_{\g\dc}\eta^{\b\dc} +
\eps^{\b\g}\eta^\dagger_{\g\dc}\eta^{\a\dc}\big)   \Big]\,, \eea
\bea\nonumber &&\hspace{-0.8cm}\bQ^{\a b} =e^{-i\pi/4}\int {\rm
d}\s\, {1\ov 2}e^{{i\ov 2}x_-}\big(- i
\eps^{\a\g}P^{b\dc}\eta_{\g\dc}^\dagger -
2\eps^{\a\g}Y^{b\dc}\eta_{\g\dc}^\dagger  -
2\eps_{\dc\dd}Y^{b\dc}\eta'^{\a\dd}~~~~~~~~~~\\\nonumber
&&~~~~~~~~~~~~~~~~~~~~~~~~~~-\eps_{\dbe\dga}P^{\a\dbe}\theta^{b\dga}-2i\eps_{\dga\drh}Z^{\a\dga}\theta^{b\drh}-2i\eps^{bc}Z^{\a\dga}\theta_{c\dga}^{\dagger\prime}\
\big)\,,~~~~~~~ \eea \bea\nonumber &&\hspace{-0.8cm}\bQ_{b\a
}^\dagger =e^{i\pi/4}\int {\rm d}\s\, {1\ov 2}e^{-{i\ov
2}x_-}\big( i \eps_{\a\g}P_{b\dc}\eta^{\g\dc} -
2\eps_{\a\g}Y_{b\dc}\eta^{\g\dc}  -
2\eps^{\dc\dd}Y_{b\dc}\eta_{\a\dd}^{\dagger\prime}~~~~~~~~~~\\\nonumber
&&~~~~~~~~~~~~~~~~~~~~~~~~~~ -\eps^{\dbe\dga}P_{\a\dbe}
\theta_{b\dga}^\dagger
+2i\eps^{\dga\drh}Z_{\a\dga}\theta_{b\drh}^\dagger
+2i\eps_{bc}Z^{\a\dga}\theta'^{c\dga}\  \big)\,,~~~~~~~ \eea
\bea\nonumber \bL^{\da\db}= \int {\rm d}\s\, \Big[{i\ov
2}\big(\eps^{\da\dc}P_{c\dc}Y^{c\db} +
\eps^{\db\dc}P_{c\dc}Y^{c\da}  \big) - {1\ov
4}\big(\eps^{\da\dc}\eta^\dagger_{\g\dc}\eta^{\g\db} +
\eps^{\db\dc}\eta^\dagger_{\g\dc}\eta^{\g\da}\big)   \Big]\,, \eea
\bea\nonumber \bR^{\dal\dbe}= \int {\rm d}\s\,\Big[ {i\ov
2}\big(\eps^{\dal\dga}P_{\g\dga}Z^{\g\dbe} +
\eps^{\dbe\dga}P_{\g\dga}Z^{\g\dal}   \big)- {1\ov
4}\big(\eps^{\dal\dga}\theta^\dagger_{c\dga}\theta^{c\dbe} +
\eps^{\dbe\dga}\theta^\dagger_{c\dal}\theta^{c\dbe}\big)\Big] \,,
\eea \bea\nonumber &&\hspace{-0.8cm}\bQ^{\dal \db}
=e^{-i\pi/4}\int {\rm d}\s\, {1\ov 2}e^{{i\ov 2}x_-}\big( - i
\eps^{\dal\dga}P^{c\db}\theta_{c\dga}^\dagger -
2\eps^{\dal\dga}Y^{c\db}\theta_{c\dga}^\dagger  -
2\eps_{cd}Y^{c\db}\theta'^{d\dal}~~~~~~~~~~\\\nonumber
&&~~~~~~~~~~~~~~~~~~~~~~~~~~ +\eps_{\b\g}P^{\b\dal}
\eta^{\g\db}+2i\eps_{\g\rho}Z^{\g\dal}\eta^{\rho\db}+2i\eps^{\db\dc}Z^{\g\dal}\eta_{\g\dc}^{\dagger\prime}\
\big)\,,~~~~~~~ \eea \bea\nonumber &&\hspace{-0.8cm}\bQ_{\db\dal
}^\dagger =e^{i\pi/4}\int {\rm d}\s\, {1\ov 2}e^{-{i\ov
2}x_-}\big(  i \eps_{\dal\dga}P_{c\db}\theta^{c\dga} -
2\eps_{\dal\dga}Y_{c\db}\theta^{d\dga}  -
2\eps_{\db\dc}Y^{c\dc}\theta_{c\dal}^{\dagger\prime}~~~~~~~~~~\\\nonumber
&&~~~~~~~~~~~~~~~~~~~~~~~~~~+ \eps^{\b\g}P_{\b\dal}
\eta_{\g\db}^\dagger-2i\eps_{\dal\dga}Z^{\b\dga}\eta_{\b\db}^\dagger
-2i\eps_{\db\dc}Z_{\g\dal}\eta'^{\g\dc}\  \big)\,,~~~~~~~ \eea and
the Hamiltonian $\bH$  and the world-sheet momentum $\bP$ up to
quadratic order in the fields are given by \bea\nonumber \bH_2
&=&\int {\rm d}\s\, \Big({1\ov 4}P_{a\da} P^{a\da} +Y_{a\da}
Y^{a\da} + Y'_{a\da}Y'^{a\da} + {1\ov 4}P_{\a\dal} P^{\a\dal}
+Z_{\a\dal} Z^{\a\dal} +Z'_{\a\dal}Z'^{\a\dal} \\\nonumber
&&~~~~~~~~~+ \eta_{\a\da}^\dagger\eta^{\a\da} + {\kappa\ov
2}\eta^{\a\da}\eta'_{\a\da} -{\kappa\ov 2}\eta^{\dagger\a\da}
\eta_{\a\da}'^\dagger + \theta_{a\dal}^\dagger\theta^{a\dal}  +
{\kappa\ov 2}\theta^{a\dal}\theta'_{a\dal} -{\kappa\ov
2}\theta^{\dagger a\dal} \theta_{a\dal}'^\dagger\Big)\,, ~~~\eea
\bea\nonumber \bP ={\Bp\ov g} =  - {1\ov g} \int\, {\rm d}\s\,
\left( P_{a\da}Y'^{a\da} + P_{\a\dal}Z'^{\a\dal} + i
\theta_{\a\da}^\dagger\theta'^{\a\da} +
i\eta_{a\dal}^\dagger\eta'^{a\dal}
 \right)\,.
\eea
Lowering the first (or raising the second)  index and omitting $e^{\pm i x_- / 2}$, one gets the following expressions for these charges  in terms of the creation and annihilation operators
\bea\nonumber
\bL_{a}{}^{b}= \int {\rm d}p\, \sum_{{\dot M}}\, {1\ov 2} \Big(\,  a^\dagger_{a {\dot M}} a^{b {\dot M}} - \eps_{ad} \eps^{bc}\, a^\dagger_{c {\dot M}} a^{d {\dot M}}\, \Big) \,,
\eea
\bea\nonumber
\bR_{\a}{}^{\b}= \int {\rm d}p\, \sum_{{\dot M}}\, {1\ov 2} \Big(\,  a^\dagger_{\a {\dot M}} a^{\b {\dot M}} - \eps_{\a\rho} \eps^{\b\g}\, a^\dagger_{\g {\dot M}} a^{\rho{\dot M}}\, \Big) \,,
\eea
\bea\nonumber
\bQ_{\a}{}^{b}= \int {\rm d}p\, \sum_{{\dot M}}\Big(\,  f_p\, a^\dagger_{\a {\dot M}} a^{b {\dot M}} -  h_p\,\eps_{\a\g} \eps^{bc}\, a^\dagger_{c {\dot M}} a^{\g {\dot M}}\, \Big) \,,
\eea
\bea\nonumber
\bQ_b^{\dagger\,\a}= \int {\rm d}p\, \sum_{{\dot M}}\Big(\,  f_p\, a^\dagger_{b {\dot M}} a^{\a {\dot M}} -  h_p\,\eps^{\a\g} \eps_{bc}\, a^\dagger_{\g {\dot M}} a^{c {\dot M}}\, \Big) \,,
\eea
\bea\nonumber
\bL_{\da}{}^{\db}= \int {\rm d}p\, \sum_{M}\, {1\ov 2} \Big(\,  a^\dagger_{M\da} a^{M\db} - \eps_{\da\dd} \eps^{\db\dc}\, a^\dagger_{M\dc} a^{M\dd}\, \Big) \,,
\eea
\bea\nonumber
\bR_{\dal}{}^{\dbe}= \int {\rm d}p\, \sum_{M}\, {1\ov 2} \Big(\,  a^\dagger_{M\dal} a^{M\dbe} - \eps_{\dal\drh} \eps^{\dbe\dga}\, a^\dagger_{M\dga} a^{M\drh}\, \Big) \,,
\eea
\bea\nonumber
\bQ_{\dal}{}^{\db}= \int {\rm d}p\, \sum_{M}(-1)^{\eps_M}\Big(\,  f_p\, a^\dagger_{M\dal } a^{M\db} -  h_p\,\eps_{\dal\dga} \eps^{\db\dc}\, a^\dagger_{M\dc} a^{M\dga}\, \Big) \,,
\eea
\bea\nonumber
\bQ_\db^{\dagger\,\dal}= \int {\rm d}p\, \sum_{M}(-1)^{\eps_M}\Big(\,  f_p\, a^\dagger_{M\db } a^{M\dal} -  h_p\,\eps^{\dal\dga} \eps_{\db\dc}\, a^\dagger_{M\dga} a^{M\dc}\, \Big) \,,
\eea
\bea\nonumber \bH_2= \int\,{\rm d}p\, \sum_{M,\dot{M}}\, \om_p\, a_{M\dot{M}}^\dagger a^{M\dot{M}} \,,\quad  \bP={\Bp\ov g}={1\ov g} \int\,{\rm d}p\, \sum_{M,\dot{M}}\, p\, a_{M\dot{M}}^\dagger a^{M\dot{M}} \,.
\eea

\subsection{Poisson brackets and the moment map}\la{momentmap}
The group ${\rm PSU(2,2|4)}$ acts on the coset space (\ref{sAdS})
by multiplication of a coset element by a group element from the
left. Fixing the light-cone gauge and solving the Virasoro
constraints, we obtain a well-defined symplectic structure
$\omega$ (the inverse of the Poisson bracket) for physical fields.
Therefore, now we are able to study the Poisson algebra of the
Noether charges corresponding to infinitesimal global symmetry
transformations generated by the Lie algebra $\psu(2,2|4)$. In the
first place we are interested in those charges which leave the
gauge-fixed Hamiltonian and, as a consequence, the symplectic
structure of the theory invariant; the corresponding subspace in
$\psu(2,2|4)$ will be called ${\cal J}$.

\medskip

Since the symplectic form $\omega$ remains invariant under the
action of ${\cal J}$, to every element $\cal M\in {\cal J}$ one
can associate a locally Hamiltonian phase flow $\xi_{\cal M}$ with
the Hamiltonian function being the Noether charge ${\rm Q}_{\cal
M}$: \bea \omega(\xi_{\cal M},\ldots)+{\rm d}{\rm Q}_{\cal M}=0\,
. \label{HE} \eea Identifying  $\psu(2,2|4)$ with its dual space,
$\psu(2,2|4)^*$, by using the supertrace operation, we can treat
the matrix ${\rm Q}$ as the {\it moment map} which maps the phase
space $(x,p,\chi)$ into the dual space to the Lie algebra:
$$
{\rm Q}:~~~(x,p,\chi)\to \psu(2,2|4)^*\,
$$
and it allows one to associate to any element ${\cal M}$ of
$\psu(2,2|4)$ a function ${\rm Q}_{\cal M}$ on the phase space.
This linear mapping from the Lie algebra into the space of
functions on the phase space is given by eq.(\ref{QM}). The
function ${\rm Q}_{\cal M}$ is a Hamiltonian function, {\it i.e.}
it obeys eq.(\ref{HE}), only if ${\cal M}\in {\cal J}$. Although
the elements of $\psu(2,2|4)$ which do not belong to ${\cal J}$
are symmetries of the gauge-fixed action, they leave neither the
Hamiltonian nor the symplectic structure invariant.

\smallskip

As is well known, eq.(\ref{HE}) implies the following general
formula for the Poisson bracket of the Noether charges ${\rm
Q}_{\cal M}$ \bea \label{PBG} \{{\rm Q}_{{\cal M}_1}, {\rm
Q}_{{\cal M}_2} \}=(-1)^{\epsilon_{{\cal M}_1}\epsilon_{{\cal
M}_2}}\str({\rm Q}[{\cal M}_1, {\cal M}_2])\, +  \texttt{C}({\cal
M}_1,{\cal M}_2) \, ,  \eea where ${\cal M}_{1,2}\in {\cal J}$.
Here $\epsilon_{\cal M}$ is the parity of a supermatrix ${\cal M}$
and $[{\cal M}_1, {\cal M}_2]$ is the graded commutator, {\it
i.e.} it is the anti-commutator if both ${\cal M}_1$ and ${\cal
M}_2$ are odd matrices, and the commutator if at least one of them
is even. The first term in the right hand side of eq.(\ref{PBG})
reflects the fact that the Poisson bracket of the Noether charges
${\rm Q}_{{\cal M}_1}$ and ${\rm Q}_{{\cal M}_2}$ gives a charge
corresponding to the commutator $[{\cal M}_1,{\cal M}_2]$. The
normalization prefactor $(-1)^{\epsilon_{{\cal
M}_1}\epsilon_{{\cal M}_2}}$ is of no great importance and it is
related to our specific choice of normalizing the even elements
with respect to the odd ones inside the matrix ${\rm Q}$. The
quantity $\texttt{C}({\cal M}_1,{\cal M}_2)$ in the right hand
side of eq.(\ref{PBG}) is the central extension, {\it i.e.} a
bilinear graded skew-symmetric form on the Lie algebra ${\cal J}$.
It Poisson-commutes with all ${\rm Q}_{\cal M}$, ${\cal M}\in
{\cal J}$. The Jacobi identity for the bracket (\ref{PBG}) implies
that $\texttt{C}({\cal M}_1,{\cal M}_2)$ is a two-dimensional
cocycle of the Lie algebra ${\cal J}$. For simple Lie algebras
such a cocycle necessarily vanishes, while for super Lie algebras
it is generally not the case. Since we consider a
finite-dimensional super Lie algebra  the central extension
vanishes if the element $\cal M$ is bosonic:  $\texttt{C}({\cal
M},\ldots)=0$.

Some comments are necessary here. As we already mentioned, the
standard feature of the light-cone closed string theory is the
presence of the level-matching constraint $p_{\rm ws}=0$. In the
off-shell theory we rather keep $p_{\rm ws}$ non-vanishing. The
light-cone Hamiltonian commutes with $p_{\rm ws}$: $\{H, p_{\rm
ws}\}=0$, {\it i.e.} $p_{\rm ws}$ is an integral of motion. The
Poisson bracket (\ref{PBG}) with the vanishing central term is
valid on-shell and it is the off-shell theory where one could
expect the appearance of a non-trivial central extension. Below we
determine a general form of the central extension based on
symmetry arguments only. The explicit evaluation of the Poisson
brackets which justifies the formula (\ref{PBG}) was discussed in
the main text.

\medskip

Let us note that formula (\ref{PBG}) makes it easy to reobtain our
results on the structure of ${\cal J}$. Indeed, from
eq.(\ref{PBG}) we find that the invariance subalgebra ${\cal
J}\subset \psu(2,2|4)$ of the Hamiltonian is determined by the
condition
$$
\{H,{\rm Q}_{\cal M}\}=\str({\rm Q}[\Sigma_+,{\cal M}])=0\, .
$$
Thus, $\cal J$ is the stabilizer of the element $\Sigma_+$ in $\psu(2,2|4)$:
$$
[\Sigma_+,{\cal M}]=0\, , ~~~~~~{\cal M}\in {\cal J}\, .
$$
Obviously, ${\cal J}$ coincides with the red-blue submatrix of
${\cal M}$ in Figure \ref{flag}. Thus, for $P_+$ being
finite\footnote{For $P_+$ finite the subalgebra which leaves
invariant both $H$ and $P_+$ coincides with the even subalgebra
${\cal J}_{\rm even}$ of ${\cal J}$. In fact ${\cal J}_{\rm even}$
is nothing else but the algebra $\alg{C}$ defined in
(\ref{linearsym}). Indeed, according to eqs.(\ref{Ham}) and
(\ref{HH}), ${\cal J}_{\rm even}$ arises as the simultaneous
solution the two equations, $[\Sigma_+,{\cal M}]=0$ and
$[\Sigma_-,{\cal M}]=0$ or, in other words, it is the centralizer
of $\Lambda(t,\phi)$ given by eq.(\ref{Lambda}). Together with
$\Sigma_{\pm}$ the algebra $\alg{C}$ comprises  the red and blue
{\it diagonal} blocks in Figure \ref{flag}.} we would obtain the
following vector space decomposition of ${\cal J}$
$$
{\cal J}=\psu(2|2)\oplus\psu(2|2)\oplus \Sigma_+\oplus \Sigma_-\,
.
$$
The rank of the latter subalgebra is six and it coincides with
that of $\psu(2,2|4)$. In the case of infinite $P_+$ the last
generator decouples.

\medskip

Now we are ready to determine the general form of the central term
in eq.(\ref{PBG}). Denote by ${\cal J}_{\rm even}\subset {\cal J}$
the even (bosonic) subalgebra of $\cal J$. It is represented by
the red and blue diagonal blocks in Figure \ref{flag}. Let $G_{\rm
even}$ be the corresponding group. The adjoint action of $G_{\rm
even}$ preserves the ${\mathbb Z}_2$-grading of $\cal J$.
Obviously, if we perform the transformation
$$
{\rm Q}\to g{\rm Q}g^{-1}\, , ~~~~~~{\cal M}\to g^{-1}{\cal M}g
$$
with an element $g\in G_{\rm even}$ the charge ${\rm Q}_{\cal M}$
remains invariant. This transformation leaves the l.h.s of the
bracket (\ref{PBG}) invariant. As a consequence, the central term
must satisfy the following invariance condition: \bea\label{invar}
\texttt{C}(g{\cal M}_1g^{-1},g{\cal M}_2g^{-1})=\texttt{C}({\cal
M}_1,{\cal M}_2)\, . \eea It is not difficult to find a general
expression for a bilinear graded skew-symmetric form on ${\cal J}$
which satisfies this condition. It is given by \bea\label{cocycle}
\texttt{C}({\cal M}_1, {\cal M}_2) = \str \bigg(\left(\varrho
{\cal M}_1 \varrho {\cal M}_2^t + (-1)^{\epsilon_{{\cal
M}_1}\epsilon_{{\cal M}_2}} \varrho{\cal M}_2 \varrho {\cal M}_1^t
\right) \Phi \bigg)\, . \eea Here
\begin{equation}
\label{Delta} \Phi = - \frac{1}{2} {\footnotesize\left(
\begin{array}{cccc}
        c_3 \, \mI_2 & 0 & 0 & 0 \\
        0 & c_1 \, \mI_2 & 0 & 0 \\
        0 & 0 &  c_4 \, \mI_2 & 0 \\
        0 & 0 & 0 & c_2 \, \mI_2
\end{array} \right)}\, ,
\end{equation}
where $\mI_2$ is the two-dimensional identity matrix and
\begin{equation}\nonumber
\varrho =  {\footnotesize\left( \begin{array}{cccc}
        \epsilon & 0 & 0 & 0 \\
        0 & \epsilon & 0 & 0 \\
        0 & 0 & \epsilon & 0 \\
        0 & 0 & 0 & \epsilon
\end{array} \right)}\,  ,
\end{equation}
where $\epsilon$ is defined in eq.(\ref{conjrule}). Note that
$\varrho$ is essentially the charge conjugation matrix. Condition
(\ref{invar}) follows from the form of the matrix $\Phi$ and the
equation
\begin{equation}\nonumber
{\mathcal J}_{\rm even}^t \varrho + \varrho{\mathcal J}_{\rm even}
= 0 \, .
\end{equation}
The coefficients $c_i$, $i=1,\ldots, 4$ can depend on the physical
fields and they are central with respect to the action of $\cal
J$:
$$
\{c_i,{\rm Q}_{\cal M}\}=0\, , ~~~~~{\cal M}\in {\cal J}\, .
$$
By using eq.(\ref{cocycle}) one can check that the cocycle
condition for $\texttt{C}({\cal M}_1,{\cal M}_2)$ is trivially
satisfied. In accordance with our assumptions, $\texttt{C}({\cal
M}_1,{\cal M}_2)$ does not vanish only if both ${\cal M}_1$ and
${\cal M}_2$ are odd.

Taking into account that $\cal J$ contains two identical
subalgebras $\psu(2|2)$ we can put $c_1=c_3$ and $c_2=c_4$. Thus,
general symmetry arguments fix the form of the central extension
up to two central functions $c_1$ and $c_2$.  Since we consider
the algebra $\psu(2|2)$, which is the real form of $\alg{
psl}(2|2)$, the conjugation rule implies that $c_1= -c_2^*$.

\section{Bibliographic remarks}
\la{sec:biblight-cone}

\small{

The phase-space light-cone gauge for strings in flat space was
introduced  in \cite{Goddard:1973qh}. It can be generalized to
strings moving in a curved background with at least one time and
one space isometry directions. If one chooses the time and space
isometries from the AdS part of the $\AdS$ background one gets the
light-cone gauge by \cite{Metsaev:2000yu}. The uniform light-cone
gauge we discuss was introduced in \cite{AF0,AFlc,AFZm},
 and belongs to the class of gauges used to study the dynamics of spinning strings in $\AdS$  \cite{KT,KT1}.

The BMN limit was introduced in the paper by Berenstein, Maldacena
and Nastase \cite{BMN}. In this limit the string sigma model on
$\AdS$ reduces to the one describing strings in the plane-wave
background \cite{Blau:2001ne,Blau:2002dy}. In the light-cone gauge
this string sigma model is a free theory of massive bosons and
fermions, and it has been analyzed in
\cite{Metsaev:2001bj,Metsaev:2002re}. The $1/P_+$ corrections to
the
 energy of string states were studied in \cite{Ryzhov}-\cite{Swanson}, \cite{FPZ}.
As was shown in \cite{AF0}, the $a=0$ uniform gauge is in fact a
non-perturbative version of the
 perturbative light-cone gauge used in \cite{Callan}-\cite{Swanson}.

 The first-order formalism for the $\AdS$ superstring model,
 the  full gauge-fixed Lagrangian and its expansion up to
 quartic order were found in \cite{FPZ}; we follow this work very closely in chapter 3.
 The reader might consult \cite{FPZ} for more details and missing derivations.

 The decompactification limit was discussed in many
 papers, see e.g. \cite{AJK}-\cite{HM}. One-soliton solutions were identified with spin chain magnons
 and named ``giant magnons" in \cite{HM}. The giant magnon solution was found in \cite{HM} by employing the conformal gauge.  The derivation of the light-cone gauge giant magnon solution and its dispersion relation in subsection \ref{magn} follows closely \cite{AFZm}.

 The two-index notation for physical fields of the light-cone model was introduced in \cite{KMRZ}. Our fields, however, differ from the ones in \cite{KMRZ} by various factors. As a result, our expressions for the supercharges in appendix \ref{app:Q} are slightly different from those in \cite{KMRZ}. Nevertheless, the T-matrix  coincides with the one computed there. The formulae for the T-matrix in subsection \ref{PSM} and in appendix \ref{app:T} are taken from \cite{KMRZ}.

The formula (\ref{Charges}) for the  $\psu(2,2|4)$ charges was
obtained in \cite{FPZ}. The centrally-extended $\su(2|2)$ algebra
was derived by using the hybrid expansion scheme in \cite{AFPZ}.
Given that the central charges retain their functional form in
quantum theory, the algebra allows one to uniquely determine the
dispersion relation. The dispersion relation implied by
eq.(\ref{Cc}) has been verified in field theory up to the fourth
order \cite{BMR} and in string theory up to the second order
\cite{KlMMZ}. The centrally-extended $\su(2|2)$ algebra coincides
with the one previously suggested in the gauge theory spin chain
context in \cite{B}. There is however no gauge theory derivation
of the centrally-extended algebra.

For the notion of the moment map  and related issues discussed in
appendix \ref{momentmap} we refer to
\cite{Souriau,Kirillov,Arnold}.

 }

\normalsize
\chapter{World-sheet S-matrix}
In the previous chapters we have demonstrated integrability of the
classical string sigma model and developed the semi-classical
quantization scheme based on the large tension expansion. The
scattering matrix of world-sheet excitations has been computed in
the Born approximation. We have also shown that in the off-shell
string theory the symmetry algebra of the light-cone Hamiltonian
coincides with two copies of the centrally extended $\psu(2|2)$
superalgebra sharing the same set of central charges.

\smallskip

Given the current lack of non-perturbative quantization schemes,
occurrence of integrability in the corresponding quantum model is
much harder to establish. Because of ultra-violet and infra-red
divergencies arising in the process of perturbative quantization,
the definition of the quantum model itself is far from obvious. At
best, it should rely on finding regularization and renormalization
schemes in the world-sheet theory which would allow one to uplift
the classical conservation laws to the quantum level. In view of
this, to make progress we will employ  a ``top-to-bottom"
approach. Namely, we will  {\it assume that our model is quantum
integrable} and then will derive the corresponding consequences.
The results obtained should obviously agree with available gauge
and string perturbative data in order to make quantum
integrability plausible. Moreover, in certain cases the results
gathered in perturbative calculations will be essentially used to
fix the structures which remain undetermined from our assumption
of quantum integrability.

\smallskip

In the decompactification limit when the circumference of the
world-sheet cylinder tends to infinity, the effective sigma model
arising on the plane is massive. The massive character of a theory
usually implies that interactions  fall off sufficiently fast with
distance, so that the concept of asymptotic states and their
scattering makes sense. Under these circumstances quantum
integrability can be understood as  the absence of particle
production and factorization of the multi-particle scattering into
a sequence of two-body events.


\vskip 0pt \noindent
\begin{figure}
\begin{minipage}{\textwidth}
\begin{center}
\includegraphics[width=0.5\textwidth]{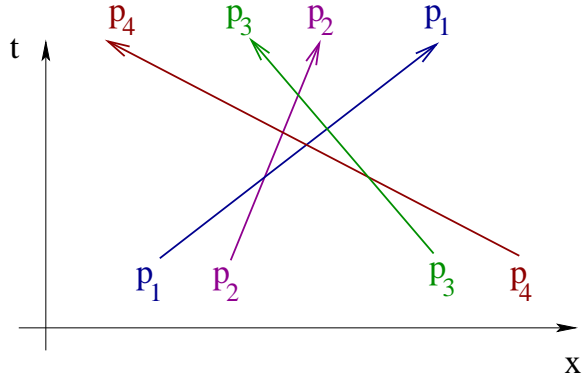}
\vskip 5pt
\parbox{5in}{\footnotesize{\caption{\label{FMP} Factorization of the multi-particle scattering.}}}
\end{center}
\end{minipage}
\end{figure}

\vspace{-0.5cm}

 In this chapter we will treat the string sigma model in the
framework of the Factorized Scattering Theory. We will show that
the symmetry principles alone lead to almost complete
determination of the exact world-sheet S-matrix and that the
latter satisfies the standard axioms of the Factorized Scattering
Theory. Besides the centrally extended $\psu(2|2)$ symmetry
algebra, an important role in our treatment will be played by
crossing symmetry which exchanges particles with anti-particles.
Compatibility of scattering with crossing symmetry will imply a
non-trivial functional equation for an overall phase of the
world-sheet S-matrix; the latter can not be constrained by other
known symmetries or by the requirement of factorization. We will
present some physically interesting solutions to this functional
equation and discuss the properties of the corresponding
world-sheet  S-matrix.

\section{Elements of Factorized Scattering Theory}
Consider scattering in a two-dimensional quantum field theory that
exhibits an infinite number of conservation laws (charges) ${\bf
q}_k$, $k=1,\ldots,\infty$, which all mutually commute. Obviously,
there exists a basis of one-particle states in which these charges
act diagonally
$$
{\bf q}_k |p\rangle = q_k(p)|p\rangle\, .
$$
If these charges are functionally independent then the
corresponding scattering theory turns out to be highly
constrained. First, the number of particles cannot change in the
collision process; particle production is absent. Second,
additivity of the conservation laws implies that
$$
\sum_{j\in {\rm in }}q_k(p_j)=\sum_{j\in {\rm out }}q_k(p_j)\,
~~~~\mbox{for~~any}~~k.
$$
Thus, the set of initial momenta is preserved under collision, the
particles are only allowed to exchange their individual momenta
and flavors, see Figure \ref{FMP}. In other words, scattering is
elastic. Finally, an infinite tower of conservation laws implies
that the multi-particle S-matrix factorizes into the product of
two-particle ones.

\smallskip

In this section we recall the basic concepts of Factorized
Scattering Theory. First, we describe the Hilbert space of the
asymptotic states as a representation carrier of the
Zamolodchikov-Faddeev (ZF) algebra; the latter is a deformed
algebra of creation and annihilation operators with defining
relations given by the scattering matrix. Second, we derive the
constraints imposed by symmetries of the Hamiltonian on the
scattering matrix. Finally, we show that the physicality
requirements on the S-matrix coincide with those which follow from
the compatibility of the ZF algebra relations.

\subsection{Zamolodchikov-Faddeev algebra}
Let $\mathscr{J}$ be the symmetry algebra of our quantum
integrable model which leaves the vacuum state $|\Omega\rangle$
invariant. Introduce a creation operator $A^\dagger_i(p)$ which
creates  a multiplet $\mathscr{V}$ of particles out of the vacuum
with momentum $p$ transforming in a linear irreducible
representation of $\mathscr{J}$. Here index $i$ labels various
states in this multiplet (the flavor index).  The hermitian
conjugate $A^i(p)$ is the vacuum annihilation operator:
\bea\nonumber A^i(p) |\Omega\rangle =0\,. \eea States in the
multiplet may have different statistics and, for this reason, it
is convenient to define parity $\eps_i$, the latter being equal to
zero or one depending on whether the value of $i$ corresponds to a
bosonic or fermionic state, respectively.

\smallskip

To describe the scattering process, we introduce the $in$-basis
and the $out$-basis of asymptotic states as \bea\nonumber
\hspace{-0.3cm} &&|p_1,p_2, \cdots , p_n
\rangle^{(in)}_{i_1,...,i_n} =
 A^\dagger_{i_1}(p_1)\cdots  A^\dagger_{i_n}(p_n)|\Omega \rangle  \,,\quad\qquad\qquad~~~~~~~ p_1>p_2>\cdots >p_n\,,\\\nonumber
\hspace{-0.3cm} &&|p_1,p_2, \cdots , p_n
\rangle^{(out)}_{i_1,...,i_n} =
(-1)^{\sum_{k<l}\gr_{i_k}\gr_{i_l}}
 A^\dagger_{i_n}(p_n)\cdots  A^\dagger_{i_1}(p_1)|\Omega \rangle  \,,\quad~~ p_1>p_2>\cdots >p_n\,.
\eea The $in$ and $out$ states are the eigenstates of the
Hamiltonian $\mathbb{H}$ of the model and the ordering of momenta
is essential. The operators $A^\dagger(p)$ should not be confused
with the fields $a^{\dagger {\rm in}/{\rm out}}(p), a^\dagger(p)$
introduced in section \ref{PSM}. In terms of the Heisenberg
creation operators the $in$ and $out$ states read as
 \bea\nonumber
 &&|p_1,p_2, \ldots , p_n \rangle^{(in)}_{i_1,...,i_n} =  a^{\dagger{\rm in}}_{i_1}(p_1)
 \cdots  a^{\dagger{\rm in}}_{i_n}(p_n)|\Omega \rangle  \,,\\\nonumber
 &&|p_1,p_2, \ldots , p_n \rangle^{(out)}_{i_1,...,i_n} =
a^{\dagger{\rm out}}_{i_1}(p_1)\cdots  a^{\dagger{\rm out}}_{i_n}(p_n)|\Omega \rangle
 \, ,
\eea where the ordering of particle momenta is the same as in the
formulae above.

\smallskip

The operators $A_i^{\dagger}$ and $A^i$ are known as the ZF
creation and annihilation operators, respectively. Contrary to
$a_i^\dagger$ and $a^i$, these operators do not satisfy the
canonical commutation relations in interacting theory. In the free
field limit the ZF operators turn into $a_i^\dagger$ and $a^i$,
which explains an extra statistics-carrying factor
$(-1)^{\sum_{k<l}\gr_{i_k}\gr_{i_l}}$ in the above formula for the
$out$-states.

\smallskip

\begin{figure}
\begin{minipage}{\textwidth}
\begin{center}
\includegraphics[width=0.8\textwidth]{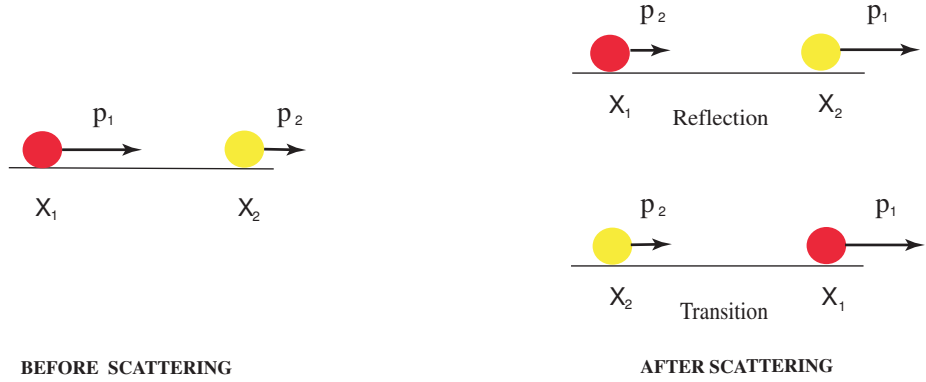}
\end{center}
\begin{center}
\parbox{5in}{\footnotesize{ \caption{\label{colorscat} In the collision process particles either keep (transition) or exchange (reflection) their momenta.
The S-matrix operates non-trivially in the flavor space.}}}
\end{center}
\end{minipage}
\end{figure}

In our new description of asymptotic states, scattering is
understood as reordering of particles (creation operators) in the
momentum space. Particles can be distinguishable, each of them
carrying a definite flavor (the value of index $i$). Then, in the
two-body collision process particles can either keep their
individual momenta, which is forward scattering (transition), or
exchange the latter (in the case of equal mass), which is backward
scattering (reflection), see Figure \ref{colorscat}. Note that the
very possibility to describe the asymptotic states and their
scattering in such a fashion  is due to \mbox{S-matrix}
factorization, as it will become apparent in a moment.

\vskip 5pt \noindent

\vspace{-1cm}

\smallskip

 According to the discussion in the previous chapter, the
$in$ and $out$ states are related by the unitary S-matrix operator
$\bS$: \bea\la{insout1} |p_1, \ldots
,p_n\rangle^{(in)}_{i_1,...,i_n}= \bS\cdot |p_1, \ldots
,p_n\rangle^{(out)}_{i_1,...,i_n}\,, \eea and one can expand
initial states on a basis of final states and vice versa. In
particular, the two-particle $in$ and $out$ states are related by
eq.(\ref{sijkl}), which  now takes the form \bea \nonumber
A^\dagger_{i}(p_1) A^\dagger_{j}(p_2)|\Omega \rangle =\bS\cdot
(-1)^{\gr_i\gr_j}A^\dagger_{j}(p_2) A^\dagger_{i}(p_1)|\Omega
\rangle= \bS^{kl}_{ij}(p_1,p_2)(-1)^{\gr_k\gr_l}
A^\dagger_{l}(p_2) A^\dagger_{k}(p_1)|\Omega \rangle\,. \eea This
formula suggests to define the new matrix elements as \bea
\label{relmatel} S^{kl}_{ij}(p_1,p_2) \equiv
\bS^{kl}_{ij}(p_1,p_2)(-1)^{\gr_k\gr_l} \, .\eea Now, by
discarding the vacuum state on both sides of the formula just
above eq.(\ref{relmatel}), we obtain the following algebra of
creation operators \bea\la{ZFu} A^\dagger_{i}(p_1)
A^\dagger_{j}(p_2) = A^\dagger_{l}(p_2)
A^\dagger_{k}(p_1)S^{kl}_{ij}(p_1,p_2)\, , \eea which is usually
referred to as the ZF  algebra.

\smallskip

Before stating the consistency conditions of these algebra
relations, it is convenient to rewrite (\ref{ZFu}) in the matrix
form. To this end, we introduce rows $E_i$ and columns $E^i$ with
all vanishing entries except the one in the $i$-th position which
is equal to the identity. The standard matrix unities are then
$E_i^{~j}=E_i\otimes E^j$ with the only non-vanishing element
equal to the identity which occurs on the intersection of the
$i$-th row with the $j$-th column. The following multiplication
rules are valid $E^kE_i^{~j}=\delta_i^k E^{j}$ and $E_i^{~j}
E_k=\delta_k^j E_{i}$ together with the product rule for the
matrix unities: $E_i^{~j} E_k^{~l}=\delta_k^j E_i^{~l}$. With this
notation at hand we can represent the ZF creation and annihilation
operators as rows and columns, respectively, \bea
\bA^{\dagger}=A_i^{\dagger}\, E^i\, , ~~~~~~ \bA=A^i\, E_i\, ,
\eea while the entities  (\ref{relmatel}) can be combined in the
following matrix \bea \label{matelement}
S(p_1,p_2)=S^{kl}_{ij}(p_1,p_2)\, E_k^{~i}\otimes E_l^{~j}\, \eea
which  is an element in ${\rm End}(\mathscr{V}\otimes
\mathscr{V})$. Thus, in the matrix notation the relations
(\ref{ZFu}) acquire the form \bea \label{ZFmatrix} \bA^\dagger_1
(p_1)\bA^\dagger_2(p_2)= \bA^\dagger_2(p_2)\bA^\dagger_1(p_1)
S_{12}(p_1,p_2)\, , \eea where $S_{12}\equiv S$, and we use the
following convention \bea\nonumber \bA^\dagger_1\,\bA^\dagger_2=
A^\dagger_i(p_1)A^\dagger_j(p_2)\,  E^i \otimes E^j\,,\quad
\bA^\dagger_2\,\bA^\dagger_1= A^\dagger_j(p_2)A^\dagger_i(p_1)\,
E^i \otimes E^j\,. \eea In what follows, if $A,B,C$ are either
columns or rows with operator entries then in the notation $A_1
B_2C_3$ the subscripts $1,2,3$ refer to the location of the
columns and rows, {\it e.g.} if $A = A^i(p_3)\, E_i\,, \
B=B_i(p_1)E^i\,,\ C=C_i(p_2)E^i$, then $A_1 B_3
C_2=A^i(p_3)B_k(p_1)C_j(p_2)E_i\otimes E^j\otimes E^k$.

This formula can be naturally supplemented by similar relations
between two annihilation operators and between creation and
annihilation operators, so that the complete algebra relations
look like \bea \label{ZFfull}
\begin{aligned}
\bA^\dagger_1 \bA^\dagger_2 &= \bA^\dagger_2 \bA^\dagger_1 S_{12}\, , \\
\bA_1 \bA_2&=S_{12} \bA_2\bA_1 \, , \\
\bA_1\bA_2^{\dagger}&=\bA_2^{\dagger}S_{21}\bA_1+\delta_{12}\, .
\end{aligned} \eea
Here $S_{21}= S^{kl}_{ij}(p_2,p_1)\,E_l^{~j} \otimes E_k^{~i}$, and $\delta_{12}=\delta(p_1-p_2)E_i\otimes E^i$,
where summation over repeated indices is assumed. In what follows we will need
the following three matrices known as the permutation matrix $P$,
the graded permutation $P^g$ and the graded identity $\mI^g$:
\bea\label{basicmatrices} P=E_i^{~j}\otimes E_j^{~i}\, ,
~~~~~P^g=(-1)^{\gr_i\gr_j}E_i^{~j}\otimes E_j^{~i}\,
,~~~~~\mI^g=(-1)^{\gr_i\gr_j}E_i^{~i}\otimes E_j^{~j}\, . \eea
The permutation matrix transforms $S_{12}$ into $S_{21}$: $PS_{12}(p,p') = S_{21}(p,p')P$.

As we have already mentioned above,  in the absence of
interactions $A^\dagger_i$ and $A^i$ become the usual bosonic
(commuting) or fermionic (anti-commuting) creation and
annihilation operators. Then, the ZF algebra relations imply that
in the free field limit the S-matrix should turn into the graded
unit matrix, {\it i.e.} into the diagonal matrix with entries $\pm
1$ depending on the statistics of the corresponding creation
operator. From this point of view the relations (\ref{ZFfull}) can
be understood as a quantization (deformation) of the free
oscillator algebra.

\smallskip
\vskip 0pt \noindent
\begin{figure}
\begin{minipage}{\textwidth}
\begin{center}
\includegraphics[width=0.8\textwidth]{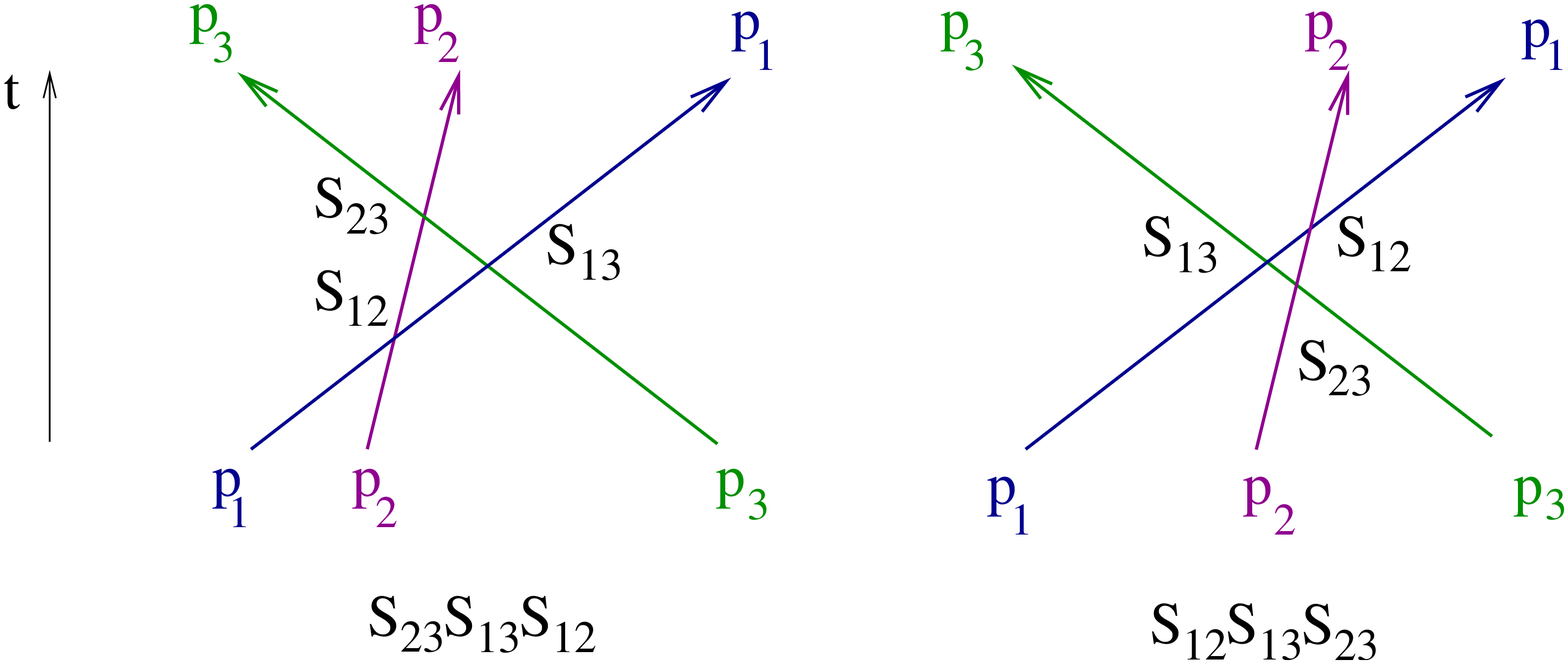}
\vskip 5pt
\parbox{5in}{\footnotesize{\caption{\label{FS3p}Factorization of the three-particle S-matrix. The result of the
three-particle scattering process does not depend on the order in
which two-particle scattering events take place.}}}
\end{center}
\end{minipage}
\end{figure}

\vspace{-0.8cm}

\subsubsection*{Yang-Baxter equation}

\noindent In the free theory the creation operators either commute
or anti-commute and, therefore, any operator monomial can be
ordered in a unique way, {\it e.g.}, by rearranging operators
according to the momentum ordering $p_1>p_2>\ldots >p_n$. It is
natural to require that this property of having a unique basis of
the lexicographically ordered monomials holds for the interacting
case as well. In the algebraic language this is known as the
Poincar\'e-Birkhoff-Witt property. Starting from any monomial
constructed from the operators $A^{\dagger}_i(p)$, we should be
able to bring it to an ordered form in a unique way by using the
defining relations (\ref{ZFmatrix}) only. Consider, for instance,
the product $\bA^\dagger_1\bA^\dagger_2\bA^\dagger_3$, where the
subscript also reflects the momentum dependence. Obviously, by
using the ZF algebra relations, this monomial can be brought to
the form $\bA^\dagger_3\bA^\dagger_2\bA^\dagger_1$ in two
different ways\footnote{Here $S_{ab}$ denotes the standard
embedding of the matrix $S(p,p')$ into the tensor product of three
spaces, {\it e.g.} $S_{13}(p,p')=S^{kl}_{ij}(p,p')\,
E_k^{~i}\otimes \mI\otimes E_l^{~j}$. Note, that in general the
momenta $p, p'$ are not attached to the indices $a,b$.}
\bea\nonumber
&&\bA^\dagger_1\bA^\dagger_2\bA^\dagger_3=\bA^\dagger_3\bA^\dagger_2\bA^\dagger_1
\, S_{12}S_{13}S_{23}\,,\\\nonumber
&&\bA^\dagger_1\bA^\dagger_2\bA^\dagger_3=\bA^\dagger_3\bA^\dagger_2\bA^\dagger_1
\, S_{23}S_{13}S_{12}\, . \eea
 If we require these two results to
coincide without imposing new (cubic) relations between ZF
operators, then the corresponding S-matrix must obey the following
equation \bea\la{YBnorm}
S_{23}(p_2,p_3)S_{13}(p_1,p_3)S_{12}(p_1,p_2)
=S_{12}(p_1,p_2)S_{13}(p_1,p_3)S_{23}(p_2,p_3)\, . \eea This is
the Yang-Baxter equation -- the fundamental equation of the
Factorized Scattering Theory.

One can show that no further constraints on the scattering matrix
arise from the ordering of higher than cubic monomials provided
the Yang-Baxter equation is satisfied. It is important to
recognize that both the left and right hand side of this equation
represent the three-particle scattering matrix, and the equation
itself is nothing else but the factorizability condition for this
S-matrix, see Figure \ref{FS3p}. Thus, the description of
scattering states in terms of ZF operators with a unique basis of
ordered monomials is only possible if the corresponding theory
exhibits  a factorizable S-matrix.

\subsubsection*{Unitarity condition}

\noindent In addition to the Yang-Baxter equation, consistency of
the ZF algebra relations imposes further requirements on the
S-matrix.

In particular, if we flip $p_1\leftrightarrow p_2$ in the ZF algebra relation
(\ref{ZFmatrix}) and then pull the permutation matrix $P$ through
its left and right hand sides, we get
$$
\bA_2(p_2)\bA_1(p_1)=\bA_1(p_1)\bA_2(p_2)S_{21}(p_2,p_1)=\bA_2(p_2)\bA_1(p_1)S_{12}(p_1,p_2)S_{21}(p_2,p_1)\,
,
$$
where the last term was obtained by applying the ZF relation
again. Thus, the S-matrix must satisfy the following property \bea
\label{unitarity} S_{12}(p_1,p_2)S_{21}(p_2,p_1)=\mI\, \eea known
as the unitarity condition.

\subsubsection{Conservation Laws}

\noindent
The fulfilment of the unitarity condition
(\ref{unitarity}) leads to the existence in the ZF algebra of a
large abelian subalgebra. Assuming for simplicity the same dispersion relation for all the particles, this subalgebra is generated by the operators \bea
\mathbb{I}_{q}=\int {\rm d}p~q(p)A_i^\dagger(p)A^i(p)\, , \eea
where $q(p)$ is an arbitrary function of particle momentum.
Indeed, applying the ZF algebra relations twice, we get \bea
\nonumber
A_i^{\dagger}(u)A^i(u)A_j^{\dagger}(p)&=&A_i^{\dagger}(u)\Big[A_k^{\dagger}(p)A^l(u)S_{jl}^{ki}(p,u)
+\delta_i^j\delta(u-p)\Big]=\\
\nonumber &=&
A_n^{\dagger}(p)A^{\dagger}_m(u)A^l(u)S_{ik}^{mn}(u,p)S_{jl}^{ki}(p,u)+
A_j^{\dagger}(p)\delta(u-p)\, . \eea
In components  the unitarity
relation (\ref{unitarity}) takes the form $S_{ik}^{mn}(u,p)S_{jl}^{ki}(p,u)=\delta_{j}^{n}\delta_{l}^{m}$, and, therefore
$$\mathbb{I}_{q}A_i^\dagger(p)=A_i^\dagger(p)(q(p)+\mathbb{I}_{q})\,,\quad
\mathbb{I}_{q}A^i(p)=A^i(p)(-q(p)+\mathbb{I}_{q})\,.$$ Thus, we
conclude  that $\mathbb{I}_{q}$ for various $q$'s do commute.
Furthermore, the formulae above imply the additivity property of
the commuting integrals
$$
\mathbb{I}_{q}~A_{i_1}^\dagger(p_1)\ldots
A_{i_n}^\dagger(p_n)|\Omega\rangle =\Big(\sum_{k=1}^n
q(p_{i_k})\Big)~A_{i_1}^\dagger(p_1)\ldots
A_{i_n}^\dagger(p_n)|\Omega\rangle\, .
$$
In particular, as a result, we get that the Hamiltonian $\bH$, the momentum operator $\bP$
and the number operator $\bN$ are given by
$$
\bH=\int {\rm d}p~\omega(p)\,A_i^\dagger(p)A^i(p)\, ,~~~~~\bP=\int
{\rm d}p~p\, A_i^\dagger(p)A^i(p)\, ,~~~~~\bN=\int {\rm d}p~\,
A_i^\dagger(p)A^i(p)\, ,
$$
where $\omega(p)$ is the dispersion relation which was assumed to
be the same for all particles from the multiplet $\mathscr{V}$.

If particles have different dispersion relations
the construction of the conservation laws admits a straightforward generalization to be discussed in due course.

\subsubsection*{Scattering and statistics}

\noindent Consider an operator $(-1)^{\mathbb{N}_F}$, where we
have introduced the following operator \bea
 {\mathbb{N}_F}= { \int {\rm
d}p~\eps_i\,A_i^\dagger(p)A^i(p)}\, . \eea Since $\eps=0$ for
bosons and $\eps=1$ for fermions, ${\mathbb{N}_F}$ is the fermion
number operator. The operator $(-1)^{\mathbb{N}_F}$ preserves the
vacuum state $(-1)^{\mathbb{N}_F}|\Omega\rangle=|\Omega\rangle$
and it defines statistics of a multi-particle state
$$
(-1)^{\mathbb{N}_F}\cdot A_{i_1}^\dagger(p_1)\ldots
A_{i_n}^\dagger(p_n)|\Omega\rangle =(-1)^{\sum_{k=1}^n \gr_{i_k}
}~A_{i_1}^\dagger(p_1)\ldots A_{i_n}^\dagger(p_n)|\Omega\rangle\,
.
$$
Since statistics of a multi-particle state cannot change under
scattering, $(-1)^{\mathbb{N}_F}$ must commute with the S-matrix
operator $\bS$. Pulling $(-1)^{\mathbb{N}_F}$ through the left and
the right hand sides of the ZF relation (\ref{ZFu}), we get
\bea\label{Sgr} (-1)^{\gr_i+\gr_j}A^\dagger_{i}(p_1)
A^\dagger_{j}(p_2)= (-1)^{\gr_k+\gr_l} S^{kl}_{ij}(p_1,p_2)
A^\dagger_{l}(p_2) A^\dagger_{k}(p_1)\,  . \nonumber \eea The last
equation leads to the following non-trivial condition for the
S-matrix elements \bea \la{grproperty}
 S^{kl}_{ij}(p_1,p_2)=(-1)^{\e_i+\e_j+\e_k+\e_l}
 S^{kl}_{ij}(p_1,p_2)\, .
\eea Obviously, this condition implies that for any non-vanishing
$S^{kl}_{ij}(p_1,p_2)$ the sum \mbox{$\e_i+\e_j+\e_k+\e_l$} is an
even number: $0,2$ or $4$. It is convenient to define the grading
matrix $\Sigma$ \bea\la{grmatrix} \Sigma=(-1)^{\gr_i}E_i^{~i}\, .
\eea Then for the matrix (\ref{matelement}) relation (\ref{Sgr})
can be cast in the form \bea [S(p_1,p_2),\Sigma\otimes \Sigma]=0\,
. \eea Thus, in the matrix language compatibility of scattering
with statistics is equivalent to commutativity of $S(p_1,p_2)$
with the matrix $\Sigma\otimes \Sigma$. It is worth pointing out
that the operator $\bN_F$ does not commute with the Hamiltonian
and, for this reason, the fermion number is not a conserved
quantity, only $(-1)^{\bN_F}$ is conserved.

\subsubsection*{ Graded S-matrix}

\noindent It is of interest to consider the following matrix \bea
S^g(p_1,p_2)=\bS^{kl}_{ij}(p_1,p_2)E_k^{~i}\otimes E_l^{~j} =
 S^{kl}_{ij}(p_1,p_2)(-1)^{\gr_k\gr_l}E_k^{~i}\otimes
E_l^{~j} \, . \nonumber\eea  By using the graded identity matrix
(\ref{basicmatrices}), the last formula can be written as \bea
S^g(p_1,p_2)=\mI^g S(p_1,p_2) \, ,\eea where $S(p_1,p_2)$ is the
matrix (\ref{matelement}). The matrix $S^g$ encodes the matrix
elements of the S-matrix operator $\bS$, but, contrary to $S$, it
does not satisfy the Yang-Baxter equation (\ref{YBnorm}). In what
follows we will refer to $S^g$ as the graded S-matrix, because it
satisfies another version of (\ref{YBnorm}) known as the graded
Yang-Baxter equation.

\smallskip

To derive the equation, we substitute in
eq.(\ref{YBnorm}) the matrix $S$ expressed via $S^g$:
$$
\mI^g_{23}S^g_{23}\,\mI_{13}^gS^g_{13}\,\mI_{12}^g
S_{12}^g=\mI^g_{12}S^g_{12}\,\mI_{13}^gS^g_{13}\,\mI_{23}^g
S_{23}^g\, .
$$
Here $S^g_{ij}$ denotes the usual embedding of the matrix $S^g$
into the product of three spaces. Now we notice that both $S$ and
$S^g$ obey the following identities \bea\la{gggYB}\begin{aligned}
&\mI_{12}^g \mI_{23}^gS_{13} = S_{13} \mI_{12}^g
\mI_{23}^g\,,\quad \mI_{12}^g\mI_{13}^g S_{23} =S_{23}\mI_{13}^g
\mI_{12}^g\,,
\\
&\mI_{12}^g \mI_{23}^gS_{13}^g = S_{13}^g \mI_{12}^g
\mI_{23}^g\,,\quad \mI_{12}^g\mI_{13}^g S_{23}^g
=S_{23}^g\mI_{13}^g \mI_{12}^g\,  \end{aligned}\eea which all
follow from eq.(\ref{grproperty}). Using these relations,
eq.(\ref{gggYB}) can be cast in the form \bea\nonumber
\underbrace{ \mI_{12}^g\mI_{13}^gS_{23}^g \mI_{13}^g
\mI_{12}^g}_{\check{S}_{23}} \underbrace{\mI_{23}^gS_{13}^g
 \mI_{23}^g}_{\check{S}_{13}}\underbrace{S_{12}^g}_{\check{S}_{12}}=
 \underbrace{S_{12}^g}_{\check{S}_{12}}\underbrace{ \mI_{23}^gS_{13}^g \mI_{23}^g }_{\check{S}_{13}}
 \underbrace{\mI_{12}^g \mI_{13}^gS_{23}^g \mI_{13}^g \mI_{12}^g}_{\check{S}_{23}} \,.
\eea We see that if we define the {\it graded embedding} of $S^g$
into the vector product of three spaces as
$$
\check{S}_{12} = S_{12}^g\,, \quad \check{S}_{13} =
\mI_{23}^gS_{13}^g \mI_{23}^g\,, \quad \check{S}_{23} =
\mI_{12}^g\mI_{13}^g S_{23}^g \mI_{13}^g \mI_{12}^g=S_{23}^g \,,
$$
we obtain the graded Yang-Baxter equation \bea
  \check{S}_{23}\check{S}_{13}\check{S}_{12}=\check{S}_{12}\check{S}_{13}\check{S}_{23}\,
  ,
\eea which looks the same as eq.(\ref{YBnorm}). Sometimes the
matrix $\check{S}$ is referred to as the graded fermionic
S-operator.

\subsection{S-matrix and its symmetries}

Now we are in position to show that the existence of a symmetry algebra of the Hamiltonian implies certain restrictions on the S-matrix.

Denote by $\hJ^\ba$ the operators which generate the symmetry
algebra $\mathscr{J}$:
$$
[\hJ^\ba,\bH]=0\, , ~~~~~~~\ba=1,\ldots, \mbox{dim}\mathscr{J}\,
.
$$
In addition to $\bH$, the symmetry generators commute with $\bP$
and $\bN$, and with all the higher conserved charges $\bI_{q}$.
The latter act diagonally in the basis of multi-particle states.

The Hilbert space created by the ZF operators carries a linear
representation of $\mathscr{J}$, and since the operators $\hJ^\ba$
commute with $\bN$ and all the higher charges they must preserve
the number of particles and the set of their momenta:
 \bea \la{strcon}\begin{aligned} &\hJ^\ba\cdot|\Omega \rangle =0\,,\\
&\hJ^\ba\cdot A^\dagger_{i}(p)|\Omega \rangle =
J^{\ba}{}^j_{i}(p)A^\dagger_{j}(p)\, |\Omega \rangle\,,\\
&\hJ^\ba\cdot A^\dagger_{i}(p_1) A^\dagger_{j}(p_2)|\Omega \rangle
= J^\ba{}^{kl}_{ij}(p_1,p_2)\, A^\dagger_{k}(p_1)
A^\dagger_{l}(p_2)|\Omega \rangle\, ,
\\ & \ldots \ldots \ldots\end{aligned}\eea Here the
tensors $J^{\ba}{}^j_{i},J^\ba{}^{kl}_{ij}$, $\ldots\,$, can be
thought of as the structure constants of the symmetry algebra in
one-particle, two-particle, etc. representations. In general these
structure constants might depend on the particle momenta. Since
$\mathscr{J}$ is a superalgebra, the generator $(-1)^{\bN_F}$
introduced in the previous section commutes with all the bosonic
algebra generators and anti-commutes with the fermionic ones \bea
(-1)^{\bN_F}\cdot \hJ^\ba=(-1)^{\gr_{\ba}}\, \hJ^\ba\cdot
(-1)^{\bN_F}\, ,\eea where $\gr_{\ba}$ is the degree of
$\hJ^{\ba}$. This leads to the selection rules for the
corresponding structure constants \bea \label{selection}
\begin{aligned}
& J^{\ba}{}^j_{i}(p)=(-1)^{\gr_{\ba}+\eps_i+\eps_j}J^{\ba}{}^j_{i}(p)\, ,\\
&
J^\ba{}^{kl}_{ij}(p_1,p_2)=(-1)^{\gr_{\ba}+\eps_i+\eps_j+\eps_k+\eps_l}J^\ba{}^{kl}_{ij}(p_1,p_2)\,
,\\
& \ldots \ldots \ldots\end{aligned}
 \eea

The crucial point is that the non-abelian symmetry algebra
$\mathscr{J}$ acting on the spectrum of the Hamiltonian implies a
non-trivial constraint on the scattering matrix. This constraint
can be derived by acting with symmetry generators $\hJ^\ba$ on the
ZF algebra relations \bea \hJ^\ba\cdot A^\dagger_{i}(p_1)
A^\dagger_{j}(p_2)|\Omega \rangle = S^{kl}_{ij}(p_1,p_2)\,
\hJ^\ba\cdot A^\dagger_{l}(p_2) A^\dagger_{k}(p_1)|\Omega
\rangle\,. \eea Recalling the formulae (\ref{strcon}), one finds
that the S-matrix elements must satisfy the following invariance
condition \bea\la{incon}
S^{mn}_{kl}(p_1,p_2)J^\ba{}^{kl}_{ij}(p_1,p_2)=
J^\ba{}^{nm}_{lk}(p_2,p_1)S^{kl}_{ij}(p_1,p_2)\,. \eea If we
combine the symmetry generator structure constants in a matrix
\bea J^\ba_{12}(p_1,p_2)\equiv J^\ba{}^{kl}_{ij}(p_1,p_2)
E^{~i}_{k} \otimes E^{~j}_{l}\,, \eea  then the invariance
condition can be written as \bea\la{incon2}
S_{12}(p_1,p_2)J^\ba_{12}(p_1,p_2) =
J^\ba_{21}(p_2,p_1)S_{12}(p_1,p_2)\,. \eea

The form of the multi-particle structure constants is determined
by the symmetry algebra of a particular model. In trivial cases,
$\mathscr{J}$ is a simple Lie superalgebra with
(momentum-independent) structure constants in the one-particle
representation
$$
[\hspace{-0.06cm}[J^{\ba},J^{\bb}]\hspace{-0.06cm}]=t^{\ba\bb{\bf
c}}J^c\, ,
$$
where $[\hspace{-0.06cm}[.,.]\hspace{-0.06cm}]$ stands for the
graded commutator
$$
[\hspace{-0.06cm}[J^{\ba},J^{\bb}]\hspace{-0.06cm}]=J^{\ba}J^{\bb}-(-1)^{\gr_{\ba}\gr_{\bb}}J^{\bb}J^{\ba}\,
.
$$
In this case the two-particle states  can be identified with the
tensor product of two one-particle states and the two-particle
symmetry generators are given by\footnote{It is worth mentioning
that eq.(\ref{2ptrep}) implies that the action of symmetry
generators on two-particle states in eq.(\ref{strcon}) satisfies
the Leibnitz rule. In general, however, this is not the case. }
\bea\la{2ptrep} J^\ba_{12} = J^\ba{}\otimes \mI + \mI^g(\mI\otimes
J^\ba{})\mI^g \, . \eea Since we work with the usual (not graded)
tensor product, the second term in the right hand side of
eq.(\ref{2ptrep}) involves the graded identity which is needed for
a proper account of the statistics.\footnote{In components, this
formula reads as $J^\ba{}^{kl}_{ij}(p_1,p_2) =
J^\ba{}^{k}_{i}(p_1)\, \delta_j^l+(-1)^{\epsilon_i\epsilon_{\ba
}}\, \delta_i^k J^\ba{}^{l}_{j}(p_2)$.}  Indeed, one can easily
check that eq.(\ref{2ptrep}) defines a representation of
$\mathscr{J}$ in the tensor product $\mathscr{V}\otimes
\mathscr{V}$. Thus, for models with momentum-independent
one-particle structure constants the invariance condition for the
S-matrix reduces to the familiar matrix equations
\bea\begin{aligned} \hspace{2cm}(J^{\ba}\otimes \mI + \mI\otimes
J^{\ba})S_{12}&=S_{12}(J^{\ba}\otimes \mI + \mI\otimes J^{\ba})\,
~~~~~~\mbox{for }~~J^{\ba}~~\mbox{bosonic} \, ,\\ (J^{\ba}\otimes
\mI + \Sigma\otimes J^{\ba})S_{12}&=S_{12}(\mI\otimes J^{\ba} +
J^{\ba}\otimes\Sigma)\, ~~~~~~\mbox{for
}~~J^{\ba}~~\mbox{fermionic} \, , \nonumber\end{aligned}\eea where
the grading matrix $\Sigma$ is defined in eq.(\ref{grmatrix}) and
we specified formula (\ref{2ptrep}) for the cases of bosonic and
fermionic algebra generators. The symmetry algebra of the
light-cone string sigma model is not of this simple type, however,
and in our subsequent analysis we have to resort to the invariance
condition (\ref{incon2}).

\smallskip
Returning to the general situation, we assume that $\mathscr{J}$
has a non-trivial center. Then, any  representation of
$\mathscr{J}$ is parametrized by the particle momentum and by the
corresponding values of the Lie algebra central elements
(charges).\footnote{The momentum $\bP$ commutes with all
$\hJ^\ba$, and therefore   is central. We prefer, however, to
separate $\bP$ from other central charges due to its special
role.} Let $J^\ba(p;c)$ be the generators of $\mathscr{J}$ in some
representation $\mathscr{V}$, where $c$ denotes a level set of the
central elements. The generators $J^\ba(p;c)$ should be thought of
as matrices depending on the parameters $p$ and $c$ but acting in
the same carrier space $\mathscr{V}$. The matrix $C$ representing
a central charge $\mathbb{C}\in \hspace{-0.2cm}\mathscr{J}$ is
 $C=c\,\mI$. Representations corresponding to various
sets of $p,c$ are inequivalent, because a transformation
$J^\ba(p;c)\to \ag J^\ba(p;c)\ag^{-1}$ cannot change the value of
the central charges.

\smallskip
Obviously, if we wish to identify $\mathscr{V}$ with a
one-particle representation in the Fock space, we have to
prescribe for $c$ some fixed value ({\it e.g.}, zero), as the
one-particle representation is characterized by the particle
momentum only and it does not involve any other continuous
parameters. The structure constants in the two-particle
representation can be then defined in a way similar to
eq.(\ref{2ptrep}) \bea\la{2ptrepgen} J^\ba_{12}(p_1,p_2) =
J^\ba{}(p_1;c_1)\otimes \mI + \mI^g(\mI\otimes
J^\ba{}(p_2;c_2))\mI^g \, . \eea For a consistent interpretation
of eq.(\ref{2ptrepgen}) as the two-particle representation, the
level sets of the first and second one-particle representations
should depend on the particle momenta $p_1,p_2$. In particular, a
non-trivial situation arises when this dependence is mutually
non-local -- $c_1$ is determined by $p_2$ and $c_2$ by $p_1$,
respectively. Plugging in eq.(\ref{2ptrep}) the matrix
representatives of $\bC$, we get \bea C_{12}=c_1 \mI\otimes \mI+
\mI^g(\mI\otimes c_2\mI)\mI^g=(c_1+c_2)\, \mI\otimes \mI\, .\eea
Of course, this formula reflects a general fact  that the value of
a central charge in a tensor product representation is given by
the sum of the values corresponding to the individual components
of this tensor  product.

\smallskip

As was established in subsection \ref{ceasu}, the  symmetry
algebra of the light-cone sigma model has the three-dimensional
center, which, in addition to $\bH$, contains the operator $\bC$
and its hermitian conjugate $\bC^{\dagger}$;  both of them are
(non-linear) functions of the momentum operator $\bP$. Thus, the
corresponding representation theory arising in the Fock space
should fit our general treatment above. Indeed, as we will show in
the forthcoming sections, the simultaneous additivity of
$\bC(\bP)$ and $\bP$ will require a realization of the
two-particle representation in the form (\ref{2ptrepgen}) with
non-trivial functions $c_1(p_2)$ and $c_2(p_1)$.

\subsection{General physical requirements}\la{gpr}
In a  physical theory the S-matrix must satisfy a number of
additional requirements reflecting analytic properties and
discrete symmetries of the corresponding Hamiltonian. In this
section we show that some of these requirements can be naturally
derived by using the ZF algebra framework. We start our discussion
with the condition of physical unitarity.

\subsubsection*{Physical unitarity}
\smallskip

Since the Hamiltonian is hermitian, the associated S-matrix
operator $\bS$ is unitary. To find the implications of this
unitarity for the two-particle S-matrix $S(p_1,p_2)$, we can use
the fact that the annihilation operators are hermitian conjugate
of the creation ones. Taking the hermitian conjugation of the
first line in eq.(\ref{ZFfull}), we get
$$
\bA_2(p_2)\bA_1(p_1)=S_{12}^{\dagger}(p_1,p_2)\bA_1(p_1)\bA_2(p_2)\,
.
$$
Changing $p_1\leftrightarrow p_2$ and pulling the permutation
$P_{12}$ through the left and the right hand side of the last
formula, we obtain
$$
\bA_1(p_1)\bA_2(p_2)=S_{21}^{\dagger}(p_2,p_1)\bA_2(p_2)\bA_1(p_1)\,
.
$$
This expression must coincide with the second line in
eq.(\ref{ZFfull}) implying the relation
$S_{21}^{\dagger}(p_2,p_1)=S_{12}(p_1,p_2)$. Using the unitarity
condition (\ref{unitarity}), this relation can be written as
 \bea \la{physicalunitarity}
S^{\dagger}(p_1,p_2) S(p_1,p_2)=\mI\,  \eea meaning that
$S(p_1,p_2)$ is a unitary matrix. This is the condition of
physical unitarity.

\subsubsection*{Parity Invariance}

\noindent As was established in subsection \ref{sect:parity}  the
Lagrangian of the world-sheet sigma model is invariant with
respect to the parity transformation $\mathscr{P}$. This
transformation acts as $\sigma\to -\sigma$ with simultaneous
multiplication of  fermions by $i$. Obviously, in the momentum
space the map $\sigma\to -\sigma$ corresponds to $p\to -p$.
Therefore, on one-particle states the action of $\mathscr{P}$ can
be naturally defined as \bea \la{parityt} \mathscr{P}\cdot
A^\dagger_{i}(p) |\Omega \rangle
=(-1)^{\sfrac{1}{2}\gr_i}A^\dagger_{i}(-p)|\Omega \rangle \, .\eea
Here $\eta_{\mathscr{P}}=(-1)^{\sfrac{1}{2}\gr_i}$ is intrinsic
parity of the particle created by $A_i^{\dagger}$. For a fermion
$\mathscr{P}^2=-1$ which reflects the double-valuedness of the
spinor representation under a rotation over an angle $2\pi$. On
multi-particle states we then have
$$
\mathscr{P}\cdot |p_1,p_2, \ldots , p_n
\rangle^{(in)}_{i_1,...,i_n} = (-1)^{\sfrac{1}{2}\sum_k
\gr_{i_k}}|-p_1,-p_2, \ldots , -p_n \rangle^{(in)}_{i_1,...,i_n}\,
.
$$
Using the representation of $in$ states in terms of the ZF
operators, we can write \bea\nonumber \mathscr{P}\cdot
A^\dagger_{i_1}(p_1)\cdots A^\dagger_{i_n}(p_n)|\Omega \rangle
=(-1)^{\sfrac{1}{2}\sum_k \gr_{i_k}}
(-1)^{\sum_{k<l}\gr_{i_k}\gr_{i_l}}A^\dagger_{i_n}(-p_n)\cdots
A^\dagger_{i_1}(-p_1)|\Omega \rangle\, , \eea where an extra
statistical factor $(-1)^{\sum_{k<l}\gr_{i_k}\gr_{i_l}}$ arises
due to the operator reordering. Now letting $\mathscr{P}$ act on
both sides of the ZF algebra
 \bea\mathscr{P}\cdot  A^\dagger_{i}(p_1)
A^\dagger_{j}(p_2) = \mathscr{P}\cdot A^\dagger_{l}(p_2)
A^\dagger_{k}(p_1)S^{kl}_{ij}(p_1,p_2)\, , \eea and pulling
$\mathscr{P}$ through, we obtain \bea\nonumber
(-1)^{\sfrac{1}{2}(\gr_i+\gr_j)+\gr_i\gr_j}A^\dagger_{j}(-p_2)
A^\dagger_{i}(-p_1) =(-1)^{\sfrac{1}{2}(\gr_k+\gr_l)+\gr_k\gr_l}
A^\dagger_{k}(-p_1) A^\dagger_{l}(-p_2)S^{kl}_{ij}(p_1,p_2)=\\
\nonumber =(-1)^{\sfrac{1}{2}(\gr_k+\gr_l)+\gr_k\gr_l}
A^\dagger_{n}(-p_2) A^\dagger_{m}(-p_1)S^{mn}_{kl}(-p_1,-p_2)
S^{kl}_{ij}(p_1,p_2) \, , \eea From here we conclude that the
matrix $S$ must obey the following condition \bea\label{parcon}
S^{mn}_{kl}(-p_1,-p_2)
S^{kl}_{ij}(p_1,p_2)(-1)^{-\gr_i\gr_j+\gr_k\gr_l+\sfrac{1}{2}(\gr_k+\gr_l-\gr_i-\gr_j)}=\delta_i^m\delta_j^n\,
. \eea Since the sum $\gr_k+\gr_l+\gr_i+\gr_j$ is an even number
and $\gr_i^2=\gr_i$, we have \bea\nonumber -
\gr_i\gr_j+\gr_k\gr_l+\sfrac{1}{2}(\gr_k+\gr_l-\gr_i-\gr_j)&=&\frac{1}{2}\big[(\gr_k+\gr_l)^2-(\gr_i+\gr_j)^2\big]=\\
\nonumber
&=&\frac{1}{2}\underbrace{(\gr_k+\gr_l-\gr_i-\gr_j)}_{\rm
even}\underbrace{(\gr_k+\gr_l+\gr_i+\gr_j)}_{\rm even}\, ,\eea
{\it i.e.} the left hand side of the last expression is also an
even number and, therefore, eq.(\ref{parcon}) reduces to
\bea\la{pinvmatrix} S(-p_1,-p_2)=S^{-1}(p_1,p_2)\, . \eea This is
the parity transformation rule for the S-matrix.

\subsubsection*{Time reversal}
In quantum field theory the time reversal operation
$\mathscr{T}:~\tau\to-\tau$ is realized by means of an
 anti-linear, anti-unitary operator $U_{\tau}$: \bea\nonumber
&& U_{\tau}\, c|\Phi\rangle =\bar{c} \, U_{\tau} |\Phi\rangle\,
,~~~~~~~~~ \langle \Psi|\Phi\rangle=\langle U_{\tau}\,
\Phi|U_{\tau}\, \Psi\rangle \eea To understand the implications of
the symmetry under time reversal, it is convenient to start with
the free field representation in terms of creation and
annihilation operators as discussed in section
\ref{subsect:quantization}. On free fields $Y^{a\da}$,
$Z^{\a\dot{\a}}$, $\theta^{a\dal}$ and $\eta^{\a\dot{a}}$ the
action of the anti-linear operator $U_{\tau}$ can be defined as
follows \bea\nonumber\begin{aligned} & U_\tau
Y^{a\da}(\s,\tau)U_\tau^{-1}=\eta_\tau\, Y^{a\da}(\s,-\tau)\, ,
~~~&&~~~ U_\tau
Z^{\a\dot{\a}}(\s,\tau)U_\tau^{-1}=\eta_\tau\,  Z^{\a\dot{\a}}(\s,-\tau)\, ,\\
& U_\tau\theta^{a\dal}(\s,\tau)U_\tau^{-1}=\eta_\tau\,
\theta^{a\dal}(\s,-\tau)\, , ~~~&&~~~
U_\tau\eta^{\a\dot{a}}(\s,\tau)U_\tau^{-1}=\eta_\tau\,
\eta^{\a\dot{a}}(\s,-\tau)\, , \end{aligned}\eea where
$\eta_{\tau}$ is intrinsic time parity which depends on the type
of a field. It is easy to see that for our string model it is
consistent to choose $\eta_\tau=1$ for all the fields, so that the
Lagrangian density (\ref{L2b}) will transform under time reversal
as
$$
U_\tau \L_2(\s,\tau)U_\tau^{-1}=\L_2(\s,-\tau)\,
$$
leaving, therefore,  the corresponding Lagrangian invariant. The
action of time reversal on creation and annihilation operators is
easy to derive by recalling the mode expansion of the
corresponding fields, {\it e.g.}, \bea\nonumber
&&Y^{a\da}(\s,\tau) = {1\ov 2\sqrt{2\pi}}\int\,\,{{\rm d}p\ov
\sqrt{\om_p}}\left( e^{ip\s-i\omega_p \tau} a^{a\da}(p) +
e^{-ip\s+i\omega_p\tau}\eps^{ab}\eps^{\da\db}
a_{b\db}^\dagger(p)\right)\\
\nonumber &&\theta^{a\dal}(\s,\tau)= {e^{-i\pi/4}\ov
\sqrt{2\pi}}\int\,{{\rm d}p\ov\sqrt{\om_p}}\,\left(
e^{ip\s-i\omega_p \tau}\,f_p\, a^{a\dal}(p) +e^{-ip\s+i\omega_p
\tau}\,h_p\,\eps^{ab}\eps^{\dal\dbe} a_{b\dbe}^\dagger(p)\right)\,
 \eea
and similarly for $Z^{\a\dot{\a}}$ and $\eta^{\a\dot{a}}$.
Applying  $U_\tau$ to these expressions, we get {\small
\bea\nonumber && \hspace{-0.7cm}U_\tau
Y^{a\da}(\s,\tau)U_\tau^{-1} ={1\ov 2\sqrt{2\pi}}\times \\
\nonumber && \hspace{-0.5cm}\times \int\,\,{{\rm d}p\ov
\sqrt{\om_p}}\left( e^{-ip\s+i\omega_p \tau} U_\tau
a^{a\da}(p)U_\tau^{-1} +
e^{ip\s-i\omega_p\tau}\eps^{ab}\eps^{\da\db} U_\tau
a_{b\db}^\dagger(p)U_\tau^{-1}\right) = Y^{a\da}(\s,-\tau)\, \eea
} and {\small \bea \nonumber && \hspace{-0.5cm}U_\tau
\theta^{a\dal}(\s,\tau)U_\tau^{-1} ={e^{i\pi/4}\ov
\sqrt{2\pi}}\times \\
\nonumber && \hspace{-0.6cm}\times \int\,{{\rm
d}p\ov\sqrt{\om_p}}\,\left( e^{-ip\s+i\omega_p \tau}\,f_p\, U_\tau
a^{a\dal}(p)U_\tau^{-1} +e^{ip\s-i\omega_p
\tau}\,h_p\,\eps^{ab}\eps^{\dal\dbe} U_\tau
a_{b\dbe}^\dagger(p)U_\tau^{-1}\right)=\theta^{a\dal}(\s,-\tau)\,
. \eea } From here we deduce the transformation law for creation
and annihilation operators \bea\la{trfree}\begin{aligned} & U_\tau
a^{a\da}(p)U_\tau^{-1}=a^{a\da}(-p) \, , ~~~&&~~~  U_\tau
a_{b\db}^\dagger(p)U_\tau^{-1}= a_{b\db}^\dagger(-p) \, ,\\
& U_\tau a^{a\dal}(p)U_\tau^{-1}=-i \,  a^{a\dal}(-p)\, , ~~~&&~~~
U_\tau a_{b\dbe}^\dagger(p)U_\tau^{-1}=i \,  a_{b\dbe}^\dagger(-p)
\, .
\end{aligned} \eea
It is interesting to note that in classical theory and before
gauge fixing, time reversal can be defined in a way similar to
parity reversal, namely, $\tau\to -\tau$ with simultaneous
multiplication of fermions $\theta$ and $\eta$ by $i$ and $-i$,
respectively. We see that in the gauge-fixed quantum theory, with
the well-defined Hamiltonian and the canonical structure, these
are fermionic creation and annihilation operators that under time
reversal are multiplied by $i$ or $-i$, rather then $\theta$ and
$\eta$.

\smallskip

Formulae (\ref{trfree}) derived for free theory suggest how to
define the time reversal operation $\mathscr{T}$ in interacting
theory. On a  one-particle state created by a ZF operator we define
an action of $\mathscr{T}$ as \bea \la{timerev} \mathscr{T}\cdot
A^\dagger_{i}(p) |\Omega \rangle
=i^{\gr_i}A^\dagger_{i}(-p)|\Omega \rangle\, . \eea Since
$\mathscr{T}$ maps $\tau\to -\tau$, it interchanges asymptotic
past and future and, for this reason, its action on multi-particle
states is given by \bea \nonumber \mathscr{T}\cdot |p_1,p_2,
\ldots , p_n \rangle^{(in)}_{i_1,...,i_n} =
(-1)^{\sfrac{1}{2}\sum_k \gr_{i_k}}|-p_1,-p_2, \ldots , -p_n
\rangle^{(out)}_{i_1,...,i_n}\, . \eea Representing $in$ and $out$
states in terms of the ZF operators, the last formula can be
written as
 \bea\nonumber \mathscr{T}\cdot
A^\dagger_{i_1}(p_1)\cdots A^\dagger_{i_n}(p_n)|\Omega \rangle
=(-1)^{\sfrac{1}{2}\sum_k \gr_{i_k}} A^\dagger_{i_1}(-p_1)\cdots
A^\dagger_{i_n}(-p_n)|\Omega \rangle\, . \eea Commuting
$\mathscr{T}$ through both sides of the ZF algebra relations
(\ref{ZFu}), one gets \bea\nonumber
(-1)^{\sfrac{1}{2}(\gr_i+\gr_j)}A^\dagger_{i}(-p_1)
A^\dagger_{j}(-p_2) =(-1)^{\sfrac{1}{2}(\gr_k+\gr_l)}
A^\dagger_{l}(-p_2) A^\dagger_{k}(-p_1)S^{*kl}_{ij}(p_1,p_2)\, ,
\eea where $S^{*kl}_{ij}$ stands for the complex conjugate of the
S-matrix element $S^{kl}_{ij}$, and we have taken into account
that $\mathscr{T}$ is an anti-unitary operator. Permuting the ZF
operators in the right hand side of the last relation, one obtains
\bea\nonumber (-1)^{\sfrac{1}{2}(\gr_i+\gr_j)}A^\dagger_{i}(-p_1)
A^\dagger_{j}(-p_2) = (-1)^{\sfrac{1}{2}(\gr_k+\gr_l)}
A^\dagger_{n}(-p_1)
A^\dagger_{m}(-p_2)S^{mn}_{lk}(-p_1,-p_2)S^{*kl}_{ij}(p_1,p_2).
\eea Thus, invariance of the theory under time reversal leads to
the following equation for the matrix elements of the S-matrix:
\bea \la{timerevcomp}
S^{*kl}_{ij}(p_1,p_2)S^{mn}_{lk}(-p_1,-p_2)(-1)^{\sfrac{1}{2}(\gr_k+\gr_l-\gr_i-\gr_j)}=\delta_i^n\delta_j^m\,
. \eea According to our discussion of the parity transform,
$$
(-1)^{\gr_i\gr_j+\gr_k\gr_l}=(-1)^{\sfrac{1}{2}(\gr_k+\gr_l-\gr_i-\gr_j)}\,
.
$$
Therefore, in the matrix form eq.(\ref{timerevcomp}) reads as
\bea\la{TRI0} \mI^g S^*_{12}(p_1,p_2)\mI^g\, S_{21}(-p_2,-p_1)=\mI
\, . \eea This is the condition on the two-particle S-matrix
implied by the  time reversal invariance.

Unitarity condition (\ref{unitarity}) in conjunction with parity
invariance (\ref{pinvmatrix}) and physical unitarity
(\ref{physicalunitarity}) allows one to rewrite the last formula
in the following form \bea \la{TRI2} S^t(p_1,p_2)=\mI^g
S(p_1,p_2)\mI^g \, ,
 \eea
that can be viewed as the consequence of the combined parity and
time reversal invariance.

\subsubsection*{Charge conjugation}
As before, we assume that particles (one-particle asymptotic
states) transform in some representation $\mathscr{V}$ of the
symmetry algebra $\mathscr{J}$. Let $\mathscr{B}$ be the bosonic
subalgebra of $\mathscr{J}$. If a theory is invariant under charge
conjugation then there are two possibilities -- either a
representation of $\mathscr{B}$ in $\mathscr{V}$ is reducible and
consists of two representations conjugate to each other  or it is
self-conjugate.

In the first case we have $\mathscr{V}=\mathscr{W}\oplus
\mathscr{W}^*$, where the first and the second components
correspond to particles and anti-particles,
respectively\footnote{The reader might have in mind, for instance,
quarks and anti-quarks which transform in fundamental and
anti-fundamental irreps of ${\rm SU(3)}$. }. If $\mathscr{D}$ is a
matrix realization of the group corresponding to $\mathscr{B}$
which acts in the space $\mathscr{W}$, then anti-particles
transform in the conjugate representations $\mathscr{W}^*$ with
the matrix realization $\mathscr{D}^*$. Note that for unitary
groups the conjugate representation coincides with the
contragradient representation: $(\mathscr{D}^{t})^{
-1}=\mathscr{D}^*$. Charge conjugation is understood as a transfer
$$C:~~~\mathscr{W}\to \mathscr{W}^*.$$
In general, $C$ belongs to the group of outer automorphisms of
$\mathscr{B}$.

In the second case, the representation $\mathscr{V}$ is
self-conjugate which means that $\mathscr{V}^*$ is equivalent to
$\mathscr{V}$. For instance, if the bosonic subalgebra of
$\mathscr{J}$ is $\su(2)$, then
$$
\mathscr{D}^*=C\, \mathscr{D}\,  C^{-1}\, ,
$$
where, according to eq.(\ref{su2conj}), $C=\epsilon$ is an
internal automorphism. Obviously, under these circumstances,
invariance under charge conjugation does not lead to any new
restrictions on the form of the two-particle S-matrix beyond those
implied by $\mathscr{J}$. This is precisely the situation we
encounter for the string sigma model.

\subsubsection*{Crossing symmetry}
So far we were considering the obvious kinematical symmetries of
the Hamiltonian. Now we introduce a new type of dynamical symmetry
which manifests itself in the scattering process as a possibility
to replace a particle with its anti-particle. In relativistic
theories this kind of symmetry is known as crossing.

Recall that in  two-dimensional Lorentz-invariant models the
particle momentum $p$ and the energy $H$ can be parametrized by a
single rapidity variable $\theta$ \bea p=\, {\rm sinh}\, \theta\,
,~~~~~~ H=\, {\rm cosh} \, \theta\,   \eea which provides a
solution to the relativistic dispersion relation \bea \la{disprel}
H^2-p^2=1\, , \eea where for simplicity we assumed a particle of
unit mass. Invariance under Lorentz transformations requires the
two-particle S-matrix to depend on the difference of the particle
rapidities: $S(p_1,p_2)=S(\theta_1-\theta_2)$.

\noindent
\begin{figure}
\begin{minipage}{\textwidth}
\begin{center}
\hspace{-2cm}\includegraphics[width=0.60\textwidth]{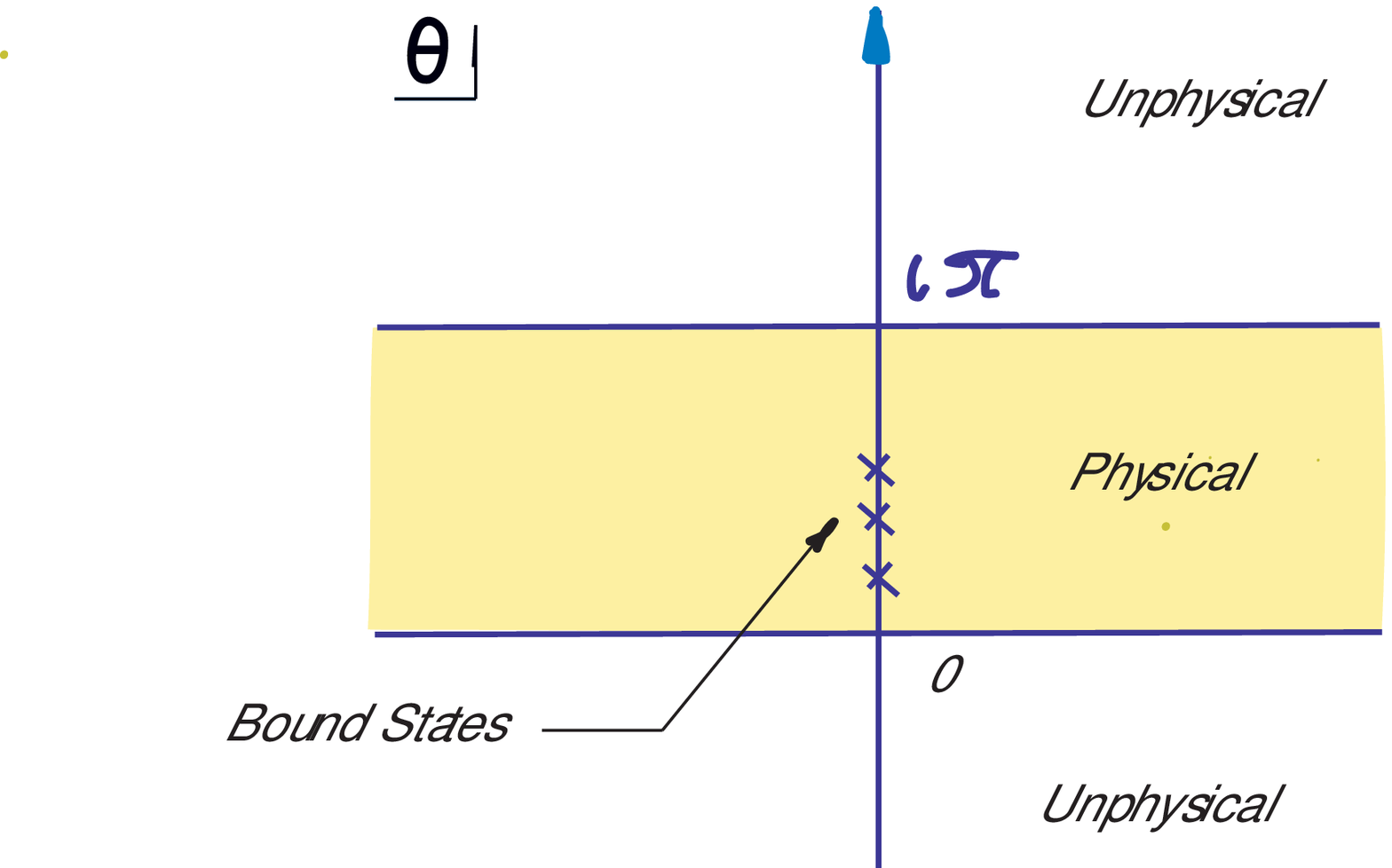}
\parbox{5in}{\footnotesize{\caption{\label{PhysStrip}Physical strip on the rapidity plane of a two-dimensional relativistic field theory.}}}
\end{center}
\end{minipage}
\end{figure}

\vskip -0.6cm To describe all states in a theory, including bound
states, the rapidity variable should be continued to the complex
plane. The already mentioned crossing symmetry transformation
corresponds to the shift $\theta\to \theta+i\pi$, because the
momentum and energy change a sign
$$
\theta\to \theta+i\pi : ~~~~~p\to -p \, ,~~~~~~~ H\to -H \, ,
$$
which signifies a transition to the corresponding anti-particle.
The difference $\theta=\theta_1-\theta_2$ takes value in the strip
$0\leq {\rm Im}\,\theta<\pi$ and $-\infty<{\rm
Re}\,\theta<\infty$, which is called the physical strip of a
relativistic field theory, see Figure \ref{PhysStrip}.

Crossing symmetry leads to further constraints on the scattering
matrix. Although the string sigma model does not have Lorentz
invariance on the world-sheet, as we will show, the corresponding
scattering theory is compatible with the assumption of crossing.

\smallskip

We will reserve a detailed discussion of crossing symmetry for
subsection 3.4.2. Here our goal will be to demonstrate that the
crossing symmetry requirement for the S-matrix naturally follows
from an additional invariance condition of the ZF algebra.

This invariance condition is related to the possibility of
exchanging in the ZF relations creation and annihilation operators
corresponding to one of the two particles. More precisely, we
define the following transformation
 \bea\la{pat} \mathbb{A}^\dagger(p)\to \mathbb{B}^\dagger(p)= \mathbb{A}^{t}(-p)
\mathscr{C}\,,\quad \mathbb{A}(p)\to
\mathbb{B}(p)=\mathscr{C}^{\dagger} \mathbb{A}^\dagger{}^{t}(-p)
\,,~~~~ \eea where $\mathscr{C}$ is a constant matrix and
superscript $t$ means transposition. We require that under this
map the ZF algebra relations (\ref{ZFfull}) for $p_1\neq p_2$
transform into themselves. More precisely, if we first replace in
the algebra relations $\mathbb{A}$ by $\mathbb{B}$ for one of the
particles, and further use the formulae (\ref{pat}) to express
$\mathbb{B}$ via $\mathbb{A}$, we should recover for $\mathbb{A}$
the same relations. Under the assumption $p_1>p_2$ the
delta-function does not contribute which makes it possible to map
by means of eq.(\ref{pat}) the exchange relations of
$\mathbb{A}(p_1)$ and $\mathbb{A}(p_2)$ to that of
$\mathbb{A}(p_1)$ and $\mathbb{A}^\dagger(p_2)$.

\smallskip

Note that flipping the sign of $p$ under (\ref{pat}) is dictated
by the compatibility of eq.(\ref{pat}) with the algebra relations
$$
\hP A^\dagger = A^\dagger( \hP + p)  \,,\quad \hP  A = A( \hP -
p)\, ,
$$
{\it i.e.} the operators ${\mathbb A}^{\dagger}$ and ${\mathbb
B}^{\dagger}$ are required to commute with $\mathbb{P}$ in the
same way.

Application of the crossing symmetry transformation to the
S-matrix requires a certain care. Crossing symmetry does not only
change the sign of $p$ but it also changes a branch of the
dispersion relation sending $H$ to $-H$. To correctly implement
the action of crossing symmetry, the S-matrix could be treated as
a function of both the particle momenta and the particle
energies: $S(p_1,H_1;p_2,H_2)$.   Of course, on any given branch the
S-matrix becomes a function of particle momenta only. Even then the S-matrix is
not a meromorphic  function of $p_i$ and $H_i$, and one still
should specify additional cuts in the $pH$-planes and choose
a proper branch. Crossing
{\it e.g.} the first particle then invokes the following
transformation
$$
S(p_1,H_1;p_2,H_2)\to S^{c_1}(-p_1,-H_1;p_2,H_2)\, ,
$$
and applying it twice one does not end up with the original
S-matrix: $\left(S^{c_1}\right)^{c_1} \neq S$.

Although below we will not specify explicitly the branch
dependence of the S-matrix, it is precisely in this sense we
understand the action of crossing on $S$. Clearly, finding an
analogue of the rapidity variable which uniformizes a given
dispersion relation and makes the S-matrix a meromorphic function
would greatly simplify the treatment of crossing symmetry as it
resolves the ambiguities of $S$ related to the choice of a branch.
For the string sigma model at hand, such a uniformization
rendering the crossing symmetric world-sheet S-matrix a
meromorphic function is unknown. We will return to this important
issue in subsection \ref{subsect: repidtorus}.

Meanwhile, we find that invariance of the ZF algebra under map
(\ref{pat}) implies the following equations
\bea\la{cros}\begin{aligned} &
\mathscr{C}_1^{-1}S_{12}^{t_1}(p_1,p_2) \mathscr{C}_1
 S_{12}(-p_1,p_2)={\mI}\, ,\\
 &
\mathscr{C}_2^{-1}S_{21}^{t_2}(p_2,p_1) \mathscr{C}_2
 S_{21}(-p_2,p_1)={\mI}\, .\end{aligned}
 \eea
Here $t_1$ and $t_2$ mean the transposition in the first and
second space, respectively, $\mathscr{C}_1 = \mathscr{C}\otimes
\mI$, $\mathscr{C}_2 = \mI\otimes \mathscr{C}$. In fact, these two
equations are equivalent: the first turns into the second after
applying the permutation and exchanging $p_1$ and $p_2$. Provided
$\mathscr{C}$ is known, eqs.(\ref{cros}) represent a further
non-trivial constraint on the S-matrix.

There is an alternative way to obtain eqs.(\ref{cros}). Without
loss of generality we assume that
$\mathscr{C}^{\dagger}\mathscr{C}=\mI$ and consider the following
singlet  (row $\times$ matrix $\times$ column)
$$
{\rm
I}(p)=\mathbb{A}^{\dagger}(-p)\mathscr{C}^{-1}\mathbb{A}^{\dagger
t }(p)\, .
$$
This operator commutes with $\mathbb{P}$ and, when applied to the
vacuum, produces a state with zero momentum. We require this
operator to have trivial scattering with all operators
$A^\dagger$:
$$
{\rm I}_1(p_1)\,
\mathbb{A}_2^\dagger(p_2)=\mathbb{A}_2^\dagger(p_2)\, {\rm
I}_1(p_1)\, .
$$
This gives \bea \nonumber
&&\hspace{-0.4cm}\underbrace{\mathbb{A}_1^{\dagger}(-p_1)\mathscr{C}_1^{-1}\mathbb{A}_1^{\dagger
t }(p_1)}_{{\rm I}_1(p_1)}\, \mathbb{A}_2^\dagger(p_2)=
\mathbb{A}_1^{\dagger}(-p_1)\mathscr{C}_1^{-1}\Big[\mathbb{A}_1^{\dagger
}(p_1)\,
\mathbb{A}_2^\dagger(p_2)\Big]^{t_1}=\\
\nonumber
&&\hspace{-0.4cm}\mathbb{A}_1^{\dagger}(-p_1)\mathscr{C}_1^{-1}\Big[\mathbb{A}_2^{\dagger
}(p_2)\, \mathbb{A}_1^\dagger(p_1)S_{12}(p_1,p_2)\Big]^{t_1}=
\mathbb{A}_1^{\dagger}(-p_1)\mathbb{A}_2^{\dagger
}(p_2)\mathscr{C}_1^{-1}S_{12}^{t_1}(p_1,p_2)\mathbb{A}_1^{\dagger
t_1}(p_1) =\\
\nonumber && ~~~~~~~~~~~~~~~~~~~~~~~~=\mathbb{A}_2^{\dagger
}(p_2)\, \underbrace{
\mathbb{A}_1^{\dagger}(-p_1)S_{12}(-p_1,p_2)\mathscr{C}_1^{-1}S_{12}^{t_1}(p_1,p_2)\mathbb{A}_1^{\dagger
t_1}(p_1)}_{{\rm I}_1(p_1)}\, , \eea {\it i.e.} we must require
$$
S_{12}(-p_1,p_2)\mathscr{C}_1^{-1}S_{12}^{t_1}(p_1,p_2)=\mathscr{C}_1^{-1}\,
,
$$
which is equivalent to eqs.(\ref{cros}). The concrete form of the
matrix $\mathscr{C}$ will be found in subsection 3.4.2.

\subsubsection*{Summary}
We conclude this section by summarizing the basic physical
requirements for the scattering matrix:
\begin{itemize}
\item {\it Generalized Physical Unitarity}
\bea\nonumber S(p_1^*,p_2^*)^\dagger\cdot S (p_1,p_2)= \mI\, \eea
\item {\it Parity Invariance}
\bea\nonumber S(-p_1,-p_2)=S^{-1}(p_1,p_2)  \eea
\item {\it Time Reversal Invariance}  \bea\nonumber S(p_1,p_2)^t =
\mI^g S(p_1,p_2)\mI^g\, \eea
\item {\it Crossing Symmetry}
\bea\nonumber S^{c_1}(p_1,p_2)\, S(-p_1, p_2)= \mI\,,\quad
S^{c_2}(p_1,p_2)\, S(p_1, -p_2)= \mI\,. \eea
\end{itemize}
Some comments are in order. For real values of momenta the
S-matrix must be unitary. If the momenta are complex, usual
unitarity is replaced by the generalized unitarity condition
above, where $p^*$ stands for the complex conjugate momentum.  The
time reversal invariance condition presented here assumes parity
invariance and physical unitarity. Finally, the crossing symmetry
relates the anti-particle-to-particle scattering matrix $S^{c_1}$
to that of particle-to-particle and it holds for the properly
normalized $S$ only.

\section{Fundamental representation of \texorpdfstring{$\su(2|2)_\cex$}{su(2|2)cex}}

\def\V{{\mathscr V}}

In this section we will describe the fundamental representation of
the centrally extended superalgebra $\su(2|2)_\cex$. For the
reader's convenience, we repeat  the Lie algebra defining
relations (see section \ref{ceasu} for notations) \bea
\label{su22abstract}\begin{aligned} &
\left[\bL_a{}^b,\bJ_c\right]=\delta_c^b \bJ_a - {1\ov 2}\delta_a^b
\bJ_c\,,\qquad~~~~ \left[\bR_\a{}^\b,\bJ_\g\right]=\delta^\b_\g
\bJ_\a - {1\ov 2}\delta^\b_\a \bJ_\g\,,
  \\
& \left[\bL_a{}^b,\bJ^c\right]=-\delta_a^c \bJ^b + {1\ov
2}\delta_a^b \bJ^c\,,\qquad
~~\left[\bR_\a{}^\b,\bJ^\g\right]=-\delta_\a^\g \bJ^\b + {1\ov 2}\delta_\a^\b \bJ^\g\,,  \\
& \{ \bQ_\a{}^a, \bQ_b^\dagger{}^\b\} = \delta_b^a \bR_\a{}^\b +
\delta_\a^\b
\bL_b{}^a +{1\ov 2}\delta_b^a\delta^\b_\a  \bH\,,  \\
& \{ \bQ_\a{}^a, \bQ_\b{}^b\} = \epsilon_{\a\b}\epsilon^{ab}~\bC\,
, ~~~~~~~~~~~~~~ \{ \bQ_a^\dagger{}^\a, \bQ_b^\dagger{}^\b\} =
\epsilon_{ab}\epsilon^{\a\b}~\bC^{\dagger} \,. \end{aligned}\eea
In section \ref{ceasu} these relations have been derived by
studying the Poisson bracket of the Noether charges of the  string
sigma-model in the light-cone gauge. It was found there that upon
going off-shell  the algebra $\su(2|2)$ receives the central
extension by two central charges $\bC$ and $\bC^{\dagger}$. To
make our treatment more general, we will for a moment assume that
$\bC$ and $\bC^{\dagger}$ are independent.

\subsection{Matrix realization}
 Introduce a basis of the
four-dimensional fundamental representation
$$
|e_M\rangle=\left\{\begin{array}{ll}  |e_a\rangle \\
|e_{\a}\rangle\, .
\end{array}\right.
$$
Here $\gr_a=0$ for $a=1,2$ and $\gr_{\a}=1$ for $\a=3,4$.  On
these basis vectors the rotation generators of
(\ref{su22abstract}) are realized as \bea \la{rot}
\begin{aligned}
&\bL_a{}^b|e_c\rangle=\delta_c^b|e_{a}\rangle-\sfrac{1}{2}\delta_a^b|e_c\rangle
~~~~~&&~~~~~ \bR_\a{}^\b |e_a\rangle=0
\\
& \bL_a{}^b|e_{\a}\rangle=0 ~~~~~&&~~~~~\bR_\a{}^\b
|e_{\gamma}\rangle=
\delta_{\gamma}^{\beta}|e_{\a}\rangle-\sfrac{1}{2}\delta_{\a}^{\beta}|e_{\gamma}\rangle
\, .
\end{aligned}
 \eea
The  supersymmetry generators will then be  represented as \bea
\label{susyrep}
\begin{aligned}
 &\bQ_\a{}^a|e_b\rangle= a\, \delta_a^b|e_{\alpha}\rangle ~~~~~&&~~~~~   \bQ_a^\dagger{}^\a|e_b\rangle=
 c\, \eps_{ab}\eps^{\a\b}|e_{\beta}\rangle\\
 &\bQ_\a{}^a|e_\beta\rangle= b\,
 \epsilon_{\a\beta}\epsilon^{ab}|e_b\rangle ~~~~~&&~~~~~
\bQ_a^\dagger{}^\a|e_\beta\rangle= d\,
 \delta_{\b}^{\a}|e_a\rangle \, .
\end{aligned}
\eea Here $a,b,c,d$ are complex numbers parametrizing
 a fundamental irrep.
One can check that the algebra relations (\ref{su22abstract}) are
satisfied provided these numbers  satisfy the following relation
\bea ad-bc=1\, . \label{conabcd}\eea The values of the central
elements are found to be \bea\label{CC} {
\bH}|e_M\rangle=(ad+bc)|e_M\rangle\, ,
~~~~~~{\bC}|e_M\rangle=ab\,|e_M\rangle\, ,
~~~~~~\bC^{\dagger}|e_M\rangle=cd\,|e_M\rangle\, . \eea In
addition, if we require this representation to be unitary, then
the parameters have to satisfy the conditions
$$
d^*=a\, , ~~~~~~~~c^*=b\, .
$$
In unitary representations, $\bH$ is hermitian and $\bC$ is the
hermitian conjugate of $\bC^{\dagger}$.

It is convenient to combine the parameters describing the set of
fundamental unitary representations into the following matrix
$$
h=\left(\begin{array}{cc} a & b \\ c & d\end{array}\right)\, .
$$
Since this matrix obeys the relation $h^{\dagger}\rho h=\rho$,
where $\rho={\rm diag}(1,-1)$ and it has unit determinant, it can
be thought of as an element of the three-dimensional ${\rm
SU}(1,1)$ group. Not all the values of the central charges are
allowed, however. Indeed,  eqs.(\ref{conabcd}) and (\ref{CC})
imply that \bea H^2-4C\bar{C}=1\, . \label{relHC}\eea This is the
so-called shortening condition which defines an atypical (short)
multiplet of $\su(2|2)_{\mathcal C}$ of dimension four. Thus, the
space of central charges corresponding to atypical
four-dimensional multiplets is parametrized by one real variable,
which is $H$, and by the phase of $C$.

\noindent
\begin{figure}
\begin{minipage}{\textwidth}
\begin{center}
\includegraphics[width=0.4\textwidth]{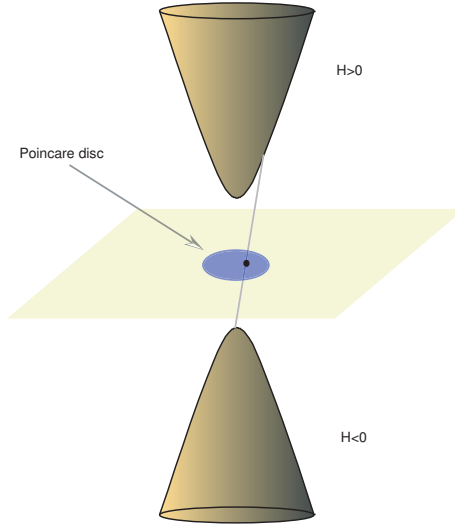}
\parbox{5in}{\caption{\label{poincare}Two branches of the dispersion relation corresponding to $H>0$ and $H<0$, respectively.
The Poincar\'e (blue) disk represents the stereographic projection
of an upper sheet on the complex plane through the origin.}}
\end{center}
\end{minipage}
\end{figure}

\vskip -0.7cm Any element of ${\rm SU}(1,1)$ gives rise to the
central charges $H$ and $C$ obeying eq.(\ref{relHC}). On the other
hand, given the charges (\ref{CC}) satisfying eq.(\ref{relHC}),
the representation parameters are not specified uniquely because a
${\rm U}(1)$ automorphism
$$
h\to \left(\begin{array}{cc} a\, e^{i\varphi} & b\, e^{-i\varphi}\\
c \, e^{i\varphi} & d \,  e^{-i\varphi}\end{array}\right)
$$
does not change the value of the charges and merely reflects a
choice of basis.  Factoring out this ${\rm U}(1)$ subgroup, we
obtain a two-sheeted hyperboloid ${\rm SU(1,1)}/{\rm U}(1)$ which
is described by eq.(\ref{relHC}). The upper sheet  $H>0$
corresponds to positive energy unitary representations,  while the
lower sheet corresponds to anti-unitary representations, see
Figure \ref{poincare}.

Finally, for the reader's convenience, we describe the
representation (\ref{rot}), (\ref{susyrep}) in terms of   $4\times
4$ matrix unities \bea\la{LRmatr}
\begin{aligned}
&L_1{}^2=E_1^{~2}\, ,   ~~~~~&&~~~~~   R_3{}^4=E_3^{~4}\, , \\
&L_2{}^1=E_2^{~1}\, ,   ~~~~~&&~~~~~   R_4{}^3=E_4^{~3} \, ,\\
&L_1{}^1=\sfrac{1}{2}(E_1^{~1}-E_2^{~2})=-L_2{}^2\, ,
~~~~~~&&~~~~~ R_3{}^3=\sfrac{1}{2}(E_3^{~3}-E_4^{~4})=-R_4{}^4\,
\end{aligned}
\eea and
\bea\la{QQdmatr}
\begin{array}{ll}
Q_3^{~1}=aE_3^{~1}+bE_2^{~4}\, ,  ~~~~~&~~~~~~
Q_1^{\dagger 3}=cE_4^{~2}+dE_1^{~3}\, , \\
Q_4^{~1}=aE_4^{~1}-bE_2^{~3}\, , ~~~~~&~~~~~~ Q_1^{\dagger
4}=-cE_3^{~2}+dE_1^{~4}\, ,
\\
Q_3^{~2}=aE_3^{~2}-bE_1^{~4}\, ,  ~~~~~&~~~~~~ Q_2^{\dagger
3}=-cE_4^{~1}+dE_2^{~3}\, ,
\\
Q_4^{~2}=aE_4^{~2}+bE_1^{~3} \, ,~~~~~&~~~~~~ Q_2^{\dagger
4}=cE_3^{~1}+dE_2^{~4}\, .
\end{array}\eea

\subsection{Outer automorphisms}
Over the complex field, the superalgebra $\su(2|2)_\cex$ admits a
group of outer automorphisms isomorphic to ${\rm SL}(2)$. This
group acts on the supercharges in the following way \bea\la{nq}
\begin{aligned}
&&\tbQ_\a{}^a =u_1\,  \bQ_\a{}^a -
u_2\,\epsilon^{ac}\,\bQ_c^\dagger{}^\g\, \epsilon_{\g\a}\,,\quad
\tbQ^\dagger_a{}^\a = v_1\, \bQ^\dagger_a{}^\a -
v_2\,\epsilon^{\a\b}\,\bQ_\b{}^b\, \epsilon_{ba}\,,\\
&&\bQ_\a{}^a =v_1\,  \tbQ_\a{}^a +
u_2\,\epsilon^{ac}\,\tbQ_c^\dagger{}^\g\, \epsilon_{\g\a}\,,\quad
\bQ^\dagger_a{}^\a = u_1\, \tbQ^\dagger_a{}^\a +
v_2\,\epsilon^{\a\b}\,\tbQ_\b{}^b\, \epsilon_{ba}\,,
\end{aligned}\eea where the coefficients $u_i,v_i$ satisfy the  condition
\bea u_1 v_1 - u_2 v_2 =1\, , \la{SL2}\eea which guarantees that
$\tbQ$ and $\tbQ^{\dagger}$ obey the same algebra relations
(\ref{su22abstract}) but with the new  central elements given by
 \bea \label{chargesn}
\begin{aligned}
&\tbH = (u_1v_1 + u_2 v_2) \bH +2u_1 v_2 \bC + 2 u_2 v_1\bC^\dagger\,,\\
&\tbC = u_1^2 \bC +u_2^2 \bC^\dagger + u_1u_2\bH \, ,\\
&\tbC^\dagger=v_1^2 \bC^\dagger +v_2^2 \bC + v_1v_2\bH  \,.
\end{aligned} \eea
The transformation parameters $u_i,v_i$ are combined into a
complex $2\times 2$-matrix
$$
\left(\begin{array}{rr} u_1 & u_2 \\
v_2 & v_1\end{array}\right)\, ,
$$
which, due to eq.(\ref{SL2}), has unit determinant. This
establishes an isomorphism of the outer automorphism group to
${\rm SL}(2)$. Restriction to the unitary representations of the
real form $\su(2|2)_\cex$ will require one to replace ${\rm
SL}(2)$ with its real form ${\rm SU(1,1)}$; the latter is defined
by imposing the following two conditions
$$
v_1^*=u_1\, , ~~~~v_2^*=u_2\, .
$$

Further, one can see that the action (\ref{chargesn}) leaves the
following combination of charges invariant
$$
\bH^2-4\bC\bC^{\dagger}\equiv \mathscr{R}^2\, .
$$
The invariant $\mathscr{R}^2$ classifies the orbits of ${\rm
SU}(1,1)$ in the space of central charges. They can be of three
types depending on the value of $\mathscr{R}^2$ -- a two-sheeted
hyperboloid for $\mathscr{R}^2>0$, a one-sheeted hyperboloid
$\mathscr{R}^2<0$ and a cone for $\mathscr{R}^2=0$. We are
interested in the $\mathscr{R}^2>0$ orbits only, because these
orbits  correspond to the positive and negative energy unitary
representations of $\su(2|2)_\cex$.

\smallskip

The outer automorphism group allows one to establish a connection
between the positive/negative energy (highest/lowest weight)
representations of $\su(2|2)_\cex$ and those of the usual
(non-extended) algebra $\su(2|2)$. Indeed, starting with an irrep
of $\su(2|2)_\cex$ characterized by some values of
$\bC,\bC^{\dagger}$, $\bH$ with $\mathscr{R}^2>0$ and choosing the
parameters $u_i,v_i$ appropriately, one can always make the
charges $\tbC$ and $\tbC^{\dagger}$  vanishing. Thus, the
transformed representation is the one for usual $\su(2|2)$ with
$\tbH$ equal to \bea\nonumber \tbH =\pm \sqrt{\bH^2-4\bC
\bC^\dagger}\, , \eea where the sign in front of the square root
correlates with the sign of $\bH$. The inverse statement is also
true: Any irreducible representation of the centrally-extended
algebra with $\mathscr{R}^2>0$ can be obtained from a
representation of the usual $\su(2|2)$ algebra with
$\bC=\bC^\dagger=0$.

\smallskip

Let us now describe in more detail the action of the outer
automorphism group on the fundamental irreps of $\su(2|2)_\cex$.
Under this action the matrix $h$  encoding the representation
parameters undergo the right shift
by an ${\rm SU(1,1)}$-matrix \bea h=\left(\begin{array}{cc} a & b \\
c & d\end{array}\right)
~~ \rightarrow ~~\left(\begin{array}{rr} u_1 & u_2 \\
v_2 & v_1\end{array}\right)  \left(\begin{array}{cc} a & b \\ c &
d\end{array}\right) \, .\eea According to the discussion above,
${\rm SU(1,1)}$ acts transitively on each sheet of the two-sheeted
hyperboloid $\mathscr{R}^2=1$. The tip of the upper sheet
corresponds to the special irrep with vanishing values for $\bC$
and $\bC^{\dagger}$, and $\bH$ equal to unity. This representation
is nothing else but the unique fundamental positive energy
representation of the non-extended algebra $\su(2|2)$.

\subsection{Parameterizations of $a,b,c,d$}

The parameters $a,b,c,d$ depend on the string  tension $g$ and the
world-sheet momenta $p$. To find their explicit dependence, we
take into account that the central charges are expressed via the
momentum $\bP$ by eq.(\ref{Cc}). Therefore, the parameters satisfy
the following relations \bea\la{abhcd}
ab={ig\ov2}\left(e^{ip}-1\right) e^{2i\xi}\,, \quad
cd={g\ov2i}\left(e^{-ip}-1\right) e^{-2i\xi}\,, \quad
H=ad+bc=2ad-1\,,~~~ \eea where $p$ is the value of the world-sheet
momentum $\bP$ on the representation.

The shortening condition (\ref{relHC}) implies that the energy
depends  on $p$ only and it leads to the following dispersion
relation for particles from the fundamental $\su(2|2)_{\mathcal
C}$ multiplet \bea H^2=1+ 4g^2\sin^2\frac{p}{2}\, . \label{disp}
\eea To simplify our further treatment, we assume that the
representation is unitary. In this case the parameters $g$, $p$,
$\xi$ are real, $H$ is positive, and  the eqs. (\ref{abhcd}),
(\ref{conabcd}) allow one to parametrize $a,b,c,d$ as
\bea\la{abcdp}\begin{aligned} a &=\eta\, e^{i\xi}\, e^{i\vp}\,,
\quad &b& =\frac{g}{2}\left(e^{ip}-1\right){i e^{i\xi}\ov\eta}\,
e^{-i\vp}\,,
\\
 d &=\eta\, e^{-{ip\ov2}}\, e^{-i\xi}\, e^{-i\vp}\,, \quad
 &c &=\frac{g}{2}\left(e^{-ip}-1\right)e^{{ip\ov2}}\,{e^{-i\xi}\ov i\eta}\, e^{i\vp}\,,
\end{aligned}
\eea where for unitary representations $\vp$ is an arbitrary real
number, and $\eta$ is expressed through the momentum $p$ and the
energy $H=\sqrt{1+ 4g^2\sin^2\frac{p}{2}}$ as follows
\bea\la{eta1} \eta = e^{ip\ov4}\, \sqrt{H+1\ov 2}\,. \eea In the
last formula the prefactor $e^{ip\ov4}$ may look rather
artificial. Nevertheless, it plays an important role in what
follows, in particular, its presence will make $\eta$  a
meromorphic function on the rapidity torus we introduce in the
next subsection.

The fundamental representation is completely determined by the
parameters $g,p,\xi$. The parameter $\vp$ just reflects a freedom
in the choice of the basis vectors $|e_M\rangle$, and in what
follows we set it to zero by proper rescaling of $|e_M\rangle$.
Then, formulae (\ref{abcdp}) render the parameters $a,b,c,d$ of
the fundamental representation as functions of the three
independent parameters  $g,p,\xi$.

\medskip

Another convenient parametrization is obtained by replacing the
momentum $p$ with two new parameters $x^+$, $x^-$.  They are
related to $p$ as \bea {x^+\ov x^-} = e^{ip}\,, \eea and satisfy
the constraint
 \bea\la{consxpxm} x^+ +{1\ov x^+}
-x^--{1\ov x^-}={2i\ov g}\,. \eea One can show that $a,b,c,d$ are
then expressed through $g$, $x^\pm$ and $\xi$ in the following way
(we set $\vp=0$)
 \bea\la{abcd}
a =\eta\, e^{i\xi}\,, \quad b =-\eta\,\, {e^{-{ip\ov2}}\ov x^-}\,
e^{i\xi}\,, \quad c =-\eta\, {e^{-i\xi}\ov x^+}\,, \quad d =\eta\,
e^{-{ip\ov2}}\,e^{-i\xi}\,,~~~ \eea where the parameter $\eta$ is
given by \bea\la{eta2} \eta = e^{ip\ov4}\, \sqrt{igx^- -
igx^+\ov2}\, . \eea With this parametrization we find that the
central charge $H$ is expressed as \bea\la{hc2} H &=& 1 + {ig\ov
x^+}- {ig\ov x^-} =  ig x^- - ig x^+ -1\, , \eea while the
remaining central charges take the form \bea C&=&{ig\ov 2}\left(
{x^+\ov x^-}-1\right)\,e^{2i\xi}\,,\quad
 \overline{C}={g\ov 2i}\left( {x^-\ov x^+}-1\right)\,e^{-2i\xi} \,.~~~~~
\eea
We will see that the S-matrix coefficients are conveniently expressed in terms of $x^\pm$.

In what follows we denote the fundamental representation as
$\V(p,\zeta)$ (or just $\V$ if the values of $p$ and $\zeta$
are not important), where $\zeta=e^{2i\xi}$.

\subsection{Rapidity torus}
\la{subsect: repidtorus}

\def\am{\mathrm{am}}
Here we would like to find an analogue of the rapidity variable
for the non-Lorentz invariant string sigma model and to understand
the action of crossing symmetry.

Our starting point is the dispersion relation (\ref{disp})  for
particles from the fundamental $\su(2|2)_{\mathcal C}$-multiplet.
This formula shows that the universal cover of the parameter space
describing the representation  is an elliptic curve. Indeed, the
eq.(\ref{disp}) can be naturally uniformized in terms of Jacobi
elliptic functions \bea\la{pez} p=2\,{\am\,z}\,,~~\quad~~
\sin{p\ov 2} = \sn(z,k)\,,~~\quad~~ H = \dn(z,k)\, , \eea where we
introduced the elliptic modulus\footnote{Our convention for the
elliptic modulus is the same as accepted in the {\it Mathematica}
program, {\it e.g.}, ${\rm sn}(z,k)={\rm JacobiSN}[z, k]$.
Throughout the paper we will often indicate only the
$z$-dependence of Jacobi elliptic functions if it cannot lead to
any confusion. } $k=-4g^2=-\frac{\lambda}{\pi^2}<0$. The
corresponding elliptic curve (the torus) has two periods
$2\omega_1$ and $2\omega_2$, the first one is real and the second
one is imaginary \bea\nonumber 2\omega_1=4{\rm K}(k)\, , ~~~~~~~~~
2\omega_2=4i{\rm K}(1-k)-4{\rm K}(k)\, ,
 \eea
where ${\rm K}(k)$ stands for the complete elliptic integral of
the first kind. The dispersion relation  is obviously invariant
under shifts of $z$ by $2\omega_1$ and $2\omega_2$. The torus
parametrized by the complex variable $z$ is the analog of the
rapidity plane in two-dimensional relativistic models.

\smallskip

In this parametrization the real $z$-axis can be called the
physical one for the original string theory, because for real
values of $z$ the energy is positive and the momentum is real due
to \bea\nonumber 1 \leq\dn(z,k)\leq \sqrt{k'}\, ,~~~~~~~~z\in
{\mathbb R}\, , \eea where $k'\equiv 1 -k$ is the complementary
modulus.
\smallskip

We further note that the representation parameters $x^{\pm}$ are expressed
in terms of Jacobi elliptic functions as \bea \la{xpxmz}
x^{\pm}=\frac{1}{2g}\Big(\frac{\cn z}{\sn z} \pm i \Big)(1+\dn z)
\, . \eea This form of $x^\pm$ follows from the requirement that
for real values of $z$ the absolute values of $x^\pm$ are greater
than unity $|x^\pm| > 1$, and the imaginary parts satisfy $\mbox{Im}(x^+)>0$  and
$\mbox{Im}(x^-)<0$.
\smallskip

\begin{table}[ht]
\begin{center}
\begin{tabular}{|c|c|c|}
\hline $z$ & $x^+$ & $x^-$ \\
\hline \hline $z+\om_1$ & $x^+$ & $x^-$ \\
\hline $z+\om_2$ & $1/x^+$ & $1/x^-$ \\
\hline $z+\sfrac{1}{2}(\om_1+\om_2)$ & $-1/x^-$ & $-x^+$ \\
\hline $z+(\om_1+\om_2)$ & $1/x^+$ & $1/x^-$ \\
\hline $z+\sfrac{3}{2}(\om_1+\om_2)$ & $-x^-$ & $-1/x^+$ \\
\hline $-z$ & $-x^-$ & $-x^+$ \\
\hline $-z+\sfrac{1}{2}(\om_1+\om_2)$ & $1/x^+$ & $x^-$ \\
\hline $-z+(\om_1+\om_2)$ & $-1/x^-$ & $-1/x^+$ \\
\hline $-z+\sfrac{3}{2}(\om_1+\om_2)$ & $x^+$ & $1/x^-$ \\ \hline
\end{tabular}\\
\vskip 0.5cm Table 1: Transformations of $x^{\pm}$ under some
shifts of $z$.
\end{center}
\end{table}
\vskip -0.2cm The transformation properties of the parameters
$x^{\pm}$ under shifts of $z$ by some fractions of the periods are
presented in the Table 1. Since both the dispersion relation and
$x^\pm$ are periodic with period $\om_1$, the range of  the real
part of $z$ can be restricted to the interval from $-\om_1/2$ to
$\om_1/2$ which corresponds to $-\pi\le p\le\pi$.

\smallskip

Now we analyze what happens to the torus in the limits
$g\to\infty$ and $g\to 0$. When $g\to\infty$ the periods exhibit
the following behavior \bea \om_1 \to {\log g\ov g}\,,\quad
\om_2\to {i\pi\ov 2g}\qquad \makebox{if}\quad g\to\infty\,. \eea
To keep the range of $\makebox{Im}( z)$ finite, we rescale $z$ as
$z\to z/(2g)$, and the momentum as $p\to p/g$. Then the dispersion
relation (\ref{disp}) acquires the standard relativistic form
(\ref{disprel}), the variable $z$ plays the role of rapidity
$\theta$ as $p = \sinh z$. As to the torus, it degenerates into a
strip with $-\pi<\makebox{Im}( z)<\pi$ and $-\infty<\makebox{Re}(
z)<\infty$. This is twice the physical strip of a relativistic
field theory.

\smallskip

In the limit $g\to 0$ the periods of the torus have the following
behavior \bea\label{gtozero} \om_1 \to \pi\,,\quad \om_2\to 2i\log
g\qquad \makebox{if}\quad g\to 0\,. \eea Thus, the torus
degenerates into the strip with $-\pi/2<\makebox{Re}( z)<\pi/2$
and $-\infty<\makebox{Im}( z)<\infty$. The limit $g\to 0$
corresponds to the one-loop gauge theory.

An important property of our parametrization of  the fundamental
representation (\ref{abcd}) is that if the parameter $e^{i\xi}$ is
a meromorphic function on the torus then all the parameters
$a,b,c,d$ are meromorphic functions as well. To show this,
 one has to resolve the branch cut ambiguities arising from the parameter
$\eta$  (\ref{eta2}).

\smallskip

This can be done in the following way. First,  the
elliptic parametrization (\ref{xpxmz}) gives
 \bea
\eta(p)=e^{\frac{i}{4}p}\, \sqrt{igx^-(p)-igx^+(p)\ov2}&=&
\frac{1}{\sqrt{2}}e^{\frac{i}{2}{\rm am}\, z}\,\sqrt{1+{\rm dn}\,
z}\,
 = \nonumber \\
&&~=\frac{1}{\sqrt{2}}\sqrt{(1+{\rm dn}\, z)({\rm
cn}\, z+i {\rm sn}\, z)}\, .\eea Second, by using the following
formulae  (recall $k = -4g^2$)
$$
1+{\rm dn}\, z=\frac{2\, {\rm dn}^2\, \frac{z}{2}}{1+4g^2\, {{\rm
sn}^4\frac{z}{2}}}\, ,\qquad {\rm cn}\, z+i\, {\rm sn}\,
z=\frac{\big({\rm cn}\, \frac{z}{2}+i \, {\rm sn}\,
\frac{z}{2}{\rm dn}\, \frac{z}{2}\big)^2}{1+4g^2\, {{\rm
sn}^4\frac{z}{2}}}\,,
$$
relating elliptic functions to those of the half argument, we can
resolve the branch cut ambiguities by means of the relation
\bea\la{etaz} e^{\frac{i}{4}p}\, \sqrt{igx^-(p)-igx^+(p)\ov2}=
\frac{{\rm dn}\, \frac{z}{2}\big({\rm cn}\, \frac{z}{2}+i \, {\rm
sn}\, \frac{z}{2}{\rm dn}\, \frac{z}{2}\big)}{1+4g^2\, {{\rm
sn}^4\frac{z}{2}}}\equiv \eta(z) \eea valid in the region
$-\frac{\om_1}{2}<{\rm Re}\, z<\frac{\om_1}{2}$ and $-\om_2/i<{\rm
Im}\, z<\om_2/i$. Finally, we notice that since
$e^{-\frac{i}{2}p}={\rm cn}\, z- i\, {\rm sn}\, z$, and  $x^\pm$
are meromorphic functions,  then the representation parameters
$a,b,c,d$ are meromorphic as well. This property greatly
facilitates the treatment of crossing symmetry.

\section{The \texorpdfstring{$\su(2|2)$}{su(2|2)}-invariant S-matrix}

Since the manifest symmetry algebra of the light-cone string
theory on $\AdS$ consists of two copies of the centrally-extended
$\su(2|2)$, the creation operators $A^\dagger_{M\dM}(p)$ carry two
indices $M$ and $\dot{M}$, where the dotted index is for the
second $\su(2|2)$.  The $n$-particle states are obtained by acting
with the creation operators on the vacuum
\begin{equation}
A^\dagger_{M_1\dM_1}(p_1)\cdots  A^\dagger_{M_n\dM_n}(p_n)|\Omega \rangle \equiv |A^\dagger_{M_1\dM_1}(p_1)\cdots  A^\dagger_{M_n\dM_n}(p_n) \rangle \,.
\end{equation}
For the purpose of this section we can think of  $A^\dagger_{M\dM}(p)$ as being a product of two creation operators $A^\dagger_{M\dM}(p)=A^\dagger_{M}(p)\times A^\dagger_{\dM}(p)$ and restrict our attention to the states created by $A^\dagger_{M}(p)$.

\subsection{Two-particle states and the S-matrix}

It is clear that a one-particle state $|A^\dagger_{M}(p)\rangle$
is identified with  the basis vector $|e_M\rangle$ of the
fundamental representation $\V(p,1)$ of $\su(2|2)_\cex$ (and we
also set $\vp=0$). Let us stress that we have to set the parameter
$\zeta$ to 1, because we use the canonical form of the central
charge $\bC$ with $\xi=0$ \bea\la{Ccc} \bC={ig\ov
2}\,(e^{i\hP}-1)\,,\quad \bC|A^\dagger_{M}(p)\rangle = {ig\ov
2}\,(e^{ip}-1) |A^\dagger_{M}(p)\rangle\, . \eea

Then the two-particle states created by $A^\dagger_{M}(p)$  should
be identified with the tensor product of fundamental
representations of $\su(2|2)_\cex$
 \bea\la{tp}
|A^\dagger_{M_1}(p_1) A^\dagger_{M_2}(p_2) \rangle \sim
\V(p_1,\zeta_1)\otimes\V(p_2,\zeta_2)\,, \eea
 equipped with the
canonical action of the symmetry generators in the tensor product.
An important observation is that the parameters $\zeta_k$ cannot
be equal to 1. The reason for that is very simple. Computing the
central charge $\bC$ on the two-particle state, we get
\bea\la{Ccn2} \bC |A^\dagger_{M_1}(p_1) A^\dagger_{M_2}(p_2)
\rangle= {ig\ov 2}(e^{i(p_1+ p_2)}-1)|A^\dagger_{M_1}(p_1)
A^\dagger_{M_2}(p_2) \rangle\, , \eea because $\hP
A^\dagger_{M}(p) = A^\dagger_{M}(p) ( \hP + p)$. On the other
hand, the value of the central charge on the tensor product of
fundamental representations is equal to the sum of their charges
\bea \bC\, \V(p_1,\zeta_1)\otimes \V(p_2,\zeta_2) = {ig\ov
2}\left( \zeta_1(e^{ip_1}-1) + \zeta_2(e^{ip_2}-1) \right)
\V(p_1,\zeta_1)\otimes  \V(p_2,\zeta_2)\,.~~~ \nonumber\eea Thus,
we must have the following identity \bea\la{iden2} e^{i(p_1+
p_2)}-1 = \zeta_1(e^{ip_1}-1) + \zeta_2(e^{ip_2}-1)\, ,\eea which
obviously cannot be satisfied if both $\zeta_1$  and $\zeta_2$ are
equal to 1. In fact, it is easy to show that there are only two
solutions to this equation for $\zeta_k$ lying on the unit circle
\bea\la{z1z2}
 \{
\zeta_1 = 1\,,\  \zeta_2=e^{ip_1}\}\,,\quad {\rm or} \quad
\{\zeta_1 = e^{ip_2}\,,\  \zeta_2=1\}\,. \eea A priori any of
these two solutions can be used  to identify a two-particle state
with the tensor product. However, the form of the S-matrix depends
on the identification, and, as we will see shortly, it is the
first solution that leads to the S-matrix which precisely agrees
with the perturbative S-matrix discussed in the previous chapter.

It is readily seen that the first solution corresponds to the
following rearrangement of the commutation relation of the central
charge $\bC$ with $A^\dagger_M(p)$ \bea\la{ca1} \bC\,
A^\dagger_M(p)= C(p)A^\dagger_M(p) +e^{ip}A^\dagger_M(p)\,\bC\,,
\eea while the second solution corresponds
 to another rearrangement of
the commutation relation \bea\la{ca2} \bC\, A^\dagger_M(p)=
C(p)A^\dagger_M(p)\,e^{i\hP} + A^\dagger_M(p)\,\bC\, . \eea The
latter has an explicit dependence on the operator of the
world-sheet momentum.

Thus, taking the first solution in (\ref{z1z2}), we see that the
invariance condition (\ref{incon2}) takes the following form for
bosonic generators $L_a{}^b$ and $R_\a{}^\b$ \bea
S_{12}(p_1,p_2)\big(J \otimes \mI + \mI \otimes J\big)=\big( J
\otimes \mI + \mI \otimes J\big) S_{12}(p_1,p_2)\,,
\la{incondb}\eea and for fermionic generators $Q_\a{}^a$  and
$Q_a^\dagger{}^\a$ \bea\nonumber
&&\hskip-0.2cm   S_{12}(p_1,p_2)\big(J(p_1;1)\otimes \mI + \Sigma\otimes J(p_2;e^{ip_1})\big)=\\
&&~~~~~~~~~~~~~~~~~~~~~~~\big( J(p_1;e^{ip_2})\otimes \Sigma +
\mI\otimes J(p_2;1) \big) S_{12}(p_1,p_2)\,,~~~\la{incondf} \eea
 where $J(p;\zeta)$
denote the structure constants matrices of the fundamental
representation parametrized by $g$, $p$ and  $\zeta = e^{2i\xi}$,
see (\ref{QQdmatr}) and (\ref{abcd}). The grading  matrix \bea
\Sigma = {\rm diag} (1,1,-1,-1) \la{GM} \eea defined in
eq.(\ref{grmatrix}) takes care of the negative sign for fermions.
These are the conditions that should be used to find the S-matrix.
With the choice of the representation parameters we made, the
resulting S-matrix satisfies the Yang-Baxter equation.

\medskip

We note that if we think of vectors from $\V(p;\zeta)$ as columns
then, as is seen from eq.(\ref{incondf}), the S-matrix can be
considered as a map \bea S_{12}(p_1,p_2): ~~~~~\V(p_1,1)\otimes
\V(p_2,e^{ip_1})\to  \V(p_1,e^{ip_2})\otimes \V(p_2,1)\,, \eea and
if we think of vectors from $\V(p;\zeta)$ as rows then the
S-matrix can be regarded as the opposite map \bea S_{12}(p_1,p_2):
~~~~~\V(p_1,e^{ip_2})\otimes \V(p_2,1)\to \V(p_1,1)\otimes
\V(p_2,e^{ip_1})
 \,.  \eea From this point of view the
action of the S-matrix corresponds to exchanging the two possible
choices of the parameters $\zeta_k$ of the two representations.
Let us stress, however, that no matter what interpretation we use,
$S_{12}(p_1,p_2)$ is a $16\times 16$ matrix acting in the
16-dimensional vector space of the two-particle states
$|A^\dagger_{M_2}(p_2) A^\dagger_{M_1}(p_1) \rangle$.

When the string coupling constant $g$ tends to infinity, the
string sigma-model becomes free, and the ZF creation operators
turn into the usual creation operators, {\it i.e.} commute or
anti-commute depending on the statistics. Therefore, in this limit
the S-matrix should be equal to the graded unity.

The S-matrix satisfying eqs.(\ref{incondb}) and (\ref{incondf})
can be easily found up to an overall scalar factor. In the
following we will give up the particle momenta $p_i=2\,{\rm am}\,
z_i$ in favor of the rapidity variables $z_i$. The invariance
condition (\ref{incondb}) that involves the bosonic algebra
generators fixes the form of the S-matrix up to ten arbitrary
coefficients \bea\la{SmatrAA}
S(z_1,z_2)=\sum_{k=1}^{10}a_k\Lambda_k \, ,\eea where
$\Lambda_1,\ldots \Lambda_{10}$ form a basis of
$\su(2)\otimes\su(2)$ invariant matrices acting in the tensor
product $\V(z_1)\otimes \V(z_2)$

\bea\nonumber~\qquad \qquad \Lambda_1 &=& E_{1111}+\frac{1}{2
   }
   E_{1122}+\frac{1
   }{2}
   E_{1221}+\frac{1
   }{2}
   E_{2112}+\frac{1
   }{2}
   E_{2211}+E_{2222}\, ,\hspace{13cm}
   \eea

 \vspace{-0.8cm}

\bea
\nonumber~\qquad \qquad
\Lambda_2 &=& \frac{1}{2}
   E_{1122}-\frac{1
   }{2}
   E_{1221}-\frac{1
   }{2}
   E_{2112}+\frac{1
   }{2} E_{2211}\, ,\hspace{13cm}
\eea

 \vspace{-0.8cm}

\bea
\nonumber~\qquad \qquad
\Lambda_3 &=& E_{3333}+\frac{1}{
   2}
   E_{3344}+\frac{1
   }{2}
   E_{3443}+\frac{1
   }{2}
   E_{4334}+\frac{1
   }{2}
   E_{4433}+E_{4444}\, ,\hspace{13cm}
\nonumber
\eea

 \vspace{-0.8cm}

\bea \nonumber~\qquad \qquad  \Lambda_4 &=& \frac{1}{2}
   E_{3344}-\frac{1
   }{2}
   E_{3443}-\frac{1
   }{2}
   E_{4334}+\frac{1
   }{2} E_{4433}\, ,\hspace{13cm}
\eea

 \vspace{-0.8cm}

\bea
\nonumber~\qquad \qquad
\Lambda_5 &=& E_{1133}+E_{1144}+E_{2233}+E_{224
   4}\, ,\hspace{13cm}
\eea

 \vspace{-0.8cm}

\bea
\nonumber~\qquad \qquad
\Lambda_6 &=& E_{3311}+E_{3322}+E_{4411}+E_{442
   2}\, ,\hspace{13cm}
\eea

 \vspace{-0.8cm}

\bea
\nonumber~\qquad \qquad
\Lambda_7 &=& E_{1324}-E_{1423}-E_{2
   314}+E_{2413}\, ,\hspace{13cm}
\eea

 \vspace{-0.8cm}

\bea \nonumber~\qquad \qquad \Lambda_8 &=&
E_{3142}-E_{3241}-E_{4132}+E_{4231}\, ,\hspace{13cm} \eea

 \vspace{-0.8cm}

\bea \nonumber~\qquad \qquad \Lambda_9 &=&
E_{1331}+E_{1441}+E_{2332}+E_{2442}\, ,\hspace{13cm} \eea

 \vspace{-0.8cm}

\bea \nonumber~\qquad \qquad \Lambda_{10} &=&
E_{3113}+E_{3223}+E_{4114}+E_{4224}\, . \hspace{13cm} \eea
 Here the symbols $E_{kilj}$ are equal to
$(-1)^{\epsilon_k\epsilon_l}\, E_k{}^i \otimes E_l{}^j$, where
$E_k{}^i \equiv  E_{ki}$ are the standard $4\times 4$ matrix
unities\footnote{Choosing $E_{kilj}\equiv E_k{}^i \otimes E_l{}^j$
will produce the corresponding graded S-matrix $S^g$.}. The
normalization of $\Lambda_i$ has been chosen in such a way that
for $a_1=a_2=\ldots =a_6=1$ and $a_7=a_8=a_9=a_{10}=0$ the matrix
$S$ coincides with the graded identity.

The unknown coefficients $a_k$ can be now determined from the
permutation relations of the S-matrix with the supersymmetry
generators. We find
 \bea
\nonumber ~\qquad\qquad  a_1=1\, ,\hspace{13cm} \eea

 \vspace{-0.6cm}

\bea
\nonumber
~\qquad\qquad
a_2=2\,\frac{(x^-_1-x^-_2) (x^+_1
   x^-_2-1)x^+_2}{(x^-_1 - x^+_2)(x^+_1 x^+_2-1)
    x^-_2}-1\,,\hspace{13cm}
 \eea

 \vspace{-0.6cm}

\bea
\nonumber
~\qquad\qquad
a_3=\frac{x^+_1-x^-_2}{x^-_1-x^+_2}\,
\frac{\tilde{\eta}_1
  \tilde{\eta}_2}{ \eta_1\eta_2} ,\hspace{13cm}
  \eea

 \vspace{-0.6cm}

\bea
\nonumber
~\qquad \qquad
 a_4=-\frac{x^+_1-x^-_2}{x^-_1-x^+_2}\,\frac{\tilde{\eta}_1
  \tilde{\eta}_2}{ \eta_1\eta_2} + 2\,\frac{
    (x^-_1-x^-_2)(x^-_1 x^+_2-1)
  x^+_1 }{ (x^-_1-x^+_2)(x^+_1 x^+_2-1)
   x^-_1} \frac{\tilde{\eta}_1
  \tilde{\eta}_2}{ \eta_1\eta_2} ,\hspace{13cm}
  \eea

 \vspace{-0.6cm}

\bea
\nonumber
~\qquad \qquad
a_5=\frac{x^+_1-x^+_2}{x^-_1-x^+_2 } \frac{\tilde{\eta}_2}{\eta _2}  ,\hspace{13cm}
\eea

 \vspace{-0.6cm}

\bea
\nonumber
~\qquad \qquad
a_6=\frac{x^-_1-x^-_2}{x^-_1-x^+_2}\frac{\tilde{\eta}_1}{\eta_1},\hspace{13cm}
 \eea

 \vspace{-0.6cm}

 \bea
 \nonumber
~\qquad \qquad
 a_7 = \frac{g}{2i}\frac{(x^-_1-x^+_1)
   (x^+_1-x^+_2)
   (x^-_2-x^+_2)}{(x^-_1-x^+_2)(x^-_1
   x^-_2-1)  }\frac{1}{\eta
   _1 \eta _2},\hspace{13cm}
\eea

 \vspace{-0.6cm}

\bea \nonumber ~\qquad \qquad a_8 =\frac{2i}{g}\frac{
   (x^-_1-x^-_2)}{  (x^-_1-x^+_2)(x^+_1 x^+_2-1)
 }\tilde{\eta}_1\tilde{\eta}_2,\hspace{13cm} \eea

 \vspace{-0.6cm}

\bea
\nonumber
~\qquad \qquad
a_9 =\frac{x^-_1-x^+_1}{x^-_1-x^+_2}\frac{\tilde{\eta}_2}{\eta_1},\hspace{13cm} \eea

 \vspace{-0.6cm}

\bea \nonumber ~\qquad \qquad
a_{10}=\frac{x^-_2-x^+_2}{x^-_1-x^+_2}\frac{\tilde{\eta}_1}{\eta_2}\,
. \hspace{13cm}
 \eea
The coefficients $a_k$ are determined up to an overall scaling
factor, and we normalize them in a canonical way by setting
$a_1=1$.  The parameters $\eta_k$ are not fixed by the invariance
condition. They are determined by imposing the generalized
unitarity condition and the Yang-Baxter equation, and are given by
the following formulae \bea\la{etak} \begin{aligned}&\eta_1 =
\eta(z_1)\, , &~~~~& \tilde{\eta}_1 =(\cn z_2+i\sn z_2)\eta(z_1) \, \\
&\tilde{\eta}_2 = \eta(z_2)\, , &~~~~&\tilde{\eta}_2 =(\cn
z_1+i\sn z_1)\eta(z_2) \,,
\end{aligned}\eea where $\eta(z)$ is defined by (\ref{etaz}).

An important property of the S-matrix (\ref{SmatrAA})  is that up
to the scalar factor it is a meromorphic function of the torus
variables $z_1,z_2$ because the parameters of all the four
representations appearing in the invariance condition
(\ref{incondf}) are meromorphic. In what follows we often refer to
the S-matrix  (\ref{SmatrAA}) with the coefficients $a_i$ given
above as to the canonical $\su(2|2)$-invariant fundamental
S-matrix.

The canonical S-matrix (\ref{SmatrAA}) satisfies all the
properties we discussed in subsection \ref{gpr}. First,  the
physical unitarity condition $S(z_1,z_2)^\dagger\cdot S(z_1,z_2) =
\mI$ for $z_1,z_2$ real can be easily checked by using the
explicit form of the coefficients $a_i$, and the hermitian
conjugation and transposition conditions \bea\nonumber
\left(\Lambda_i\right)^\dagger =
\left(\Lambda_i\right)^t=\Lambda_i\,,\quad i=1,\ldots ,6\,;\qquad
\left(\Lambda_7\right)^\dagger =
\left(\Lambda_7\right)^t=-\Lambda_8\,,\ \
\left(\Lambda_9\right)^\dagger =
\left(\Lambda_9\right)^t=\Lambda_{10}\,. \eea Moreover, with the
choice (\ref{etak}) of $\eta_i$, the S-matrix also satisfies the
generalized unitarity condition $ S(z_1^*,z_2^*)^\dagger\cdot S
(z_1,z_2)= \mI\,, $ and it is also a graded-symmetric matrix $
S^t(z_1,z_2)=\mI^g S(z_1,z_2)\mI^g$. The latter property implies
that the coefficients $a_i$ satisfy the following relations
\bea\la{rel112} &&a_7(z_1,z_2) = a_8(z_1,z_2)\,,\quad
a_9(z_1,z_2)=a_{10}(z_1,z_2)\,,~~~~~~~~ \eea which, in fact,
reduces the number of independent coefficients to seven.

If $p_1=p_2$ or, equivalently, $z_1=z_2$ the canonical S-matrix
becomes the permutation matrix. As will become clear in the next
section, the world-sheet S-matrix reduces at this special point to
minus the permutation due to the scalar factor which tends to -1.

We further note that the form of the structure constants matrices
$J(p;\zeta)$ allows one to
 determine the commutation relations  of the symmetry operators with the creation and annihilation operators.
 It is convenient to use the matrix notations, {\it i.e.} to combine $A^\dagger_M$ and $A_M$ into a row and column, respectively, and the symmetry algebra structure constants  of the one-particle representation (\ref{abcd}) with $\xi=0$ into matrices $L_a{}^b$, $R_\a{}^\b$, $Q_\a{}^a$ and $Q_a^\dagger{}^\a$, see (\ref{LRmatr}).
 Then,  the commutation relations for the centrally-extended algebra $\su(2|2)_{\cal C}$  can be
written in the following simple form of the braided
(anti)-commutators \bea\la{JAs}\begin{aligned} \bL_a{}^b\,
\bA^\dagger(p) - \bA^\dagger(p)\,\bL_a{}^b =&\,
\bA^\dagger(p)\,L_a{}^b \,,~~~~\\ \bR_\a{}^\b \, \bA^\dagger(p)-
\bA^\dagger(p)\,\bR_\a{}^\b
=&\,\bA^\dagger(p)\,R_\a{}^\b \,,\\
  \bQ_\a{}^a \, \bA^\dagger(p)-e^{ip/2}\bA^\dagger(p)\,\Sigma\, \bQ_\a{}^a
=&\,\bA^\dagger(p)\,Q_\a{}^a(p)\,,\\ \bQ_a^\dagger{}^\a \,
\bA^\dagger(p)- e^{-ip/2}\bA^\dagger(p)\,\Sigma\,
\bQ_a^\dagger{}^\a =&\,\bA^\dagger(p)\, Q_a^\dagger{}^\a(p)\,.
\end{aligned}\eea Thus, the braiding factors in (\ref{JAs}) are
the exponents $e^{\pm ip/2}$. This form of the commutation
relations is the one that usually appears in models with non-local
charges.

It is worthwhile to notice that  the form of the two-particle
structure constant matrices appearing in the invariance condition
(\ref{incondf}) allows us to reformulate the problem by using the
Hopf algebra language, see appendix (\ref{hopf}) for detail.

\subsection{Multi-particle states}
Multi-particle states created by $A^\dagger_{M}(p)$  are
correspondingly identified with the tensor product of fundamental
representations of $\su(2|2)_{\cal C}$ \bea\la{tpm}
|A^\dagger_{M_1}(p_1)\cdots  A^\dagger_{M_n}(p_n) \rangle \sim
\V(p_1,\zeta_1)\otimes \cdots \otimes \V(p_n,\zeta_n)\,, \eea
equipped with the canonical action of the symmetry generators in
the tensor product, and the parameters  $\zeta_k$ have to satisfy
the following identity \bea\la{iden} e^{i(p_1+\cdots + p_n)}-1 =
\sum_{k=1}^n\zeta_k(e^{ip_k}-1)\,. \eea In general, there are many
different solutions to this equation. In our case, however, the
choice of $\zeta_k$ is fixed by the commutation relations
(\ref{JAs}) and (\ref{ca1}). One can easily see that the only
solution compatible with (\ref{JAs}) is
 \bea
\zeta_1=1\,,\quad \zeta_2=e^{ip_1}\,,\quad \ldots\,, \quad
\zeta_{n-1}=e^{i(p_1+\cdots +p_{n-2})}\,,\quad
\zeta_n=e^{i(p_1+\cdots +p_{n-1})}\,.~~~ \eea
It is clear that the multi-particle
S-matrix just maps the vector space with this choice of
$\zeta_k$ to the (isomorphic) space with the following choice of
$\zeta_k$
\bea
\zeta_1=e^{i(p_2+\cdots +p_n)}\,,\quad \zeta_2=e^{i(p_3+\cdots
+p_n)}\,,\quad \ldots\,, \quad \zeta_{n-1}=e^{ip_n}\,,\quad
\zeta_n=1\,,~~~ \eea
because the second choice obviously corresponds to the order of the ZF creation operators in the $out$-state.

Due to the integrability of the model the multi-particle S-matrix
factorizes into a product of two-particle ones, and the
consistency condition for the factorizability  is equivalent to
the Yang-Baxter equation.

\section{Crossing symmetry}

\subsection{World-sheet S-matrix and dressing phase}

The canonical $\su(2|2)$-invariant fundamental S-matrix  can be
used to find the  corresponding $\su(2|2)\oplus\su(2|2)$ invariant
world-sheet S-matrix which describes the scattering of fundamental
particles in the light-cone string sigma model. To this end, one
should multiply the tensor product of two copies of the canonical
S-matrix by a scalar factor so that the resulting matrix would
satisfy an equation imposed by crossing symmetry. Thus, the
world-sheet S-matrix describing the scattering of two fundamental
particles is of the form \bea\la{saa} {\cal S} (z_1,z_2) =
S_0(z_1,z_2)\, S(z_1,z_2)\, \check\otimes \, S(z_1,z_2)\,. \eea
The tensor product in (\ref{saa}) is unusual and it takes care of
various signs which arise due to factorization of the ZF creation
operators. For the graded S-matrix these signs were determined in
subsection \ref{sec:Sfact}, see eq. (\ref{Sfact}). Taking into
account that the graded S-matrix is equal to the product of  the
graded identity and the S-matrix (\ref{saa}), one finds the
indexed version of eq.(\ref{saa}) \bea\la{Saafact} S_{M\dM,
N\dN}^{P\dP,Q\dQ} (z_1,z_2)= (-1)^{\eps_{\dM}\eps_N +
\eps_{P}\eps_{\dQ}}\,S_0(z_1,z_2)\,S_{MN}^{PQ}(z_1,z_2)
\dot{S}_{\dM\dN}^{\dP\dQ}(z_1,z_2)\,. \eea

Since $\Lambda_1$ is the only $\su(2)\oplus\su(2)$ invariant
matrix which contains the term \mbox{$E_1{}^1\otimes E_1{}^1$},
the S-matrix component
$$ S_{1\dot{1}, 1\dot{1}}^{1\dot{1},1\dot{1}}  (z_1,z_2)  \equiv S_{\su(2)}  (z_1,z_2)=
S_0(z_1,z_2)\,a_1(z_1,z_2)^2
$$
describes the scattering of particles in the $\su(2)$ sector of
the theory.  Since we have set $a_1$ equal to unity, the scalar
factor $S_0$ in eq.(\ref{saa}) is simply equal to the S-matrix of
the $\su(2)$ sector \bea\nonumber S_0(z_1,z_2) =
S_{\su(2)}(z_1,z_2)\,. \eea Thus, assuming integrability and
preservation of classical symmetries at the quantum level, we
conclude that the S-matrix in the $\su(2)$ sector encodes the full
dynamics of the model as its form cannot be fixed by kinematical
symmetries. This S-matrix was determined by using various indirect
arguments involving both string and gauge theory considerations
which will be discussed in Part II of the review. Here we will
present the resulting expression
\bea S_{\su(2)}(z_1,z_2) = e^{i
a(p_2 \om_1-p_1 \om_2)}\,{x_1^+\ov x_1^-} {x_2^-\ov x_2^+}\,
\frac{1}{\s(x_1^\pm,x_2^\pm)^2}\, \frac{u_1-u_2-\frac{2i
}{g}}{u_1-u_2+\frac{2i}{g}}\,.\quad\la{ssu2} \eea In
eq.(\ref{ssu2})  the spectral parameters $u_k$ are expressed in
terms of $x_k^\pm$ as follows \bea\nonumber u_k= {1\ov
2}\left(x^+_k +{1\ov x_k^+} +x_k^- +{1\ov x_k^-} \right)\,, \eea
and in terms of the $u$-parameters the last term in  (\ref{ssu2})
is nothing else but the S-matrix of the Heisenberg spin chain. It
exhibits a pole at $u_1-u_2=-\frac{2i}{g}$ which corresponds to a
bound state of two fundamental particles from the $\su(2)$ sector,
as we will show in Part II.

The first factor in (\ref{ssu2}) depends on $a$ which is the
parameter of the uniform light-cone gauge (\ref{ulc}), and
\bea\om_i = \sqrt{1+4g^2\sin^2{p_i\ov 2}}\nonumber\eea is the
energy of the $i$-th particle. Under crossing both $p$ and
$\omega$ change sign and, as a consequence, the gauge-dependent
factor solves the homogeneous crossing equation. Without loss of
generality, in what follows we set $a=0$.

\smallskip
The gauge-independent function $\s(x^\pm_1,x^\pm_2)$ is called the
dressing factor, and it is often written in the exponential form
 $\s(x^\pm_1,x^\pm_2) =
e^{i\theta(x^\pm_1,x^\pm_2)}$.  Here
 the dressing phase
 \bea
\theta(x_1^+,x_1^-,x_2^+,x_2^-) = \sum_{r=2}^\infty
\sum_{\textstyle\atopfrac{s>r}{ r+s={\rm odd}}
}^\infty c_{r,s}(g)\Big[
q_r(x_1^\pm)q_{s}(x_2^\pm) - q_r(x_2^\pm)q_{s}(x_1^\pm)
\Big]\, \la{phase}
 \eea
  is a two-form on the vector
space of conserved charges $q_r(x^\pm)$ \bea \la{localconcharge}
q_r(x_k^-,x_k^+) &=& {i\ov r - 1}\left[ \left({1\ov
x^+_k}\right)^{r - 1} - \left({1\ov x^-_k}\right)^{r - 1}\right]\,
. \eea The coefficients $c_{r,s}(g)$ are nontrivial real functions
of the string tension and they admit an asymptotic large $g$
expansion \bea c_{r,s}(g) = g\, \sum_{n=0}^\infty {1\ov g^{n}}\,
c_{r,s}^{(n)} \,,\quad g\gg 1\,, \la{phexp}\eea where the
numerical coefficients $c_{r,s}^{(n)}$ can be determined from
string sigma model perturbative computations.  The leading order
coefficients $c_{r,s}^{(0)}$ and the functional form (\ref{phase})
of the dressing phase were found by discretizing the finite-gap
integral equations which describe the spectrum of classical
spinning strings. The result is \bea c_{r,s}^{(0)}= {1\ov 2}
\de_{r+1,s}\,. \la{afsc}\eea

The leading coefficients (\ref{afsc}) are already enough to relate
the exact world-sheet S-matrix we discuss here and the tree-level
S-matrix computed in chapter 2. First, we construct the graded
version of the exact S-matrix: ${\cal S}^g(p_1,p_2)=\mI^g{\cal
S}(p_1,p_2)$. Second, we rescale the particle momenta $p_i\to
p_i/g$ and take the limit $g\to \infty$. One then finds that the
leading term in the strong coupling expansion of the exact
(graded) world-sheet S-matrix is the identity, while the
subleading one reproduces precisely the perturbative S-matrix of
chapter 2.

Returning to the discussion of the dressing phase, the subleading
coefficients in (\ref{phexp}) were fixed by analyzing the one-loop
corrections to energies of circular spinning strings, and turn out
to be \bea c_{r,s}^{(1)}= -{2\ov \pi}{(r-1)(s-1)\ov
(r+s-2)(s-r)}\,, \quad r+s={\rm odd}\,. \eea The requirement that
the dressing factor satisfies the crossing symmetry equations
leads to the following proposal for all the coefficients
$c_{r,s}^{(n)}$ \bea \la{phasestrong}
c_{r,s}^{(n)}=\frac{(-1)^n\zeta(n)}{2\, \pi^n \Gamma(n-1)
}(r-1)(s-1)
\frac{\Gamma[\frac{1}{2}(s+r+n-3)]\Gamma[\frac{1}{2}(s-r+n-1)]}{\Gamma[\frac{1}{2}(s+r-n+1)\Gamma[\frac{1}{2}(s-r-n+3)]}\,,~~
\eea where we use that $ r+s={\rm odd}$.

 The coefficients $c_{r,s}(g)$ also admit a convergent small $g$ expansion
\bea c_{r,s}(g) =g\, \sum_{n=r+s-3}^\infty g^{n}\,
\tilde{c}_{r,s}^{(n)} \,,\quad g< {1\ov2}\,. \eea where the
numerical coefficients $ \tilde{c}_{r,s}^{(n)}$ can, in principle,
be extracted from anomalous dimensions of primary operators of the
perturbative gauge theory. The first nonvanishing coefficient
$\tilde{c}_{2,3}^{(2)}$ requires an involved four-loop
perturbative computation and it appears to be \bea
\tilde{c}_{2,3}^{(2)} = -{\z(3)\ov2}\,. \eea The remaining
coefficients were conjectured by assuming analytic continuation,
and are given by \bea  \la{phaseweak}
\tilde{c}_{r,s}^{(n)}=\frac{\cos(\sfrac{1}{2}\pi
n)(-1)^{s+n}2^{-n}\zeta(1+n)\Gamma(2+n)\Gamma(1+n)(r-1)(s-1)}{\Gamma[\frac{5+n-r-s}{2}]
\Gamma[\frac{3+n+r-s}{2}]\Gamma[\frac{3+n-r+s}{2}]
\Gamma[\frac{1+n+r+s}{2}]} \,, \eea where we use again that
$r+s={\rm odd}$. This formula shows that the coefficients are
nonvanishing for, and only for, even  $n$.

\subsection{Crossing equations}

Here we will come back to the issue of crossing symmetry, which is
essentially  related to the existence of the two branches of the
dispersion relation, the one corresponding to unitary
representations with $H>0$ and the other corresponding to
anti-unitary ones with $H<0$.

\smallskip

Recall that on the upper sheet of the hyperboloid (\ref{relHC})
the variable $z$ takes values $-\om_1/2\le z\le \om_1/2$. Shifting
$z$ by half of the imaginary period, we find \bea\label{crossz}
\begin{aligned} H(z)&\to H(z+\omega_2)=\dn(z+\omega_2,k)=-\dn(z,k)=-H(z)\,  , \\
p(z)&\to p(z+\omega_2)=2\am(z+\omega_2)=-2\am(z)=-p(z)\, .
\end{aligned} \eea
Thus, under this transformation the positive energy branch of the
dispersion relation transforms into the negative one; both the
Hamiltonian and the momentum change their sign. Thus, the map
$z\to z+\omega_2$ is the analogue of the crossing symmetry
transformation in two-dimensional relativistic field theories. In
what follows we regard $z$ as a complex variable and refer to
eq.(\ref{crossz}) as the crossing transform.

\smallskip

Let $M\equiv M(H,C)$ be a matrix realization of a fundamental
unitary irrep of $\su(2|2)_{\cex}$ characterized by the central
charge values $H$ and $C$. Consider now the following map (``minus
supertransposition")
$$
M\to -M^{st}\, .
$$
Obviously, under this map the central charge values change their
signs. Moreover, $-M(H,C) ^{st}$ is an irrep of $\su(2|2)_{\cex}$,
but with exactly the opposite values of the central charges. In
particular, if $M(H,C)$ belongs to the positive branch of the
dispersion relation, then $-M(H,C) ^{st}$ is on the negative
branch.

\smallskip

There are two transformations acting on the space of central
charges: the first one is the crossing transform which essentially
interchanges the positive and negative sheets between themselves,
the second one is an outer automorphism belonging to ${\rm
SU}(1,1)$. By combining these two, one is always able to transform
$(-H,{-\rm C})$ into $(H,C)$. To understand this issue, we recall
that in the elliptic parametrization both $H$ and $C$ are
functions of $z$. Under the crossing transform, the Hamiltonian
and the momentum change sign, which is, however, not always the
case for $C(z)$. In fact, we find that
$$
C(z+\omega_2)=\frac{i}{2}g(e^{-ip}-1)e^{2i\xi}=-e^{-ip}C(z)\, ,
$$
where we assume that $\xi$ is independent of $z$.

On the other hand, one should recall a ${\rm U}(1)$ automorphism
(a part of the outer ${\rm SU}(1,1)$ automorphism group), which
acts on the super- and central charges as
$$
Q(z)\to e^{i\rho} Q(z)\, , ~~~~~~C(z)\to e^{2i\rho}C(z)\, .
$$
Thus, if we pick  the ${\rm U }(1)$-automorphism obeying
the condition
$$
e^{2i\rho}C(z+\omega_2)=-e^{2i\rho}e^{-ip}C(z)=-C(z)\, ,
$$
{\it i.e.}, $e^{i\rho}=e^{i\frac{p}{2}}$, then after applying the
combined crossing and ${\rm U}(1)$ transformations, an original
irrep with $(-H,- C)$ will receive the central charges
$(H,C)$ and, for this reason, must be equivalent to $M(H,C)$. In
other words, \bea\la{intrel}
-\hat\rho(M(z+\omega_2))^{st}=\mathscr{C}M(z)\mathscr{C}^{-1}\, ,
\eea where $\mathscr{C}$ is an intertwining matrix and $\hat\rho$
denotes the action of the ${\rm U}(1)$-automorphism. In
particular, specifying eq.(\ref{intrel}) for the kinematical
generators we get \bea \la{rel1}
&&\mathscr{C}\,L_a^b=- L_b^a\, \mathscr{C}\,,\\
\nonumber &&\mathscr{C}\,R_\a^\b=-R_\b^\a\, \mathscr{C}\, , \eea
where we have taken into account that $(L_a^b)^t = L_b^a$, and
$(R_\a^\b)^t =R_\b^\a$. These relations fix the form of
$\mathscr{C}$ up to two coefficients
$$
\mathscr{C}=
\left(
\begin{array}{cc}
c_1\, \s_2 &  0\\
0 & c_2\,  \s_2\\
 \end{array}
 \right)\, ,
$$
where $\s_2$ is the Pauli matrix.
It is clear that only the ratio $c_1/c_2$ matters, and in what follows we  set $c_1=1$ for definiteness.  Then,
specification of eq.(\ref{intrel}) for the supersymmetry
generators gives
\bea\la{Qcr1}\begin{aligned}
e^{i\frac{p}{2}}Q_\a{}^a(z+\omega_2)^{st}
=-\mathscr{C}Q_\a{}^a(z)\mathscr{C}^{-1}\, , \\
e^{-i\frac{p}{2}}Q_a^\dagger{}^\a(z+\omega_2)^{st}
=-\mathscr{C}Q_a^\dagger{}^\a(z)\mathscr{C}^{-1} \, .
\end{aligned}\eea
The transformed supersymmetry generators can be easily found by
using the following relations \bea x^\pm(z+ \om_2)=1/x^\pm(z)\,,
\quad \eta(z+ \om_2) = {i\ov x^+(z)}\eta(z)\,.~~~ \eea One can
further show that the matrix $\mathscr{C}$ is given
by\footnote{Essentially, $\mathscr{C}$ is a product of the charge
conjugation and the parity transform matrices:
$$
\mathscr{C}=-i^{1/2}\left(\begin{array}{cc} \epsilon & 0
\\ 0 & \epsilon \end{array}\right)\left(\begin{array}{cc} i^{1/2}\mI_2
& 0
\\ 0 & i^{-1/2}\mI_2 \end{array}\right)\, ,
$$
where $\epsilon$ is defined in eq.(\ref{conjrule}).
 }
 \bea\la{Ccr} \mathscr{C}= \left(
\begin{array}{cc}
\s_2 &  0\\
0 & i\,  \s_2\\
 \end{array}
 \right)\, .
\eea It is worthwhile mentioning that the representation  obtained
by shifting $z$ in the opposite direction is related to the
original one through the matrix $\mathscr{C}^{-1}$
\bea\la{Qcr2}\begin{aligned}
e^{i\frac{p}{2}}Q_\a{}^a(z-\omega_2)^{st}
=-\mathscr{C}^{-1}Q_\a{}^a(z)\mathscr{C}\, , \\
e^{-i\frac{p}{2}}Q_a^\dagger{}^\a(z-\omega_2)^{st}
=-\mathscr{C}^{-1}Q_a^\dagger{}^\a(z)\mathscr{C}\, ,
\end{aligned}\eea
because $\eta(z- \om_2) = -{i\ov x^+(z)}\eta(z)$.

\medskip

To derive the crossing equations, we  use eqs.(\ref{Qcr1}),
(\ref{Qcr2}) that relate the contragradient  representation to the
original one, and the invariance conditions (\ref{incondb}),
(\ref{incondf}).  Taking the transpose of (\ref{incondb}) with
respect to the first  factor in the tensor product of two
matrices, and using the relations (\ref{rel1}), we get that the
matrix $ \mathscr{C}^{-1}_1S_{12}^{t_1} \mathscr{C}_1$ is
$\su(2)\oplus\su(2)$-invariant, {\it i.e.} it commutes with the
bosonic generators.

Next, we rewrite eq.(\ref{incondf}) in the following form \bea
\nonumber  S_{12}(z_1,z_2)\big[J_1(z_1;1) + \Sigma_1
J_2(z_2;e^{ip(z_1)})\big]=\big[ J_1(z_1;e^{ip(z_2)}) \Sigma_2 +
J_2(z_2;1) \big] S_{12}(z_1,z_2)\, ,
\\
\la{incondf2} \eea where the subscripts $1,2$ indicate the
embedding of the matrices into the tensor product:
$J_1(z;\z)\equiv J(z;\z)\otimes \mI \,,\ J_2(z;\z)\equiv
\mI\otimes J(z;\z)$. Then, we take the transpose of
eq.(\ref{incondf2}) with respect to the first  factor in the
tensor product of two matrix spaces, and use the relations
(\ref{Qcr2}), (\ref{Qcr1})
 written in the form\footnote{Here we have taken into
account that $M^{st}=M^t\,\Sigma$ and also indicated a possible
dependence of the supersymmetry generators  on the parameter $\z$.
} \bea\la{Qcr2b}
\begin{aligned}
&Q_\a{}^a(z;\z)^{t}
=-e^{\frac{ip(z)}{2}}\mathscr{C}Q_\a{}^a(z-\omega_2;\z)\mathscr{C}^{-1}\,
\Sigma\, ,~~\\ &Q_\a{}^a(z;\z)^{t}
=-e^{\frac{ip(z)}{2}}\mathscr{C}^{-1}Q_\a{}^a(z+\omega_2;\z)\mathscr{C}\,
\Sigma\, , \end{aligned}\eea and similar formulae for
$Q_a^\dagger{}^\a$. By using the first formula in (\ref{Qcr2b}),
after a simple computation, we find that the matrix $
\mathscr{C}_1S_{12}^{t_1}(z_1+\om_2,z_2) \mathscr{C}_1^{-1}$
satisfies the same invariance conditions as the matrix
$S_{12}^{-1}(z_1,z_2)$ and, therefore, the two matrices can differ
only by a function of $z_1,z_2$. The crossing symmetry condition
is just a statement that an $\su(2|2)$-invariant S-matrix could be
multiplied by a scalar factor such that these two matrices become
equal to each other \bea\la{crrel1}
 \mathscr{C}_1S_{12}^{t_1}(z_1+\om_2,z_2) \mathscr{C}_1^{-1} = S_{12}^{-1}(z_1,z_2)\,.
\eea In the same way,  transposing eq.(\ref{incondf2}) with
respect to the second  factor, we derive the second crossing
equation
 \bea\la{crrel2}
 \mathscr{C}_2S_{12}^{t_2}(z_1,z_2-\om_2) \mathscr{C}_2^{-1} = S_{12}^{-1}(z_1,z_2)\,.
\eea
The crossing equations impose important restrictions on the form of the S-matrix  scalar factor. We find, in particular, that the S-matrix in the $\su(2)$ sector should satisfy the following crossing symmetry equations
\bea\la{crosseqSu2}\begin{aligned}
S_{\su(2)}(z_1,z_2)\,S_{\su(2)}(z_1+\om_2,z_2)  &=
f(x_1^\pm,x_2^\pm)^2\,,\\
S_{\su(2)}(z_1,z_2)\,S_{\su(2)}(z_1,z_2-\om_2)  &=
f(x_1^\pm,x_2^\pm)^2\,,
\end{aligned}\eea
where the function $f(x_1^\pm,x_2^\pm)$ is defined by
\bea\label{funf} f(x_1^\pm,x_2^\pm)
=\frac{(x_1^--x_2^-)\Big(1-\frac{1}{x_1^-x_2^+}\Big)}{(x_1^+-x_2^-)\Big(1-\frac{1}{x_1^+x_2^+}\Big)}
\eea where the variables $x_i^\pm$ should be expressed through
$z_i$ by using eq.(\ref{xpxmz}).

\smallskip
 These equations together with the formula (\ref{ssu2}) can be used to
 derive the crossing equations for the dressing factor $\s$.
 In fact,
the simplest form of these equations arises  for a function
$\Sigma(z_1,z_2)$ which differs from $\s$ by the extra factor
${x_1^-\ov x_1^+} {x_2^+\ov x_2^-}$ entering in eq.(\ref{ssu2})
\bea\la{dressn} \Sigma(z_1,z_2) = \left({x_1^-\ov x_1^+} {x_2^+\ov
x_2^-}\ \right)^{1\ov 2}\, \s(x_1^\pm,x_2^\pm)\,. \eea It is not
difficult to show that $\Sigma(z_1,z_2)$ should satisfy the
following crossing equations\footnote{The second equation in
(\ref{crosseqg}) follows from the first one by using the unitarity
condition $\Sigma(z_1,z_2)\Sigma(z_2,z_1)=1$.}
\bea\la{crosseqg}\begin{aligned}
\Sigma(z_1,z_2)\,\Sigma(z_1+\om_2,z_2)  &=
h(x_1^\pm,x_2^\pm)\,,\\
\Sigma(z_1,z_2)\,\Sigma(z_1,z_2-\om_2)  &=
h(x_1^\pm,x_2^\pm)\,,
\end{aligned}\eea where the function $h(x_1^\pm,x_2^\pm)$ is given
by \bea\nonumber h(x_1^\pm,x_2^\pm)=
\frac{(x_1^--x_2^+)\Big(1-\frac{1}{x_1^-x_2^-}\Big)}{(x_1^+-x_2^+)\Big(1-\frac{1}{x_1^+x_2^-}\Big)}\,
\,. \eea It is important to notice that the function
$h(x_1^\pm,x_2^\pm)$ obeys the following identities \bea
h(1/x_1^\pm,x_2^\pm)h(x_2^\pm,x_1^\pm) = 1\,,\quad h(x_1^\pm,
1/x_2^\pm)h(x_2^\pm,x_1^\pm) = 1\,, \eea which are incompatible
with the assumption that the dressing factor is both unitary and
meromorphic function of $z_i$. Since unitarity is a physical
requirement, the dressing factor cannot be a meromorphic function
of the torus rapidity variables.


\section{Appendix}

\subsection{Monodromies of the S-matrix }
The canonical $\su(2|2)$-invariant fundamental S-matrix is defined
on a product of two rapidity tori. As such, it exhibits certain
monodromy properties under shifts of rapidity variables by certain
fractions of the real and imaginary periods of the torus.

By using the explicit form (\ref{SmatrAA}), one finds
 \bea
\begin{aligned}
 S(z_1+2\om_1,z_2)& =\Sigma_1\,
S(z_1,z_2)\Sigma_1=\Sigma_2\, S(z_1,z_2)\Sigma_2\, , \, \\
 S(z_1+2\om_2,z_2)& =\Sigma_1\, S(z_1,z_2)\Sigma_1=\Sigma_2\,
S(z_1,z_2)\Sigma_2\, .
\end{aligned}\label{mon2w} \eea
Hence, the S-matrix has the same monodromies over real and
imaginary cycles and it is a periodic function on a double torus
with periods $4\omega_1$ and $4\omega_2$. Here
$\Sigma_1=\Sigma\otimes \mI$ and  $\Sigma_2= \mI\otimes\Sigma$,
where $\Sigma$ is given by eq.(\ref{GM}). The element $\Sigma$ is
in the center of the group ${\rm SU}(2)\times {\rm SU}(2)$. We
recall that compatibility of scattering with statistics implies
that  \bea [S(z_1,z_2),\Sigma\otimes \Sigma]=0\,
.\label{commSigma}\eea

Now we establish the monodromy properties with respect to shifts
by half-periods. Under the shift by the real half-period we get
\bea
\begin{aligned}
S(z_1+\om_1,z_2)&=\big(V\otimes\Sigma\big)\,
S(z_1,z_2)\big(V^{-1}\otimes \mI\big)\, , \\
S(z_1,z_2+\om_1)&=\big(\Sigma\otimes V^{-1}\big)\,
S(z_1,z_2)\big(\mI\otimes V\big)\,
\end{aligned}\eea
and, as a consequence, \bea \nonumber
S(z_1+\om_1,z_2+\om_1)=(\Sigma\otimes\Sigma)(V\otimes
V^{-1})S(z_1,z_2)(V^{-1}\otimes V) \, .\eea Here $V={\rm
diag}\big(e^{\frac{i\pi}{4}},e^{\frac{i\pi}{4}},e^{-\frac{i\pi}{4}},e^{-\frac{i\pi}{4}}\big)
$. Thus, up to multiplication by $\Sigma\otimes \Sigma$, under the
simultaneous shift of the rapidity variables by the real
half-period the S-matrix undergoes a similarity transformation
with $V\otimes V^{-1}$.

The shift by the imaginary half-period is the crossing
transformation which has been already discussed in section 3.4.
For completeness, we present it here for the S-matrix
(\ref{SmatrAA}) \bea
\begin{aligned}
\mathscr{C}_1^{-1}S^{t_1}(z_1,z_2)\mathscr{C}_1 &=
\frac{1}{f(z_1,z_2)}S^{-1}(z_1+\omega_2,z_2) \, , \\
\mathscr{C}_2^{-1}S^{t_2}(z_1,z_2)\mathscr{C}_2 &=
\frac{1}{f(z_1,z_2)}S^{-1}(z_1,z_2-\omega_2) \, ,
\end{aligned}\eea
where the function $f(z_1,z_2)\equiv f(x_1^{\pm},x_2^{\pm})$ is
defined in (\ref{funf}). Combining the last formulae and using the
parity invariance of the S-matrix, we further find that
\bea\la{doublecross}
\mathscr{C}_1^{-1}\mathscr{C}_2^{-1}S^{t}(z_1,z_2)\mathscr{C}_1\mathscr{C}_2
=S(z_1+\omega_2,z_2+\omega_2)\, , \eea where
$S^{t}(z_1,z_2)=S^{t_1,t_2}(z_1,z_2)$. Here we have used that
$f(z_1,z_2)f(-z_1-\omega_2,-z_2)=1$. On the other hand, on can
independently verify that\bea S^{t}(z_1,z_2)= {\mathscr
C}_1{\mathscr C}_2 S(z_1,z_2){\mathscr C}_1^{-1}{\mathscr
C}_2^{-1}={\mathscr C}_1^{-1}{\mathscr C}_2^{-1}
S(z_1,z_2){\mathscr C}_1{\mathscr C}_2\, . \la{anothert}\eea These
two expressions for the transposed S-matrix are compatible due to
the fact that $\mathscr{C}^2=\Sigma$. Eq.(\ref{anothert}) together
with eq.(\ref{doublecross}) implies that the S-matrix remains
invariant under the simultaneous shift of $z_1$ and $z_2$ by
$\omega_2$: \bea\la{Sm12} S(z_1+\om_2,z_2+\om_2)=S(z_1,z_2)\, .
\eea Finally, we note that the time reversal invariance and
eq.(\ref{anothert}) lead to another commutativity property \bea
[S(z_1,z_2), \mI^g(\mathscr{C}\otimes\mathscr{C})]=0\, .
\label{commC}\eea We remark that for the S-matrix (\ref{SmatrAA})
both equations, (\ref{commSigma}) and (\ref{commC}), are trivially
satisfied without invoking the explicit form of the coefficients
$a_i$.

The monodromic properties of the S-matrix together with
generalized physical unitarity allow one to consistently define an
elliptic analog of the ZF algebra (the ZF algebra on the rapidity
torus). We however will not consider it here.

\subsection{One-loop S-matrix} \la{oneloopS}
\la{oneloop} Here we describe the properties of the ``one-loop"
S-matrix which is obtained from the S-matrix (\ref{SmatrAA}) upon
taking the limit $g\to 0$. We continue to work in the elliptic
parametrization introduced in subsection 3.2.4. According to
eq.(\ref{gtozero}), in this limit Jacobi elliptic functions
degenerate into the corresponding trigonometric ones and we find
the following trigonometric S-matrix:

{\footnotesize
 \bea
  \label{S1loop}
 \begin{aligned}
 S(z_1,z_2)\, =\, &
\Big(E_{1}^{1}\otimes E_{1}^{1}+E_{2}^{2}\otimes
E_{2}^{2}+E_{1}^{1}\otimes E_{2}^{2}+E_{2}^{2}\otimes
E_{1}^{1}\Big) &\\
 +\frac{2i } {\cot z_1-\cot z_2-2
i}\, &\Big(E_{1}^{1}\otimes E_{2}^{2}+E_{2}^{2}\otimes
E_{1}^{1}-E_{1}^{2}\otimes E_{2}^{1}-E_{2}^{1}\otimes
E_{1}^{2}\Big)\, &
\\
  -e^{-i(z_1-z_2)}\frac{\cot z_1-\cot z_2+2 i}{\cot z_1-\cot
z_2-2
 i}\,  & \Big(E_{3}^{3}\otimes E_{3}^{3}+E_{4}^{4}\otimes
E_{4}^{4}+E_{3}^{3}\otimes
E_{4}^{4}+E_{4}^{4}\otimes E_{3}^{3}\Big) &\\
  + e^{-i(z_1-z_2)}\frac{2i} {\cot z_1-\cot z_2-2 i}\, &
\Big(E_{3}^{3}\otimes E_{4}^{4} +E_{4}^{4}\otimes
E_{3}^{3}-E_{3}^{4}\otimes E_{4}^{3}-E_{4}^{3}\otimes
E_{3}^{4}\Big) &\\
 + e^{-iz_1}\frac{\cot z_1-\cot z_2}{\cot z_1-\cot z_2-2
i}\, &\Big(E_{1}^{1}\otimes E_{3}^{3}+E_{1}^{1}\otimes
E_{4}^{4}+E_{2}^{2}\otimes E_{3}^{3}+E_{2}^{2}\otimes
E_{4}^{4}\Big) &\\
  + e^{iz_2}\frac{\cot z_1-\cot z_2}{\cot z_1-\cot z_2-2
i}\, &\Big(E_{3}^{3}\otimes E_{1}^{1}+E_{4}^{4}\otimes
E_{1}^{1}+E_{3}^{3}\otimes E_{2}^{2}+E_{4}^{4}\otimes
E_{2}^{2}\Big) &\\
  -  e^{-\sfrac{i}{2}(z_1-z_2)}\frac{2i } {\cot
z_1-\cot z_2-2 i}\,   &\Big(E_{1}^{3}\otimes
E_{3}^{1}+E_{1}^{4}\otimes E_{4}^{1}+E_{2}^{3}\otimes
E_{3}^{2}+E_{2}^{4}\otimes E_{4}^{2} \Big) &\\
 -  e^{-\sfrac{i}{2}(z_1-z_2)}\frac{2i } {\cot z_1-\cot
z_2-2 i}\,  &\Big(E_{3}^{1}\otimes E_{1}^{3}+E_{4}^{1}\otimes
E_{1}^{4}+E_{3}^{2}\otimes E_{2}^{3}+E_{4}^{2}\otimes E_{2}^{4}
\Big) &\, .
\end{aligned}
\eea }

\vskip 0.2cm \noindent The relations between the $z$-variable, the
momentum and the rescaled rapidity $u\to gu$  transform in the
limit $g\to 0$ into \bea p=2z\, , ~~~~u=\cot z=\cot\frac{p}{2}\, .
\eea Surprisingly enough, this S-matrix cannot be written in the
difference form, {\it i.e.} as a function of one variable being
the difference of a properly introduced spectral parameter. By
construction, this S-matrix satisfies the Yang-Baxter equation
\bea\label{YB}
S_{23}(z_2,z_3)S_{13}(z_1,z_3)S_{12}(z_1,z_2)=S_{12}(z_1,z_2)S_{13}(z_1,z_3)
S_{23}(z_2,z_3) \, ,\eea as one can also verify by direct
calculation. On the other hand, at one-loop there is another
``canonical" S-matrix which is a linear combination of the graded
identity and the usual permutation:
 \bea S^{\rm
can}_{12}=\frac{u_1-u_2}{u_1-u_2-2 i}\mI^g_{12}
-\frac{2i}{u_1-u_2-2i}P_{12} \, . \label{Schain}\eea This S-matrix
satisfies the same Yang-Baxter equation (\ref{YB}).

\smallskip
 It appears that two S-matrices, (\ref{S1loop}) and
(\ref{Schain}), are related by the following transformation
\bea\nonumber S^{\rm
can}(z_1,z_2)=U_2(z_1)\Big[V_1(z_1)V_2(z_2)S_{12}(z_1,z_2)V_1^{-1}(z_1)V_2^{
-1}(z_2)\Big]U_1^{-1}(z_2)\, , \eea where we have introduced the
diagonal matrices \bea\nonumber U(z)&=&{\rm
diag}(1,1,e^{iz},e^{iz})\, ,\\ \nonumber
 V(z)&=&{\rm
diag}(e^{i\frac{z}{4}},e^{i\frac{z}{4}},e^{-i\frac{z}{4}},e^{-i\frac{z}{4}})
\, . \eea The transformation by $V$ is a ``gauge" transformation
which always preserves the Yang-Baxter equation. On the other
hand, transformation by $U$ is a twist that generically transforms
the usual Yang-Baxter equation into the twisted one and vice
versa. Note also that the twist $U$ does not belong to the
symmetry group ${\rm SU }(2)\times {\rm SU}(2)$ of the ``all-loop"
S-matrix.

\smallskip

To understand why at one loop the Yang-Baxter equation is
preserved under the twisting, we first write the Yang-Baxter
equation for $S^{\rm can}$ by using\footnote{The gauge
transformation by the matrix $V$ decouples from the Yang-Baxter
equation.} eq.(\ref{Schain}) \bea \nonumber
&&U_3(z_2)S_{23}U_2^{-1}(z_3)U_3(z_1)S_{13}U_1^{-1}(z_3)U_2(z_1)S_{12}U_1^{-
1}(z_2)=\\
&&~~~~~~~~=U_2(z_1)S_{12}U_1^{-1}(z_2)U_3(z_1)S_{13}U_1^{-1}(z_3)U_3(z_2)S_{
23}U_2^{-1}(z_3)\, , \eea which can be reshuffled as follows
\bea\nonumber
&&U_3(z_2)S_{23}U_2(z_1)U_3(z_1)S_{13}U_1^{-1}(z_3)U_2^{-1}(z_3)S_{12}U_1(z_
2)=\\
&&~~~~~~~~=U_2(z_1)U_3(z_1)S_{12}U_1^{-1}(z_2)S_{13}U_3(z_2)S_{23}U_1^{-1}(z
_3)U_2^{-1}(z_3)\, . \eea It is clear now that we will get the
usual Yang-Baxter equation for $S$ provided it obeys the following
relation \bea\label{1loopU} [S,U\otimes U]=0\, , \eea where $U$ is
an {\it arbitrary diagonal matrix}. One can easily verify that
both S-matrices, (\ref{S1loop}) and (\ref{Schain}), do indeed
satisfy this relation. At higher orders in $g$ the relation
(\ref{1loopU}) does not hold anymore. The corresponding all-loop
S-matrix (\ref{SmatrAA}) satisfies only a weaker condition
\bea\label{allloopS} [S,G\otimes G]=0\, , ~~~~~~~~G\in {\rm SU
}(2)\times {\rm SU}(2)\, ,\eea which is nothing else but the
invariance condition. As a consequence, the Yang-Baxter equation
is preserved by the twist only at the one-loop order.

\subsection{Hopf algebra interpretation} \la{hopf} In
section 3.3 we have determined the commutation relations of the
$\su(2|2)$ symmetry algebra generators with the ZF operators. This
allowed us to define the action of this symmetry algebra in the
multi-particle states constructed by successive application of
creation operators. An alternative way to define this action is to
use the concept of a Hopf algebra.

Let $\mathcal{A}$ be a vector space over complex numbers. Consider
the following two maps \bea \Delta:~~\mathcal{A}\to
\mathcal{A}\otimes\mathcal{A}\, , ~~~~~~~\epsilon:~~\mathcal{A}\to
{\rm complex~numbers}\, . \eea If these maps satisfy the relations
\bea \la{CA}
\begin{aligned}
& ({\rm id}\otimes \Delta)\circ \Delta =(\Delta\otimes {\rm
id})\circ \Delta \, ,\\
& ({\rm id}\otimes \epsilon)\circ \Delta ={\rm
id}=(\epsilon\otimes {\rm id})\circ \Delta \, ,
\end{aligned}
\eea then $\mathcal{A}$ is called a coalgebra.
 Accordingly, the map $\Delta$ is called the coproduct (or comultiplication) of $\mathcal{A}$ and $\epsilon$
 is the counit of $\mathcal{A}$.
A bialgebra $\mathcal{A}$  is both a unital associative algebra
and a coalgebra such that $\Delta$ and $\epsilon$ are algebra
homomorphisms, and multiplication $\mu$ and identity $\mI$ are
coalgebra homomorphisms. The fact that $\Delta$ and $\epsilon$ are
algebra homomorphisms is expressed as
 $$
\Delta(ab)=\Delta(a)\Delta(b)\, ,
~~~~~\epsilon(ab)=\epsilon(a)\epsilon(b)\, , ~~~~a,b
\in\mathcal{A} \, ,
 $$
Finally, a Hopf algebra is a bialgebra equipped with a bijective
map $S:~~\mathcal{A}\to\mathcal{A}$, called antipode, obeying the
following relations
$$
\mu(S\otimes{\rm id})\circ\Delta=\mI\circ\epsilon=\mu({\rm
id}\otimes S)\circ \Delta\, .
$$

Let now $\mathcal{A}$ be a unitary graded associative algebra
generated by even rotation generators $\bL_a{}^b$, $\bR_\a{}^\b$,
the odd supersymmetry generators $\bQ_\a{}^a$,
$\bQ_a^\dagger{}^\a$ and two central elements $\bH$ and $\bP$
subject to eqs.(\ref{su22abstract}). The central charges $\bC$ and
$\bC^{\dagger}$ are expressed via $\mathbb{P}$ by means of
eqs.(\ref{Cc}).

In what follows we make use of the graded tensor product, that is
for any algebra elements $a,b,c,d$
$$
(a\,\hat{\otimes}\, b)(c\,\hat{\otimes}\,
d)=(-1)^{\epsilon_b\epsilon_c}(ac\,\hat{\otimes}\, bd)\, ,
$$
where $\epsilon_a=0$ if $a$ is even and $\epsilon_a=-1$ if $a$ is
odd.

Now we are ready to supply $\mathcal{A}$ with the structure of a
Hopf algebra. We define the following coproduct
 \bea
\nonumber \Delta({\bJ}) &=&{\bJ}\,\hat{\otimes}\, \mI +
\mI\,\hat{\otimes}\, {\bJ}\, \quad ~~~~{\rm for\ any\ even\
generator}\,,\\\la{coprod} \Delta(\bQ_\a{}^a)
&=&\bQ_\a{}^a\,\hat{\otimes}\, \mI +
e^{\frac{i}{2}\hP}\,\hat{\otimes}\, \bQ_\a{}^a\,,\\\nonumber
\Delta(\bQ_a^{\dagger}{}^\a)
&=&\bQ_a^{\dagger}{}^\a\,\hat{\otimes}\, \mI +
e^{-\frac{i}{2}\hP}\,\hat{\otimes}\, \bQ_a^{\dagger}{}^\a\, , \eea
the counit \bea \epsilon(\mI)=1\, ,
~~~~\epsilon(\bJ)=\epsilon(\bQ_\a{}^a)=\epsilon(\bQ_a^{\dagger}{}^\a)=0\,
\eea and the antipode \bea S(\bJ)=-\bJ\, ,
~~~~S(\bQ_\a{}^a)=-e^{-\frac{i}{2}\bP}\, \bQ_\a{}^a\, ,
~~~S(\bQ_a^{\dagger}{}^\a)=-e^{\frac{i}{2}\bP}\,
\bQ_a^{\dagger}{}^\a\,  \eea and $S(\mI)=1$.
 The reader can easily verify that with
these definitions all Hopf algebra axioms are satisfied. For
instance, we compute\footnote{Since all elements here are even we
can use the usual tensor product. }
$$
\Delta(\bC)=\frac{ig}{2}(e^{\hP\otimes \mI+\mI\otimes
\hP}-\mI\otimes \mI)=\frac{ig}{2}(e^{\hP}\otimes
e^{\hP}-\mI\otimes \mI)=\bC\otimes \mI+e^{i\hP}\otimes \bC\, .
$$
On the other hand, \bea \nonumber &&\hskip -0.8cm
\{\Delta\bQ_\a{}^a,\Delta\bQ_\b{}^b\}=
\{\bQ_\a{}^a\,\hat{\otimes}\, \mI +
e^{\frac{i}{2}\hP}\,\hat{\otimes}\, \bQ_\a{}^a\,
,\bQ_\b{}^b\,\hat{\otimes}\, \mI +
e^{\frac{i}{2}\hP}\,\hat{\otimes}\, \bQ_\b{}^b\,\}=\\
\nonumber &&~~~~~\{\bQ_\a{}^a,\bQ_\b{}^b\}\otimes
\mI+e^{i\hP}\otimes
\{\bQ_\a{}^a,\bQ_\b{}^b\}=\epsilon_{\a\b}\epsilon^{ab}(\bC\otimes
\mI+e^{i\hP}\otimes \bC )=\epsilon_{\a\b}\epsilon^{ab}\Delta \bC\,
,
 \eea
{\it i.e.} $\Delta$ is indeed an algebra homomorphism.

Let us show that the coproduct agrees with the form of the
two-particle structure constants appearing in (\ref{incondf}). Let
$\mV$ be a vector space of the fundamental representation of
$\cA$. This space has a natural grading; the corresponding grading
matrix is given by $\Sigma$. The action of, say, supersymmetry
generators $\bQ_\a{}^a$ on the tensor product $\mV\otimes \mV$ is
given by application of the coproduct (\ref{coprod}) \bea
\Delta(\bQ_\a{}^a)\cdot v\otimes u &=&
(\bQ_\a{}^a\,\hat{\otimes}\, \mI  +
e^{\frac{i}{2}\hP}\,\hat{\otimes}\, \bQ_\a{}^a)\cdot v\otimes u
\\\nonumber &=&  \bQ_\a{}^a\cdot v\otimes u +
\Sigma\, e^{\frac{i}{2}\hP}\cdot v\otimes \bQ_\a{}^a\cdot u\, ,
\eea where $v\otimes u$ is an element of $\mV\otimes \mV$. Now one
can recognize that the two-particle representation coincides with
the one appearing on the left hand side of (\ref{incondf}).

The action of the Hopf algebra operations on the algebra
generators depends on the  chosen bases. Recall that $\mathcal{A}$
admits an automorphism
$$
\bQ\to e^{i\xi}\bQ\, ,~~~~~~\bC\to e^{2i\xi}\bC\, ,
$$
where $\xi$ might be a non-trivial function of the central
charges. For the choice $\xi=-\frac{1}{4} \bP$ the central charges
$\bC$ and $\bC^{\dagger}$ take the form (\ref{Cc2}), and they
become real and coincide. The action of the coproduct on the
redefined supercharges takes the most symmetric form \bea
\la{coprod1} \begin{aligned}\Delta(\bQ_\a{}^a)
&=\bQ_\a{}^a\,\hat{\otimes}\, e^{-\frac{i}{4}\hP} +
e^{\frac{i}{4}\hP}\,\hat{\otimes}\, \bQ_\a{}^a\,,\\\nonumber
\Delta(\bQ_a^{\dagger}{}^\a)
&=\bQ_a^{\dagger}{}^\a\,\hat{\otimes}\, e^{\frac{i}{4}\hP} +
e^{-\frac{i}{4}\hP}\,\hat{\otimes}\, \bQ_a^{\dagger}{}^\a\, .
\end{aligned}\eea In the new basis the antipod becomes trivial for
any algebra element \bea S(\bJ)=-\bJ\, ,
~~~~S(\bQ_\a{}^a)=-\bQ_\a{}^a\, , ~~~S(\bQ_a^{\dagger}{}^\a)=-
\bQ_a^{\dagger}{}^\a\, . \eea The only drawback of this algebra
basis is that with $\bC$ real a basis of the corresponding
fundamental representation cannot depend meromorphically on the
torus variable $z$.

Our final comment  concerns permutation relations (\ref{JAs}). We
observe that they can be cast in  the usual (anti)-commutator form
by redefining the supersymmetry generators $\bQ_\a{}^a$ and
$\bQ_a^\dagger{}^\a$ in the following way \bea\la{redq}
\bQ_a{}^\a\to \bQ_a{}^\a\, e^{i\hP/2} \,,\quad
\bQ_a^\dagger{}^\a\to \bQ_a^\dagger{}^\a\, e^{-i\hP/2}\,. \eea
Relations  (\ref{JAs}) for the redefined supersymmetry charges
take the form of the (anti)-commutators
\bea\la{JAs2b}\begin{aligned}
 \bQ_\a{}^a\, \bA^\dagger(p) -\bA^\dagger(p)\,\Sigma\,
\bQ_\a{}^a &=\bA^\dagger(p)\,Q_\a{}^a(p)\, e^{-i\hP/2} \,,\\
\bQ_a^\dagger{}^\a \,\bA^\dagger(p)-\bA^\dagger(p)\,\Sigma\,
\bQ_a^\dagger{}^\a &=\bA^\dagger(p)\,Q_a^\dagger{}^\a(p)\,
e^{i\hP/2} \,.\end{aligned}\eea The only difference with the
standard relations is the appearance of the operator $e^{\pm
i\hP/2}$ in the right hand side of eqs.(\ref{JAs2b}). As in our
discussion above, redefinition (\ref{redq}) changes the
momentum-dependence of the central charge $\bC$: \bea \bC \to
{ig\ov2}\,(e^{i\hP}-1) e^{-i\hP} =   {ig\ov2}\,(1 - e^{-i\hP})\,,
\eea and, therefore, the boundary conditions for the light-cone
coordinate $x_-$. Obviously, it does not change the form of the
S-matrix if one keeps track of the additional phases because the
redefined supercharges also commute with the S-matrix.

\section{Bibliographic remarks}
{\small The Factorized Scattering Theory has been developed in
\cite{ZZ}. For important applications the reader may consult
\cite{Lukyanov:1993pn,Lukyanov1}. The ZF algebra has been
introduced in \cite{ZZ,Fad}. Its various properties and
representation theory have been extensively discussed in the
literature, see {\it e.g.} \cite{Mintchev:1999tc,Ragoucy:2001zy}.
Our exposition of the Factorized Scattering Theory and its
application to the string sigma model follows closely
\cite{AFZzf,AFtba}.

The exact dispersion relation (\ref{disp}) has been conjectured in
\cite{BDS}. In this work the local conserved charges
(\ref{localconcharge}) has been introduced as the ``higher-loop"
generalization of conserved charges of the Heisenberg model.

The $\psu(2|2)$-invariant S-matrix has been obtained in \cite{B}
by exploiting the corresponding invariance condition. This
condition severely constraints its matrix structure but does not
fix it uniquely. In general, the S-matrix depends on a few
parameters \cite{B2}, which reflects the freedom of choice of a
two-particle basis, and, as the result, it satisfies a twisted
version of the Yang-Baxter equation. In a physical theory the
S-matrix must be unique (up to unitary transformations). In
two-dimensional integrable models it must satisfy the condition of
factorized scattering, {\it i.e.} the Yang-Baxter equation. This
requirement partially fixes the two-particle basis and the
corresponding S-matrix \cite{AFZzf} leaving the possibility to
perform momentum-dependent transformations of one-particle states.
An additional requirement of generalized physical unitarity, or,
equivalently, of physical unitarity of the S-matrix of the mirror
model, leads to a unique matrix expression \cite{AFtba} up to
constant transformations of the one-particle basis. Then, the only
undetermined piece of the S-matrix is an overall normalization
(the scalar factor). The graded S-matrix obtained in chapter 3 is
the inverse of the graded version of $S$ found in
\cite{AFZzf,AFtba}. This is done to get an agreement with the
perturbative S-matrix of chapter 2, the latter was computed by
using the standard field-theoretic prescriptions.

The idea that the overall scalar factor can be constrained by
requiring the world-sheet scattering matrix to satisfy an analogue
of crossing symmetry has been put forward in \cite{Janik}, where
also a functional equation for this factor implied by crossing
symmetry has been derived. In fact, in relativistic integrable
models compatibility of scattering with crossing symmetry is a
standard requirement \cite{ZZ,Bernard:1990ys,Bernard:1992mu}. In
this respect, a peculiarity of the string sigma model lies in the
absence of the two-dimensional Lorentz invariance on the
world-sheet. In the last chapter we exhibited three different
faces of crossing symmetry: crossing symmetry as an additional
invariance condition for the ZF algebra \cite{AFZzf}, crossing
symmetry as a requirement of trivial scattering of the singlet
state \cite{B2} (see also \cite{Ahn:2008df}) and, finally,
crossing symmetry as the particle-to-antiparticle transform
\cite{Janik}.

The representation theory of the centrally extended $\su(2|2)$
algebra has been studied in \cite{B2}, where, in particular,
conditions leading to the multiplet shortening have been
determined and the outer automorphism group ${\rm SL}(2)$ has been
identified. The rapidity torus has been introduced in
\cite{Janik}, although our uniformization (the same uniformization
has been also used in \cite{B2}) for the dispersion relation in
terms of elliptic functions is different from that in
\cite{Janik}. Table 1 representing the transformation properties
of $x^{\pm}(z)$ under shifts of $z$ by some fractions of the
periods is taken from \cite{B2}.

The most non-trivial part of the overall scalar factor of the
world-sheet S-matrix is the dressing phase. Its functional form in
terms of local conserved charges of the model was conjectured in
\cite{AFS} by discretizing the finite-gap solutions \cite{KMMZ} of
the classical string sigma model. The most general functional form
of the dressing phase compatible with integrability
\cite{Beisert:2005wv} is given by eq.(\ref{phase}).

Further progress in determination of the dressing phase relied on
comparison of the energies of spinning strings at the classical
and the one-loop level \cite{FT,Frolov:2003qc},
\cite{Frolov:2003tu}-\cite{SchaferNameki:2006gk} with those
obtained by solving the asymptotic Bethe ansatz equations
\cite{Beisert:2003xu}-\cite{Gromov:2005gp}. The general method for
determining the one-loop correction to the dressing phase has been
developed in \cite{Beisert:2005cw} and used to obtain the one-loop
correction to the coefficient $c_{2,3}$ in eq.(\ref{phase}). This
approached has been further applied to completely determine the
dressing phase at one loop \cite{HL,Freyhult:2006vr}. The same
results were later derived by using the algebraic curve techniques
\cite{Gromov:2007cd,Gromov:2007ky}.

Two known orders in the strong coupling expansion of the dressing
phase \cite{AFS,HL,Freyhult:2006vr} were shown to solve  the
functional equation implied by crossing symmetry \cite{AF06}.
Formula (\ref{phasestrong}) that encodes an all-order asymptotic
solution for the dressing phase was obtained in \cite{BHL} by
exploiting its functional form (\ref{phase}) together with the
crossing equation. Opposite to the strong coupling expansion,
gauge theory perturbative expansion of the dressing factor is in
powers of $g$ and it has a finite radius of convergence.  A
proposal (\ref{phaseweak}) leading to the exact dressing factor
has been put forward in \cite{BES} and it passed several very
non-trivial tests \cite{RTT}-\cite{BMR}, \cite{Cachazo:2006az}. A
check that this exact dressing phase obeys the crossing equation
for finite values of $g$, {\it i.e.} not in the asymptotic sense,
is currently lacking, however.

As was discussed at the beginning of chapter 3, quantum
integrability  is a plausible but yet unproven property of the
string sigma model. To reveal it, one has to demonstrate the
absence of particle production and factorization of multi-particle
scattering. This important question has been investigated in
\cite{Puletti:2007hq,Hentschel:2007xn}, where factorization has
been shown to hold at leading orders in the strong coupling
expansion.

The monodromy properties of the $\su(2|2)$-invariant S-matrix have
been established in \cite{AFtba}, where an elliptic analog of the
ZF algebra has been introduced. The one-loop limit of the S-matrix
and its relation to the canonical S-matrix built out of the graded
identity and the permutation has been also analyzed there. The
Hopf algebra structure discussed in the appendix and in
\cite{AFZzf} seems to be equivalent (up to a twist and some
redefinitions of the supersymmetry generators and the central
elements) to the one studied in \cite{Gomez:2006va,Plefka:2006ze}.

}
\vskip 1cm

\subsubsection{Acknowledgement}
We thank F. Alday, N. Beisert, J. Plefka, R. Roiban, M.
Staudacher, A. Tseytlin and M. Zamaklar for enjoyable
collaborations. We are also grateful to  N. Dorey, P. Dorey, D.
Hofman, R. Janik, V. Kazakov, C. Kristjansen, M. de Leeuw, J.
Maldacena, J. Minahan  and K. Zarembo  for many valuable
discussions. We also thank  M. de Leeuw, E. Quinn and A. Torrielli
for the careful reading of the manuscript and M. de Leeuw for
helping with figures.

\noindent The work of G.~A. was supported in part by the RFBI
grant 08-01-00281-a, by the grant NSh-672.2006.1, by NWO grant
047017015 and by the INTAS contract 03-51-6346. The work of S.F.
was supported in part by the Science Foundation Ireland under
Grant No. 07/RFP/PHYF104.



\end{document}